\documentclass[useAMS,usenatbib]{mn2e}

\usepackage{times}
\usepackage{graphicx}
\usepackage{subfigure}
\usepackage{amsmath}

\newcommand{\gsim}{\mathrel{\hbox{\rlap{\lower.55ex \hbox {$\sim$}}
                   \kern-.3em \raise.4ex \hbox{$>$}}}}
\newcommand{\lsim}{\mathrel{\hbox{\rlap{\lower.55ex \hbox {$\sim$}}
                   \kern-.3em \raise.4ex \hbox{$<$}}}}

\title[Diversity and properties of protostellar discs]{On the diversity and statistical properties of protostellar discs}
\author[M. R. Bate]{Matthew R. Bate$^{1}$\thanks{E-mail:
mbate@astro.ex.ac.uk}\\
$^{1}$ School of Physics and Astronomy, University of Exeter, Stocker
Road, Exeter EX4 4QL  
}

\date{\today}
\begin{document}
\maketitle
\begin{abstract}
We present results from the first population synthesis study of protostellar discs.  We analyse the evolution and properties of a large sample of protostellar discs formed in a radiation hydrodynamical simulation of star cluster formation.  Due to the chaotic nature of the star formation process, we find an enormous diversity of young protostellar discs, including misaligned discs, and discs whose orientations vary with time.  Star-disc interactions truncate discs and produce multiple systems.  Discs may be destroyed in dynamical encounters and/or through ram-pressure stripping, but reform by later gas accretion.  We quantify the distributions of disc mass and radii for protostellar ages up to $\approx 10^5$~yrs.  For low-mass protostars, disc masses tend to increase with both age and protostellar mass.  Disc radii range from of order ten to a few hundred au, grow in size on timescales $\lsim 10^4$~yr, and are smaller around lower-mass protostars.  The radial surface density profiles of isolated protostellar discs are flatter than the minimum mass solar nebula model, typically scaling as $\Sigma \propto r^{-1}$.  Disc to protostar mass ratios rarely exceed two, with a typical range of $M_{\rm d}/M_* = 0.1-1$ to ages $\lsim 10^4$~yrs and decreasing thereafter.  We quantify the relative orientation angles of circumstellar discs and the orbit of bound pairs of protostars, finding a preference for alignment that strengths with decreasing separation.  We also investigate how the orientations of the outer parts of discs differ from the protostellar and inner disc spins for isolated protostars and pairs.
\end{abstract}

\section{Introduction}
\label{introduction}

Discs around young stars are a natural result of angular momentum conservation and energy dissipation during protostellar collapse.  Their existence was implied by the near coplanarity of the planets in the Solar System which led to the development of the Nebular Hypothesis for the formation of the Sun and its planets in the 18th century, contributed to by Emmanuel Swedenborg, Immanuel Kant, and Pierre-Simon Laplace \cite[see][for a review]{Koerner1997}.  Thus, discs around young stars were hypothesised long before they were observed.  Early observational evidence for discs around young stars came in the form of excess infrared emission in spectral energy distributions.  This excess over that expected from the star's spectrum alone implied the existence of a reservoir of dust and gas surrounding the young star, and the form of the excess could be well explained by disc-like structures \citep[see the reviews of][]{BecSar1993,BecSar1996}.  Direct observations of protostellar discs were very rare \cite[e.g.][]{Beckwithetal1984} prior to the advent of the Hubble Space Telescope (HST).  The fiducial 0.1~arcsecond resolution offered by HST led to dozens of resolved scattered light and silhouette images of discs (see \citealt{OdeWenHu1993,McCODe1996} and the review of \citealt{McCStaClo2000}).  Since then, sub-arcsecond imaging both from space and the ground has resulted in a wealth of observations of discs around young stars \citep[see the review of][]{Watsonetal2007}, including millimetre observations that can measure masses and kinematics \cite[e.g][]{Beckwith_etal1990, DutGuiSim1994, Saitoetal1995, Dutreyetal1996, Dutreyetal1998, Duvertetal1998}.  Most direct observations to date, however, are of the discs of so-called Class II \citep{Lada1987, AndWarBar1993} or T-Tauri type stars \citep[e.g.][]{Andrewsetal2009, Andrews_etal2010}.  These have little envelope material left and tend to have masses much less than the masses of the stars.  Often called protoplanetary discs, these discs presumably provide the initial conditions for planet formation, but they tell us less about the earlier processes of stellar growth and disc formation \citep[see the reviews by][]{Armitage2011, WilCie2011}.  

On the theoretical side, numerical simulations of star formation have produced discs for more than 30 years \citep[e.g.][]{Tscharnuter1975,Boss1987}.  Fixed-grid calculations employed central sink cells \citep{BosBla1982} to enable the calculation to be followed beyond formation of a single stellar object.  Later, the invention of sink particles  \citep*{BatBonPri1995} allowed non-axisymmetric calculations to be followed well beyond one initial cloud free-fall time to study disc formation in multiple stellar systems (i.e.\ both circumstellar and circum-multiple discs).  Since then, hydrodynamical star formation calculations have tended to fall into one of two categories -- those that follow the collapse of individual molecular cloud cores to form single stars or small multiple systems and resolve discs reasonably well (e.g. to au scales), or those that study the formation of many protostars in molecular clouds but do not resolve most protostellar discs \citep[e.g.][]{KleBurBat1998,BonBat2002,KruKleMcK2012}.  Only the calculations of Bate \citep*[e.g.][]{BatBonBro2002b,BatBonBro2003, BatBon2005, Bate2009a, Bate2009b, Bate2012} and Offner (\citealt*{OffKleMcK2008}; \citealt{Offneretal2009}) have tried to bridge these two regimes by using small sink particles ($\lsim 10$~au).  Even then, individual protostars are only followed for $\lsim 10^5$ yrs, so it is difficult to compare the discs in these calculations with discs around Class II objects (which tend to have ages $\gsim 10^6$ yrs; \citealt{Evansetal2009}).  It should be noted, however, that observational classification of young stars into different Classes comes about primarily from the distribution of dust around the young star (in particular the relative masses of the envelope and disc) rather than the actual age.  As shown by \cite{Kurosawaetal2004} and \cite{Offneretal2012}, in hydrodynamical simulations even protostars with ages $< 10^5$ yrs can appear as Class II or Class III objects if they are not sufficiently embedded in cloud material (e.g.\ due to dynamical ejection from the cloud).

However, with the burgeoning of sensitive interferometers with sub-arcsecond resolution working at (sub-)millimetre wavelengths (e.g.\ the Plateau de Bure Interferometer (PdBI), the Combined Array for Research in Millimeter-wave Astronomy (CARMA), the Submillimetre Array (SMA), the Atacama Large Millimetre Array (ALMA), and the Karl G. Jansky Very Large Array (VLA)), we can now peer inside star-forming molecular cloud cores to examine the very youngest protostellar discs.  
Some early results did not find discs \citep[e.g.][]{Mauryetal2010}, but recently discs with radii $>30$~au have been found around both Class 0 and I objects  \citep[e.g.][]{Lee_etal2009,ChoTatKan2010,Tobinetal2012, MurLai2013, Yen_etal2013,Codella_etal2014, Ohashi_etal2014,Harsono_etal2014,Tobinetal2015, Aso_etal2015,LeeHwaLi2016,SeguraCox_etal2016,Lee_etal2017c,Yen_etal2017,Aso_etal2017}.  One hierarchical triple system even displays the sort of disc morphology that would be expected if the wider component had recently formed via the fragmentation of a circumbinary disc \citep{Tobinetal2016}.  Class 0 objects are thought to have typical lifetimes of $\approx 10^5$~yrs \citep{Evansetal2009}.
 
Therefore, we now stand on the brink of having large samples of discs at the Class 0 and I stages of star formation from both observations and numerical simulations which can be compared.  This paper is the first to examine the properties of a large sample of discs ($>100$) from a hydrodynamical simulation of star cluster formation.  \cite{Offner_etal2010} studied $\sim 10$ discs from \cite{Offneretal2009}, but their main interest was in whether multiple systems formed primarily by core fragmentation or disc fragmentation rather than in the properties of the discs per se.  The radiation hydrodynamical calculation from which we extract our sample of discs was published by \cite{Bate2012} who studied the statistical properties of the protostars (i.e.\ mass distribution, multiplicity, and the properties of multiple system) and found good agreement with the statistical properties of Galactic stars.  But the paper included little discussion of the protostellar discs.  The calculation employed sink particles with accretion radii of 0.5~au and, therefore, resolves discs down to radii of a few au.  The discs display a huge diversity in morphology, mass, radius, and in how they evolve with time, which this paper attempts to summarise.  It is hoped that this may aid the interpretation of future observations of young discs, and allow the process of comparing the statistical properties of observed and numerical samples of discs to begin.  We also note that knowing the statistical properties of protoplanetary discs is one of the major bottlenecks in understanding planet formation \citep[e.g.][]{MorRay2016}

In Section 2, we briefly summarise the method and initial conditions that were used to carry out the calculation.  In Section 3, we highlight the diversity of the protostellar discs produced during the calculation and how they evolve.  In Section 4, we discuss the statistical properties of the discs, including their masses, radii, and the orientations of discs in binary systems, and how they evolve with time and vary with protostellar mass.  In Section 5, we compare the numerical results with observed samples of discs and discuss future research directions. Finally, we draw our conclusions in Section 6.

\section{Method}
\label{sec:method}

The calculation discussed in this paper was original published in \cite{Bate2012}.
In Sections \ref{hydro} and \ref{initialconditions} we give a brief summary of the method used to perform the
calculation -- a more detailed description may be found in the original paper.  
Section \ref{sec:characterisation} gives a detailed description of the method we used to extract the discs from the hydrodynamical calculation and characterise their properties.

\subsection{The radiation hydrodynamical calculation}
\label{hydro}

The calculation was performed using the three-dimensional smoothed particle
hydrodynamics (SPH) code, {\tt sphNG}, based on the original 
version of \citeauthor{Benz1990} 
(\citeyear{Benz1990}; \citealt{Benzetal1990}), but substantially
modified using the methods described in \citet{BatBonPri1995}, \citet{PriMon2007},
\citet*{WhiBatMon2005}, \citet{WhiBat2006} and 
parallelised using both OpenMP and MPI.

Gravitational forces between particles and a particle's 
nearest neighbours are calculated using a binary tree.  
The smoothing lengths of particles varied in 
time and space and were set such that the smoothing
length of each particle 
$h = 1.2 (m/\rho)^{1/3}$ where $m$ and $\rho$ are the 
SPH particle's mass and density, respectively
\cite[see][for further details]{PriMon2007}.  The SPH equations were 
integrated using a second-order Runge-Kutta-Fehlberg 
integrator \citep{Fehlberg1969} with individual time steps for each particle
\citep{BatBonPri1995}.
To reduce numerical shear viscosity, the
\cite{MorMon1997} artificial viscosity was employed 
with $\alpha_{\rm_v}$ varying between 0.1 and 1 while $\beta_{\rm v}=2 \alpha_{\rm v}$
\citep[see also][]{PriMon2005}.

The calculation employed two temperature (gas and radiation) radiative transfer in the flux-limited
diffusion approximation \citep{WhiBatMon2005, WhiBat2006}.
The gas and dust temperatures were assumed to be the same.  Taking solar metallicity 
gas, the opacity was set to be the maximum of 
the interstellar grain opacity tables of \citet{PolMcKChr1985} and, at higher 
temperatures when the dust has been destroyed, the gas opacity
tables of \citet{Alexander1975} (the IVa King model)  \citep[see][for further details]{WhiBat2006}.
The gas equation of state had hydrogen and helium mass fractions
of $X=0.70$ and $Y=0.28$, respectively.
The contribution of metals to the equation of state is neglected. 

The calculation followed the hydrodynamic collapse of each protostar through the first core
phase and into the second collapse (which begins at densities of
$\sim 10^{-7}$~g~cm$^{-3}$) due to molecular hydrogen dissociation \citep{Larson1969}.
However, due to the decreasing size of the time steps, sink particles
\citep{BatBonPri1995} were inserted when the density exceeded
$10^{-5}$~g~cm$^{-3}$.  This density is 
just two orders of magnitude before the
stellar core begins to form (density $\sim 10^{-3}$~g~cm$^{-3}$) and the associated free-fall time
is only one week.

A sink particle is formed by 
replacing the SPH gas particles contained within $r_{\rm acc}=0.5$ au 
of the densest gas particle in region undergoing second collapse 
by a point mass with the same mass and momentum.  Any gas that 
later falls within this radius is accreted by the point mass 
if it is bound and its specific angular momentum is less than 
that required to form a circular orbit at radius $r_{\rm acc}$ 
from the sink particle.  Thus, gaseous discs around sink 
particles can only be resolved if they have radii $\gsim 1$ au.
Sink particles interact with the gas only via gravity and accretion.
There is no gravitational softening between sink particles.
The angular momentum accreted by a sink particle is recorded
but plays no further role in the calculation.
The sink particles used in the calculation discussed in this
paper did not contribute radiative feedback 
\citep[see][for a detailed discussion of this limitation]{Bate2012}.

Sink particles were permitted to merge if they
passed within 0.01 au of each other (i.e., $\approx 2$~R$_\odot$).
However, no mergers occurred during the calculation.

\subsection{Initial conditions and resolution}
\label{initialconditions}

For a full description of the initial conditions, see \cite{Bate2012}.  Briefly, the initial 
conditions consisted of an initially uniform-density molecular cloud containing 
500 M$_\odot$ of molecular gas, with a radius of 0.404 pc (83300 au) giving an initial density of 
$1.2\times 10^{-19}$~g~cm$^{-3}$ and an initial free-fall time of the cloud of 
$t_{\rm ff}=6.0\times 10^{12}$~s or $1.90\times 10^5$ years.  The initial temperature was 10.3 K.  
Although the cloud was uniform in density, we imposed an initial 
supersonic `turbulent' velocity field in the same manner
as \citet*{OstStoGam2001} and \cite{BatBonBro2003}.  
We generated a divergence-free random Gaussian velocity field with 
a power spectrum $P(k) \propto k^{-4}$, where $k$ is the wavenumber 
on a $128^3$ uniform grid and the velocities of the particles were interpolated from the grid.  
The velocity field was normalised so that the kinetic energy 
of the turbulence was equal to the magnitude of the gravitational potential 
energy of the cloud, giving an initial root-mean-square (rms) Mach number of the turbulence, ${\cal M}=13.7$.

The calculation used $3.5 \times 10^7$ SPH particles to model the cloud.  This resolution is sufficient to resolve the local Jeans mass throughout the calculation, which is necessary to model fragmentation of collapsing molecular clouds correctly (\citealt{BatBur1997, Trueloveetal1997, Whitworth1998, Bossetal2000}; \citealt*{HubGooWhi2006}).  More recently, there has been much discussion in the literature about the resolution necessary to resolve fragmentation in isolated gravitationally unstable discs (\citealt{Nelson2006,MerBat2011,MerBat2012,HopChr2013,Rice_etal2014,YouCla2015,YouCla2016,LinKra2016}; \citealt*{TakTsuInu2016,BaeKlaKra2017,DenMayMer2017}; \citealt{Klee_etal2017}).  As yet, there is no consensus as to the resolution that is necessary and sufficient to capture fragmentation of such discs.  Moreover, the gravitationally unstable discs that form in the calculation discussed in this paper are usually accreting rapidly, rather than evolving in isolation.  Rapid accretion can be important for driving fragmentation \citep{Bonnell1994,BonBat1994a,Hennebelle_etal2004, KraMatKru2008, Kratter_etal2010}, adding another complication.  \cite{KraLod2016} provide a recent review of gravitational instabilities in circumstellar discs.  The fact that the criteria for disc fragmentation is not well understood should be kept in mind as a caveat throughout this paper.

\subsection{Method of disc characterisation}
\label{sec:characterisation}

As will be seen in Section \ref{sec:diversity}, the protostellar discs produced by the hydrodynamical calculation are continually evolving due to a variety of different processes.  Therefore, to examine the disc properties statistically, we extract their properties many times during the calculation.  Specifically, we extract the disc properties from snapshots of the calculation at intervals of 0.0025~$t_{\rm ff} $ (i.e. every 476 yrs) for each protostar.  This gives 11831 instances of circumstellar discs around 183 protostars (with protostars that former earlier in the calculation contributing more instances).

\subsubsection{Circumstellar discs}
\label{sec:csdisc}

For each protostar (i.e.\ sink particle), we sort the SPH gas particles (and other sink particles) by distance from the sink particle.  Beginning with the closest SPH particle, we consider this particle to be part of the disc of the protostar if it has not already been assigned to another disc and the instantaneous ballistic orbit of that particle has an apastron distance less than 2000 au and an eccentricity $e < 0.3$.  The sensitivity of the results to the chosen upper limit on the eccentricity is explored in Section \ref{sec:ecctest}.  If the particle satisfies these criteria, its mass is added to that of the system and the position and velocity of the centre of mass of the system are computed.  The test is then repeated using these quantities for the next SPH particle.  We only consider particles out to a distance of 2000 au.  This distance was chosen empirically as it is larger than the apparent radius of any disc.  

If a sink particle is discovered when moving to more and more distant particles (e.g. the protostar being considered is part of a binary system, or there is a passing protostar), then the identity of the sink particle companion is recorded, and the determination of the circumstellar disc mass is terminated; it cannot include any particles more distant than the first neighbouring protostar.  Throughout this paper, we refer to protostars that do not have a companion within 2000 au as being `isolated'.  Note that a protostar may be a single protostar but not an isolated protostar if it has a protostar closer than 2000 au, but the two protostars are not bound to each other (the masses of the circumstellar discs are included when determining whether two protostars are bound; see the next section).

We have found empirically that the above algorithm generally results in sensible disc extraction of circumstellar disc properties from snapshots of the simulation.  However, separating the `disc' from the `envelope' of protostars is difficult (both theoretically and observationally) and sometimes the above algorithm identifies low-mass `discs' with very large radii.  These are not really discs, they are just parts of the infalling envelope.  To deal with this, we exclude any `circumstellar disc' for which the mass is $<0.03$~M$_\odot$ (i.e.\ $<2100$ SPH particles) and the radius that contains 63\% (see below for the origin of this value) of this mass is $>300$~au -- this is essentially a cut based on the disc's mean surface density.  We also exclude any `discs' for which the radius containing 63\% of the disc's mass is greater than three times the radius that contains 50\% of the mass.  These two cuts reduce the number of instances of circumstellar discs that are used in the analysis by 4.6\% (giving 11281 instances).  The number of instances of circumstellar discs around isolated protostars is reduced by 9.9\% (to give 2186 instances), with more than half of these excluded systems having ages $<3000$~yrs.

For each disc, we measure the radii that contain 2, 5, 10, 20, 30, 40, 50, 63.2, 70, 80, 90, 95, and 100 percent of the total disc mass.  These multiple radii can be used to measure the disc's surface density profile by fitting power-law radial surface density profiles.  Observers \citep[e.g.][]{Andrews_etal2010, Tazzari_etal2017} often fit discs assuming a truncated power-law surface density profile
\begin{equation}
\Sigma(r) = \Sigma_{\rm c} \left(\frac{r}{r_{\rm c}}\right)^{-\gamma} \exp\left[ - \left(\frac{r}{r_{\rm c}}\right)^{(2-\gamma)} \right],
\label{eqn:sigma}
\end{equation}
where $r_{\rm c}$ is the characteristic or cut-off radius of the disc and $\gamma$ is the power-law radial radial density profile in the bulk of the disc.
  This surface density profile is based on models of viscously evolving discs in which the kinematic viscosity scales as $\nu \propto r^\gamma$ \citep{LynPri1974, Hartmann_etal1998}.  Equation \ref{eqn:sigma} only gives sensible profiles for $\gamma < 2$, since for values greater than two, the exponential cut-off radius, $r_{\rm c}$, actually becomes an {\em inner} cut-off and outside of this radius the surface density profile falls off very steeply.  For $\gamma=2$, the exponential term becomes unity so the density profile becomes a pure power law (and thus the disc mass never converges).  We also note that for $\gamma<0$ the disc has an inner hole.  Interestingly, {\em for all} $\gamma < 2$ the characteristic radius, $r_{\rm c}$, is {\em always equal} to the radius that contains a fraction $(1-1/e)$ of the total disc mass (i.e.\ 63.2\%).  Therefore, if the disc is well described by equation \ref{eqn:sigma}, the characteristic radius, $r_{\rm c}$, can be obtained simply by measuring the radius containing 63.2\% of the total disc mass (rather than fitting the analytic profile).  But measuring this radius alone also provides a measure of the radius that is more general -- even for discs that have very different surface density profiles, it will still give a sensible measure of the disc radius.  Thus, throughout this paper, when quoting a disc radius, we use the radius containing 63.2 percent of the total mass.  Since many of the discs are perturbed, we found that using a larger value (particularly above 80\%) can give values that appear qualitatively too large when examining images of the discs. We note that \cite{Tripathi_etal2017} used a different surface density model to fit their observed discs and adopted an effective radius containing 68\% of the flux to characterise their discs.  They noted that using values from 50\% to 80\% made little difference to their analysis, and this is also true of the analysis presented below.  Thus, direct comparison of the disc radii that we report throughout this paper can also be made with \cite{Tripathi_etal2017} because the disc radii obtained using mass fractions of 63.2 or 68\% usually differ by less than 10\% percent.

\begin{figure*}
\centering \vspace{-0.5cm} \hspace{0cm}
    \includegraphics[width=8.5cm]{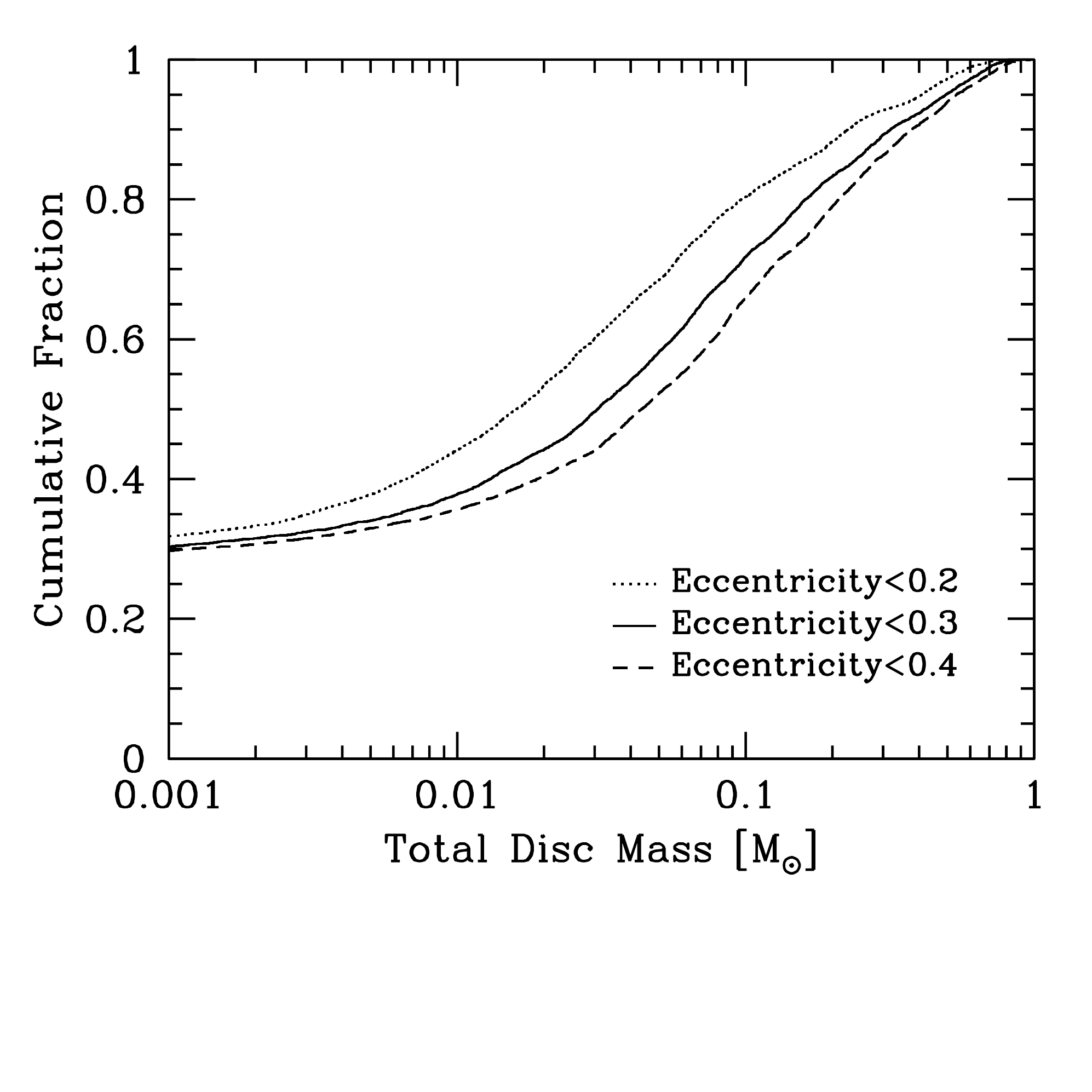}      \vspace{0.0cm}
    \includegraphics[width=8.5cm]{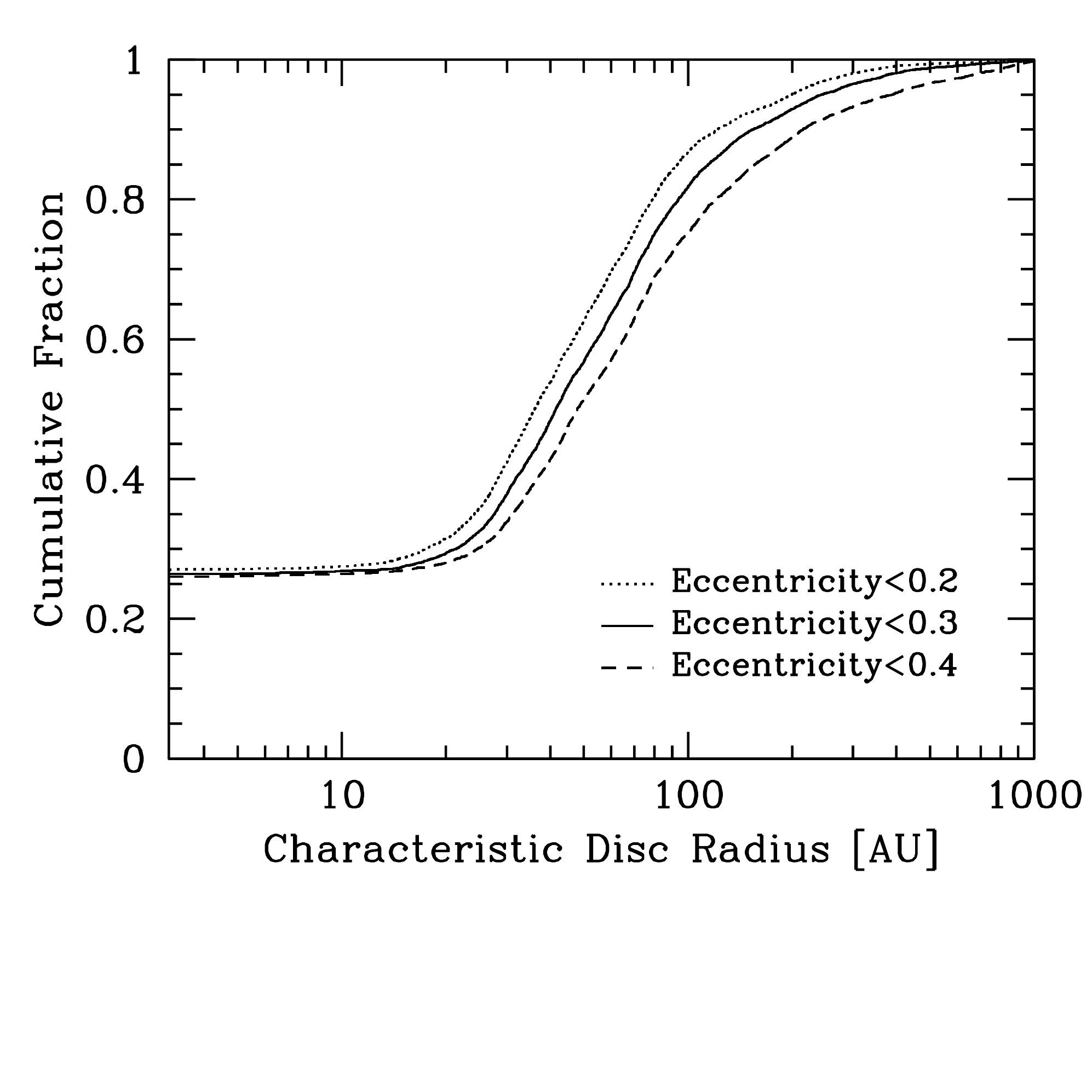}      \vspace{-1.5cm}
\caption{The cumulative distributions of disc mass (left) and characteristic radius (right) of the protostellar systems that are obtained when using different values of the upper limit for the eccentricity, $e$, that an SPH particle may have in order to be considered to be part of a disc.  Our default criterion is that particles with ballistic orbits with $e<0.3$ are considered to be in a disc.  Using a lower cut off obviously results in discs being assigned lower masses.  It also tends to result in slightly smaller discs.  Using the difference between the distributions obtained using $e<0.3$ and $e<0.4$ as an indication of the uncertainty in the disc extraction process, the typical uncertainty of our disc masses is $\pm 40$\% and for the characteristic disc radii it is $\pm 20$\%.}
\label{eccentricitytest}
\end{figure*}

\subsubsection{Circum-multiple discs}
\label{sec:cmdisc}

Characterising the properties of circumstellar discs (i.e.\ discs surrounding a single star) is relatively straighforward.  However, many of the protostars are located in bound multiple systems.  The discs in these systems are usually much more complex (see Section \ref{sec:diversity}).  For example, in a binary protostellar system there can be two circumstellar discs and a circumbinary disc, and these may not be aligned with each other, or with the binary's orbital plane.  Higher-order systems are even more complex!  With dozens of multiple complex systems, (relative) brevity demands that we limit our analysis to gross properties.  First, as in \cite{Bate2012}, we only consider single, binary, triple, and quadruple systems. For example, if a septuple system that is composed of a quadruple system bound to a triple system, the quadruple system and triple systems are treated as separate systems. Most very high-order systems are unstable or will have dynamical encounters with other protostars and undergo dynamical decay relatively quickly anyway.  Second, for the overall population of these systems, we limit our analysis to the total disc mass and a characteristic radius of the disc material.  But for bound pairs of protostars (either binaries or components of hierarchical higher order systems), we also examine the alignments of the circumstellar discs, protostellar spins, and the orbital plane of the pair.

Our method of extracting the discs of a multiple system from the hydrodynamical simulation is as follows.  Firstly, we need to find bound protostellar systems.  We use the same method as \cite{Bate2009a}, except that the disc mass is also included in the analysis (as opposed to only the mass of the sink particle).  After extracting the circumstellar discs of all protostars (Section \ref{sec:csdisc}), we have the total mass and the centre of mass location and velocity of all the protostars and their circumstellar discs.  These are denoted as nodes.  We then search for the pair of nodes that have the lowest total energy (kinetic plus potential energy) that are also mutual nearest neighbours.  This pair is grouped into a new node (a pair) and the total mass and centre of mass location and velocity of the node are determined.  The two nodes that formed the new node are removed from the list of active nodes.

We then need to extract the circum-multiple disc of this new node.  We do so only for nodes that are composed of 2--4 protostars.  A binary may have a circumbinary disc.  A triple that is composed of a pair and a third component on a wider orbit may have both a circumbinary disc surrounding the pair, and a circumtriple disc (in addition to the three circumstellar discs).  A quadruple can be composed either of two pairs, or a triple with a fourth component on a wider orbit.  In the former case, there may be two circumbinary discs and a circum-quadruple disc, while in the latter case there may be a circumbinary disc, a circumtriple disc, and a circum-quadruple disc (in either case there are up to 7 discs).

The method for extracting the disc of a multiple system is similar to that of extracting a circumstellar disc.  SPH particles (and sink particles) are sorted by distance from the centre of mass of the node.  Again, beginning with the closest particle, the particle is considered to be part of the circum-multiple disc if it has not already been assigned to a disc and its instantaneous ballistic orbit around the node (using its centre of mass location and velocity) has an apastron distance less than 2000 au and an eccentricity $e<0.3$.  If it satisfies these criteria, its mass is added to that of the circum-multiple disc of the node and the total mass of the node and the position and velocity of the centre of mass of the node are updated.  The test is then repeated for the next SPH particle, out to a distance of 2000 au.  If a sink particle is encountered which is not one of the components of the node, the determination of the disc is terminated.

Once the circum-multiple disc of the new node has been extracted, the above process is repeated on the list of active nodes until there are no pairs of nodes that have a negative total energy and are mutual nearest neighbours.  The final list of protostellar systems is provided by traversing the list of all nodes from the highest-order system to the lowest-order systems, only writing out the data for nodes that contain four or fewer protostars and whose components have not been previously written out (e.g.\ a triple is not written out if it is a component of a quadruple system; a binary is not written out of it is a component of a quadruple system or a triple system).

The total disc mass of the system is easy to compute; it is simply the total mass of all of the different discs that have been extracted for the system (e.g. for a triple, there are three circumstellar discs, one circumbinary disc, and one circumtriple disc, some of which may have little or no mass).  However, there are many ways that a characteristic disc radius may be determined for a multiple system.  The method we have chosen is based on combining the radial information of each disc in the system.  As for the circumstellar discs, for each circum-multiple disc we record the radii that contain 2, 5, 10, 20, 30, 40, 50, 63.2, 70, 80, 90, 95 and 100\% of the disc mass.  To determine a characteristic disc radius for a multiple system we loop over all of its component discs starting from the smallest of these radii, keeping a cumulative sum of the mass contained within each part of the disc.  For example, consider a binary system in which the radii containing 2\% and 5\% of the disc masses are 5 au and 8 au for the primary's circumstellar disc, 4 au and 6 au for the secondary's circumstellar disc, and 100 au and 120 au for the circumbinary disc.  The cumulative sum first takes 2\% of the secondary's disc mass (because this is at the smallest radius), then 2\% of the primary's disc mass, then (5-2)=3\% of the secondary's disc mass, then 3\% of the primary's disc mass, etc.  All of the circumstellar disc mass will generally be added in before the circumbinary mass starts to be added because the inner radius of the circumbinary disc is expected to be larger than separation of the binary.  The characteristic disc radius for the system is set as the radius at which the cumulative sum first exceeds 63.2\% of the total disc mass.  This algorithm is easily applied to a system of arbitrary order, and it provides a reasonable value for the size of a disc system.  If a close binary has low-mass circumstellar discs and a comparatively massive circumbinary disc, the characteristic radius will lie within the circumbinary disc as one would expect.  If a binary has no circumbinary disc or one that is significantly less massive than the combined mass of the circumstellar discs, then the characteristic radius will be related to the mass distribution of the circumstellar discs.  In this case, if the two circumstellar discs are identical, then the characteristic disc radius of the system is the same as it is for each of the circumstellar discs individually.  If the two circumstellar discs are very different, the value of the characteristic radius of the system will lie somewhere in between the two characteristic radii of the individual discs.  If there is only one circumstellar disc (e.g.\ a circumprimary disc), then the characteristic disc radius of the system is the same as that of the circumprimary disc, as expected.  We note, however, that this does mean that the characteristic radius of the discs in a system will not lie in the disc that surrounds the entire system (e.g.\ a circumbinary disc for a binary system, or a circum-triple disc for a triple system) unless the circum-system disc contains more than $\approx 37$\% of the total disc mass.  The bottomline is that there is no simple way to define a characteristic disc size for a multiple system, and if comparisons of disc sizes in multiple systems are going to be made in the future, extreme care must be taken to ensure that a like-for-like comparison is made.

Finally, when analysing protostellar systems we make the same cuts as for circumstellar discs.  We exclude any circum-multiple disc for which the total mass is $<0.03$~M$_\odot$ (i.e.\ $<2100$ SPH particles) and the radius that contains 63\% of this mass is $>300$~au.  We also exclude any discs for which the radius containing 63\% of the disc's mass is greater than three times the radius that contains 50\% of the mass.  These two cuts reduce the number of instances of discs of protostellar systems that are used in the analysis by 9.2\% (to give 6388 instances).  The number of instances of circumstellar discs around single protostars is reduced by 10.5\% (to give 3845 instances).  Note that the number of instances of single protostars is almost twice the number of instances of isolated protostars.  This is because, at various times, a significant number of protostars have other protostars that are nearby, but unbound.

\subsubsection{Sensitivity of the disc extraction to the eccentricity limit}
\label{sec:ecctest}

The disc extraction algorithm defines an SPH particle as belonging to a disc if its ballistic orbit (relative to a protostar or protostellar system and the other particles that have previously been identified as belonging to its disc(s)) has an eccentricity $e<0.3$.  This number has been chosen empirically based on examining the some of the discs that are extracted using the algorithm.  Using $e<0.3$ is a compromise between the algorithm identifying too much of the infalling `envelope' as `disc' (which may happen if the limiting eccentricity is too high) and not picking up the all of the disc (which may happen if the limiting eccentricity is too low).  Particles in a disc may have significant eccentricities if the disc itself is eccentric, or if the disc is gravitationally unstable and displays spiral shocks.  To give the reader confidence that the chosen algorithm does a sensible job of extracting the discs, in Appendix \ref{appendixA}, we provide some examples of the discs that are extracted from the calculation.  

In Section \ref{sec:stats}, the statistical properties of the discs are discussed.  The main properties of the discs that are analysed are their masses and their characteristic radii.  Using a lower limit for the maximum eccentricity a particle can have to be considered to be part of a disc necessarily results in lower disc masses, while using a higher limit always gives a greater mass.  To quantify the typical level of uncertainty in the values of the disc masses and characteristic radii, in Fig.~\ref{eccentricitytest} we show the cumulative distributions of the disc masses and characteristic disc radii of protostellar systems that are obtained using three different upper limits on the eccentricity: $e<0.2$, $e<0.3$, and $e<0.4$.  By eye, using $e<0.1$ or $e<0.2$ usually excludes a significant number of particles that clearly should constitute part of a disc.  Indeed, the left panel of Fig.~\ref{eccentricitytest} shows that using $e<0.2$ results in disc masses that are typically a factor of two less massive than using $e<0.3$.  Conversely, although if we use $e<0.4$ the algorithm may pick up a few more particles for eccentric or strongly gravitationally unstable discs, it will also often include more of the infalling envelope than is desirable.  From Fig.~\ref{eccentricitytest}, using $e<0.4$ increases the typical disc mass by $\approx 40$\% compared to using $e<0.3$.  We take this as the typical level of uncertainty of the disc masses obtained in the rest of this paper (i.e. $\pm 40$\%).

The values of the characteristic disc radii are less dependent on the upper value of the eccentricity that is used by the extraction algorithm than the disc masses (right panel of Fig.~\ref{eccentricitytest}).  Using the $e<0.2$ and $e<0.4$ distributions to estimate the typical uncertainty in the characteristic radii, we conclude that the characteristic disc radii obtained using $e<0.3$ have an uncertainty of $\pm 20$\%.  These uncertainties are small enough that they do not impact any of the conclusions of this paper, but they should be kept in mind throughout Sections \ref{sec:stats} and \ref{sec:discuss}.

\begin{figure*}
\centering \vspace{0cm} \hspace{0cm}
    \includegraphics[width=17cm]{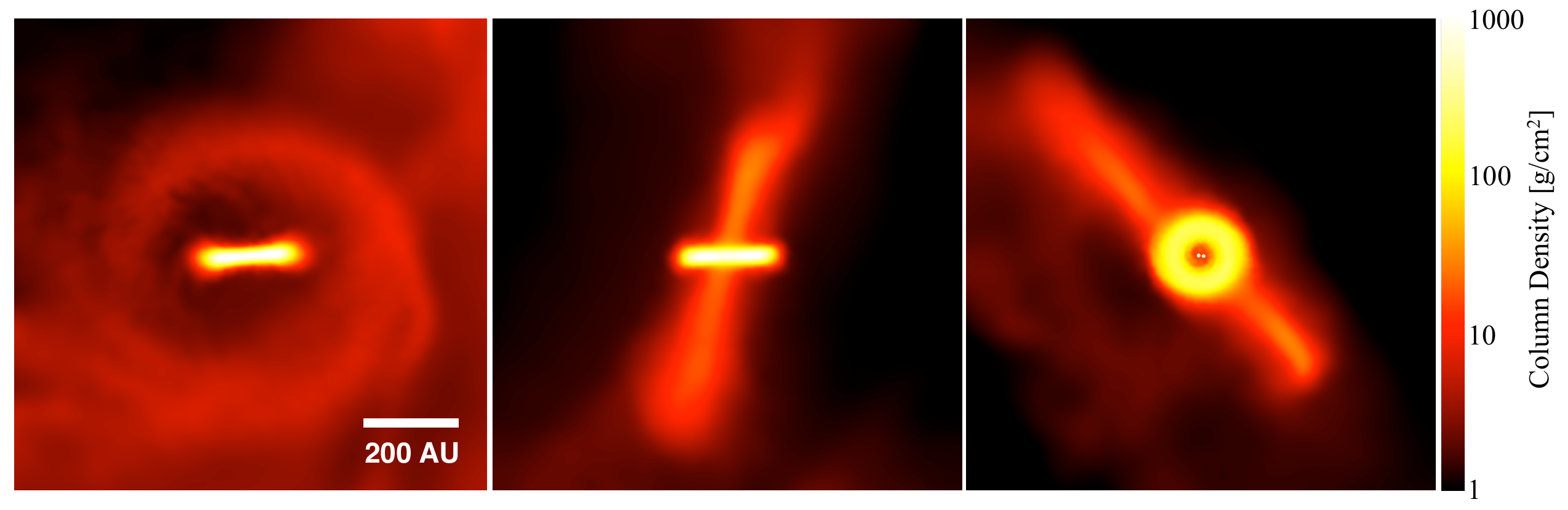} \vspace{-0.25cm}
\caption{A binary protostellar system (protostars 6 and 13) with a circumbinary disc which is misaligned.  It consists of an inner disc extending from radius, $r \approx 25-120$~au, and an outer disc extending from $r \approx 200-350$~au. The inner and outer discs are misaligned by approximately 75 degrees.  Each panel shows the system at the same time ($t=1.12~t_{\rm ff}$), but from three different angles. Sink particles are plotted as white filled circles that have radii 4 times larger than the actual sink particle accretion radius.  }
\label{misaligned}
\end{figure*}

\
\section{The diversity of discs}
\label{sec:diversity}

The cloud was evolved to $t=1.20~t_{\rm ff}$ (228,300~yrs), by which time 88.2~M$_\odot$ of gas (17.6\%) had produced 183 protostars, with a mean mass of 0.48~M$_\odot$ and a median mass of 0.21~M$_\odot$.  In this paper, the protostars are numbered by the order in which they formed (i.e. sink particles were inserted).  The first protostar formed at $t=0.73~t_{\rm ff}$, so when the calculation was stopped the oldest protostar had an age of 90,000~yrs.  At this time, 36 objects had masses less than the brown dwarf limit (taken to be 0.075~M$_\odot$).  The mass distribution of the protostars was consistent with the parameterisation of the observed Galactic IMF given by \cite{Chabrier2005}.  At the end of the calculation, the protostars were arranged as 84 single protostars and 40 multiple systems, the latter consisting of 28 binaries, 5 triples, and 7 quadruple systems (systems of higher-order were not treated as bound because they were deemed likely to undergo further dynamical evolution).  The final parameters of the 40 multiple systems are provided in Table 3 of \cite{Bate2012}.  The multiplicity was found to be a strongly increasing function of primary mass, consistent with observations.  For example, only 3 of the 38 very-low-mass systems (primary masses $<0.1~{\rm M}_\odot$) were binaries at the end of the calculation, but of the 15 systems with primary masses $>1.2~{\rm M}_\odot$, 11 were multiple systems.  See \cite{Bate2012} for further details on the mass function and the statistical properties of the multiple systems.

Although the calculation resolved many protostellar discs, \cite{Bate2012} included very little analysis of protostellar discs. \citeauthor{Bate2012} examined the distribution of closest encounters, finding that by the end of the calculation half of the protostars had been involved in encounters closer than $\approx 10$~au.  For binary systems, he also investigated the relative orientations of the spins of the sink particles and their orientations relative to their orbits.  When a gas particle is accreted by a sink particle, its linear momentum is added to that of the sink particle and its angular momentum relative to the sink particle is added to the spin angular momentum of the sink particle \citep{BatBonPri1995}.  The sink particle spin plays no further role in the calculation, but the spin can be thought of as representing the combined angular momentum of the protostar and its inner disc (size scales $\lsim 0.5$~au).  \citeauthor{Bate2012} found a strong tendency for sink particle spin alignment in binaries with separations $\lsim 40$~au, and a similar preference for alignment between sink particle spins and binary orbits for close binaries.  However, the resolved discs themselves were ignored (size scales $\gsim 1$~au).

There are two main reasons that \cite{Bate2012} omitted discussion of the resolved discs.  Firstly, many of the discs undergo dramatic dynamical evolution during the calculation.  Therefore, it is not sufficient simply to discuss the distribution of disc properties at the end of the calculation -- the evolution of the population must be studied. Secondly, the discs in the calculation are not very well resolved.  The calculation had a mass resolution of 70,000 SPH particles per solar mass, so even a relatively massive disc of $M_{\rm d} = 0.1~{\rm M}_\odot$ only contains 7000 SPH particles.  Moreover, since SPH is a Lagrangian method, the effective spatial resolution of the SPH method decreases with decreasing mass.  Thus, care is required when interpreting the properties of the protostellar discs.  To investigate the effects of the limited numerical resolution on disc evolution, in Appendix \ref{appendixB} we present results from two simple star formation calculations performed at different resolutions.  Generally, the properties of more massive discs are more reliable than those of low-mass discs.  Based on the resolution testing from the simple calculations, we find that discs modelled by $\gsim 2000$ SPH particles (i.e. discs with masses $\gsim 0.03$~M$_\odot$) should be well modelled in terms of their total mass and characteristic radius for the ages of the protostars produced in the cluster formation calculation (i.e. the typical error in their masses should be less than the typical uncertainty of $\approx 40$\% that arises from the disc extraction algorithm).  Below this mass, the disc masses are likely to be significantly underestimated and the radii of isolated discs are likely to be overestimated.  To relate this to observational classifications, this means that the properties of discs of Class 0 and I objects will be more reliable than those of Class II objects.   

Despite the limitations, the calculation contains a wealth of information on the types of discs that may be formed in a dense, interactive, star-forming environment, and on the dynamical processes that may drive the evolution of very young protostellar discs. This is the topic of this paper.  

The calculation produces a large population of discs with diverse properties.  This is best appreciated by watching the animation that is included with the Supporting Information that accompanies this paper.  The animation shows a mosaic of 183 animations, each of which displays a region with dimensions of $400\times 400$ au centred on one of the protostars (sink particles) that is produced during the simulation.  Clearly we are unable to discuss the evolution of the discs around all of these individual protostars in detail in this paper.  Instead, we begin by highlighting a few dramatic cases of discs produced during the calculation.

\subsection{A circumbinary disc with misaligned inner and outer components}
\label{sec:misaligned}

In Fig.~\ref{misaligned} we show the disc that existed at $t=1.12~t_{\rm ff}$ around the binary system consisting of protostars 6 and 13 (numbered by the order in which they formed).  This binary formed via a star-disc encounter between protostars 6 and 13 that occurred at $t=0.87~t_{\rm ff}$.  Prior to this, each of the two protostars had discs with masses of $M_{\rm d} \approx 0.1~{\rm M}_\odot$ and radii of 20 and 40 au, respectively.  After binary formation, further accretion quickly produced a massive circumbinary disc with a radius of 100 au and a mass ranging from $0.3-0.4$~M$_\odot$.  The binary then evolved in relative isolation for $\approx 0.1~t_{\rm ff}$, during which its components grow in mass from 0.6 and 0.1 M$_\odot$ to 1.0 and 0.5 M$_\odot$, respectively.  

At $t=0.98~t_{\rm ff}$, the binary started capturing further cloud material, but the angular momentum of this material was almost completely misaligned with the angular momentum of the existing circumbinary disc and binary, producing the dramatic misaligned disc system depicted in Fig.~\ref{misaligned}.
This misaligned disc survived until $t\approx 1.15~t_{\rm ff}$ (i.e. for $\approx 0.15~t_{\rm ff} \approx 30,000$~yr), when an encounter with protostar 10 and several other protostars caused the accretion of much of the disc material onto the binary and stripped away the rest. The encounter resulted in a high-order ($>4$ protostars) multiple system that persisted until the end of the calculation.  During the phase with the misaligned disc, protostars 6 and 13 increased in mass from $1.0$ to 1.1~M$_\odot$ and $0.5$ to 0.75~M$_\odot$, respectively.  By the end of the calculation, the two stars had grown via the accretion of the circumbinary disc material to masses of 1.9 and 1.3~M$_\odot$, respectively, and their semi-major axis was 1.2 au.

We note that recent hydrodynamical simulations of discs around black holes that are misaligned with the spin of the black hole, or circumbinary discs that are strongly misaligned with the binary's orbital plane show that these discs may tear into discrete precessing rings of gas \citep{Nixon_etal2012,NixKinPri2013}.  Although such dynamical evolution may result in similar disc structures to those found in Fig.~\ref{misaligned}, this system formed by accretion of gas with different angular momenta, not by disc tearing.  Since both mechanisms can result in similar protostellar disc structures, care must be taken in the interpretation of any similar systems found in future observations.

\subsection{Misaligned circumstellar discs in multiple systems}
\label{sec:misalign_csdiscs}

\begin{figure}
\centering \vspace{0cm} \hspace{0cm}
    \includegraphics[width=8.5cm]{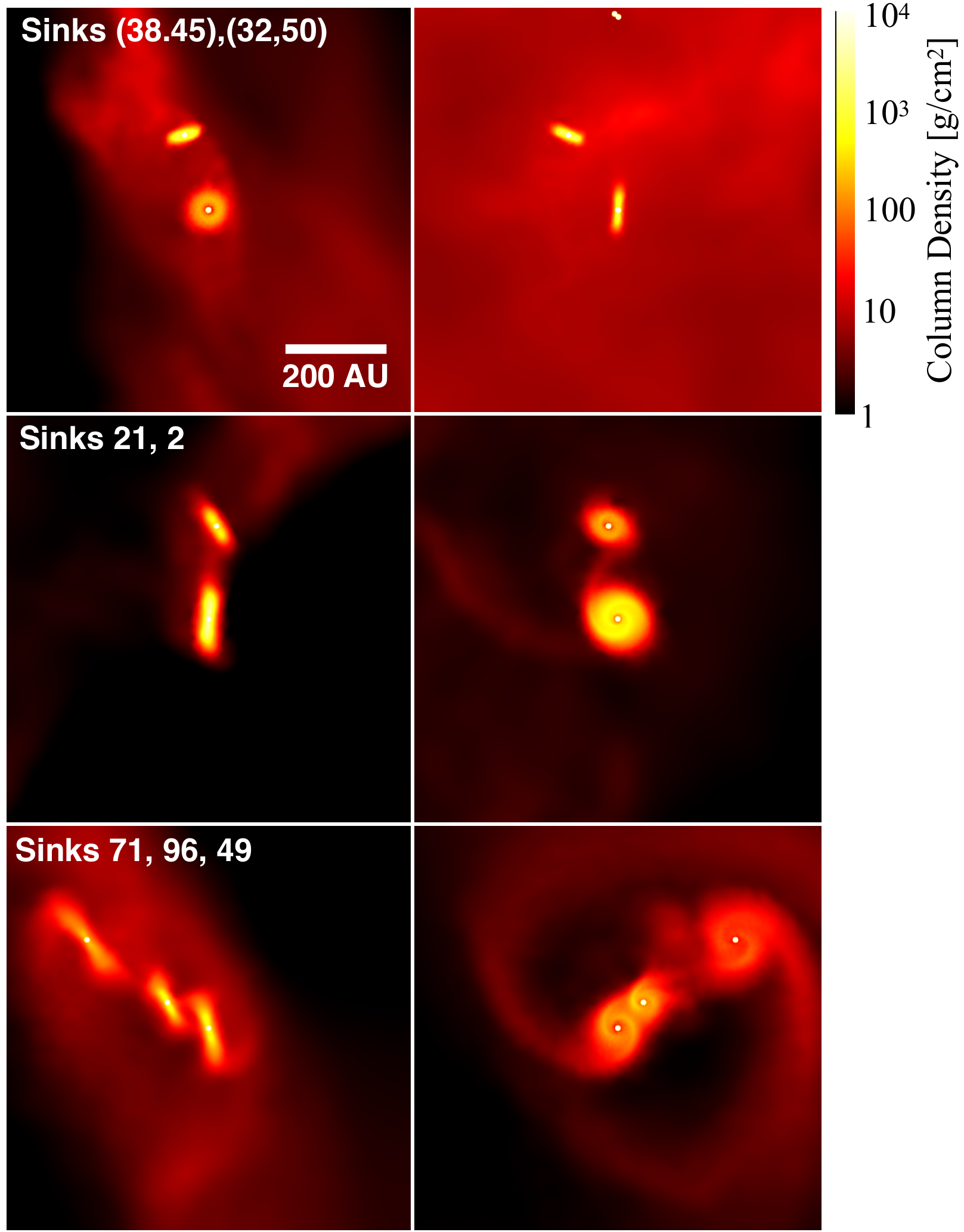} \vspace{-0.25cm}
\caption{Three examples of misaligned circumstellar discs in binary or multiple systems.  Two images (left and right) give perpendicular views of each system. Top row: a quadruple system consisting of two tight pairs separated by 200-au, with each pair surrounded by a circumbinary disc  ($t=1.18~t_{\rm ff}$).  The discs are inclined by $\approx 80$ degrees to one another.   Centre row: a 200-au binary with two circumstellar discs inclined at 44 degrees to one another ($t=1.07~t_{\rm ff}$).  Bottom row: A triple system with three circumstellar discs that are only moderately misaligned  (the left-most and right-most discs are misaligned by 22 degrees, $t=1.20~t_{\rm ff}$). Sink particles are plotted as white filled circles that have radii 10 times larger than the actual sink particle accretion radius.  Sink particles are numbered in order of the formation, and within each panel the their numbers are given listed according to their position in the images, from top to bottom.}
\label{csdiscs}
\end{figure}

The disc discussed in the previous section is a unique case of a circumbinary disc whose disc plane differs between the inner and outer regions of the disc.  A more common type of multiple-system disc found in the simulation is where two or more components of a multiple system each has a separate circumstellar or circumbinary disc.  Sometimes these discs are misaligned with each other; other times they are close to being coplanar.  In Fig.~\ref{csdiscs} we give three clear examples.  The first is a quadruple system consisting of two very tight pairs, separated by $\approx 200$~au, each with a $\approx 50$~au radius circumbinary disc.  The two discs have a relative orientation angle of $\approx 80$ degrees.  The second is a binary system with a separation of $\approx 180$~au in which each component has a circumstellar disc, with radii of $\approx 70$~au and $\approx 40$~au.  These discs are misaligned by $\approx 45$ degrees.  This binary formed with the aid of a star-disc encounter and earlier in the evolution of the pair, the two discs were perpendicular to each other.  The final example is a triple system in which all three components have resolved circumstellar discs that are close to being coplanar, but are still misalignment up to $22$ degrees.

Observationally, there are plenty of examples of misaligned discs in wide binaries and some in higher-order multiple systems.  Early evidence for such systems came from the observation that spins of binary stars are frequently misaligned with the binary's orbit \citep{Weis1974,Guthrie1985}, with \cite{Hale1994} finding a preference for alignment for binary separations $\lsim 30$~au and random uncorrelated stellar rotation and orbital axes for wider systems.  Misaligned jets from protostellar systems \citep{DavMunEis1994,Lee_etal2016} and inferred jet precession \citep{Eisloffeletal1996} also provided indirect evidence of misaligned discs.   Polarimetry can also be used to study disc alignment \citep{MonMenDuc1998, Jensenetal2004,WolSteHen2001,MonMenPer2006}.  However, we now have a growing list of directly imaged misaligned discs in wide ($\gsim 100$~au) Class II systems \citep{Koresko1998,Stapelfeldtetal1998,Kang_etal2008,Ratzka_etal2009,Roccatagliata_etal2011,JenAke2014,Salyk_etal2014,Williams_etal2014}, including the recently observed system Ophiuchus SR24 that appears to have two discs misaligned by $\approx 108^\circ$ \citep{FerZapGab2017}.  Evidence for misalignment is also starting to be found in both closer and younger multiple systems.  The Class II triple system TWA 3 \citep{Kellogg_etal2017} consists of a spectroscopic (35-day) binary with a circumbinary disc and a disc-less low-mass companion star at $\approx 50$~au, with evidence that the disc and the orbits are misaligned by at least $30^\circ$.   \citep{Lee_etal2017} has reported a Class I M-dwarf binary with a separation $\sim 1000$~au that has two circumstellar discs misaligned by $\approx 70^\circ$. \citep{Brinch_etal2016} studied gas kinematics in the young 74-au binary protostar IRS 43 that has two Keplerian circumstellar discs and a circumbinary disc.  They find the circumstellar discs may be significantly misaligned with each other ($\gsim 60^\circ$), and with the binary's orbit.

\begin{figure*}
\centering \vspace{0cm} \hspace{0cm}
    \includegraphics[width=17cm]{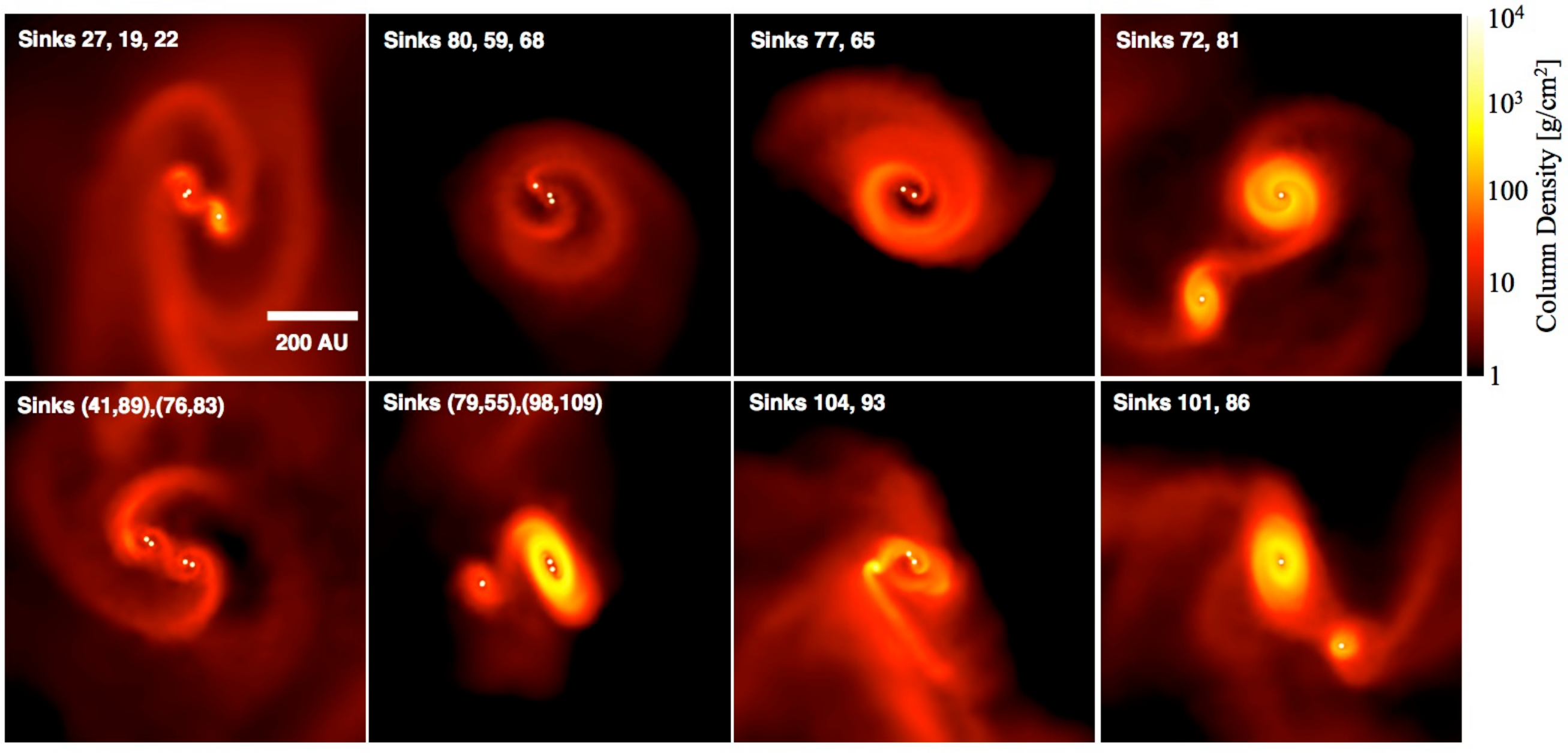} \vspace{0cm}
\caption{Eight examples of discs in binary or higher-order multiple systems.  The case with sink particles 104 and 93 is shown just before the disc fragments to form sink 134 and the morphology is very similar to the ALMA image of L1448 IRS3B published by Tobin et al. (2016).  Sink particles are plotted as white filled circles that have radii 10 times larger than the actual sink particle accretion radius.  Sink particles are numbered in order of the formation, and within each panel the their numbers are given listed according to their position in the images, from top to bottom.}
\label{multdiscs}
\end{figure*}

It has become fashionable over the past couple of decades to refer to `turbulence' when discussing the origins of systems where discs and/or orbits are misaligned (e.g. gravoturbulent fragmentation -- \citealt{Jappsen_etal2005}; turbulent fragmentation -- \citealt{Offner_etal2010}).  Such misaligned systems do naturally form in turbulent cloud simulations.  \cite{Bate2009a,Bate2012} present statistics on the misalignment angles of orbits in triple systems, and the misalignments between sink particle spins and orbits in binary systems, both of which display some similar trends to observed systems. Frequently these systems are produced by two objects forming separately initially, and subsequently evolving into a closer bound system.  The two objects may initially marginally bound to each other but on highly eccentric orbits or completely unbound, but in either case may become more tightly bound through accretion and/or star-disc encounters (see Section \ref{sec:stardisc}).   \cite{Offner_etal2016} show that binaries resulting from turbulent fragmentation have randomly orientated angular momentum, and that partial misalignment persists even after inward orbital migration.

However, it is important to recognise that it is not {\em necessary} to have turbulence to produce such systems.  For example, a binary with circumstellar discs whose axes are misaligned with the binary's orbital axis can be produced in a {\em laminar} core simply by having misalignment between the orientation of the initial density structure and the angular momentum vector(s) in the dense core.  Following such an idea, \cite{Bonnell_etal1992} produced binary systems with discs that were misaligned with the binary's orbit by having cylindrical (i.e.\ filamentary) clouds that rotated about an arbitrary axis.  \cite{Pringle1989} referred to non-linear density structure in molecular clouds as leading to `prompt fragmentation', since the seeds for fragmentation were already present in the initial conditions prior to collapse.  The distinction between appealing to fully developed turbulence versus non-linear density structure may be important since the velocity dispersion {\em within} dense molecular clouds cores is typically subsonic and independent of scale \citep{Goodman_etal1998,Caselli_etal2002} and there is observational evidence that dense cores may be kinematically distinct from the large clouds in which they are embedded \citep{Pineda_etal2010,Hacar_etal2016}.

\subsection{Circumbinary and circum-multiple discs}
\label{sec:cmdiscs}

With binary or higher-order multiple systems it is common in the simulation for circumstellar, circumbinary, and/or circum-multiple discs to exist simultaneously.  There are more than 30 examples of such discs visible in the simulation at various times.  Eight examples of these are displayed in Fig.~\ref{multdiscs}.  Four of these are binary systems.  System (77,65) shows a large $\approx 200$~au circmbinary disc around a $\approx 25$-au binary.  This system would be expected to have circumstellar discs as well, but these are poorly resolved in the simulation.  

Systems (72,81) and (101,86) are both wide binaries (separations $>200$~au) with two well resolved circumstellar discs and small amounts of circumbinary material.  Qualitatively, these systems are similar to the Class I system L1551 NE, in which a binary with projected separation of 70 au has two circumstellar discs and a 300-au circumbinary disc with strong spiral arms \citep{Takakuwa_etal2012, Takakuwa_etal2017}.  Although the two examples we give here are each approximately twice as large in physical scale as L1551 NE, their morphological structure of two circumstellar discs with high surface densities, and the strongly-perturbed circumbinary disc with a low surface density and streams feeding the circumstellar discs is very similar.  We note that \cite{Takakuwa_etal2017} suggest that the circumstellar discs of L1551 NE may be misaligned with each other and with the circumbinary disc due to the differing position angles of their major axes.  The two circumstellar discs in system (72,81) are misaligned by 40 degrees, and those in system (101,86) are misaligned by 68 degrees.  However, we caution that care must be taken when using the position angles of discs to infer misalignment since, as can be seen in the image of system (72,81), the smaller (circumsecondary) disc is eccentric, its eccentricity varies with time (see the animation), and the discs may contain spiral arms which may also complicate the determination of the disc's major and minor axes.

System (104,93) is shown just before its circumbinary disc fragmented to form a third protostar (number 134).  The geometry of this system is very similar to the recent ALMA image of the triple protostar L1448 IRS3B \citep{Tobinetal2016}.  The spatial size of the system is about half that of L1448 IRS3B, with projected separations of $\approx 30$ and $\approx 90$ au compared to the separations of the observed system of $61$ and $183$ au, respectively.  Similarly, the masses are lower.  The observed close pair have a combined mass of $\approx 1$~M$_\odot$ \citep{Tobinetal2016}, while each component of the close pair in the simulated system has a mass of $\approx 0.2$~M$_\odot$ at the time of disc fragmentation.  In the 10,000 yrs following the formation of the third protostar, the stellar masses of the pair each grew to $\approx 0.2$~M$_\odot$, while the third component grew to $\approx 0.15$~M$_\odot$.  The estimated mass of the third component in the observed system is $\approx 0.09$~M$_\odot$.  The total disc mass remained around $\approx 0.1$~M$_\odot$ during this time due to ongoing accretion from the cloud, whereas in L1448 IRS3B the total disc mass is estimated to be $\approx 0.3$~M$_\odot$.

The four other systems in  Fig.~\ref{multdiscs} are higher-order multiples -- two triples and two quadruples.  The two triples both consist of a close pair and a wider component, and large circum-triple discs with strong spiral arms.  The circumstellar and circumbinary discs are better resolved in system ((19,22),27) than in system ((59,68),80).  The two triples both consist of two tight pairs separated by $\approx 150-200$~au.  System (41,89),(76,83) displays both circumbinary discs and a large circum-quadruple disc with strong spiral arms.  System (79,55),(98,109) has two resolved circumbinary discs, but there is little circum-quadruple material.

There are not many resolved observations of circumbinary discs to date.  The first were of GG Tau \citep{DutGuiSim1994,GuiDutSim1999} and UY Auriga \citep{Duvertetal1998}, and these are still the best examples.  The edge-on disc of HH30 apparently contains a binary \citep{Guilloteau_etal2008}.  There are also some well known unresolved circumbinary discs such as V4046 Sgr \citep{Byrne1986, SteGah2004}, UZ Tau E \citep{MatMarMag1996,Martin_etal2005}, DQ Tau \citep{Mathieu_etal1997}, and RX J0530.7-0434 \citep{Covino_etal2001}, with GW Ori \citep{MatAdaLat1991} actually being a close triple system \citep{Berger_etal2011}.  With improved resolution, more resolved systems should be expected in the future.

\begin{figure*}
\centering \vspace{0cm} \hspace{0cm}
    \includegraphics[width=17cm]{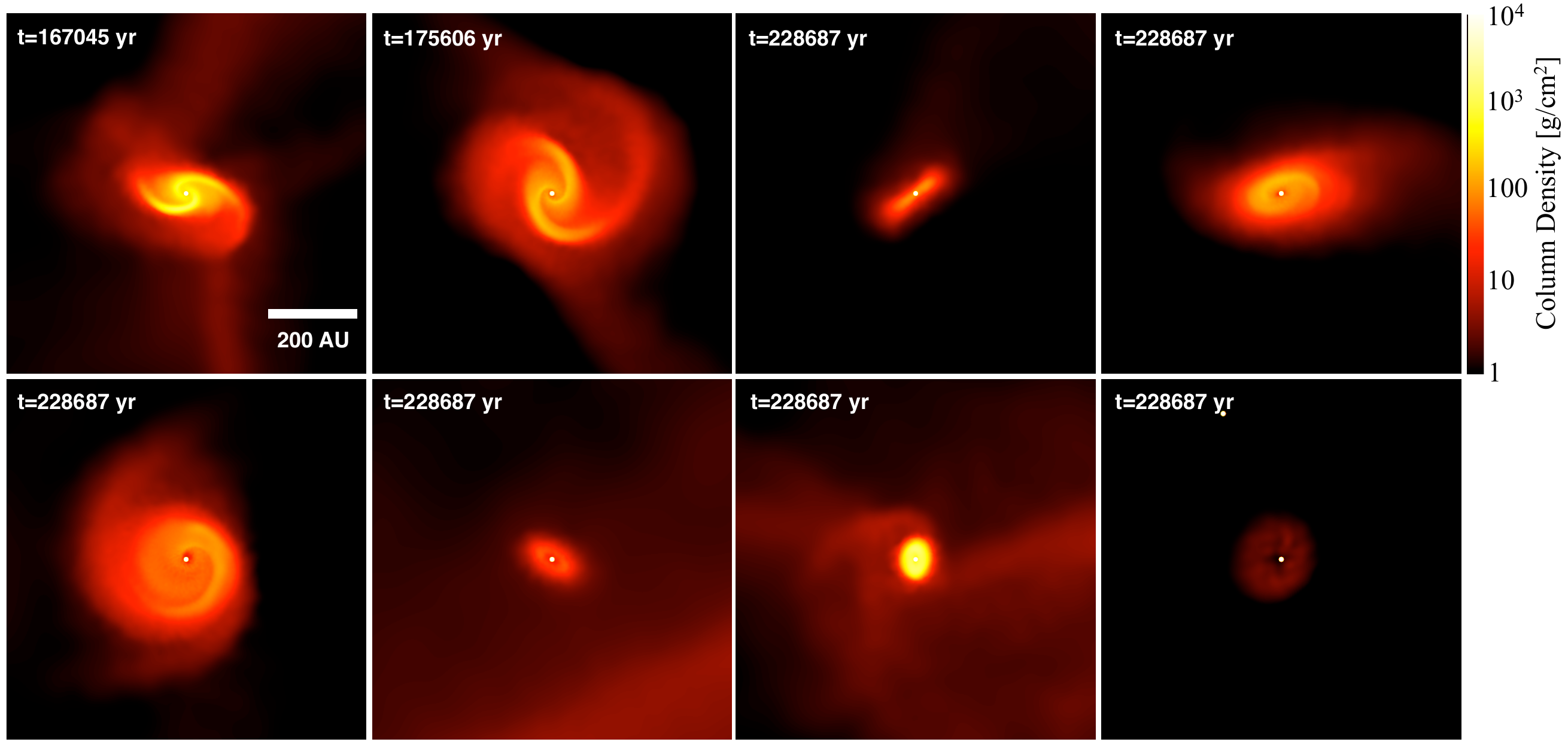} \vspace{0cm}
\caption{Examples of eight of the circumstellar discs around single protostars from the calculation.  The first two cases demonstrate gravitational instabilities in young massive discs soon after they have formed (protostar numbers 2 at $t=0.88~t_{\rm ff}$ and 4 at $t=0.92~t_{\rm ff}$).  The remaining 6 panels show discs at the end of the calculation around protostars 53, 85, 99, 119, 136, and 141, respectively.  The discs have a wide variety of radii and masses.  Sink particles are plotted as white filled circles that have radii 10 times larger than the actual sink particle accretion radius. }
\label{single}
\end{figure*}

\subsection{Discs around single stars}
\label{sec:single}

In the above sections, we have illustrated the variety of the discs found in multiple systems.  However, there are also a lot of discs around single stars.  At various times in the calculation there are more than four dozen single protostars with resolved discs.  Not all of these remain single to the end of the calculation, and even for those that do, not all of the resolved discs survive to the end of the calculation due to various processes which will be discussed in the following section.

In Fig.~\ref{single} we display snapshots of eight discs around single protostars.  Many single protostars have large ratios of disc mass to stellar mass soon after they form.  Consequently, these discs display strong spiral arms because they are gravitationally unstable.  Some of these fragment (see Section \ref{sec:fragment}), but others are stable enough to avoid fragmentation and transport mass and angular momentum rapidly via gravitational torques from the spiral arms \cite[e.g.][]{LynKal1972,Paczynski1978,LinPri1987,LauBod1994}.  In the first two panels of Fig.~\ref{single} we give examples of such massive discs, those around protostar numbers 2 and 4.  At the times shown, protostar 2 had a mass of 0.20~M$_\odot$ and its disc mass was 0.25~M$_\odot$; protostar 4 had a mass of 0.23~M$_\odot$ and its disc mass was 0.27~M$_\odot$.

At the end of the calculation, the discs around the single protostars have a wide range of properties.  The remaining 6 panels of Fig.~\ref{single} show some of them.  These protostars have masses of 0.18, 0.26,  0.29, 0.11, 0.50, and 0.17~M$_\odot$, respectively, while their discs have masses of 0.03, 0.25, 0.33, 0.02, 0.38, and 0.004~M$_\odot$, respectively.  The disc radii are approximately 60, 100, 100, 50, 30, 70 au in radius, respectively, where in each case this is the radius containing 63.2\% of the disc mass.   Since the SPH particles have masses of 1/70000~M$_\odot$ each, the latter of these discs only contains $\approx 280$ SPH particles, which is why it is so faint in the image.

Recent observations have detected a number of spiral waves in circumstellar discs. Examples of spiral waves in Class II objects and transition discs include: AB Aur \citep{Hashimoto_etal2011}, MWC 758 \citep{Grady_etal2013, Benisty_etal2015}, SAO 206462 \citep{Muto_etal2012, Garufi_etal2013, Stolker_etal2016}, HD 100546 \citep{Boccaletti_etal2013, Avenhaus_etal2014, Currie_etal2015, Garufi_etal2016, Follette_etal2017}, HD 100453 \citep{Wagner_etal2015}, AK Sco \citep{Janson_etal2016}, Elias 2-27 \citep{Perez_etal2016}.  \cite{Alves_etal2017} have presented observations of the Class I object BHB07-11 with a dense 80-au radius disc surrounded by a lower density disc extending to $\approx 300$~au that has spiral structure.  The Class 0 triple protostar L1448 IRS 3B also has spiral structures \cite{Tobinetal2016}.  In the absence of more information it is difficult to know whether observed spiral structure is generated by a companion \citep[as in the case of HD 100453;][]{Wagner_etal2015, Dong_etal2016, Benisty_etal2017} or disc self gravity.  However, the Class II object Elias 2-27 which has a clear `grand design' spiral \citep{Perez_etal2016} is a strong candidate for a disc in which the spiral structure is driven by disc self-gravity \citep{Tomida_etal2017, Meru_etal2017}.  Similarly, it has been argued that the triple system L1448 IRS 3B was recently formed by disc fragmentation (see Sections \ref{sec:cmdiscs} and  \ref{sec:fragment}), in which case the disc must have been strongly self-gravitating.

\begin{figure*}
\centering \vspace{0cm} \hspace{0cm}
    \includegraphics[width=17cm]{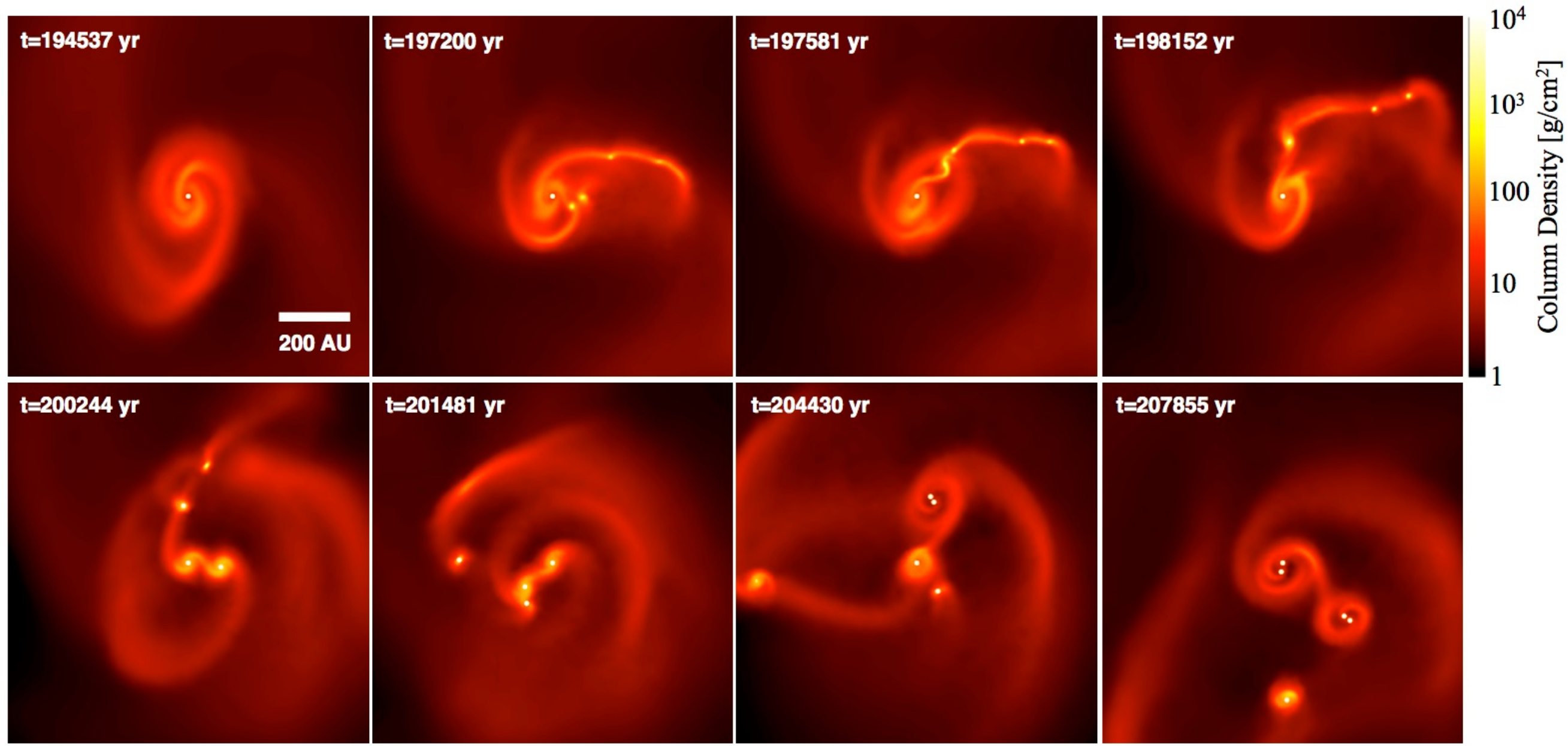} \vspace{0cm}
\caption{Time sequence showing the fragmentation of the massive disc around sink particle 41.  In panels 2--4 three potential fragments merge into a single object before it collapses to a stellar core (sink number 76).  Two more fragments at the top right of the 4th panel eventually collapse to stellar cores (sink numbers 83, 89) and these pair up with 76 and 41, respectively to produce a quadruple system consisting of two pairs: (41,89),(76,83).  The 5th sink (number 135, visible in the last two panels) is eventually ejected from the system.  Sink particles are plotted as white filled circles that have radii 10 times larger than the actual sink particle accretion radius. }
\label{fragmentation41}
\end{figure*}

\begin{figure*}
\centering \vspace{0cm} \hspace{0cm}
    \includegraphics[width=17cm]{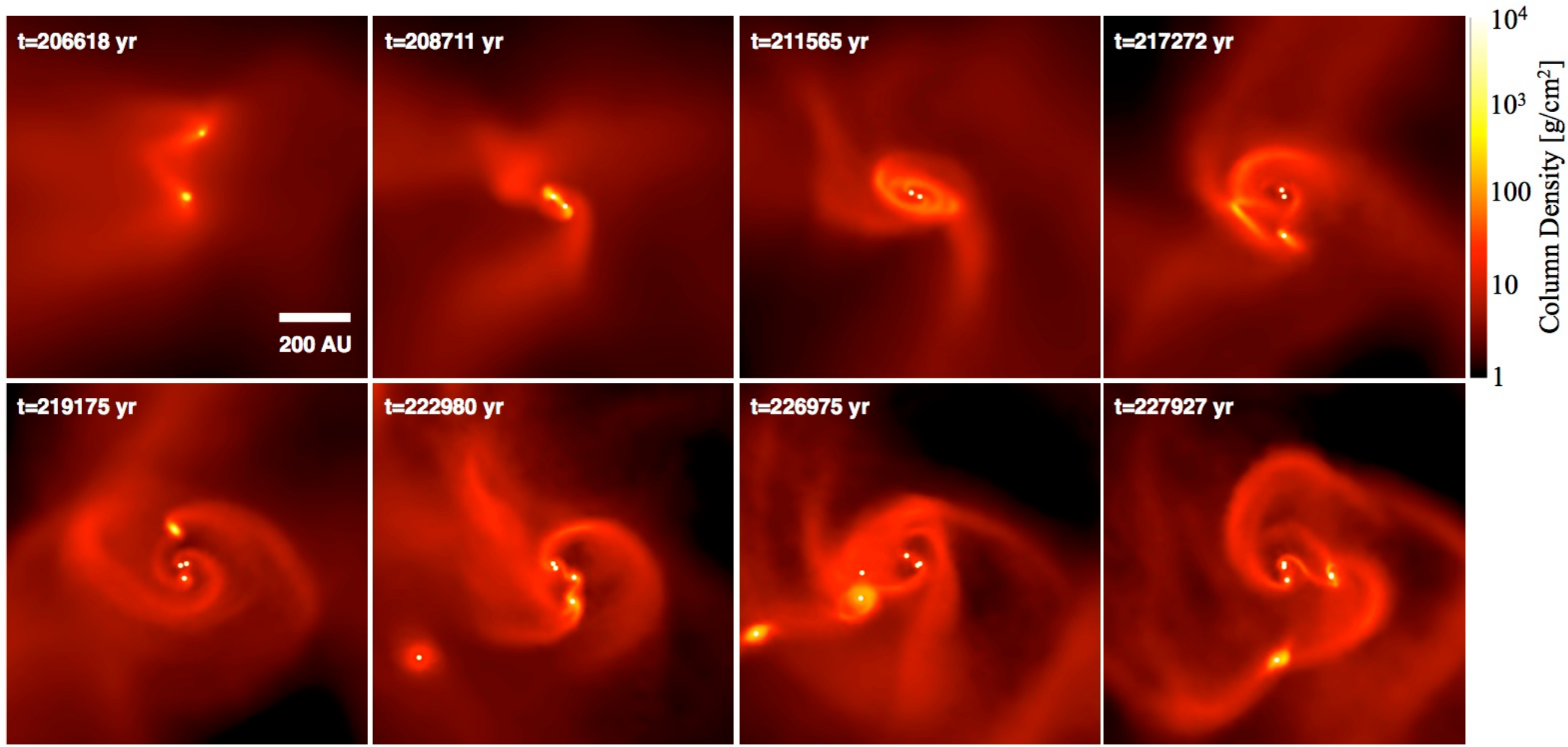} \vspace{0cm}
\caption{Time sequence showing the fragmentation of the massive disc around sink particle 122.  In panels 1--3, two protostars (sink numbers 122, 123) form separately but bound, undergoing a star-disc encounter to form a tight binary with a circumbinary disc. This disc fragments to produce a triple (panels 4 \& 5; sink number 145), and again to produce sink number 159 (panel 5).  Sink number 150 forms separately and falls into the system, colliding with the disc around sink 159 (panel 7) to produce a tight binary companion to the triple.  The widest companion in panels 7 and 8 (sink number 180) formed in the disc just before the calculation was stopped. Sink particles are plotted as white filled circles that have radii 10 times larger than the actual sink particle accretion radius. }
\label{fragmentation122}
\end{figure*}

\section{Dynamical evolution of discs}

As we have seen in the previous sections, the discs in the protostellar systems have diverse morphologies, due both to their formation in a turbulent, chaotic environment, and due to gravitational interactions with companions or even the self-gravity of the discs.

However, the discs also evolve with time.  Self-gravitating discs transport mass and angular momentum via gravitational torques \citep{LauBod1994} and may also fragment \citep{Bonnell1994,BonBat1994a}.  Gravitational interactions between binaries and circumbinary discs or higher-order multiples and circum-multiple discs can lead to orbital decay \citep{Artymowiczetal1991}.  The discs form from the collapse and accretion of gas from the molecular cloud, and in many cases this continues to the end of the simulation.  Conversely, discs can accrete gas \citep{MoeThr2009,Scicluna_etal2014,Wijnen_etal2016,Wijnen_etal2017} or suffer from ram-pressure stripping as they pass through density cloud material \citep{Wijnen_etal2016}.  Star-disc interactions can also strip away or truncate discs \citep{ClaPri1991b}, and/or energy loss during a star-disc interaction can produce binaries or high-order multiple systems from protostars that were previously unbound \citep*{ClaPri1991a,HalClaPri1996}.  Finally, even if none of these processes play a significant role in disc evolution, the numerical simulations have some shear viscosity and this will lead to viscous evolution of the discs \citep{LynPri1974}.   Examples of all these evolutionary processes can be seen during the simulation (see the animation in the Supporting Information that accompanies this paper).  In the following sections, we briefly discuss these further and, in some cases, give examples.

\begin{figure*}
\centering \vspace{0cm} \hspace{0cm}
    \includegraphics[width=17cm]{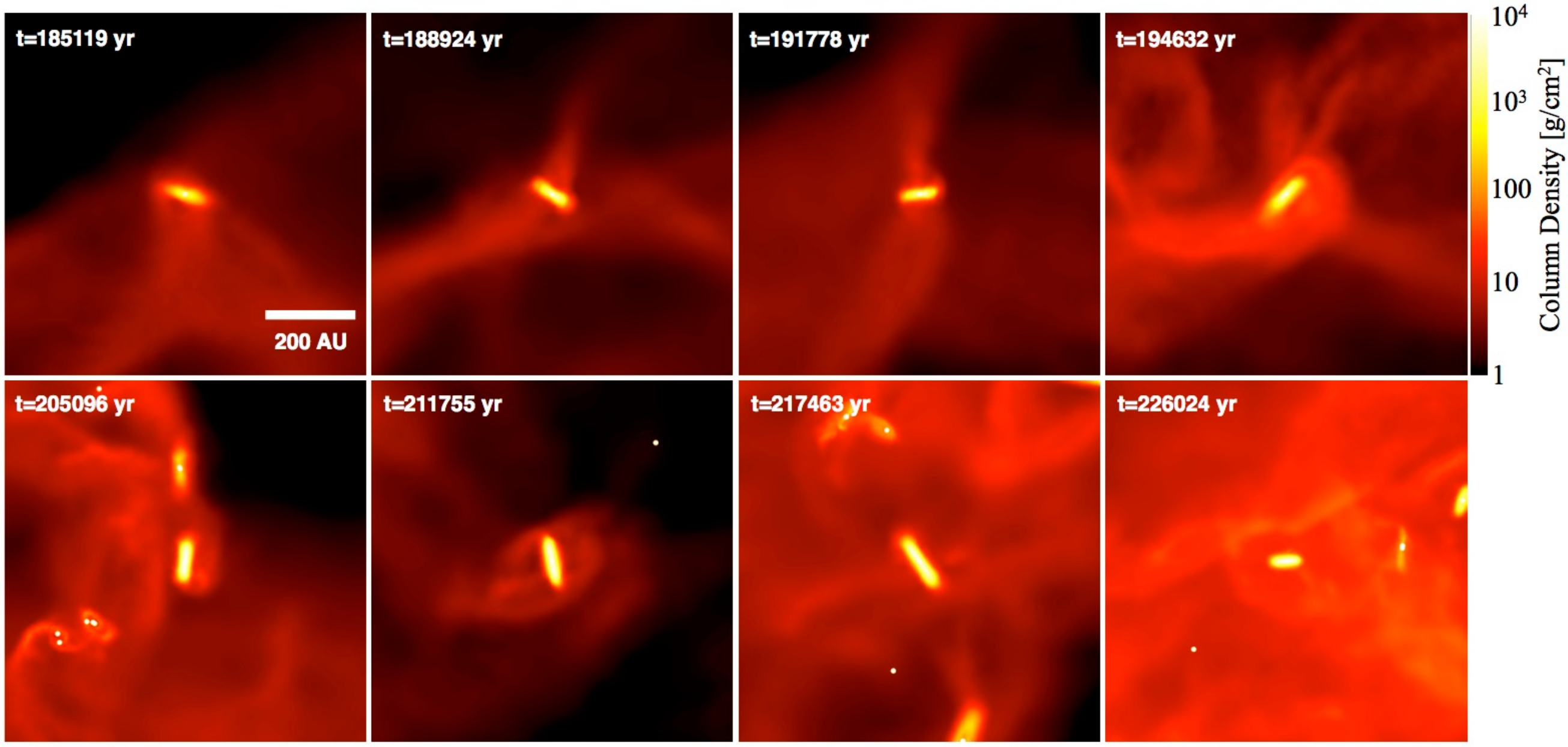} \vspace{0cm}
\caption{Time sequence showing the variation of the orientation of the disc around sink particle 40 due to accretion of gas with different angular momentum.  Between panels 2 and 8 the angular momentum vector of the disc rotates by more then 180 degrees.  Sink particles are plotted as white filled circles that have radii 10 times larger than the actual sink particle accretion radius. }
\label{variable}
\end{figure*}

\subsection{Disc fragmentation}
\label{sec:fragment}

Although gravitational fragmentation of massive discs is not as common in calculations that include radiative transfer \citep[e.g.][]{Bate2012} as in calculations that use a barotropic equation of state \citep[e.g.][]{Bate2009a}, there are ten discs that undergo fragmentation in the calculation.  All but four of these produce multiple fragments (one produces 6 fragments, another produces 5, two produce 3 fragments, and two produce 2 fragments), so together 25 protostars are formed by disc fragmentation (i.e.\  about 1/7 of the total number of protostars).  The fragmentation of the circumbinary disc of system (104,93) to produce a third protostar (number 134) which has a very similar  morphology to the Class 0 system L1448 IRS 3B \citep{Tobinetal2016} was discussed in Section \ref{sec:cmdiscs}.  In this section, we give two other examples.

In Fig.~\ref{fragmentation41} we show a time sequence of the evolution of the massive disc surrounding protostar number 41.  In the first panel, the mass of the protostar is 0.07~M$_\odot$ while the disc mass is 0.17~M$_\odot$.  The gravitationally unstable disc has strong spiral arms.  In the second panel, four fragments are forming, but in the third and fourth panels three of these merge into a single object, while two further fragments forms in the outer parts of the largest arm.  This shows the importance of not replacing gas fragments with sink particles until just before a stellar core would be formed in reality (see Section \ref{sec:method}).  If these fragments had been replaced by sink particles earlier, the fragmentation would have been artificially enhanced.  The fragment resulting from the triple merger does then undergo the second collapse phase and is replaced by a sink particle (protostar number 76) producing a binary (fifth panel).  The two outer fragments also collapse and are replaced by sink particles (protostar numbers 83 and 89).  Protostar 83 forms a tight pair with protostar 76, while the other forms a tight pair with protostar 41, resulting in a hierarchical quadruple system (panels 6--8).  In the meantime a further protostar has formed from the largest arm, resulting in a pentuple system.

As the second example, in Fig.~\ref{fragmentation122} we show a time sequence of the evolution of the massive disc surrounding the binary system composed of protostar numbers 122 and 123.  These form from two separate, but nearby, condensations (first panel) and quickly form a binary which accretes a circumbinary disc (second and third panels).  This disc is gravitationally unstable (disc mass $\approx 0.15$~M$_\odot$, protostellar masses 0.12, 0.10~M$_\odot$, respectively, at $t=216,000$~yrs) and fragments to produce two additional protostars which arrange into a hierarchical triple system with a fourth outer component (panels 5--7).  The subsequent evolution is complicated by the infall of protostar number 150, which formed separately from the system and a mutual star-disc encounter with protostar number 159 produces a tight pair which is bound to the triple.  Meanwhile an additional protostar has formed from a gravitationally unstable arm of the circum-mulitple disc (panels 7 and 8), resulting in a sextuple system overall.

\subsection{Evolution of disc orientation}

After a protostellar system has formed, it can continue to accrete further gas from the cloud.  Since the cloud is turbulent, the orientation of the angular momentum of this additional gas relative to the protostellar system may be very different from the orientation of the angular momentum that originally produced the system.  In Section \ref{sec:misaligned} we saw how this could also produce a disc in which the inner and outer parts of the disc had different orientations.  However, in the simulation discussed in this paper, a much more common affect is that substantial accretion can re-orientate the plane of a disc.   \cite{BatLodPri2010} investigated how the accretion of such material may lead to stellar spins being misaligned with planetary orbital planes, potentially explaining observations of misaligned exoplanet systems \citep[see also][]{Fielding_etal2015}.  

There are at least ten examples in the simulation of disc orientations being changed by accretion.  In Fig.~\ref{variable}, we show a time sequence of one of these -- the disc surrounding the single protostar, number 40.  Between the first two panels, it can be seen that accretion rotates the disc plane clockwise in the figure by about 20 degrees.  Then the effect reverses, and most of the remainder of the simulation, the disc plane rotates anticlockwise.  Between the second panel and the last panel, the angular momentum vector of the disc rotates by approximately 220 degrees!  During the period from 195,000 to 223,000 yrs, the disc mass remains between 0.5 and 0.7 M$_\odot$ but the mass of the star increases from 0.4 to 2.5 M$_\odot$. This clearly demonstrates that the orientation of protostellar discs can be altered dramatically by accretion in such a chaotic environment.  Such reorientation would also be expected to alter the direction of a protostellar jet \citep[see also][]{BatLodPri2010}.

\begin{figure*}
\centering \vspace{0cm} \hspace{0cm}
    \includegraphics[width=17cm]{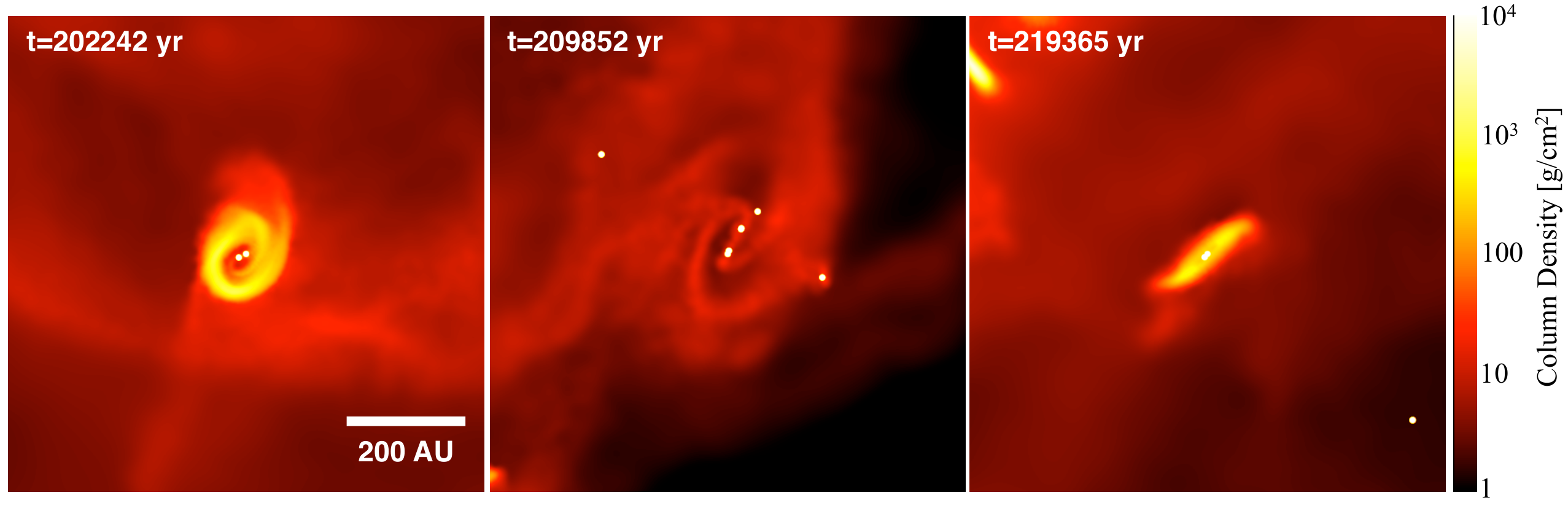} \vspace{-0.25cm}
\caption{Time sequence showing the stripping and reforming of a disc around protostar number 37 (located at the exact centre of each panel).  Initially protostar 37 forms a binary with protostar number 48 and the pair are surrounded by a circumbinary disc.  This disc is stripped away via dynamical encounters with several other protostars.  During these encounters, protostar 48 is unbound and replaced by protostars 25 and 26, forming a tight triple system.  This triple system then accretes new material from the molecular cloud, producing a circumtriple disc.  Sink particles are plotted as white filled circles that have radii 10 times larger than the actual sink particle accretion radius.}
\label{reform}
\end{figure*}

\subsection{Disc erosion and discs renewed by accretion}

Many protostars in the simulation have their discs eroded or truncated either by ram pressure stripping as they quickly move through dense molecular cloud material, or when they have dynamical encounters with other protostars.  There are at least two dozen examples of such disc erosion which can be seen in the animation.  In some of these a smaller, resolved disc survives, but in many the discs are stripped away completely due to the finite numerical resolution of the calculations.  In reality, the cases in the calculation in which the discs are stripped entirely would be expected to retain small, low-mass discs.  However, with sink particle accretion radii of 0.5~au and the SPH resolution length scaling with density, $\rho$, as $h \propto \rho^{-1/3}$, discs with radii $\lsim 10$~au are not usually resolved in the calculation.  As mentioned in Section \ref{sec:single} when discussing the last panel of Fig.~\ref{single}, a disc mass of 0.004~M$_\odot$ corresponds to only 280 SPH particles and is clearly not very well resolved.  In Appendix \ref{appendixB}, we also show that discs that are modelled by $\lsim 2000$ particles are likely to suffer some numerical viscous evolution over the typical timescales modelled in the calculation, and for those modelled by $\lsim 500$ particles this evolution is likely to be significant.

There are a few cases in the calculation of discs being eroded, and then new discs being accreted from the molecular cloud.  An example of this is shown in Fig.~\ref{reform}.  In this case, the original disc is destroyed during dynamical encounters with other protostars, and a new disc (with a different orientation) is later accreted from the molecular cloud.

Accretion by a disc passing through an ambient medium and ram-pressure truncation of circumstellar discs has been studied in detail by \cite{MoeThr2009}, and \cite{Wijnen_etal2016,Wijnen_etal2017}.   \cite{Wijnen_etal2017b} also study the effects of disc reorientation as a protostellar disc travels through an ambient medium.  \citep{Wijnen_etal2017c} find that face-on accretion and ram pressure stripping are more important for setting disc radii than dynamical encounters when the total mass in stars is $<30$\%.  However, this assumes a `smooth' (non-clustered) stellar distribution.  In the simulation studied here, protostars tend to be formed in small groups (either in filament fragmentation or disc fragmentation, or both).  Because of this, {\em both} dynamical interactions and ram-pressure stripping are very important in truncating and stripping discs (even though at the end of the calculation less than 20\% of the mass is in protostars).  Furthermore, because other forming protostars are embedded in dense gas, both effects can occur during a single encounter.  Examining the evolution of all 183 protostars, we find that dynamical encounters alone are responsible for stripping approximately 26 discs, ram-pressure stripping alone is responsible for stripping approximately 7 discs.  Another 18 discs are stripped by a {\em combination} of ram-pressure stripping and encounters with other protostars.  Thus, both processes are important.

\begin{figure*}
\centering \vspace{-0.5cm} \hspace{0cm}
    \includegraphics[width=8.5cm]{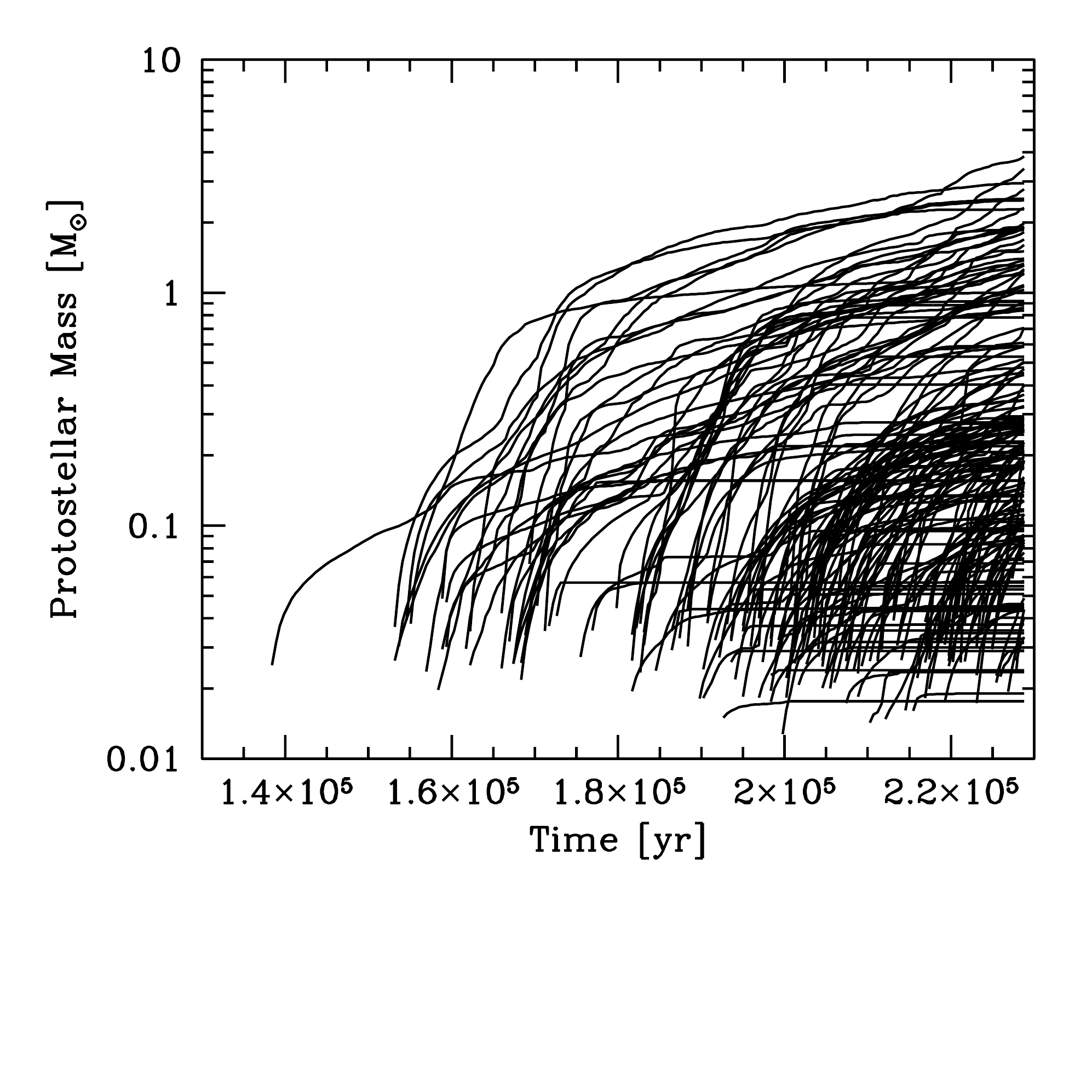} \vspace{0cm}
    \includegraphics[width=8.5cm]{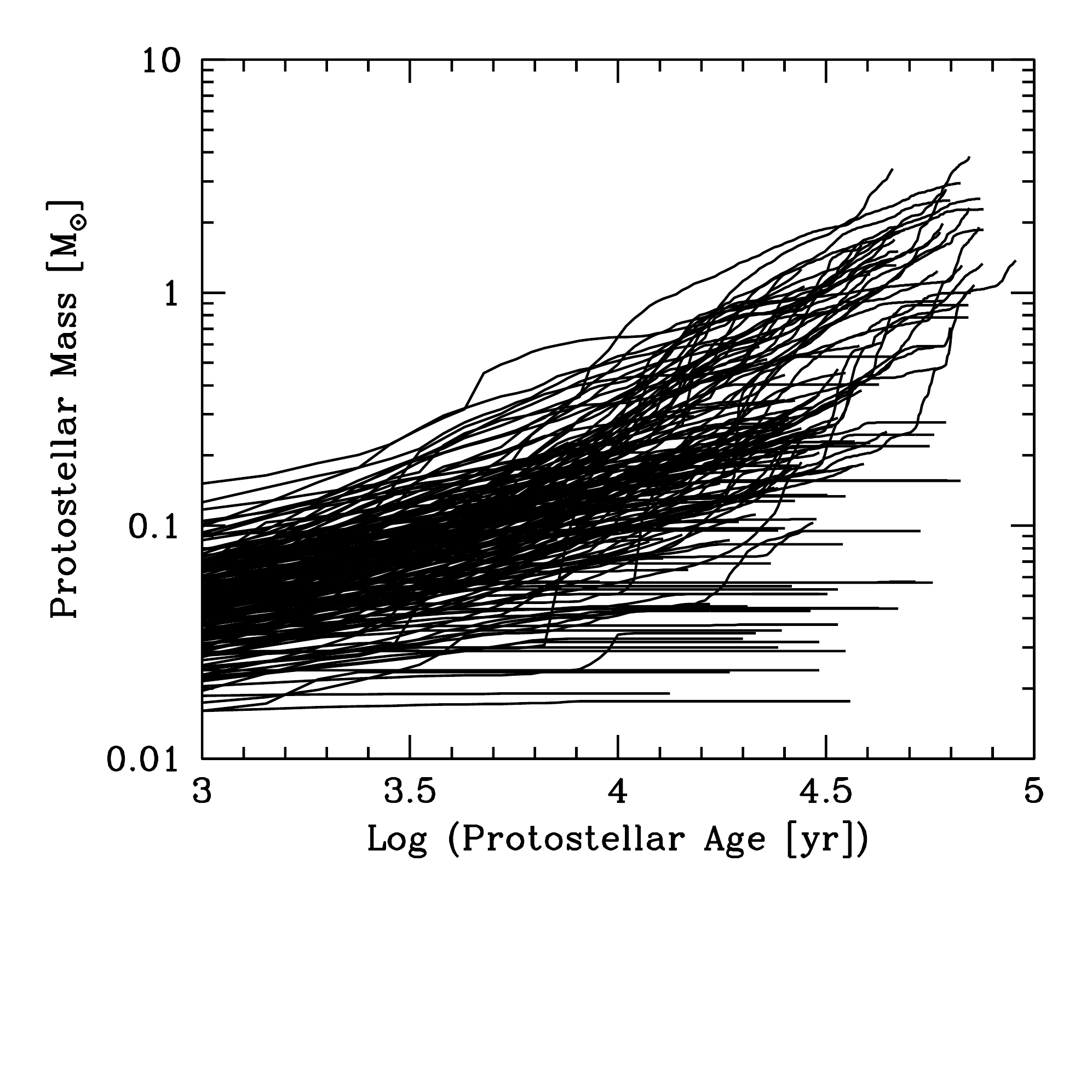} \vspace{-1.5cm}
\caption{The time evolution of the masses of all of the protostars (sink particles) formed in the calculation.  As is expected, the stellar masses increase monotonically with time, and there is greater dispersion when plotting the mass versus absolute time (left panel; linear time axis) than when plotting mass versus the time since protostar was formed, i.e. the age of the protostar  (right panel; logarithmic time axis).  }
\label{starmass}
\end{figure*}

\subsection{Star-disc encounters and orbital decay}
\label{sec:stardisc}

Star-disc encounters are very common in the calculation,  More than four dozen can be counted by looking at the animation.  One example was discussed in Section \ref{sec:fragment} and is illustrated in Fig.~\ref{fragmentation122} (panels 6, 7 and 8).  Star-disc encounters are frequently involved in forming binary systems (32 cases) or higher-order multiple systems (at least 14 cases) from protostars which form in separate, but nearby, condensations in the highly-structured molecular cloud.  The close binary in Fig.~\ref{misaligned} was formed this way.  After producing bound systems from two unbound protostars, there is usually rapid decay of the orbital separation and eccentricity as the binary transfers angular momentum and energy to the dissipative gas, often producing a circumbinary disc.  \cite*{BatBonBro2002b} argued that orbital decay from interactions with circumbinary or circum-multiple discs (in addition to dynamical interactions and accretion) are crucial for producing close binary systems (separations $\lsim 10$~au) which cannot form via direct fragmentation since the typical sizes of first hydrostatic cores are $\approx 5$~au in radius \citep{Larson1969}.

\cite{ClaPri1991a} studied star-disc capture rates in young stellar groups and clusters and found that the rates were too low to provide an important binary formation mechanism.  However, their study examined virialised stellar groups with stellar densities and velocity dispersions typical of nearby star-forming regions.  It does not apply to the earlier stage of the fragmentation of highly-structured or turbulent molecular gas.  Both numerical simulations \citep{BatBonBro2003} and recent observations \citep{Andre_etal2007,Foster_etal2015,Rigliaco_etal2016,Sacco_etal2017} find the typical velocity dispersions in dense molecular gas, from which protostars form, are much lower than (typically $\approx 1/3$) the velocity dispersions of young stars.  \cite{BatBonBro2003} attributed the larger velocity dispersion of stars to gravitational interactions between stars after they had formed (e.g. dynamical interactions with binaries and the break up of multiple systems).  Prior to this, the low velocity dispersion of the molecular gas means that protostars frequently form in separate condensations that are either marginally unbound or marginally bound to each other.  It is then common for these objects to undergo relatively slow star-disc encounters in which the two objects become bound, or the orbits of already bound objects become tighter, less eccentric, and the system changes its orbital orientation.  Discs that are misaligned with the orbit of such a binary are a natural outcome of this process \citep[e.g.][]{Offner_etal2016}, and if the discs in the simulation were better resolved they would likely be warped.  \cite{MoeBal2006} performed well resolved hydrodynamical simulations of star-disc encounters, examining the torquing of the disc and its reorientation.

\begin{figure}
\centering \vspace{-0.5cm} \hspace{0cm}
    \includegraphics[width=8.5cm]{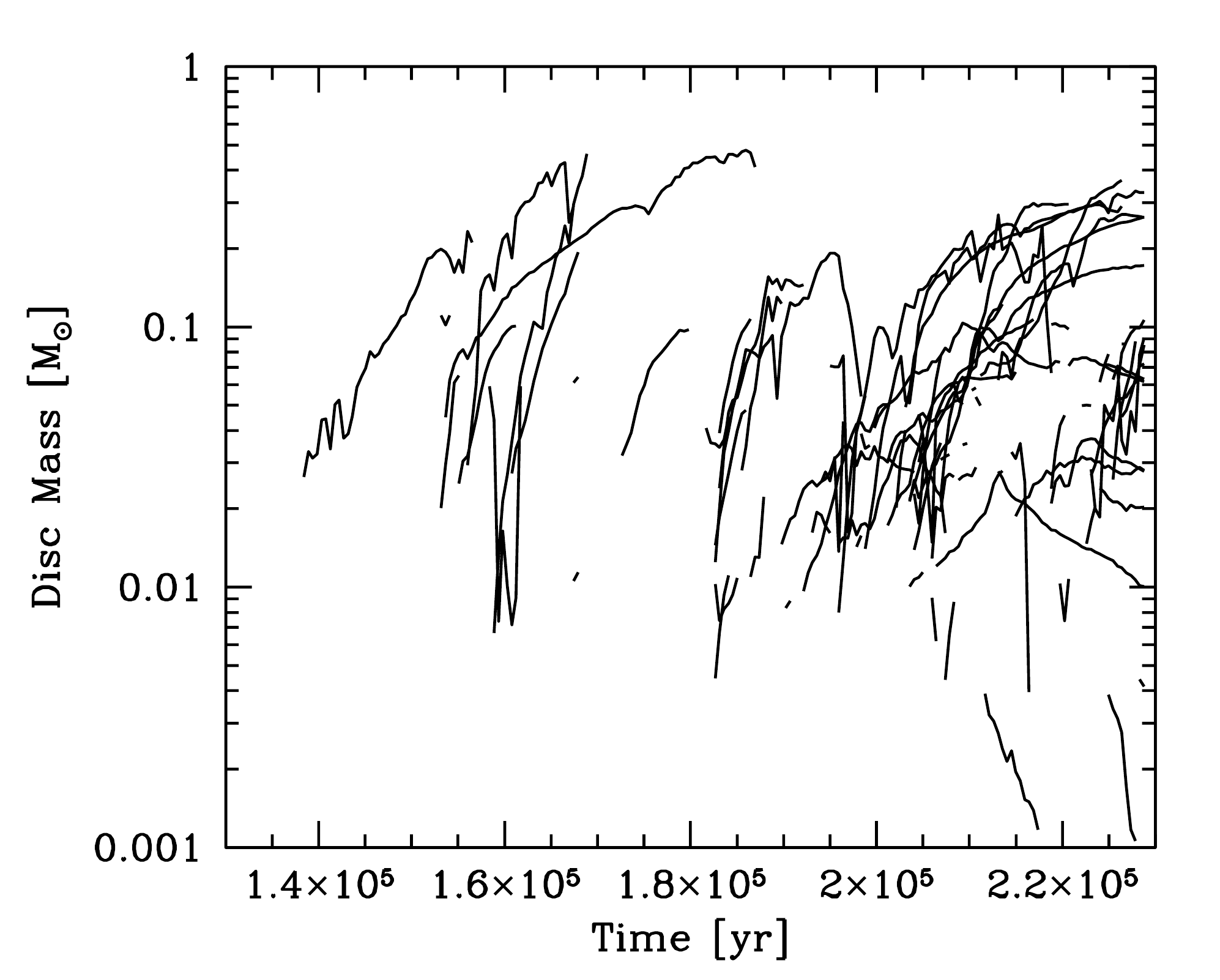} \vspace{0.0cm}
    \includegraphics[width=8.5cm]{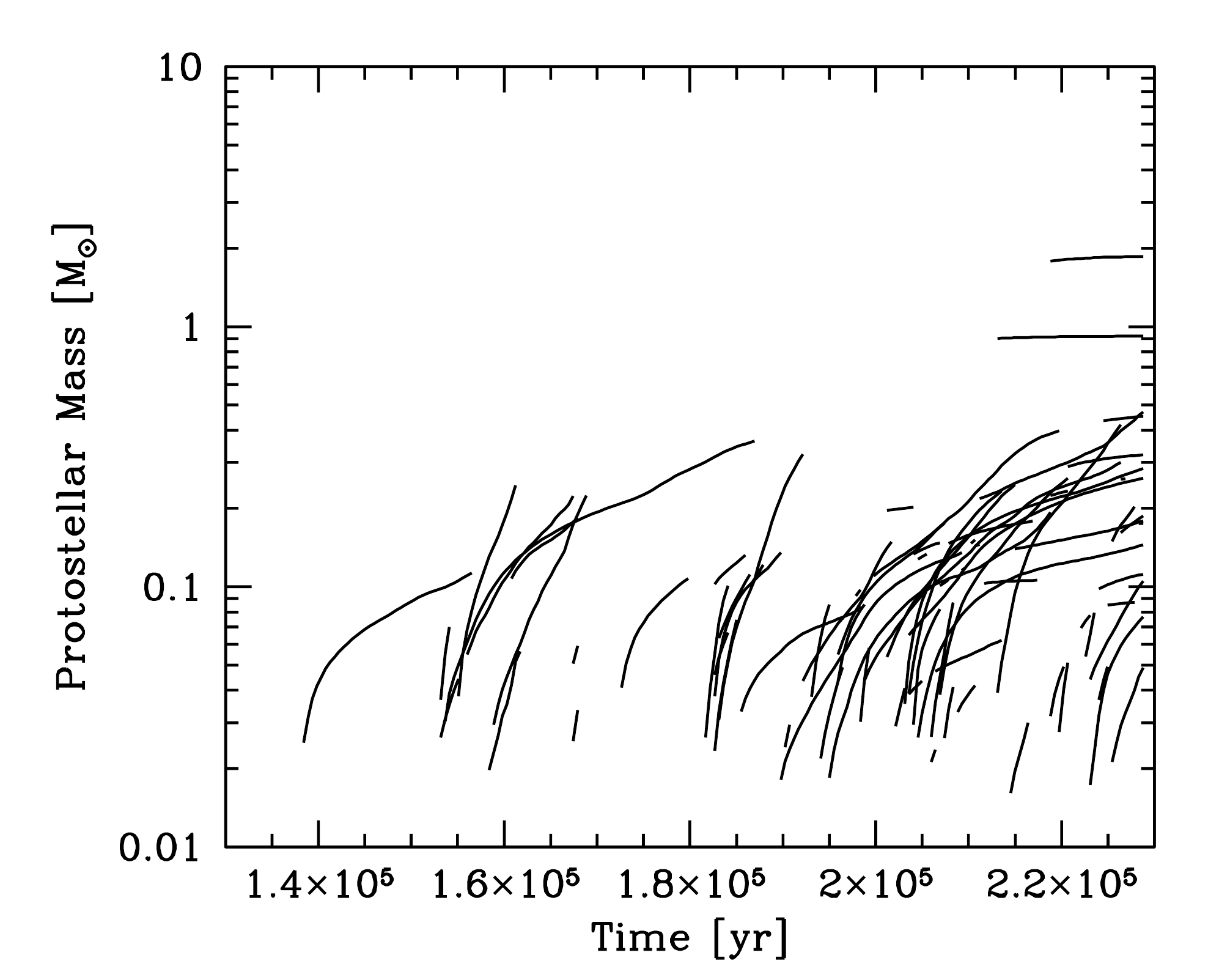} \vspace{0.0cm}
    \includegraphics[width=8.5cm]{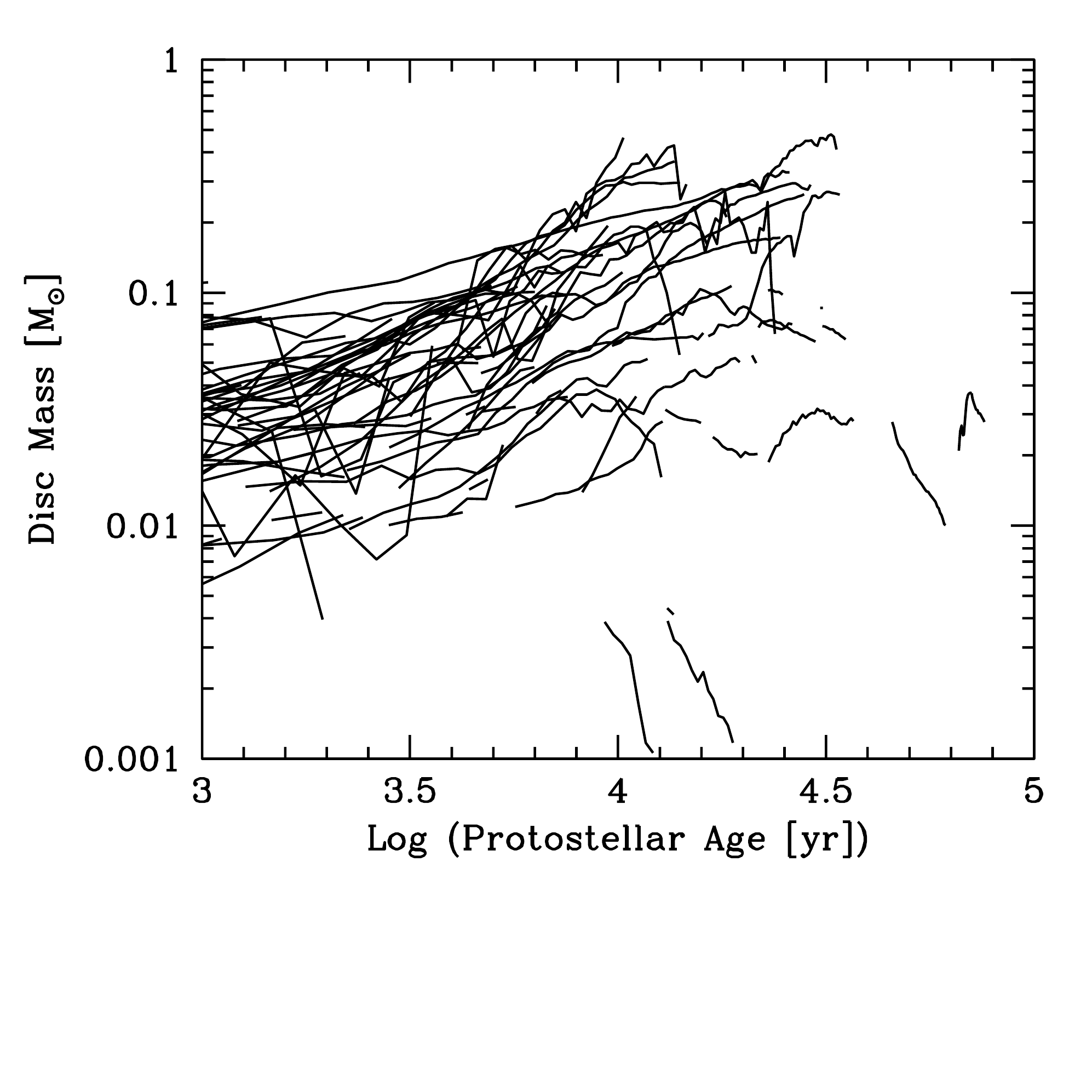} \vspace{-2cm}
\caption{The time evolution of the disc masses of isolated protostars (sink particles) during the calculation.  The top panel gives the disc mass versus the (linear) time in the calculation, while the bottom panel gives the disc mass versus the age of the protostar (using a logarithmic time scale).  For comparison with the top panel, the middle panel provides the time evolution of the protostellar mass (i.e. sink particle mass) of the isolated protostars whose disc masses are given in the top panel.  Each line represents the evolution of the disc around a particular isolated protostar.  Lines may stop and start, for example, if the protostar becomes part of a multiple system, or is expelled from a multiple system, respectively.}
\label{discmass}
\end{figure}

\section{The statistical properties of the discs}
\label{sec:stats}

In this section, we give an overview of the statistical properties of the discs and how they evolve with time.  This is difficult because, as seen in the previous section, there many different types of discs, and they are continually evolving through self-gravity, accretion, ram-pressure stripping, and interactions with other protostars.  

In the following sections, we first discuss the properties of `isolated' protostars, which we define as those without companions closer than 2000 au.  There are 2186 instances of isolated discs.  Note that these are defined as being isolated in that particular snapshot.  They may have been members of multiple systems or suffered close encounters in the past, or they may become members of multiple systems later in the calculation.  This policy is consistent with what an observer would see -- they only know whether a protostar is currently isolated and cannot tell what may have happened in the past or what may happen in the future.  However, if a protostar has had an encounter with another object it is likely to have affected its disc.  Therefore, we also consider the disc properties of the subset of protostars that have never had another protostar closer than 2000 au. After discussing isolated protostars, we discuss the statistical properties of discs in protostellar {\em systems} (i.e. both discs around single protostars, and those found in bound multiple systems).

To put the disc properties in context, in Fig.~\ref{starmass} we provide graphs of the time evolution of the mass of each protostar (i.e.\ sink particle).  The left panel shows mass as a function of simulation time using a linear time axis, while the right panel shows the mass as a function of the age of each protostar (i.e.\ the time since a sink particle was inserted) using a logarithmic time axis.  As discussed in \cite{Bate2012}, over the first $10^4$ yrs of the life of a protostar, the typically protostellar accretion rate in the calculation is $1.5 \times 10^{-5}$~M$_\odot$~yr$^{-1}$, with a dispersion of 0.37 dex.  Some protostars obviously stop accreting (flat lines).  This typically occurs when the protostars are involved in dynamical encounters that expel them from the dense molecular gas, or increase their velocities so that they cannot accrete cloud material at a significant rate \citep[since the Bondi-Hoyle accretion rate is inversely proportional to speed cubed;][]{BonHoy1944}.

\subsection{Discs of isolated protostars}
\label{sec:isolated}

We examine the distributions of disc mass, disc radii, and disc surface density profile for isolated protostars (those without other protostars within 2000 au).  We begin by examining how the disc mass depends on the time in the simulation, and also on the age of the protostar.  The former is more like an observer would see (i.e. a mixture of protostars at different ages), while the latter allows us to investigate how disc properties depend on the age of the protostar.

In the top panel of Fig.~\ref{discmass}, we plot the disc masses of isolated protostars versus time.  Each continuous line gives the evolution of the disc mass for a particular protostar.  Individual tracks may be short or may stop and start because the protostar may not be isolated for very long, or it may change from being isolated to not isolated or visa versa.  The middle panel of Fig.~\ref{discmass} gives the mass evolution of the protostars whose disc masses are plotted in the top panel.  It is clear that almost all of the isolated protostars have masses between 0.02 and 0.4 M$_\odot$.  This is because although many protostars initially form as isolated objects, they do not remain isolated.  Also, as is observed, stars that are more massive are more likely to have companions \citep[see][for an extensive discussion of stellar multiplicity]{Bate2012}.  This means that we have essentially no information on the disc properties of isolated protostars with masses $M_* \gsim 0.4$~M$_\odot$.

Plotting protostellar quantities as functions of simulation time makes it difficult to study how protostellar properties evolve with age because at any particular time there is a mixture of protostars with different ages.  This produces broad dispersions of properties at any particular time (e.g.\ Figs. \ref{starmass} and \ref{discmass}).   This should be born in mind by observers since they have no choice but to look at a star-forming region at a particular time, and ages of individual protostars usually cannot be reliably determined \citep{HarCasKen1997,TouLivBon1999,BarChaGal2009}.

\begin{figure}
\centering \vspace{-0.5cm} \hspace{0cm}
    \includegraphics[width=8.5cm]{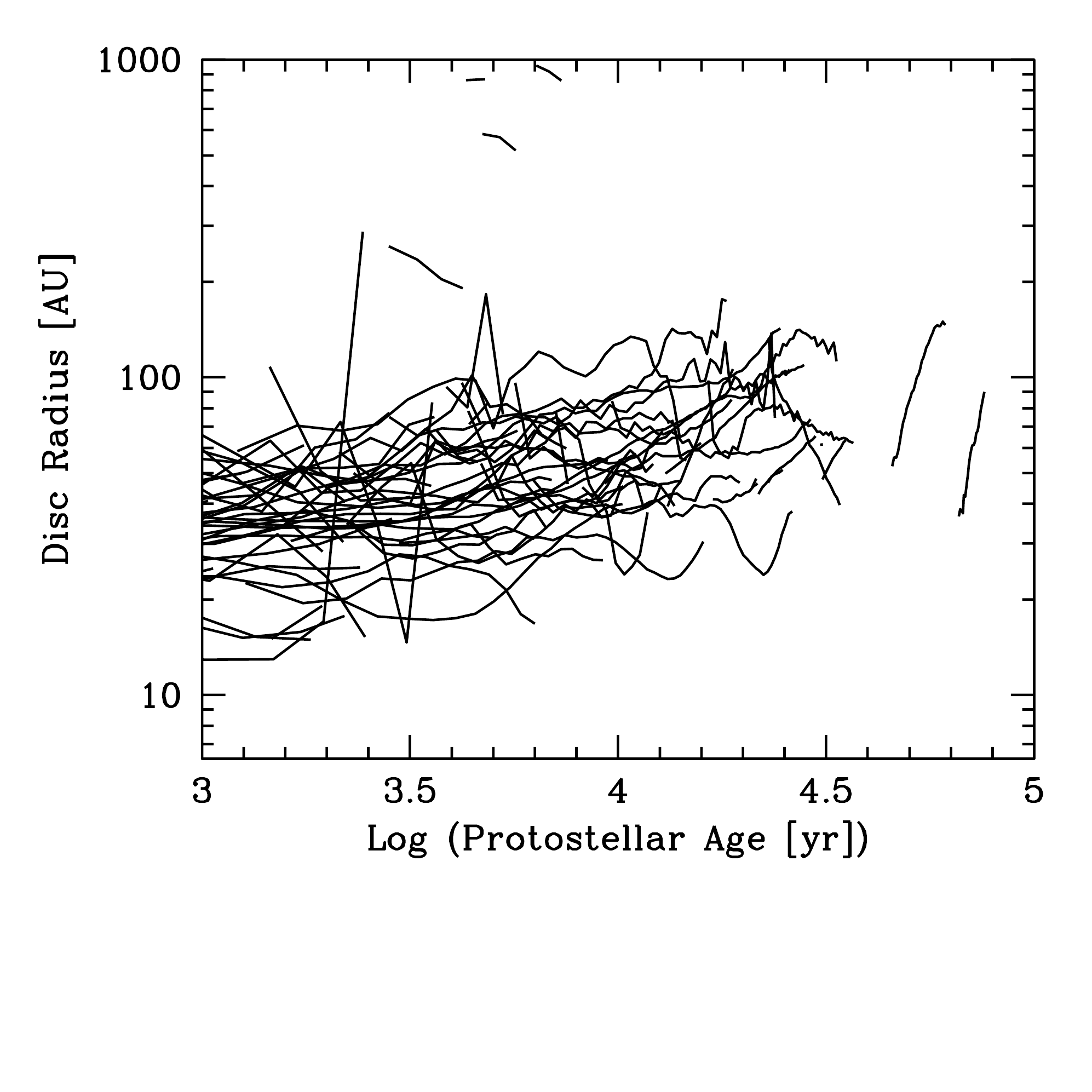} \vspace{-2cm}
\caption{The evolution of the disc radii of isolated protostars (sink particles) as a function of age.  Each line represents the evolution of the disc around a particular protostar.  Lines may stop and start, for example, if the protostar becomes part of a multiple system, or is expelled from a multiple system, respectively.}
\label{discrad}
\end{figure}

\begin{figure}
\centering \vspace{-0.5cm} \hspace{0cm}
    \includegraphics[width=8.5cm]{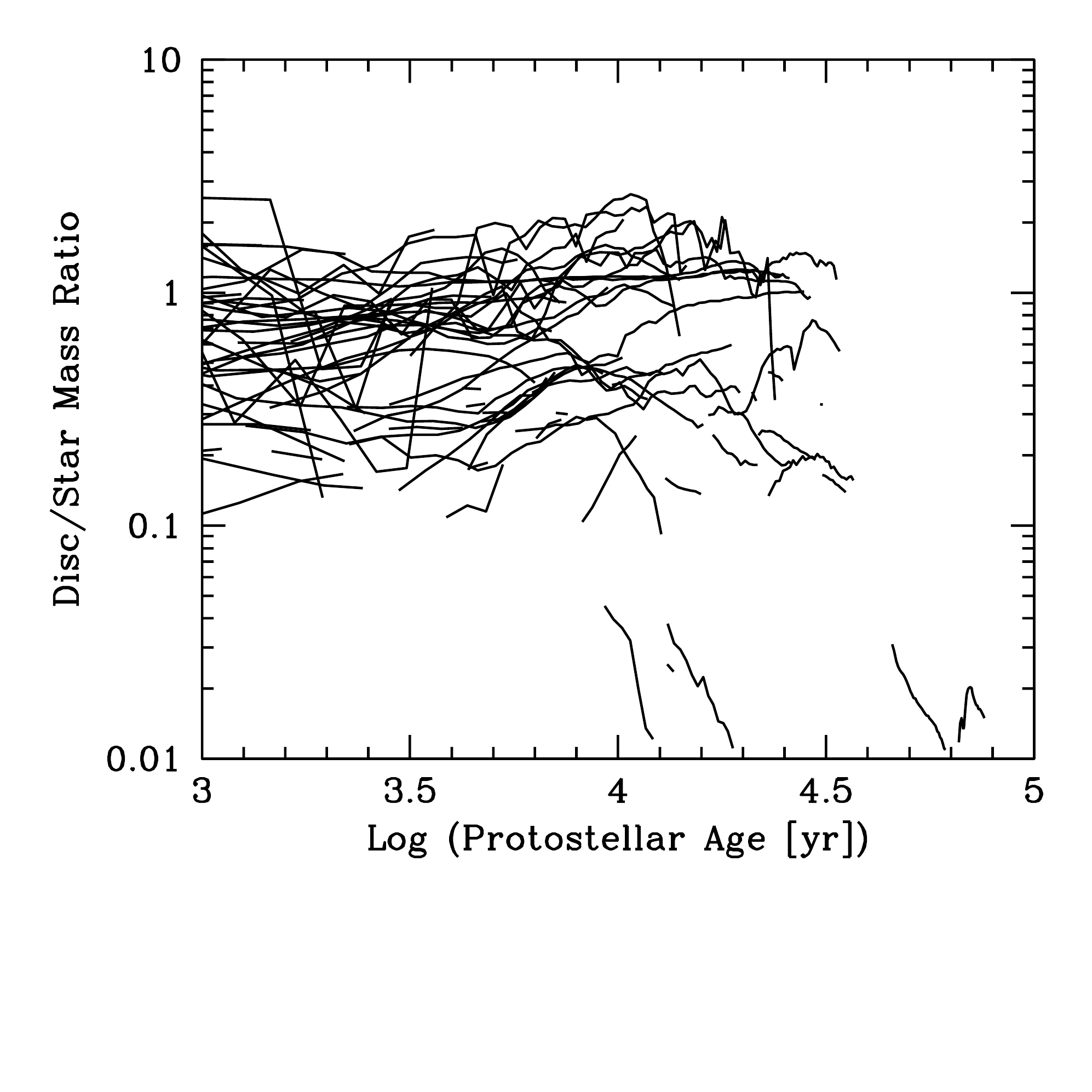} \vspace{-2cm}
\caption{The evolution of the ratio of the disc mass to protostellar (sink particle) mass as a function of age for isolated protostars.  Each line represents the evolution of the disc around a particular protostar.  Lines may stop and start, for example, if the protostar becomes part of a multiple system, or is expelled from a multiple system, respectively.}
\label{discratio}
\end{figure}

\begin{figure*}
\centering \vspace{-0.1cm} \hspace{0.0cm}
    \includegraphics[width=5.8cm]{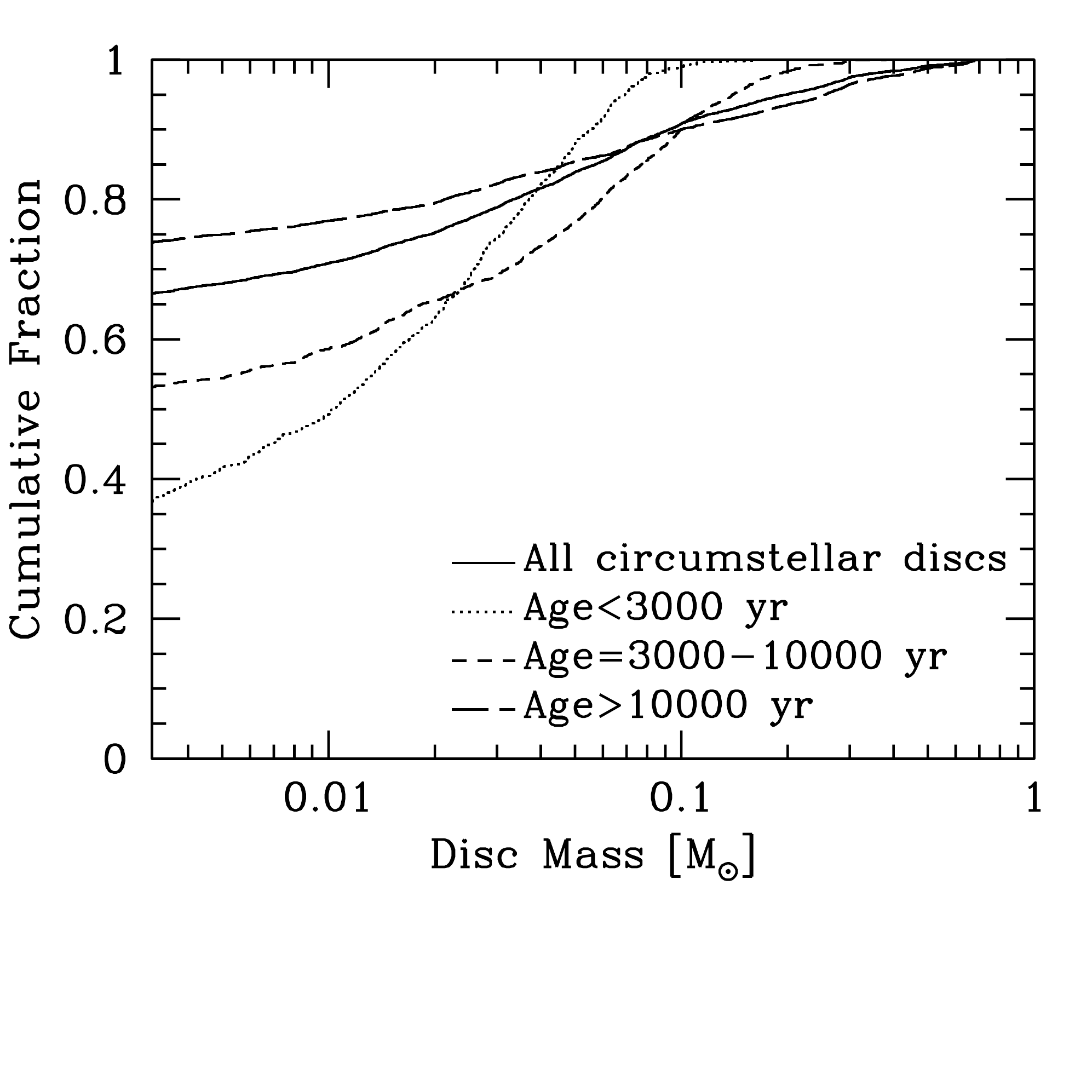} 
    \includegraphics[width=5.8cm]{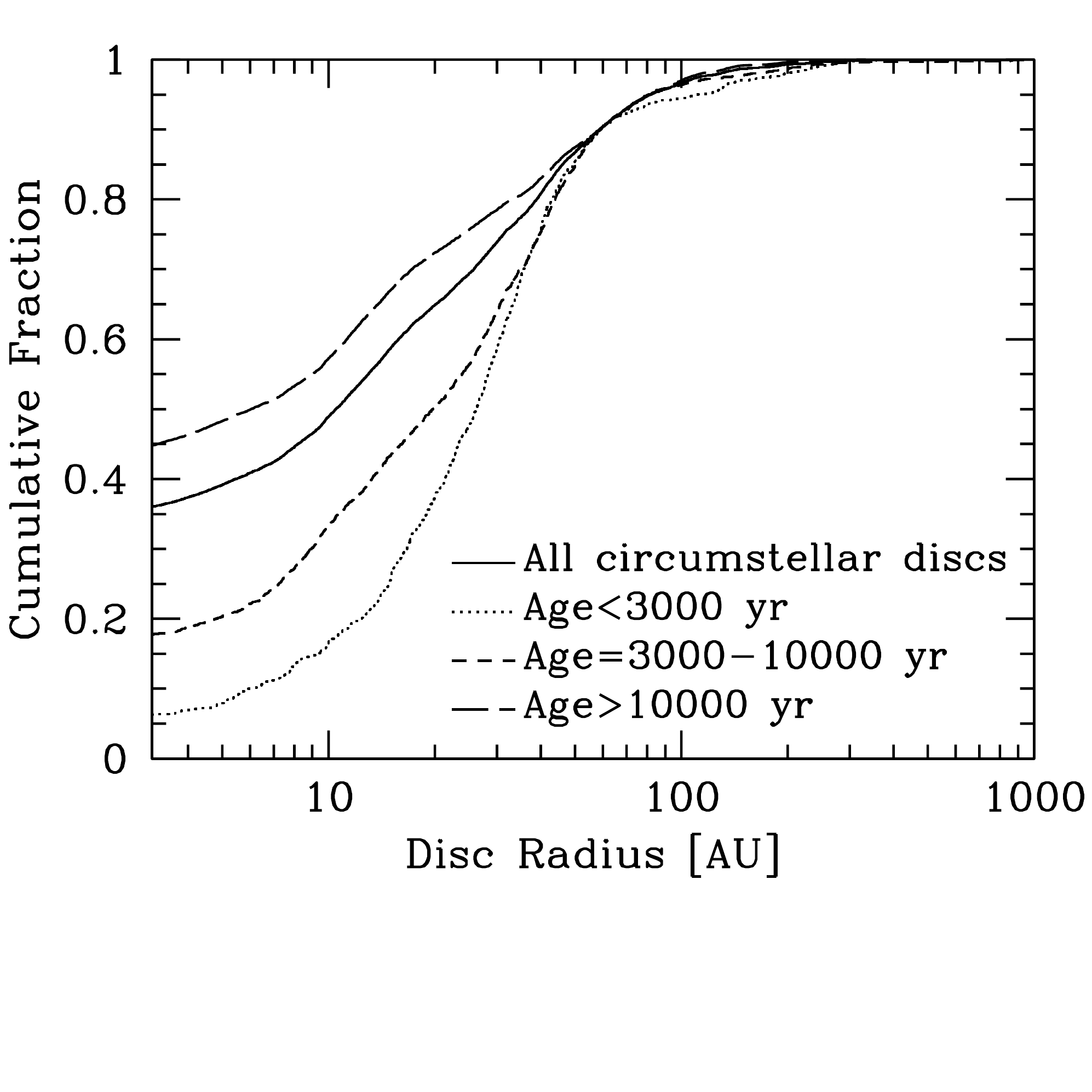}
    \includegraphics[width=5.8cm]{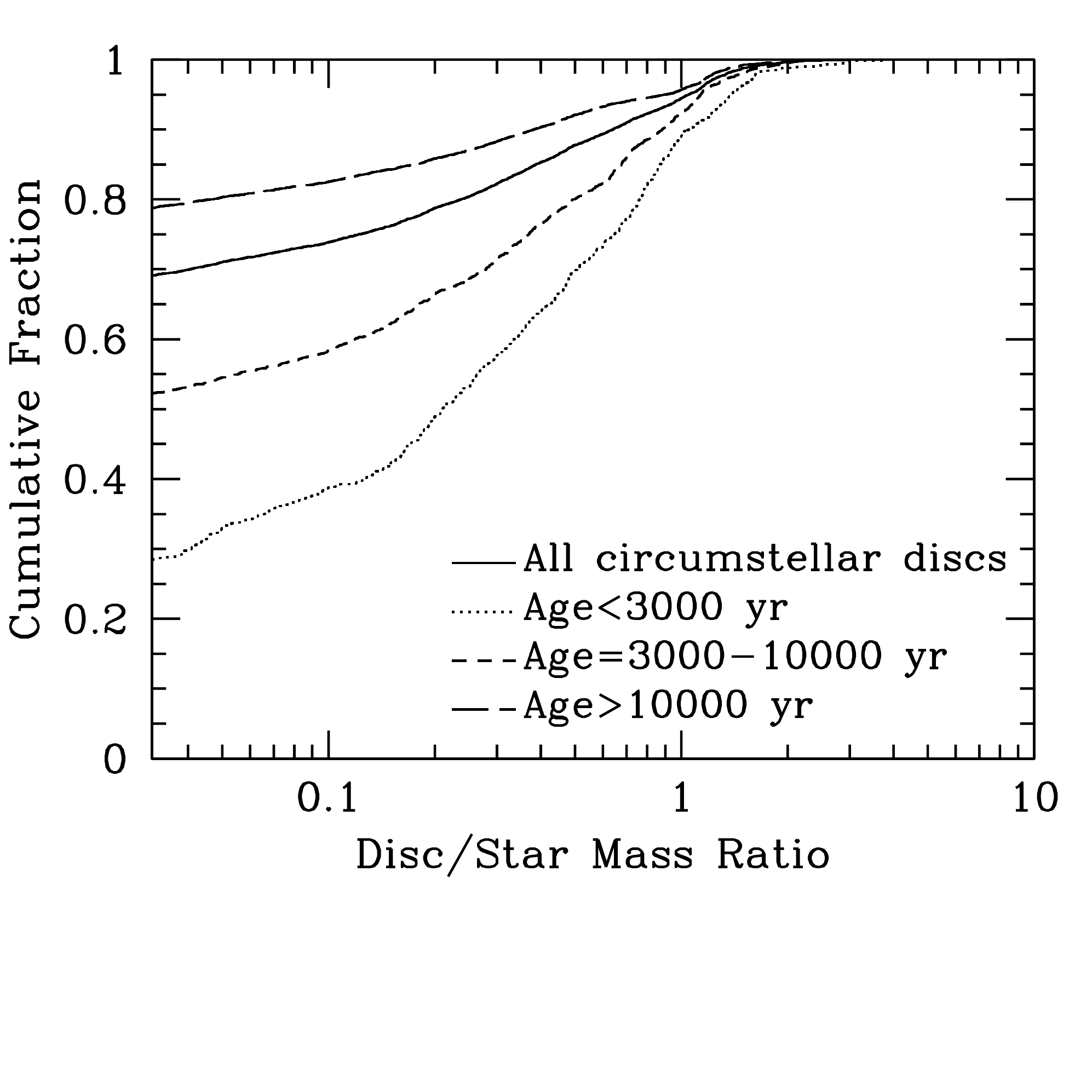} 
    \includegraphics[width=5.8cm]{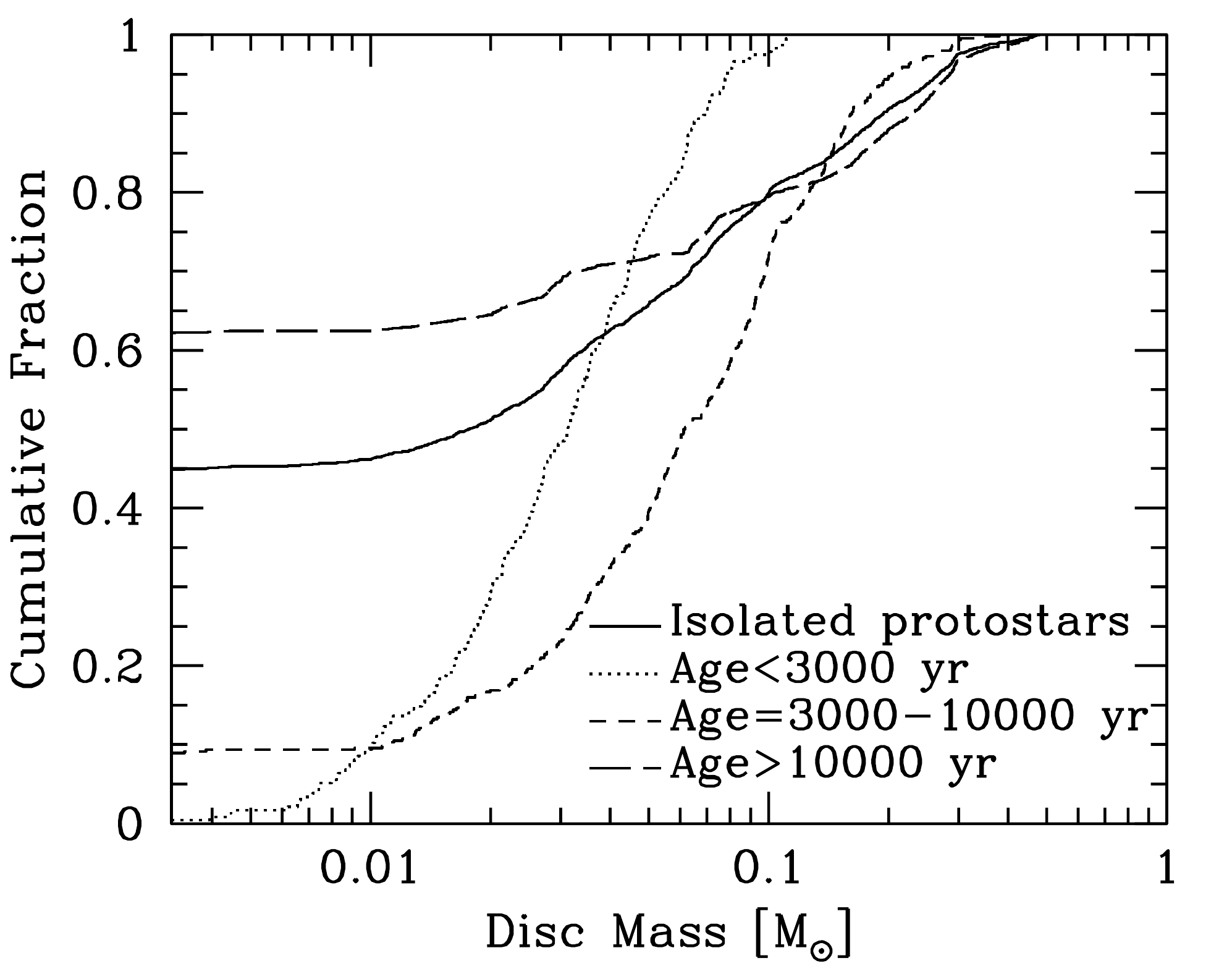} 
    \includegraphics[width=5.8cm]{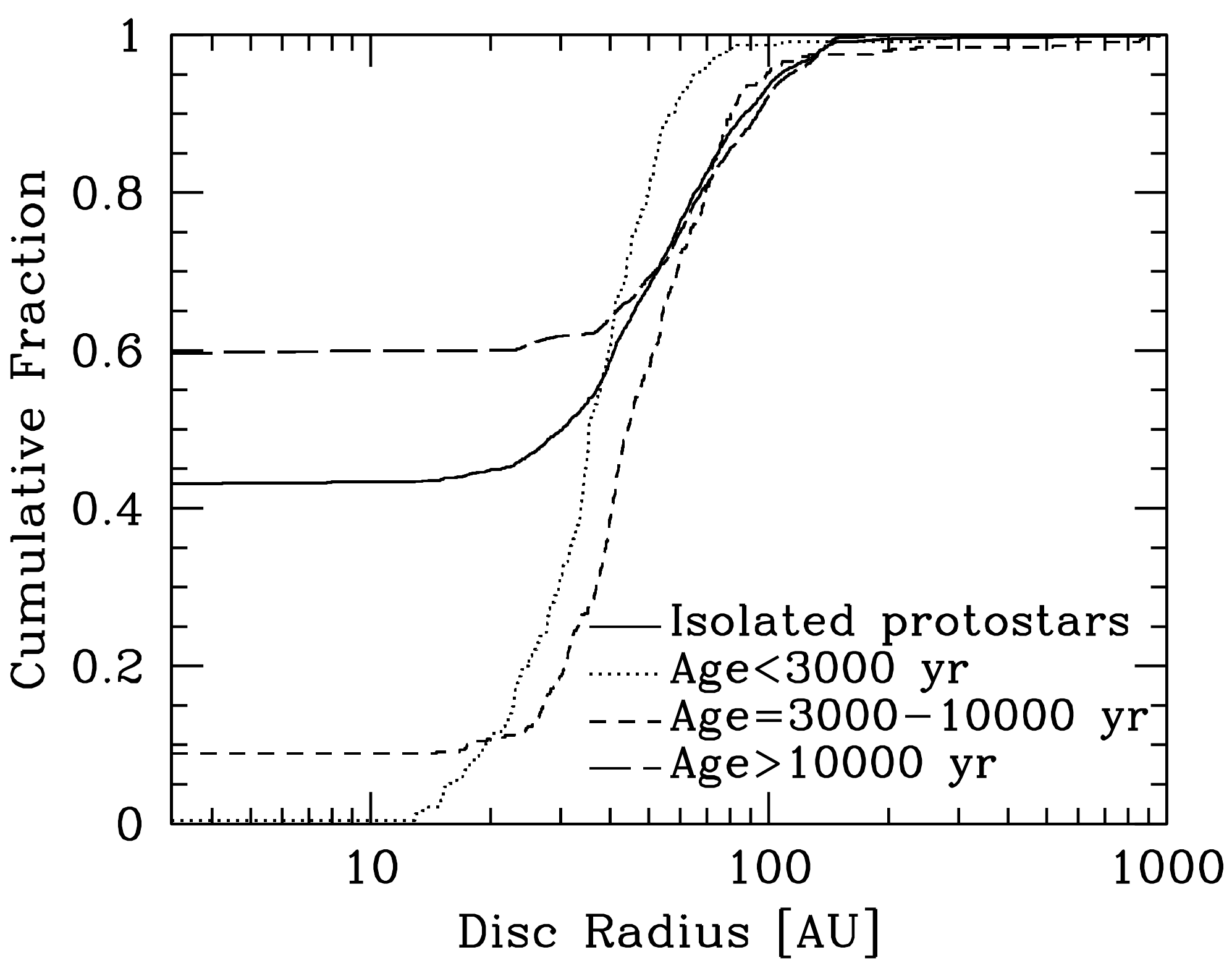}
    \includegraphics[width=5.8cm]{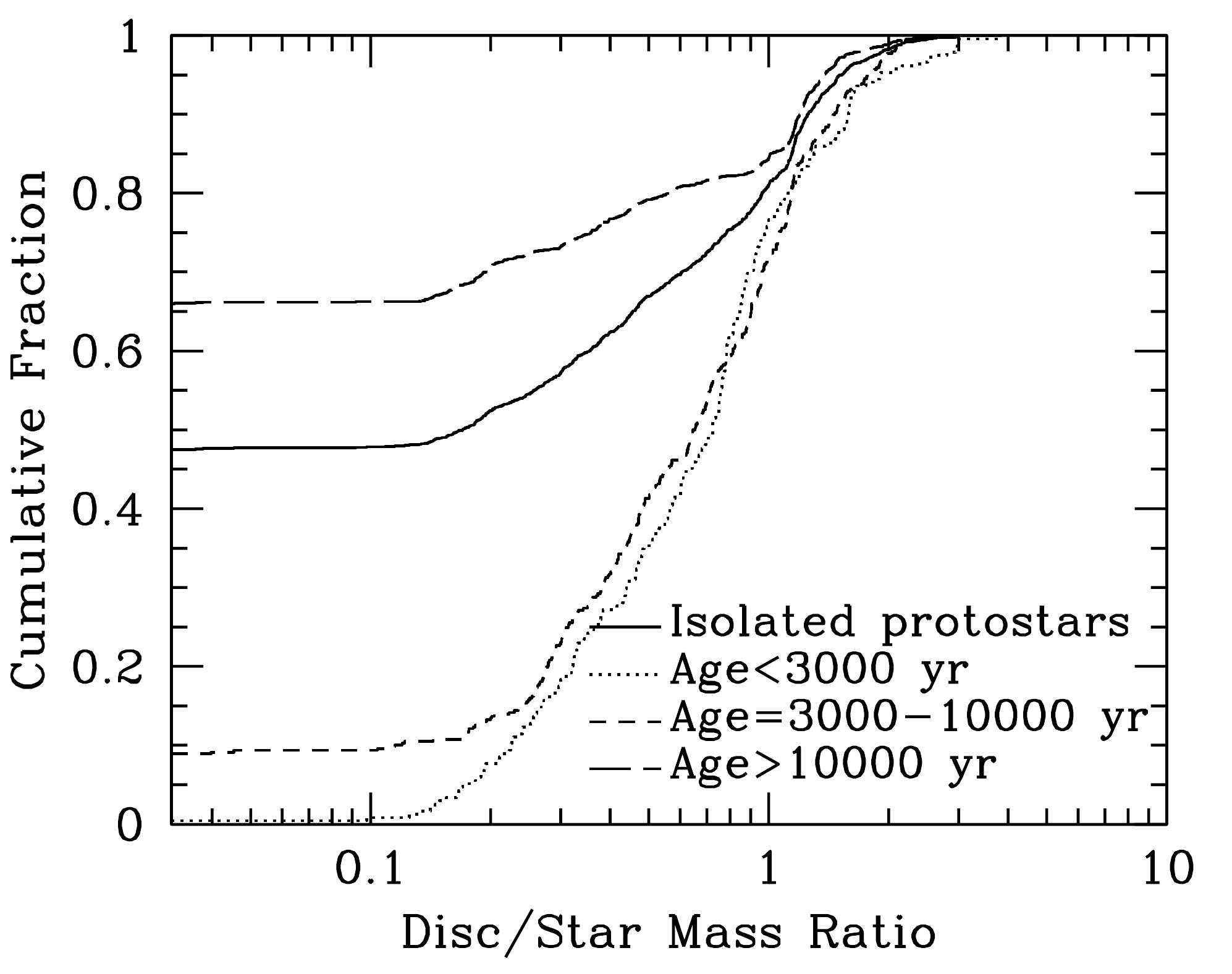} 
    \includegraphics[width=5.8cm]{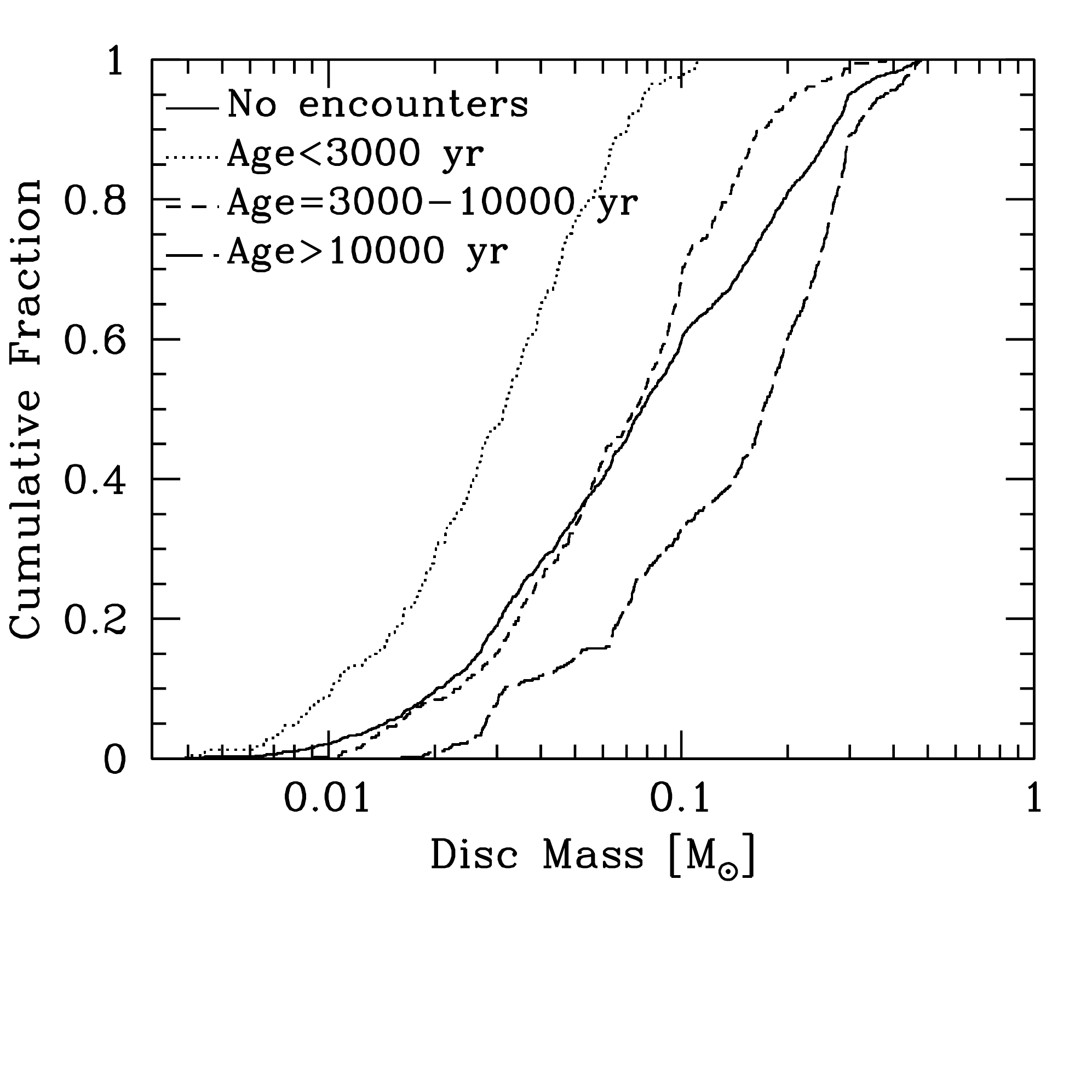} 
    \includegraphics[width=5.8cm]{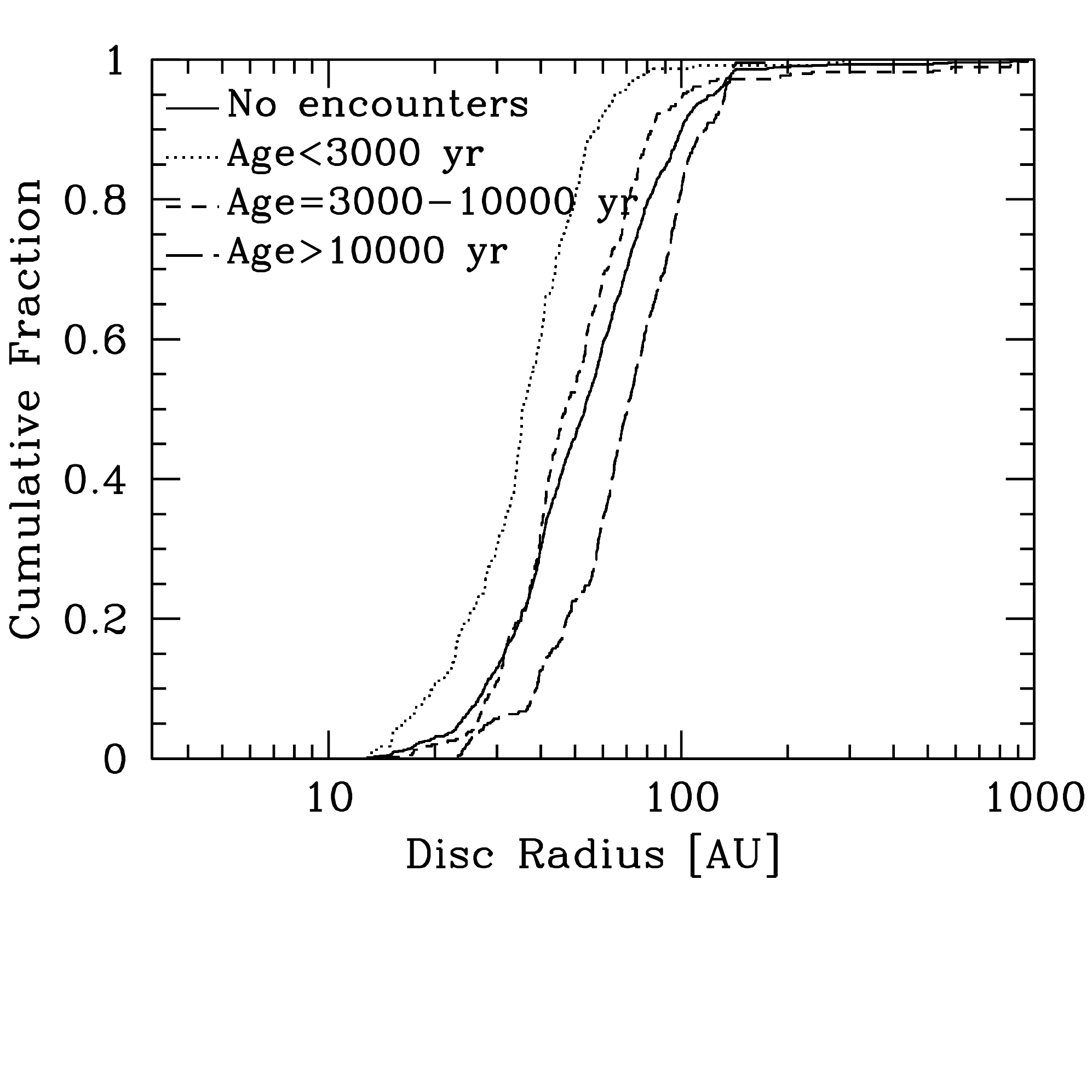}
    \includegraphics[width=5.8cm]{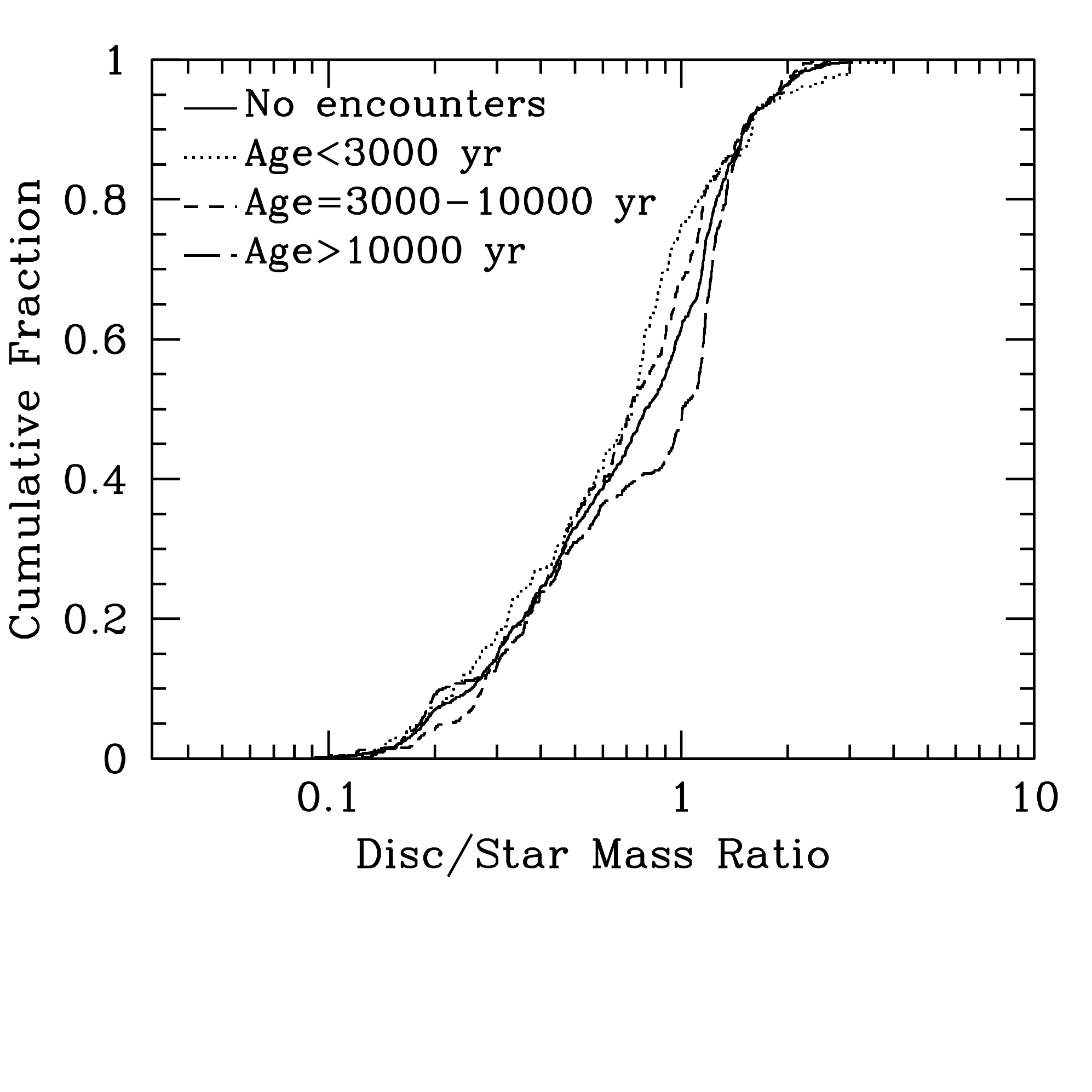} 
    \vspace{-1.0cm}
\caption{The cumulative distributions of disc mass (left), radius (centre), and the disc to stellar mass ratio (right) for circumstellar discs (i.e. discs around each individual protostar; top row), discs of isolated protostars (middle row), and discs of single protostars that have never had another protostar within 2000 au (bottom row).  The solid lines give the distributions for protostars of all ages.  The dashed, long-dashed, and dot-dashed lines give the distributions for protostars in three age ranges: $0-3000$~yrs, $3000-10000$~yrs, and $>10000$~yrs, respectively.  As the single protostars age, the median disc mass and radius increases, but the disc to star mass ratio distribution remains approximately constant.  As the isolated protostars age, the median disc mass and radius of those with resolved discs increases, but a growing fraction of isolated stars also have no resolved disc due to dynamical encounters and the ejection of protostars from multiple systems.  Considering all circumstellar discs, the fractions of protostars without resolved discs is even higher than for the isolated protostars, due to the interactions with companions. }
\label{disc_cumulative}
\end{figure*}

\begin{figure*}
\centering \vspace{-0.4cm} \hspace{0.0cm}
    \includegraphics[width=5.8cm]{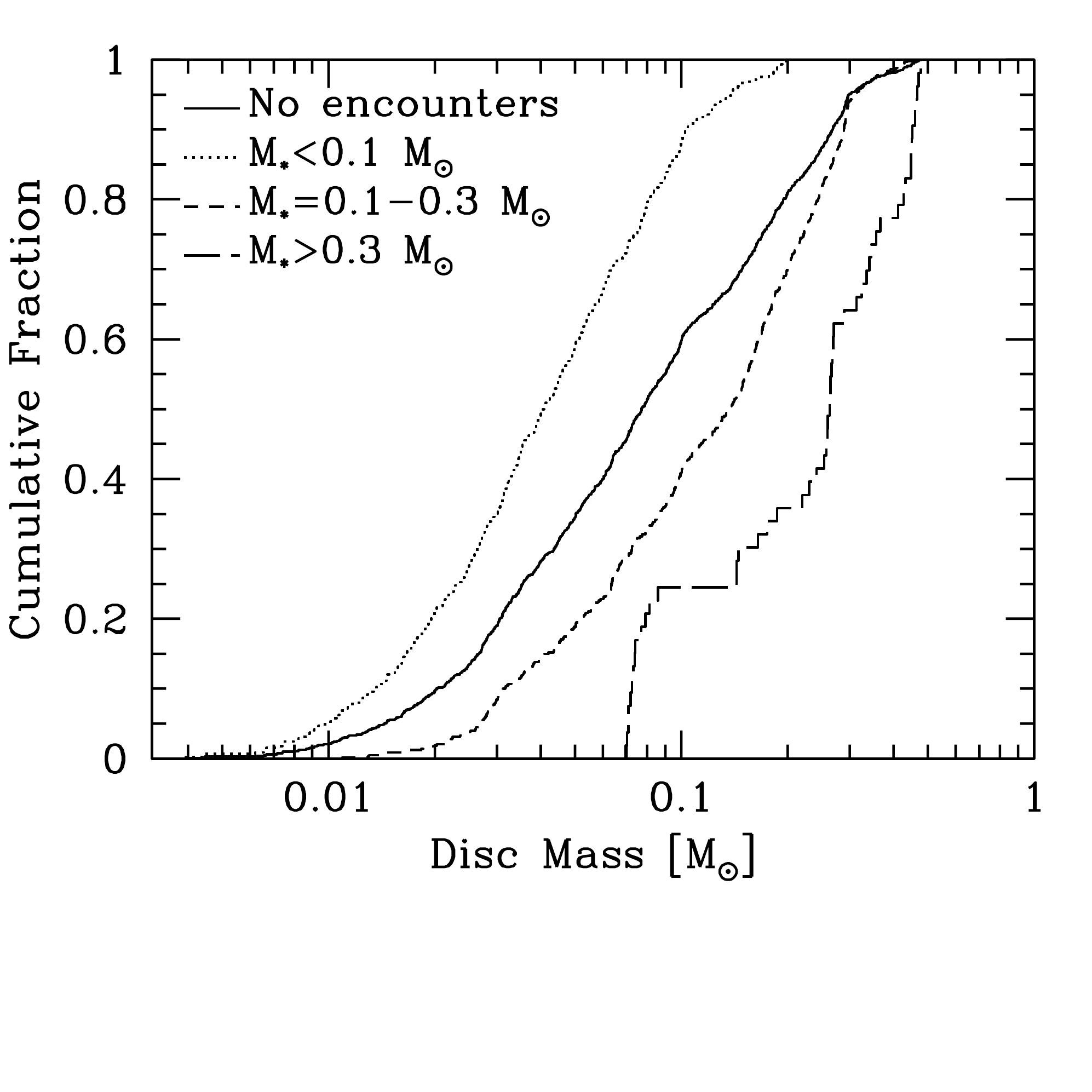} 
    \includegraphics[width=5.8cm]{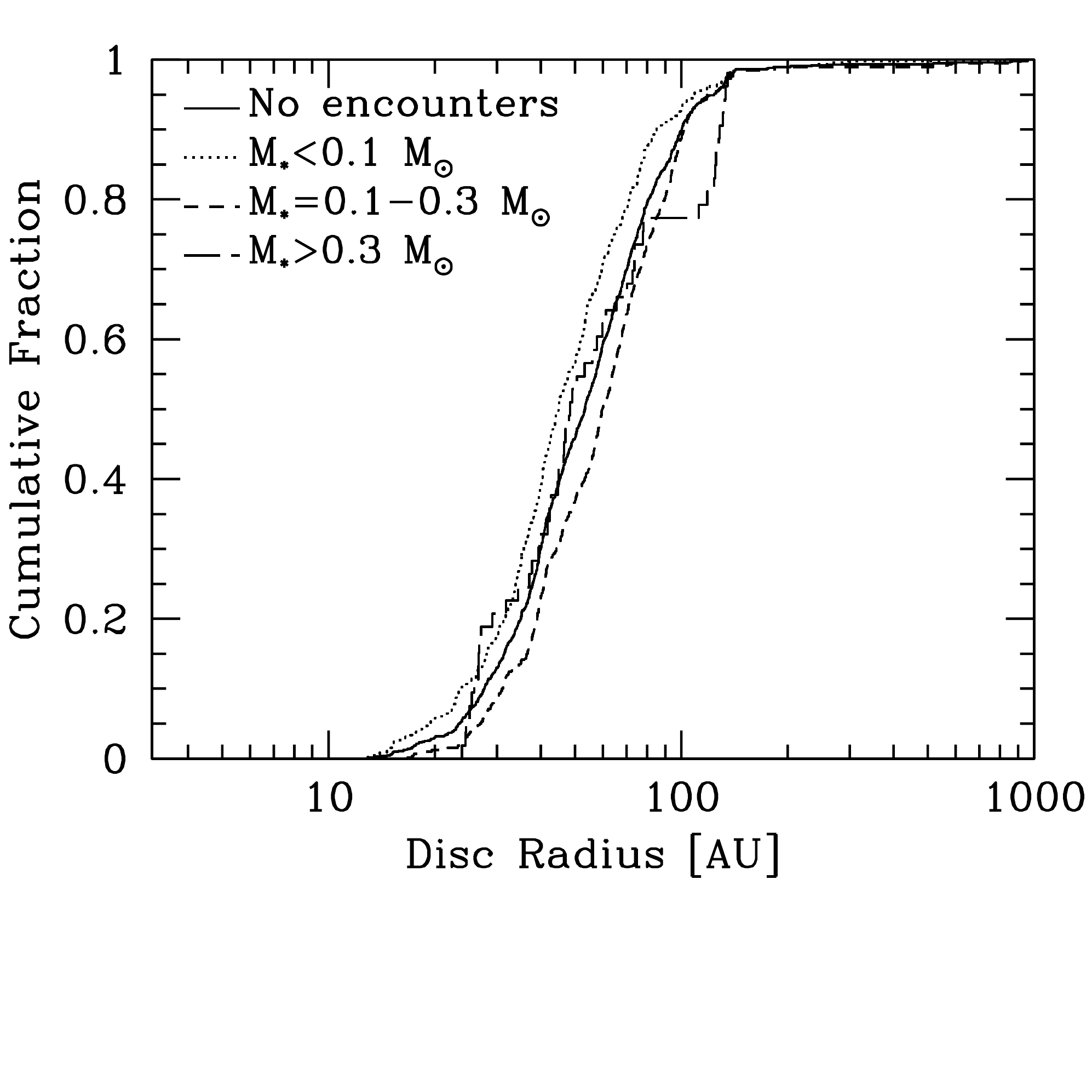}
    \includegraphics[width=5.8cm]{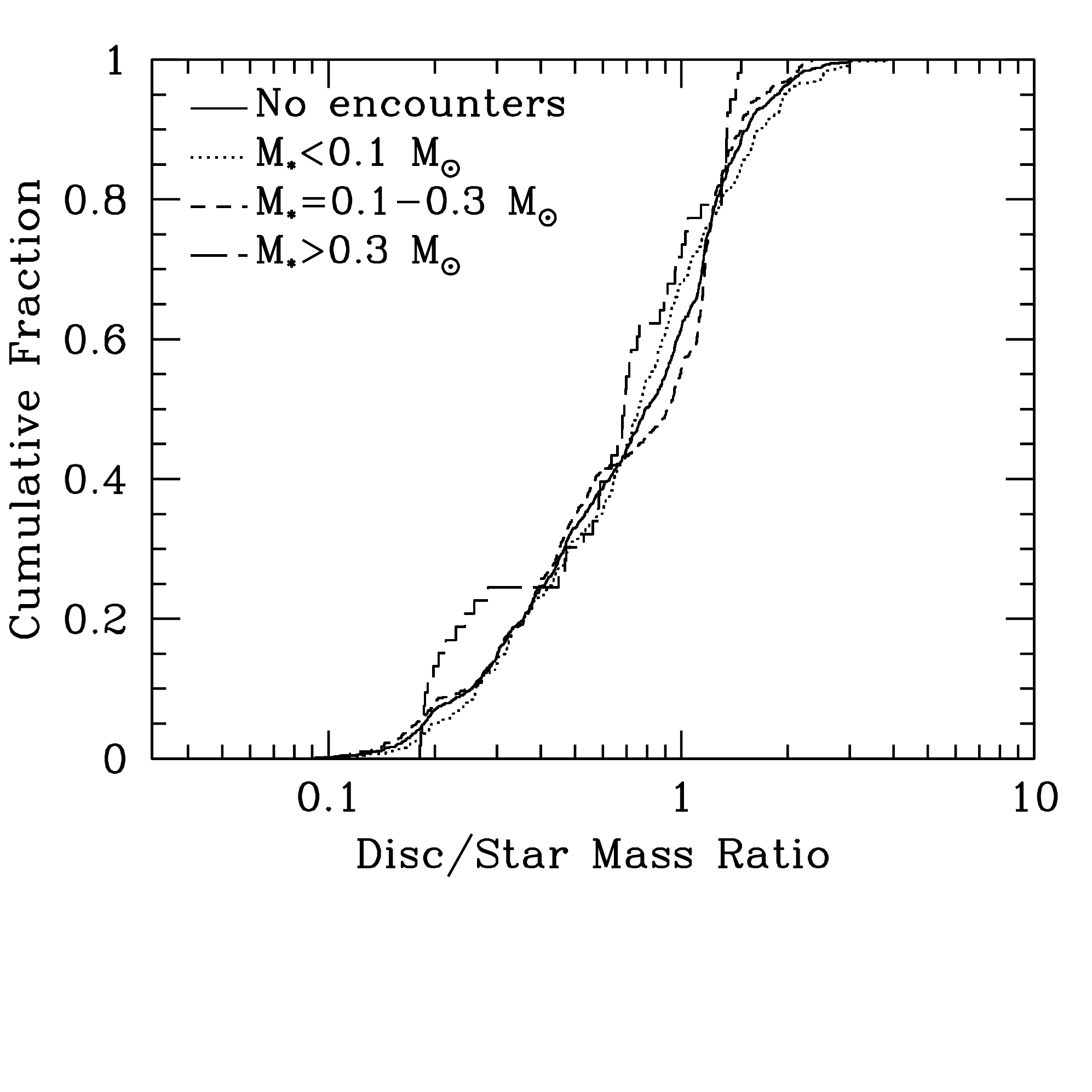} 
    \vspace{-1.0cm}
\caption{The cumulative distributions of disc mass (left), radius (centre), and the disc to stellar mass ratio (right) for circumstellar discs of single protostars that have never had another object within 2000 au.  The solid lines give the distributions for all protostellar masses.  The dashed, long-dashed, and dot-dashed lines give the distributions for single protostars in three mass ranges: $<0.1$~M$_\odot$, $0.1-0.3$~M$_\odot$, and $>0.3$~M$_\odot$, respectively.  Single protostars with greater masses have substantially more massive discs, such that the disc to star mass ratio distribution is essentially independent of protostellar mass.  More massive protostars also tend to have slightly larger discs, but the effect is weak. }
\label{disc_noencounter_mass}
\end{figure*}

However, from hydrodynamical calculations we can determine protostellar ages, so from this point on we will discuss how properties depend on age.  In the bottom panel of Fig.~\ref{discmass}, we plot the disc masses of isolated protostars versus their age.  Here it is clear that the disc masses of isolated protostars increase with age from $\approx 0.03$~M$_\odot$ at $10^3$ yrs old to $\approx 0.10$~M$_\odot$ at $10^4$ yrs old, with a dispersion of $\approx 0.3$ dex.  The most massive discs exceed 0.4~M$_\odot$.  Note that there are few isolated protostars older than 30,000 yrs when the calculation is stopped.  Also, after $\approx 10^4$~yrs, it is clear that the disc masses around some protostars rapidly decline (due to accretion, encounters with other protostars, and/or ram-pressure stripping).  Some protostars also suddenly become isolated as they are ejected from multiple systems and these usually have low disc masses.

In Fig.~\ref{discrad} we plot the disc radii of isolated protostars versus age.  As with disc mass, there is a general trend for disc radii to get larger with time.  They range from radii of 10--50~au at $10^3$ yrs old to 20--100~au at $10^4$ yrs old.  Note that even though the calculation does not treat magnetic fields (which could provide angular momentum transport by magnetic braking), the disc radii are not unusually large.  We will discuss this further in Section \ref{sec:discuss}.

In Fig.~\ref{discratio} we plot the ratio of the disc mass to the protostellar mass (i.e. sink particle mass) versus age for all isolated protostars.  Generally, the lines are relatively flat, indicating that the ratio of the disc mass to the stellar mass is relatively constant.  For protostellar ages less than $\approx 10^4$~yrs, the ratios range from $\approx 0.1-2$, indicating that self-gravity will be important for the evolution of many discs (as seen in Sections \ref{sec:single} and \ref{sec:fragment}).  Many disc/star mass ratios still exceed 0.1 beyond ages of $10^4$~yrs, but some low-mass discs (with disc/star mass ratios $<0.1$) also appear.  Again, if protostars are ejected from multiple systems and these usually have low disc masses.

In Fig.~\ref{disc_cumulative}, we give the cumulative distributions of circumstellar disc masses, radii, and star to disc mass ratios.  The top row of panels gives the distributions for all discs containing only one protostar, i.e.\ circumstellar discs (including those that are components of multiple systems).  The second row of panels gives the equivalent distributions but for isolated protostars only.  The bottom row of panels gives the distributions for protostars that have never had an encounter within 2000 au. In each case, we also give the distributions obtained by limiting the samples to protostellar age ranges of $<3000$~yrs, $3000-10000$~yrs, and $>10000$~yrs.

From these cumulative distributions we draw similar conclusions as we did from Figs.~\ref{discmass} to \ref{discratio}.  First, the masses of resolved discs tend to increase with age, for circumstellar discs in general, for those surrounding isolated protostars, and for those that have never had encounters.  From this point on we will often refer to {\em resolved} discs, which we define as those that have masses $M_{\rm d}\approx 0.01$~M$_\odot$ (i.e. they are modelled by more than 700 SPH particles).  It is clear that, for protostars that have not had encounters, the typical (median) mass of their discs increases with age.  But when a significant fraction of the protostars no longer have resolved discs (i.e.\ considering isolated protostars or all circumstellar discs) we cannot be sure whether median disc mass of the population increases or not.  For example, in the top left panel of Fig.~\ref{disc_cumulative} for ages $>10000$~yrs, the cumulative line passes through 0.78 at $M_{\rm d}\approx 0.01$~M$_\odot$ and rises to unity, so the median value for protostars that have resolved discs is when the cumulative fraction is equal to $(0.78+1.0)/2 = 0.89$ and the associated disc mass is $M_{\rm d}\approx 0.08$~M$_\odot$.  However, at this age, the vast majority of systems (78\%) do not have resolved discs, so we cannot determine the median disc mass for all protostars in this age range. Second, the disc radii tend to increase with age for discs around isolated protostars (with the clearest trend being seen for protostars that have not had encounters), but this is {\em not} apparent for all circumstellar discs.  Circumstellar discs in multiple systems have their outer radii limited by gravitational interactions with companions \citep{ArtLub1994}.  Third, the distribution of resolved disc to protostar mass ratios tends not to evolve significantly with age.  This is true regardless of whether we examine all circumstellar discs, those around isolated protostars, or those that have never had encounters.  It is very clear for the protostars that have never had encounters. But even for the more diverse populations, the disc to protostar mass ratios almost all lie in the range $M_{\rm d}/M_*= 0.1 - 2$.

Fig.~\ref{disc_cumulative} also gives us information that the earlier figures cannot show -- information on the fractions of protostars without resolved discs.  Isolated protostars with the youngest ages ($<3000$~yrs) essentially all have resolved discs, with radii typically ranging from $r_{\rm c} \approx 10-70$~au and masses ranging from $M_{\rm d} \approx 0.01-0.1$~M$_\odot$.  But for both older isolated protostars and protostars in multiple systems, a significant number do not have resolved circumstellar discs.  Comparing the distributions for the isolated protostars and those that have never had encounters, it is clear that encounters with other protostars are primarily responsible for producing protostars without resolved discs (as opposed to ram-pressure stripping, or numerical viscous evolution).

Similarly, comparing the top panels of Fig.~\ref{disc_cumulative} with the middle row of panels, it is also clear that protostars in multiple systems are much less likely to have resolved circumstellar discs than isolated protostars.  A trend of lower disc fractions for multiple systems is also apparent observationally (\citealt*{JenMatFul1994,JenMatFul1996,OstBec1995,AndWil2005};\citealt{Harris_etal2012}).  \cite{Harris_etal2012} find that the incident rate of detectable disc emission for stars in multiple systems is half that of single stars in Taurus.  These trends are, no doubt, largely due to dynamical interactions between the protostars truncating the discs \citep{ArtLub1994} and the differential accretion rates of protostars in multiple systems \citep{BatBon1997, Bate2000}.  However, in the hydrodynamical calculation, numerical viscous evolution also plays a role (see Appendix \ref{appendixB}).  Viscously evolving circumstellar discs in multiple systems are likely to be replenished less quickly than those in isolated systems because of the presence of the companion, and the disc around the secondary is expected to evolve faster \citep*{ArmClaTou1999}.  Since the numerical viscosity increases with decreasing disc mass in SPH calculations, low-mass discs will evolve much quicker than is realistic and will drain away.

For the isolated protostars, the fraction without resolved discs increases to $\approx 10$\% for ages $3000-10000$~yrs and $\approx 60$\% for ages $>10000$~yrs.  Note that this does not necessarily mean that most isolated protostars lose their resolved discs during the calculation because later in the calculation many protostars become isolated when they are ejected from multiple systems.  Many of these either wouldn't have had resolved circumstellar discs before they were lost, or else their discs may have been lost during the break up of the multiple system.  However, regardless of the origin, it does mean that many old protostars (ages $>10000$~yrs) do not have resolved circumstellar discs.

Finally, for this section, in Fig.~\ref{disc_noencounter_mass}, we also investigate the dependence of disc properties on protostellar mass for protostars that have never had encounters closer than 2000 au. From the distributions of disc to protostellar mass ratios, it is clear that although there is a distribution of these mass ratios, the distribution does not depend on the protostellar mass and the typical disc mass scales linearly with the mass of the protostar.  In each protostellar mass range, the disc masses range over $\approx 1.5$ dex. By contrast, the disc characteristic radii have a smaller range ($\approx 20-150$ au) and there is less dependence on protostellar mass (the median disc radius for $M_*<0.1$~M$_\odot$ is $\approx 40$~au, while for $M_*>0.3$~M$_\odot$ is $\approx 60$~au.

\subsection{Radial surface density profiles of the discs of isolated protostars}
\label{sec:radial}

If the radial surface density distribution of a disc can be described as $\Sigma(r) \propto r^{-\gamma}$, then the disc mass contained within radius $r$ scales as $M_{\rm d}(r) \propto r^{2-\gamma}$ ($\gamma<2$).  Therefore, performing a least squares linear regression on $\log(M_{\rm d})$ vs $\log(r)$ can be used to obtain the best fitting value of $\gamma$ for a disc.  In the analysis that follows, we perform linear regressions on the values of the disc radii that contain various percentages of the total disc mass.  The maximum radius used for the fits is that containing 80\% of the disc mass.  The last 20\% of the disc mass often stretches to large radii and is not indicative of the distribution of the bulk of the mass.  We vary the minimum radius used in the fits, using values of 2, 10, 30, 40, and 50\%.  We limit our analysis to discs with $M_{\rm d} > 0.05~{\rm M}_\odot$ (3500 SPH particles).  Similar results are obtained using a lower disc mass limit of $M_{\rm d} > 0.03~{\rm M}_\odot$ (2100 SPH particles), but dropping the limit of $M_{\rm d} > 0.01~{\rm M}_\odot$ (700 SPH particles) results in a large number of almost constant surface density discs, as may be expected for discs that are poorly resolved.  The resulting sample includes 372 instances of isolated discs around 39 protostars.

\begin{figure}
\centering \vspace{0cm} \hspace{0cm}
    \includegraphics[width=8.5cm]{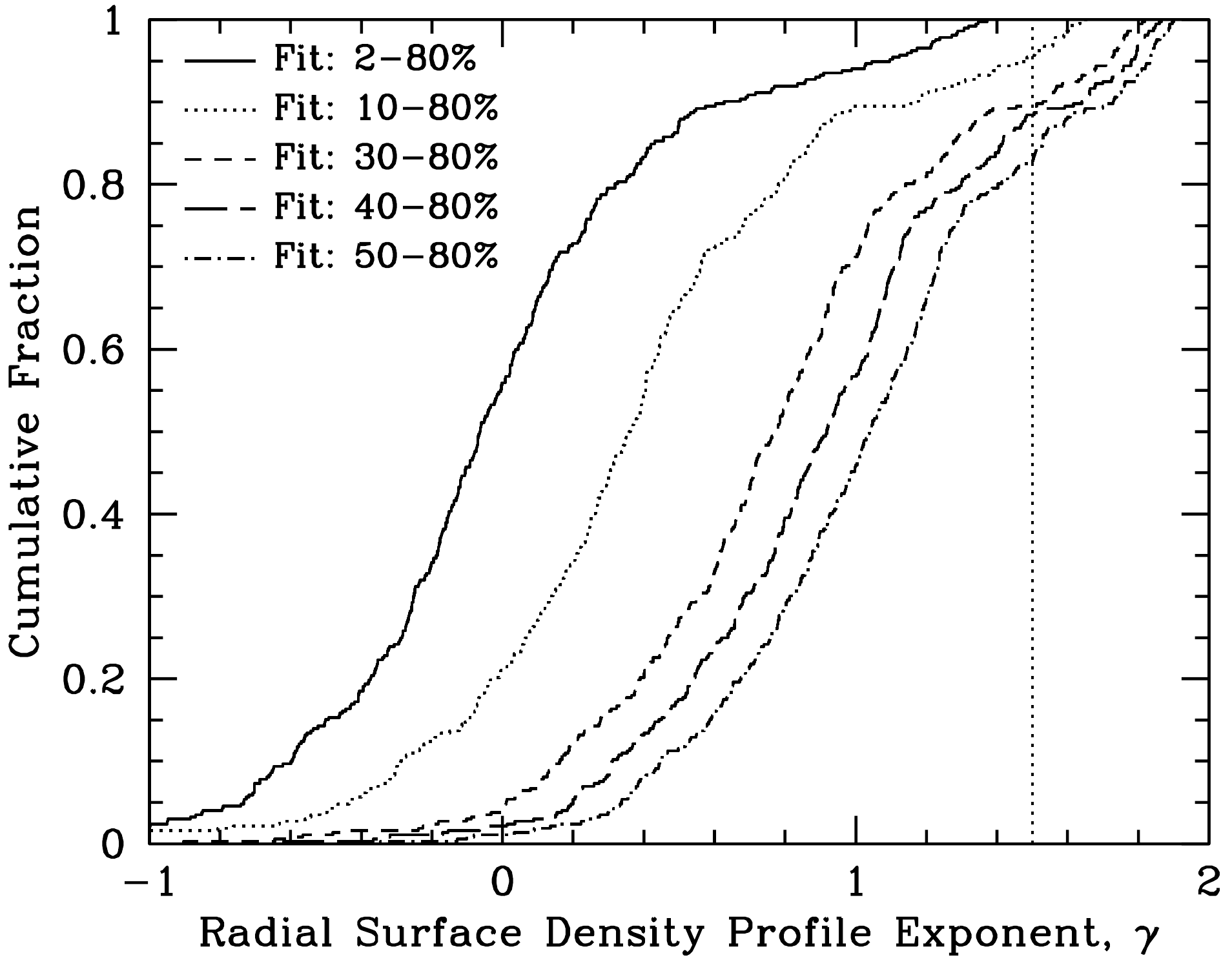} \vspace{-0.25cm}
\caption{The cumulative distributions of radial surface density profile exponents, $\gamma$, (where $\Sigma(r) \propto r^{-\gamma}$) for isolated protostars with disc masses $>0.05~{\rm M}_\odot$.  The values of $\gamma$ obtained depend on the ranges of the radii being fit.  We give several distributions, computed using the radii containing $2-80$, $10-80$, $30-80$, $40-80$, and $50-80$\% of the total disc mass.  The fact that varying the inner radius significantly affects the fits indicates that the discs are not well fit by power-law radial surface density profiles.  In particular, the inner parts of the discs have flatter radial profiles than the outer parts (or have holes). This is largely numerical in origin.  The vertical dotted line gives the value of $\gamma$ for the Minimum Mass Solar Nebula model.  Regardless of how our fits are computed, almost all of our discs have flatter surface density profiles than the MMSN model.}
\label{fig:surfdens}
\end{figure}

In Figure \ref{fig:surfdens}, we plot the cumulative distribution of exponents, $\gamma$, for discs around isolated protostars (i.e.\ over all ages).  The distributions do not vary much when particular age ranges are used.  It is clear from the figure that the distribution of the exponent depends on the minimum value of the radius that is used in the fitting (which depends on the minimum percentage of the total disc mass).  Fundamentally, this indicates that the surface density profiles are not well fit by power-laws.  As the minimum radius is decreased, the typical exponent decreases.  This means that the inner parts of the discs typically have flatter density profiles than the outer parts.  When including the inner-most radii (i.e.\ fits ranging from 2--80\% or 10--80\% of the disc mass) the value of $\gamma$ is often negative.  This indicates a surface density that {\em increases} with increasing radius (i.e.\ a hole in the inner disc).  Although young inner holes have been found recently in discs that are thought to be young (e.g.\ HL Tau:  \citealt{ALMA_etal2015}; WL 17: \citealt{SheEis2017}), in the simulations analysed in this paper the inner holes are certainly numerical, due to the sink particle accretion radius.  Firstly, no mass within the accretion radius is resolved, inevitably producing a $\approx 1$~au hole in the disc mass distribution.  But even outside the accretion radius, the inner edge of the disc is eroded by the accretion radius because we make no attempt to include sink particle boundary conditions \citep{BatBonPri1995}.  Therefore, the most reliable fits exclude both the inner and outer regions of the discs (e.g.\ those using from $30-80$ to $50-80$ of the enclosed disc mass).  These typically have $\gamma \approx 1$.

For a steady-state constant-alpha disc $\Sigma(r) \propto r^{q-3/2}$ \citep*[e.g.][]{FraKinRai2002}, where the mid-plane temperature scales as $T(r) \propto r^{-q}$.  For a marginally Toomre-stable disc (i.e.\ $Q(r)=1$), it is expected that $\Sigma(r) \propto r^{-q/2-3/2}$.  Typically $q \approx 3/4$, so the expected values of $\gamma$ range from $\approx 0.75 - 1.1$. When fitting the exponent $\gamma$ for radii containing $30-80$\% of the total disc mass (or $40-80$ or $50-80$\%) most of the values of $\gamma$ lie within this range.  It is interesting to note that the value of $\gamma$ for the Minimum Mass Solar Nebula (MMSN) model is $\gamma=3/2$ \citep{Weidenschilling1977, Hayashi1981}.  Regardless of the mass range used for the fitting, almost all of our discs have flatter radial density profiles than the MMSN,  although it is important to note that we are considering gas rather than solids.

\begin{figure*}
\centering \vspace{0.0cm} \hspace{0.0cm}
    \includegraphics[width=5.8cm]{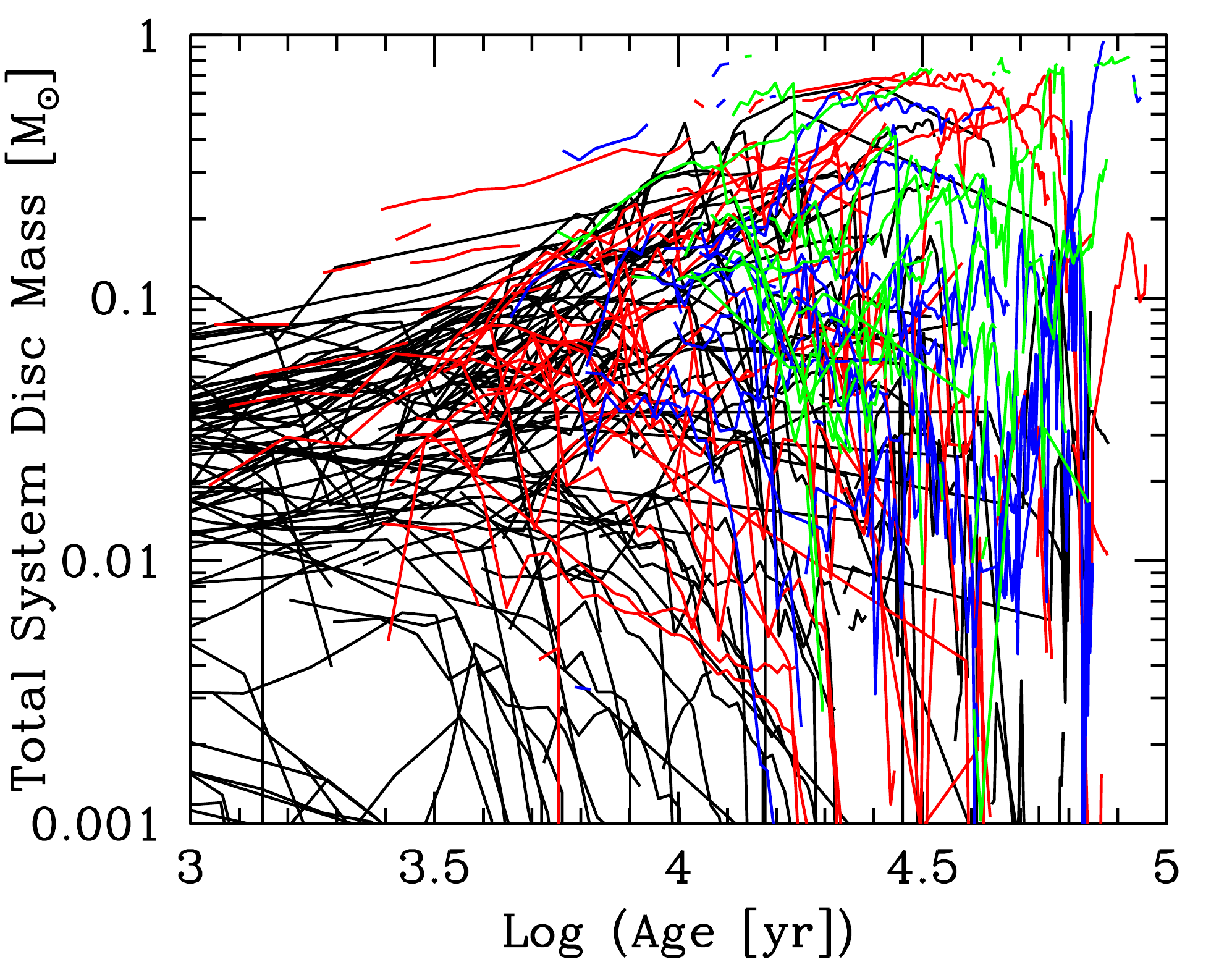}
    \includegraphics[width=5.8cm]{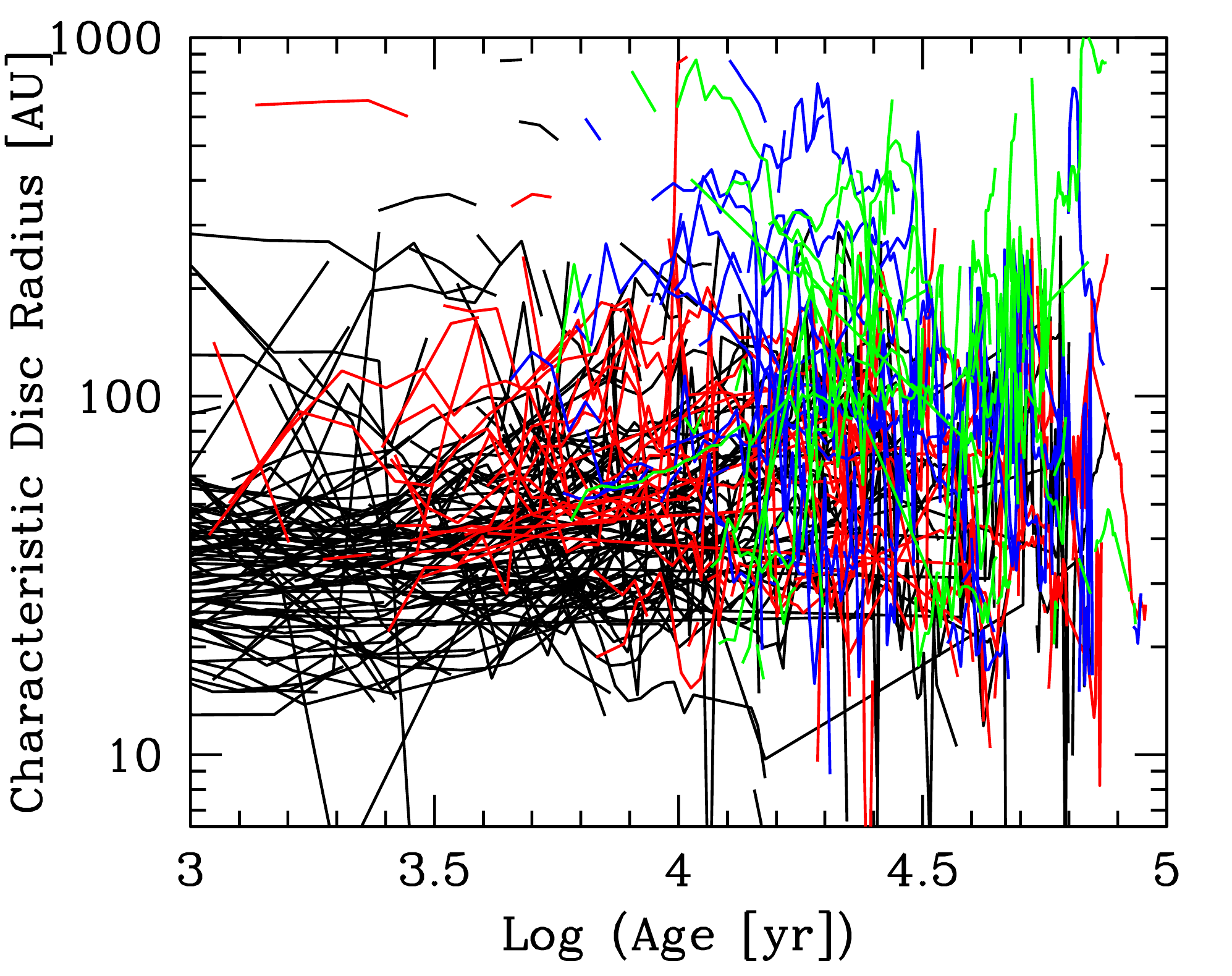} 
    \includegraphics[width=5.8cm]{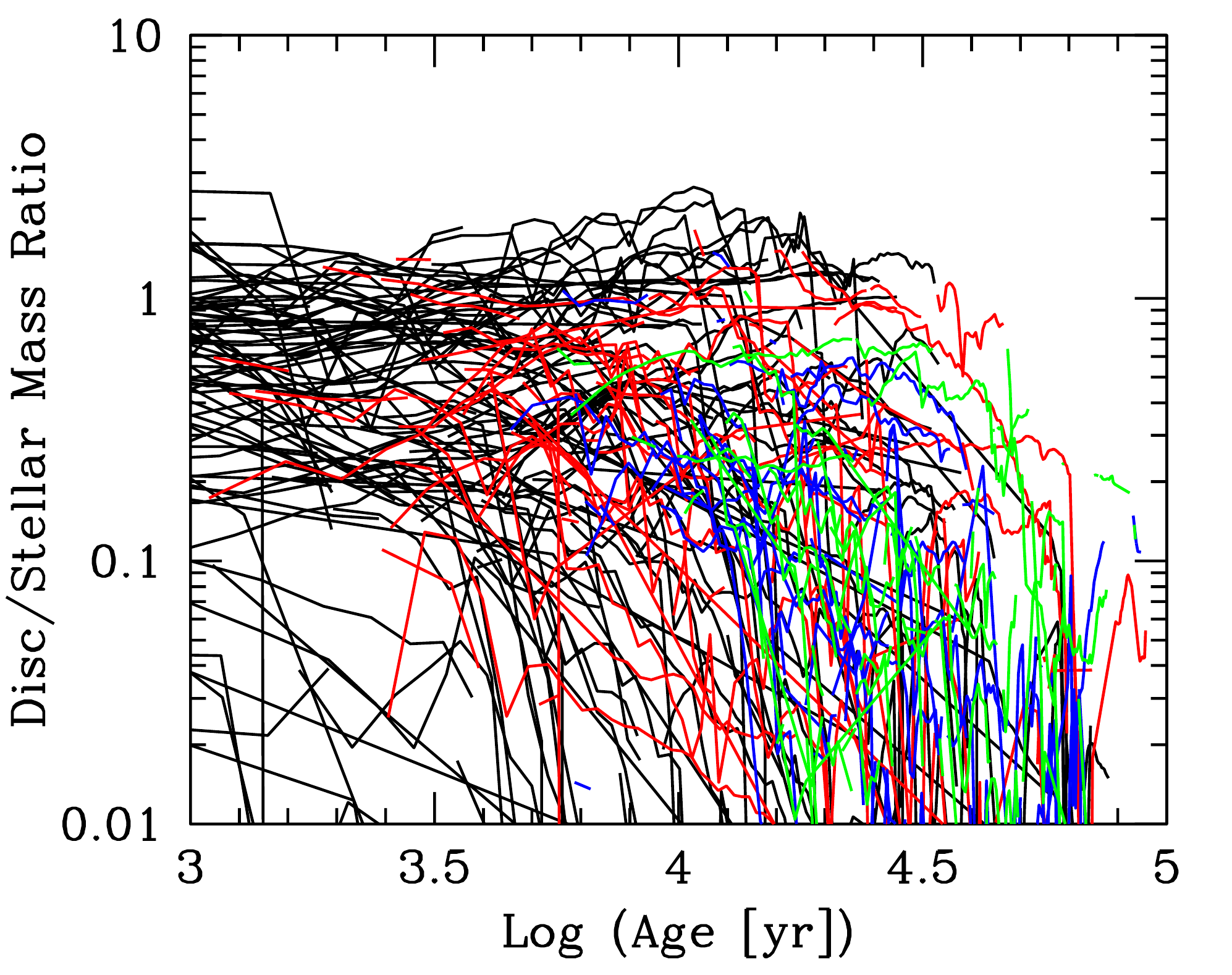}        
    \includegraphics[width=5.8cm]{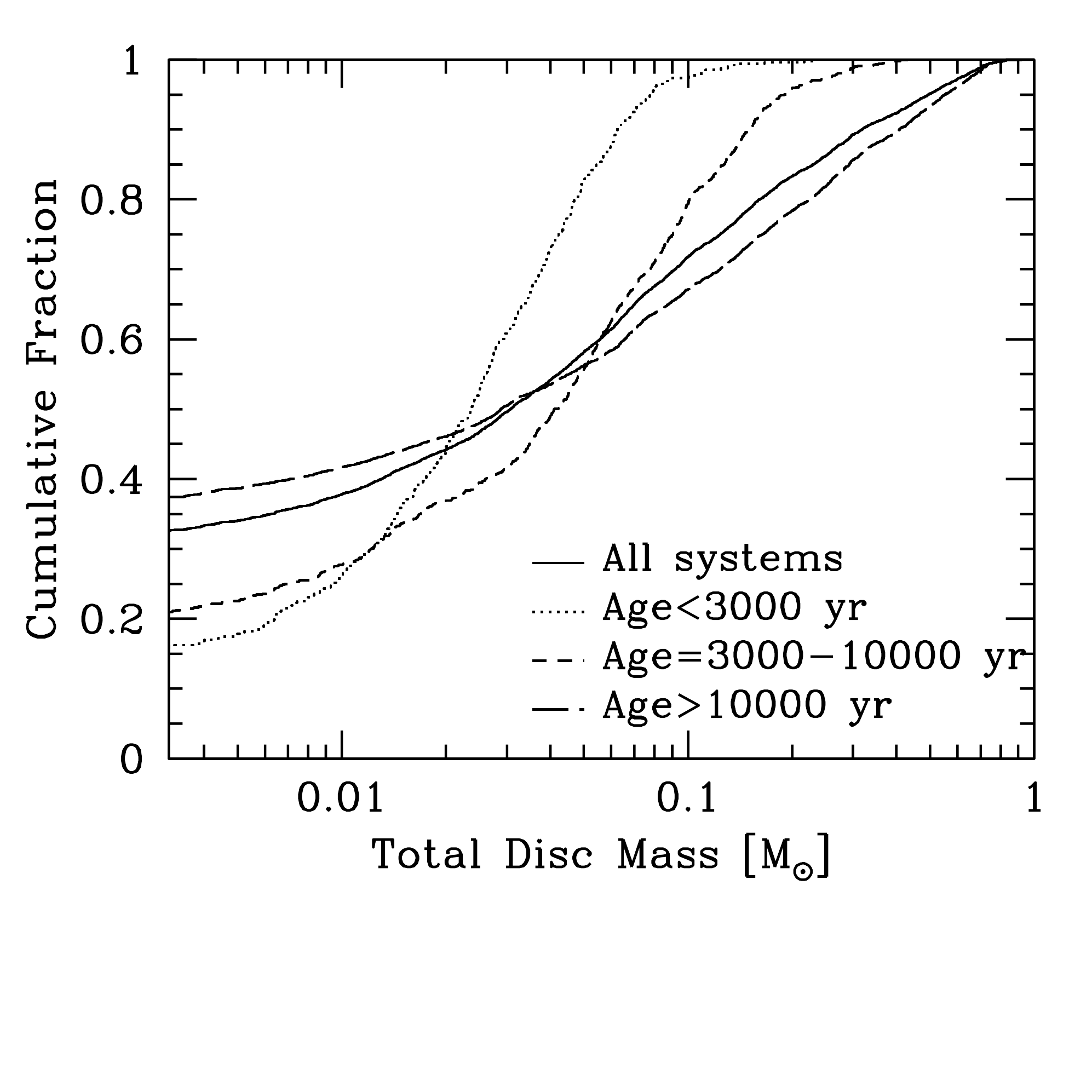} 
    \includegraphics[width=5.8cm]{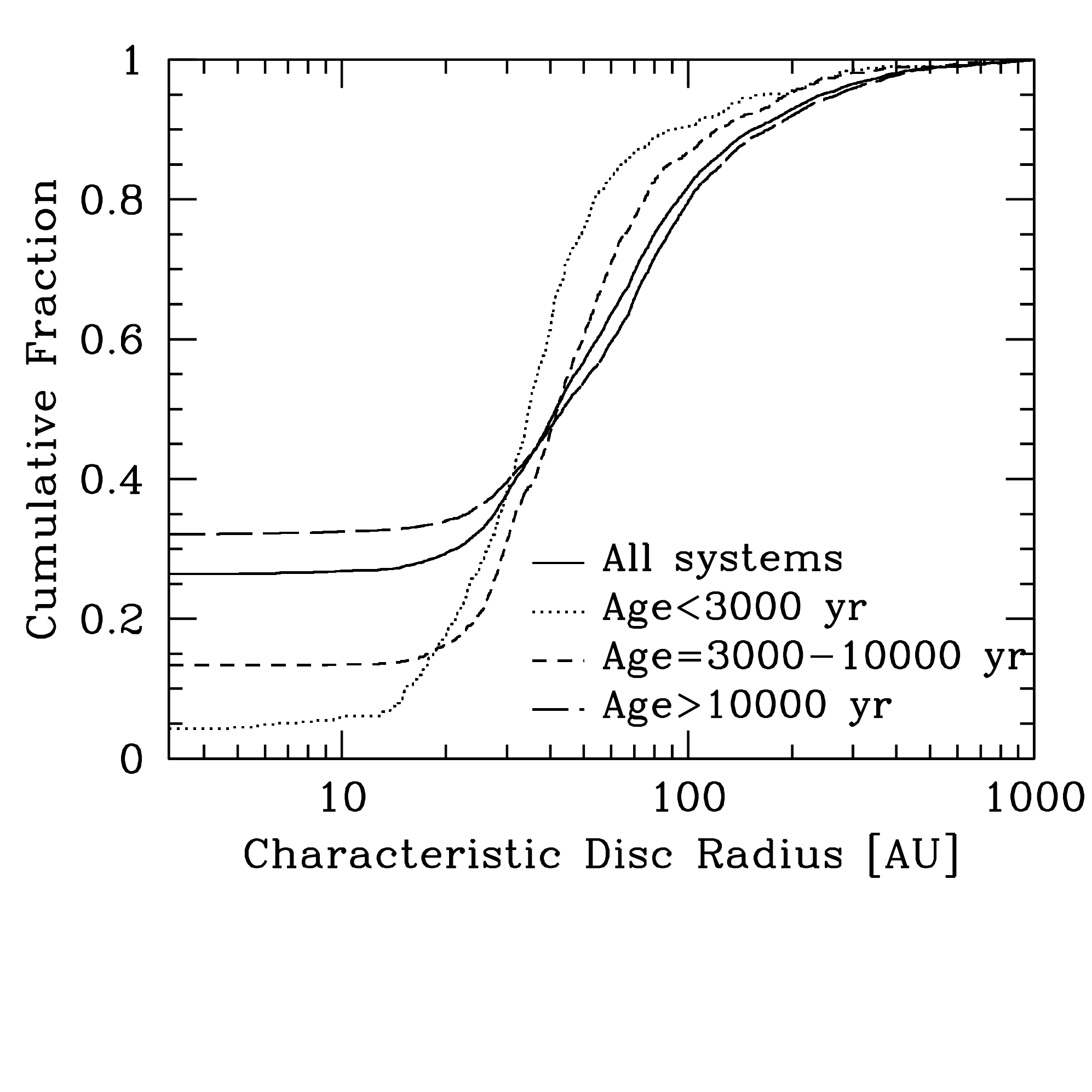} 
    \includegraphics[width=5.8cm]{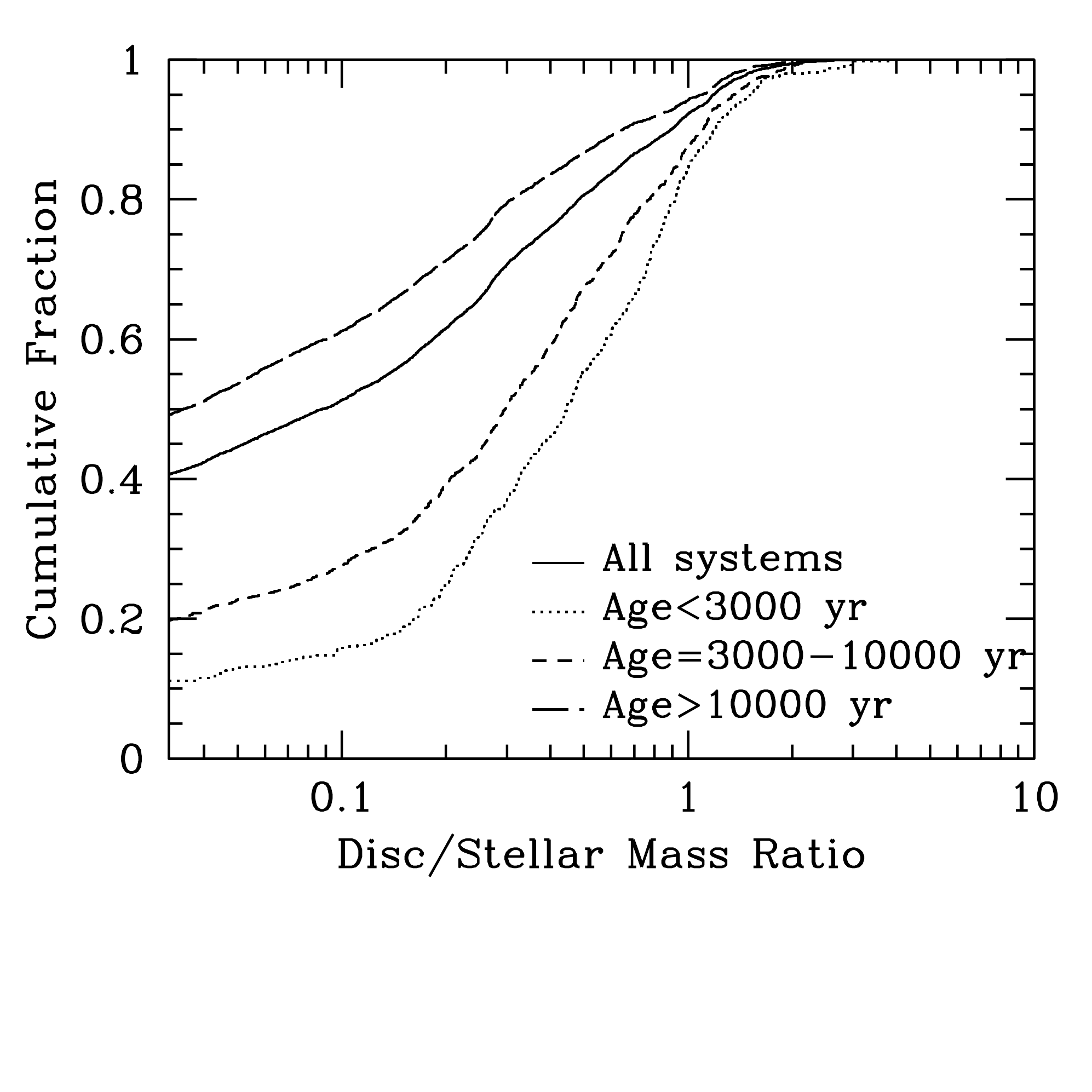} 
\vspace{-1cm}
\caption{The evolution of the total disc masses (left), characteristic disc radii (centre), and the ratio of total disc mass to total stellar mass (right) of the protostellar systems versus their age.  In the upper panels, each line represents the evolution of the disc(s) of a particular system, which may be a single protostar or a bound multiple protostellar system.  The colours denote the order of the system: single (black), binary (red), triple (blue), or quadruple (green).  Lines may stop and start if the components of a system change.  For example, if a single protostar becomes bound to a binary system (e.g. via star-disc capture), then the lines for both the single protostar and the binary will stop, and a new line will appear that represents the evolution of the new triple system.  In the lower panels, we give the cumulative distributions for all protostellar systems, and for systems in three ages ranges: $<3000$~yrs, $3000-10000$~yrs, and $>10000$~yrs. Disc masses and radii typically increase until ages of $10^4$~yrs and the ratio of disc to stellar mass is approximately constant.  Beyond this age, the disc masses tend to stabilise and some resolved discs are lost, so the ratio of disc to stellar mass tends to decline.}
\label{disc_systems_age}
\end{figure*}

\begin{figure*}
\centering \vspace{0.0cm} \hspace{0.0cm}
    \includegraphics[width=5.8cm]{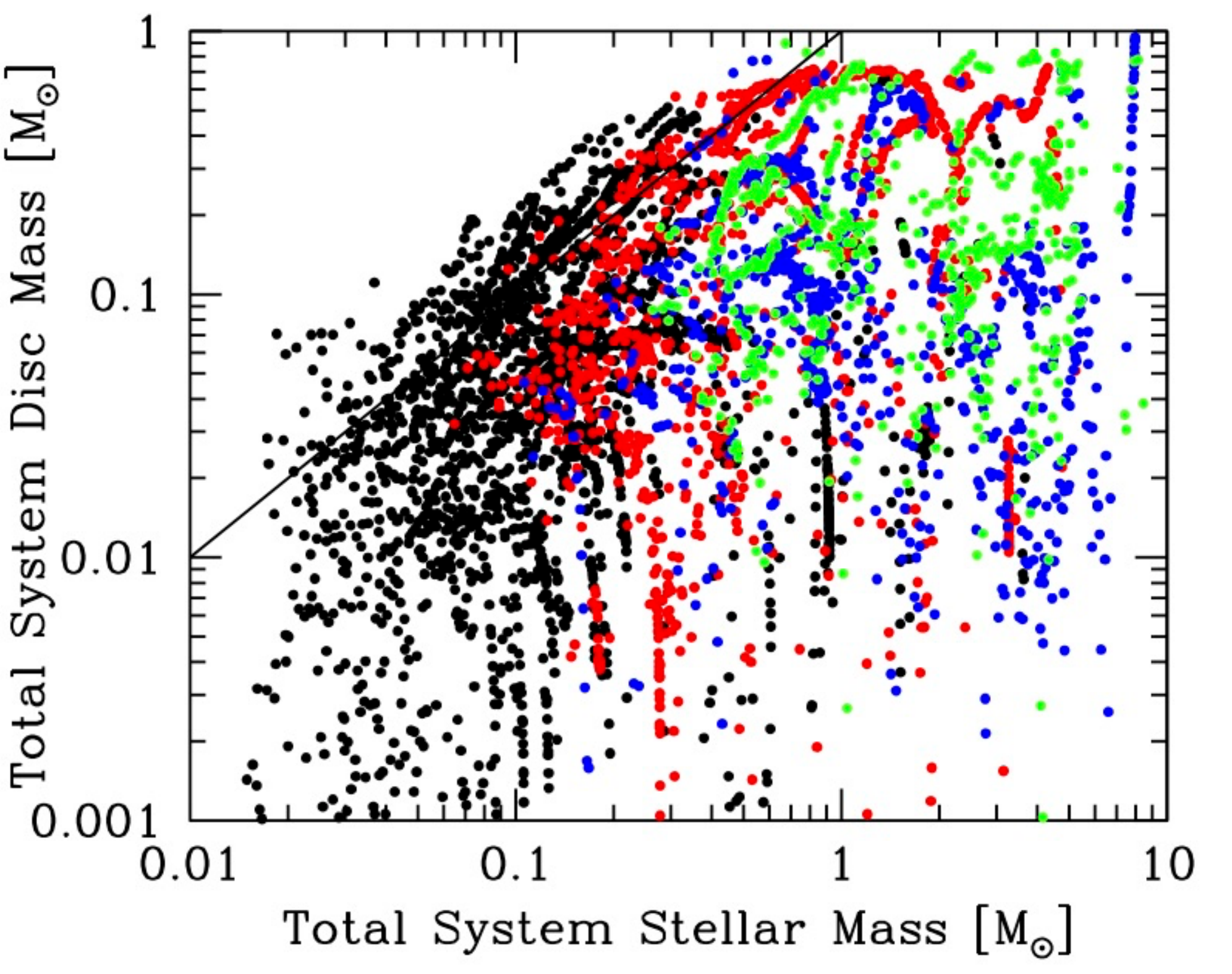} 
    \includegraphics[width=5.8cm]{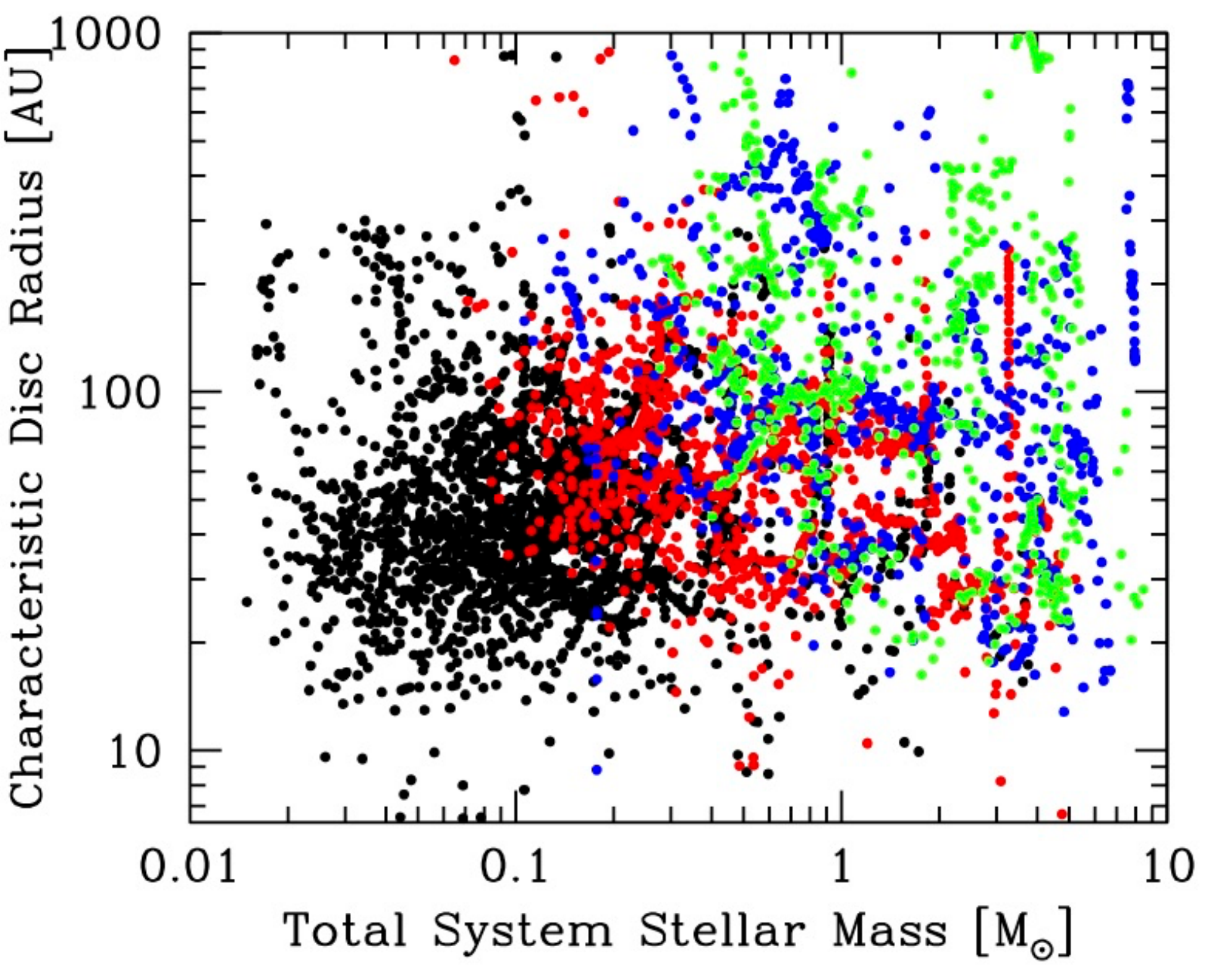} 
    \includegraphics[width=5.8cm]{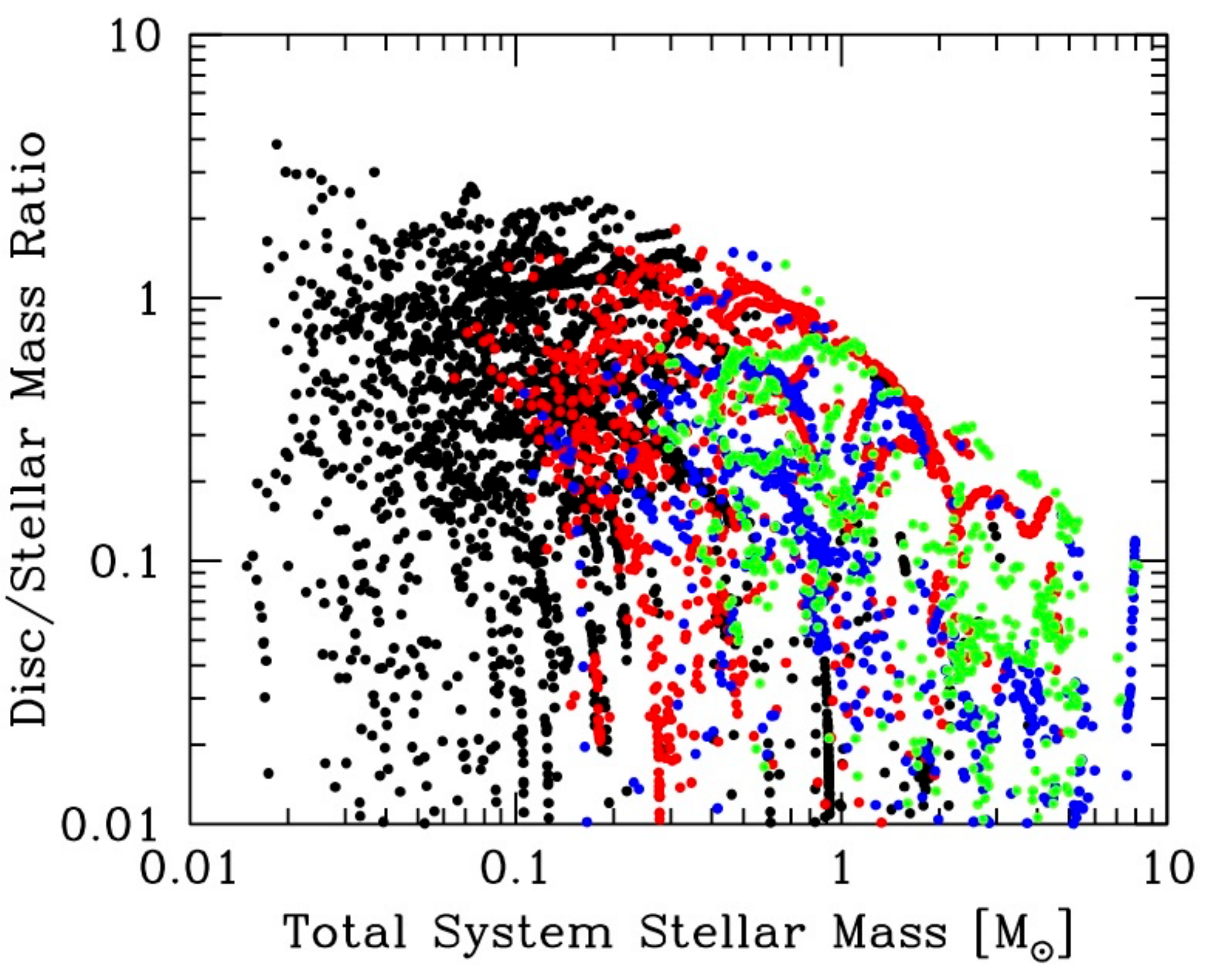} 
    \includegraphics[width=5.8cm]{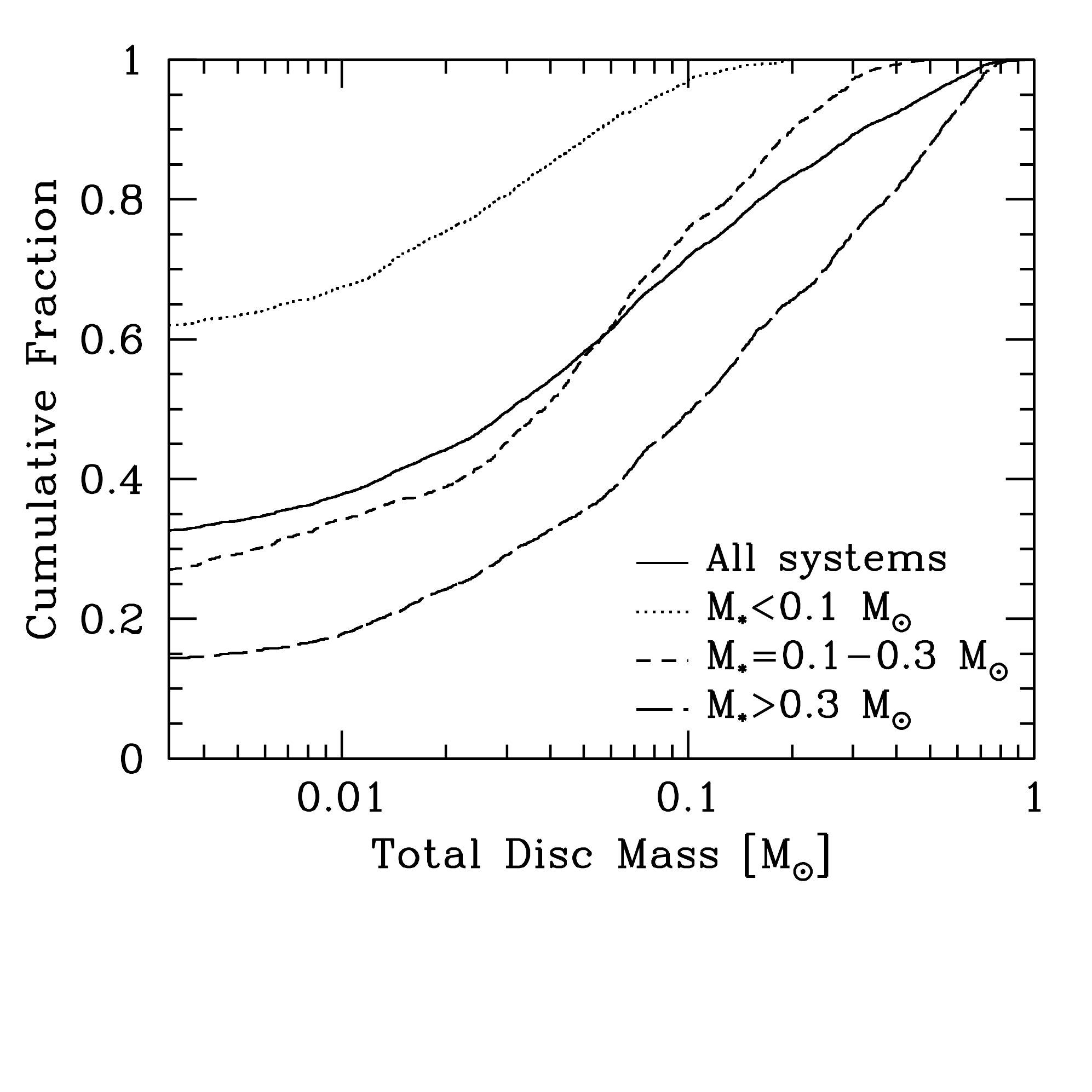}
    \includegraphics[width=5.8cm]{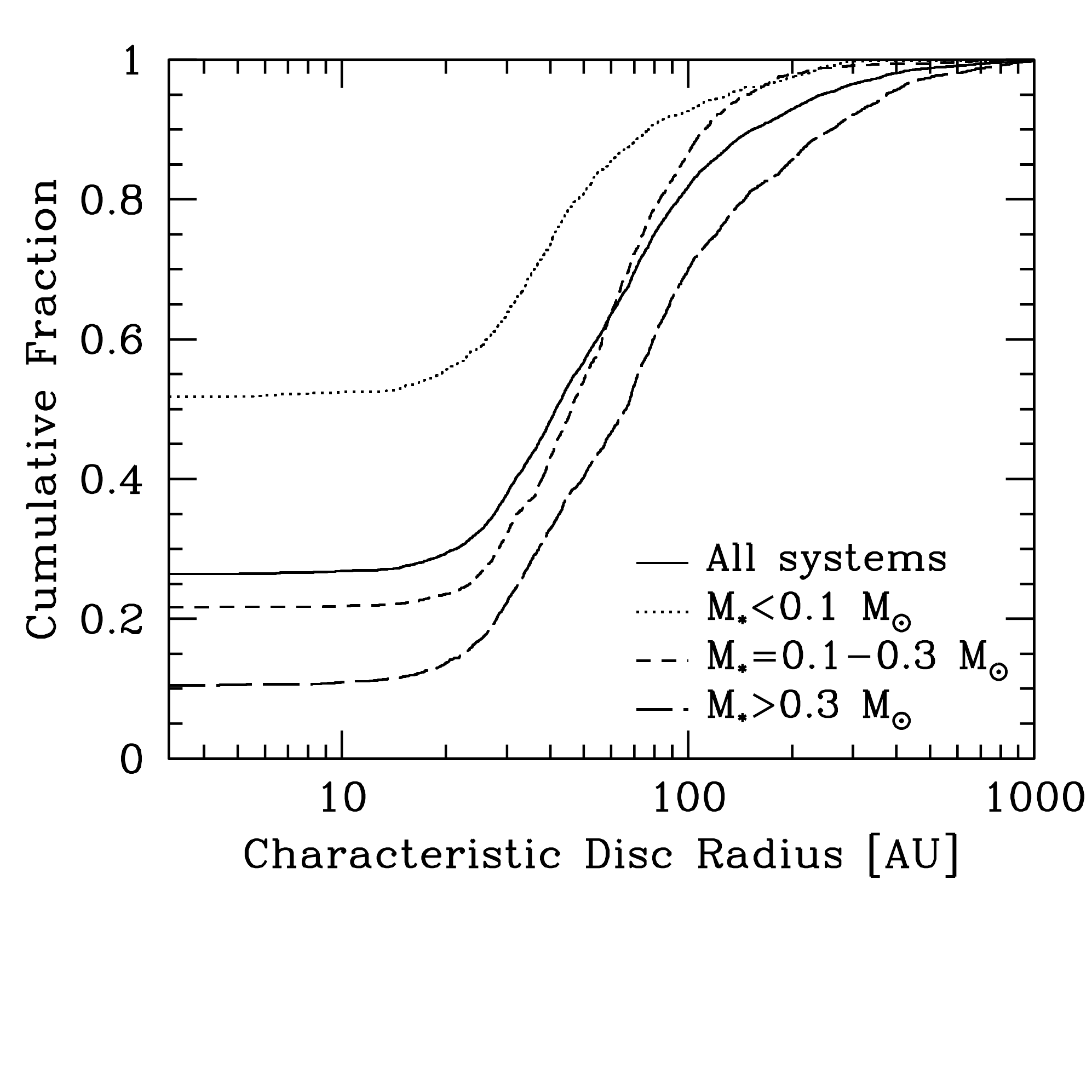} 
    \includegraphics[width=5.8cm]{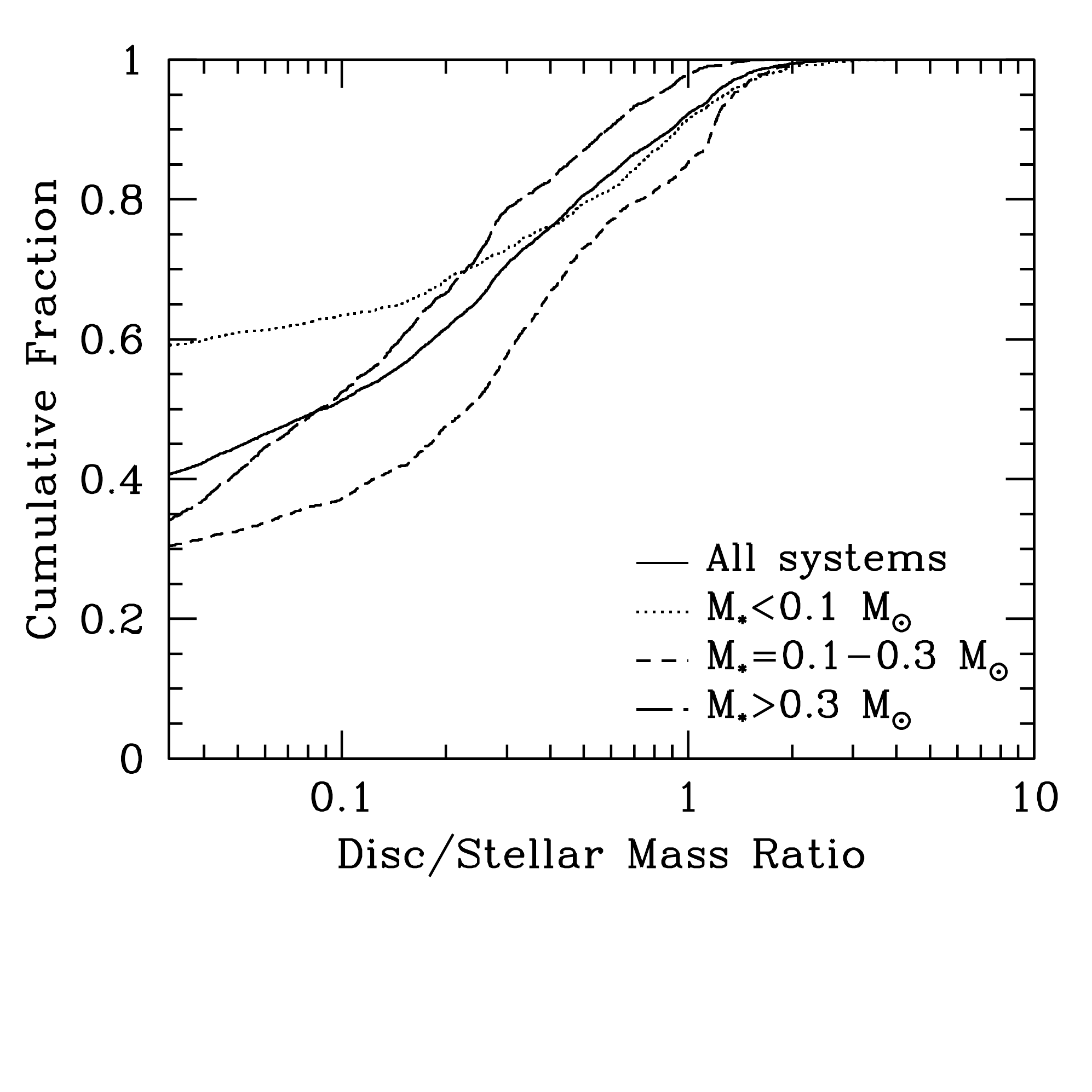} 
    \vspace{-1cm}
\caption{The total disc masses (left), characteristic disc radii (centre), and the ratio of total disc mass to total stellar mass (right) of the protostellar systems versus their total stellar mass.  In the upper panels, each dot represents an instance of disc(s) of a particular system, which may be a single protostar or a bound multiple protostellar system.  A single system may be represented by many dots that give the state of the system at different times.  The colours denote the order of the system: single (black), binary (red), triple (blue), or quadruple (green).  A particular system may be represented by many instances (taken at different times).  In the lower panels, we give the cumulative distributions for all protostellar systems, and for systems in three stellar mass ranges: $<0.1$~M$_\odot$, $0.1-0.3$~M$_\odot$, and $>0.3$~M$_\odot$. Disc masses tend to be greater for more massive protostellar systems.  Discs around very low mass ($<0.1$~M$_\odot$) are noticeably smaller than those around more massive systems.  For more massive systems, the typical disc size does not depend strongly on the total stellar mass, but the largest discs tend to be found around some of the most massive systems, and these tend to be multiple systems.  The typical ratios of disc to stellar mass tend to be highest for  systems with intermediate masses ($0.1-0.3$~M$_\odot$) and they tend to decline strongly with increasing mass for $>0.3$~M$_\odot$. }
\label{disc_systems_mass}
\end{figure*}

\subsection{The discs of stellar systems (single and multiple)}

In this section, we discuss the statistical properties of bound stellar systems (i.e. both single protostars and bound multiple systems).  The discs in such systems may include circumstellar, circumbinary and/or circum-multiple discs.  

In what follows, we take the age of a multiple protostellar system to be the age of the oldest protostar in the system, regardless of when the system became a multiple system.  Our initial analysis is similar to that presented in Section \ref{sec:isolated} for isolated protostars, but this time we only consider the total disc mass and the system's characteristic disc radius (as defined in Section \ref{sec:cmdisc}).  We do not attempt to measure the radial surface density profiles as we have seen from Section \ref{sec:diversity} that the discs in multiple systems have complex morphologies.  However, for bound pairs of protostars, we also examine the relative orientations of their circumstellar discs, sink particle spins, and orbit.

\subsubsection{Disc masses and radii}
\label{sec:sys_mass_rad}

In the left panels of Fig.~\ref{disc_systems_age}, in the top panel we plot total disc masses versus age for all protostellar systems, while in the bottom panel we provide the cumulative distributions of total disc mass for all systems and for three different age ranges ($<3000$~yrs, $3000-10000$~yrs, and $>10000$~yrs).  As with isolated protostars, the disc masses generally increase with age from $\sim 0.03$~M$_\odot$ at $10^3$ yrs to $\sim 0.1$~M$_\odot$ at $10^4$ yrs.  But the dispersion of the disc masses is much greater than for isolated protostars.  Whereas only a few isolated protostars had disc masses $M_{\rm d}>0.3$~M$_\odot$ at ages $>10^4$ yrs, many multiple systems have total disc masses that exceed $0.3$~M$_\odot$ at ages $10^4-10^5$~yrs.  The most massive total disc mass now approaches 1~M$_\odot$.  There are also a significant number of systems with disc masses $M_{\rm d}<0.01$~M$_\odot$.   As with the isolated protostars, the disc masses of some protostellar systems rapidly decline (due to accretion, dynamical evolution and/or ram-pressure stripping).

In the middle panels of Fig.~\ref{disc_systems_age}, we plot the characteristic radii of the discs versus age for all protostellar systems, and cumulative distributions of disc radius.  As with isolated protostars, the typical disc size tends to increase with age, but only by a factor of two or so.  The median characteristic disc radius of resolved discs is $\approx 30$~au at ages $<3000$~yrs and $\approx 60$~au at ages $>10^4$~yrs.  At ages $<3000$ yrs, the vast majority of discs have characteristic radii ranging from 10--60~au, while at ages of $>10^4$ yrs about 20\% of discs have characteristic radii exceeding 100 au.  The largest discs tend to be found around multiple systems.

In the right panels of Fig.~\ref{disc_systems_age}, we plot the ratio of the total disc mass to the total protostellar mass (i.e. sink particle mass) versus age for all protostellar systems, and the corresponding cumulative distributions.  As with the isolated protostars, until $\approx 10^4$~yrs, the lines are relatively flat, indicating that the ratio of the total disc mass to the total stellar mass is relatively constant and the ratio lies in the range $0.1-2$ for the vast majority of systems.  Beyond $10^4$~yrs, the typical value of the ratio declines.  There will be many reasons for this decline, including accretion (e.g. driven by gravitational torques in self-gravitating discs), fragmentation, and ram-pressure stripping.  None of the systems have total disc masses exceeding their total stellar mass beyond ages of 40,000 yrs, but some still have ratios $>0.1$ until $\approx 10^5$~yrs (the calculation is stopped when the oldest system has an age of $90,000$~yrs).  Significant numbers of low-mass discs (with disc/star mass ratios $M_{\rm d}/M_*<0.1$) have appeared by ages of $\approx 3000$ yrs and by $10^4$~yrs about half of systems have ratios $M_{\rm d}/M_*<0.1$.  If protostars are ejected from multiple systems, these usually have low disc masses.  Multiple systems tend to have lower ratios than single stars.  In some cases this will be due to massive discs fragmenting to produce the multiple system, thus leaving a lower disc mass to stellar mass ratio than before the fragmentation occurred.  Another contributing factor is dynamical clearing (and accretion) of disc material in a multiple system.

The protostellar systems we study are very young and many are accreting rapidly.  Therefore, as they age, they are also become more massive.  Because of this it is not possible to separate the evolution of discs with time from the dependence of disc properties on stellar mass -- disc properties depend on both age and stellar mass.

In the left panels of Fig.~\ref{disc_systems_mass}, we plot the total disc mass versus the total protostellar mass (that is, the total sink particle mass) for different instances of all protostellar systems, and the cumulative distributions of total disc mass for the instances of the systems and, separately, for three different ranges of total stellar mass ($M_*<0.1$~M$_\odot$, $0.1 \le M_* < 0.3$~M$_\odot$, and $M_*\ge 0.3$~M$_\odot$).  Systems that have a greater total mass tend to have a higher multiplicity (as is observed).  Disc masses tend to be greater for more massive protostellar systems up until $M_* \approx 0.5$~M$_\odot$.  Beyond this mass there is no strong trend in the disc masses.  Lower mass systems have fewer resolved discs -- more than half of the very-low-mass (VLM; $M_*<0.1$~M$_\odot$) systems do not have resolved discs.

In the middle panels of Fig.~\ref{disc_systems_mass}, we plot the characteristic disc radius versus the total stellar mass for different instances of the protostellar systems, and we give the corresponding cumulative distributions.  Discs of VLM systems tend to be a factor of two smaller than those around systems with masses $0.1 \le M_* < 0.3$~M$_\odot$, and three times smaller than systems with $M_*\ge 0.3$~M$_\odot$.  For systems with masses $M_* \gsim 0.5$~M$_\odot$, the typical disc size does not depend strongly on the total stellar mass, but the largest discs tend to be found in multiple systems.  

In the right panels of Fig.~\ref{disc_systems_mass}, we plot the ratio of the total disc mass to the total protostellar mass (i.e. sink particle mass) versus the total stellar mass for different instances of the protostellar systems, and we provide the corresponding cumulative distributions.  The typical ratios of disc to stellar mass tend to be highest for systems with intermediate masses ($0.1-0.3$~M$_\odot$).  For lower masses, many protostars do not have resolved discs.  For systems of higher mass, as we noted above, the total disc mass becomes independent of the total protostellar mass, so the disc to star mass ratios tend to decline roughly inversely proportional to the total protostellar mass.

\begin{figure*}
\centering \vspace{-0.5cm} \hspace{0cm}
    \includegraphics[width=8.5cm]{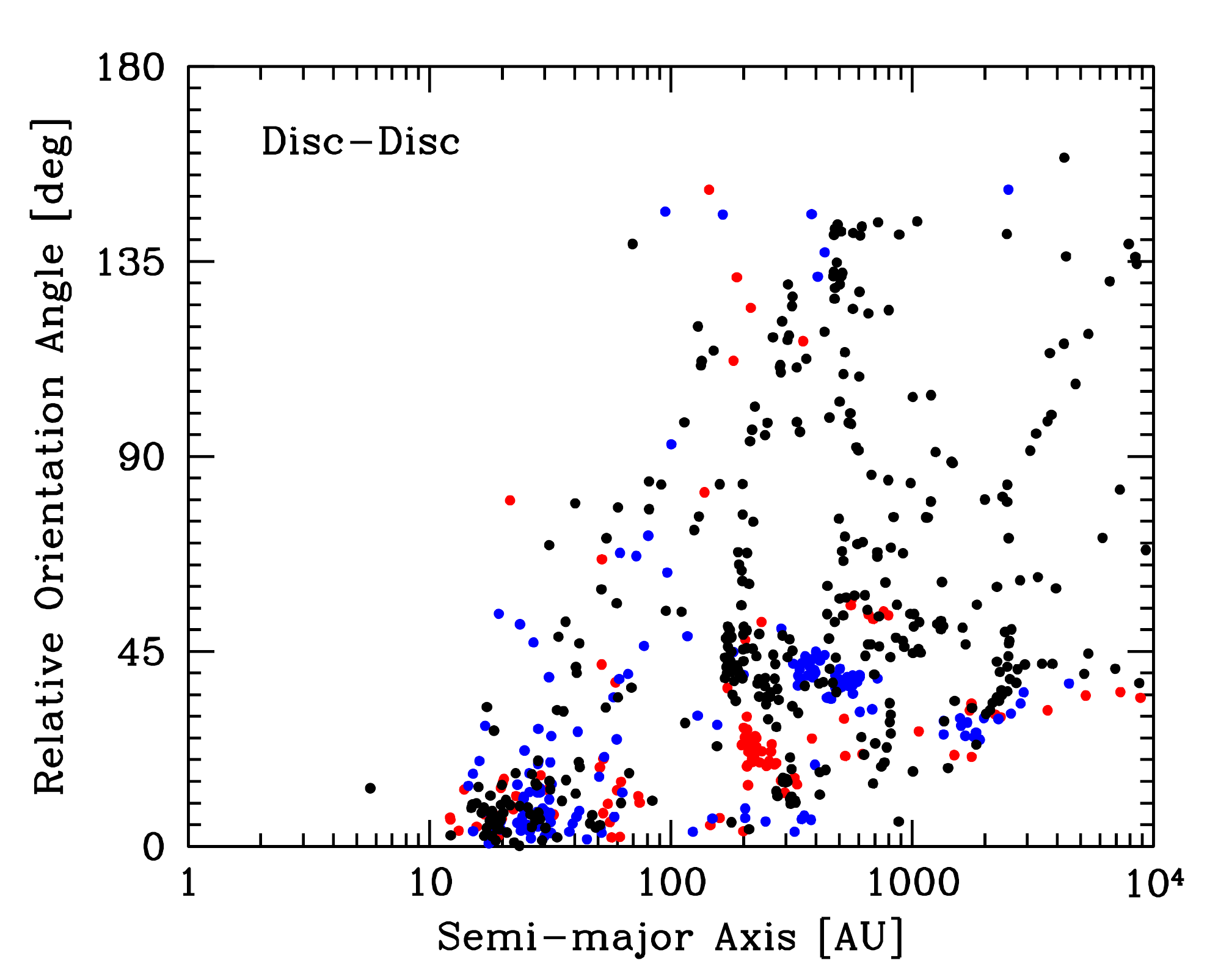} \vspace{0cm}
    \includegraphics[width=8.5cm]{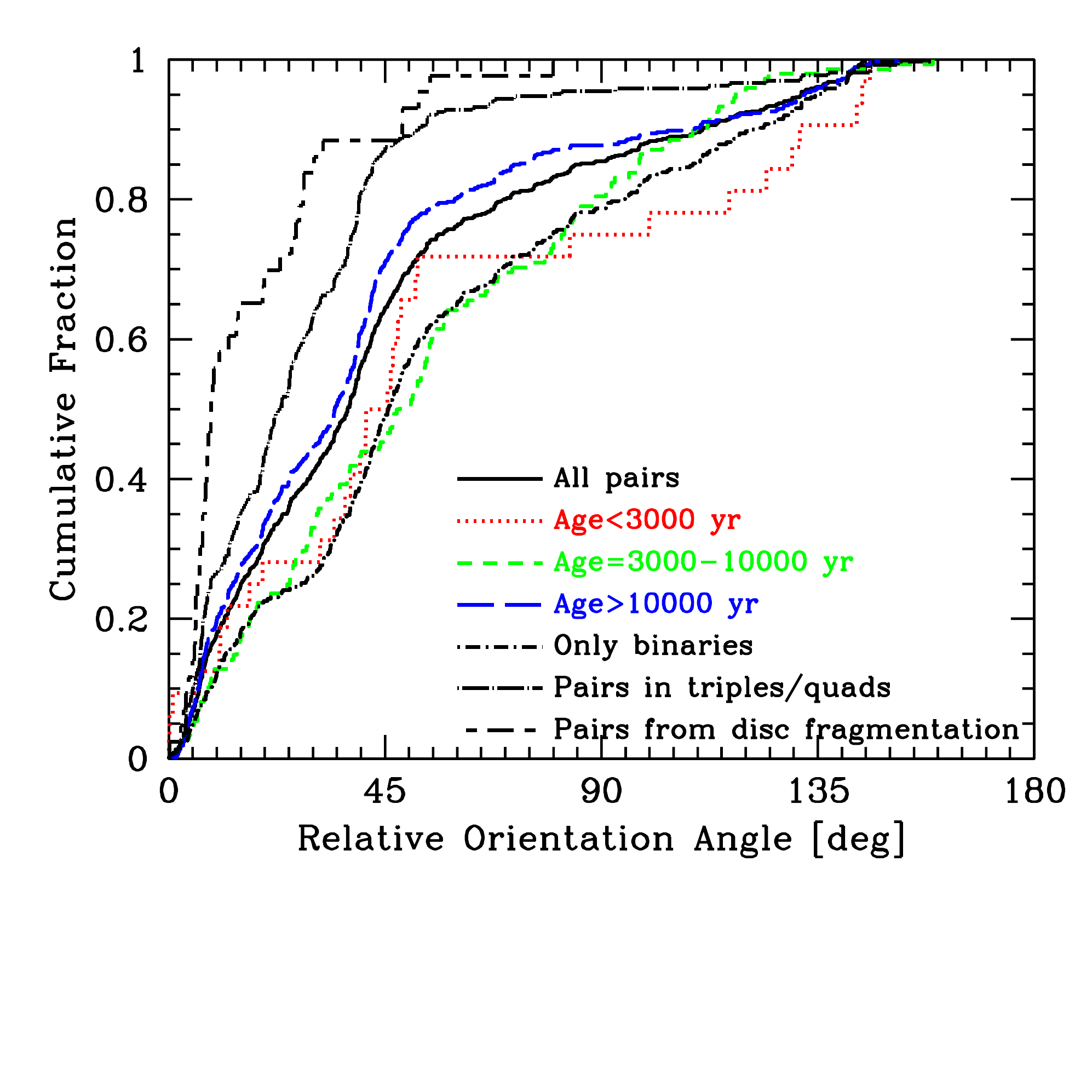} \vspace{0cm}
    \includegraphics[width=8.5cm]{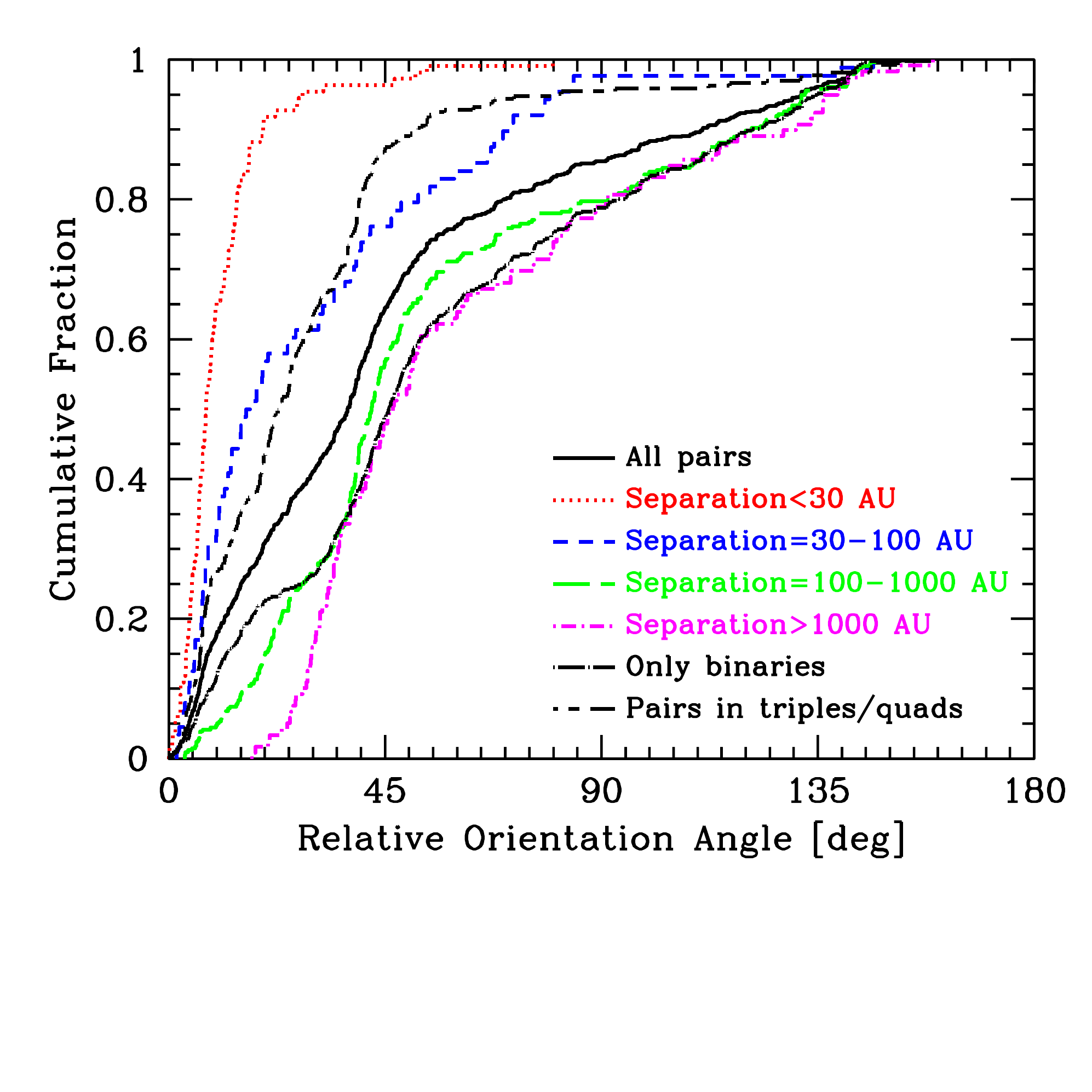} \vspace{0cm}
    \includegraphics[width=8.5cm]{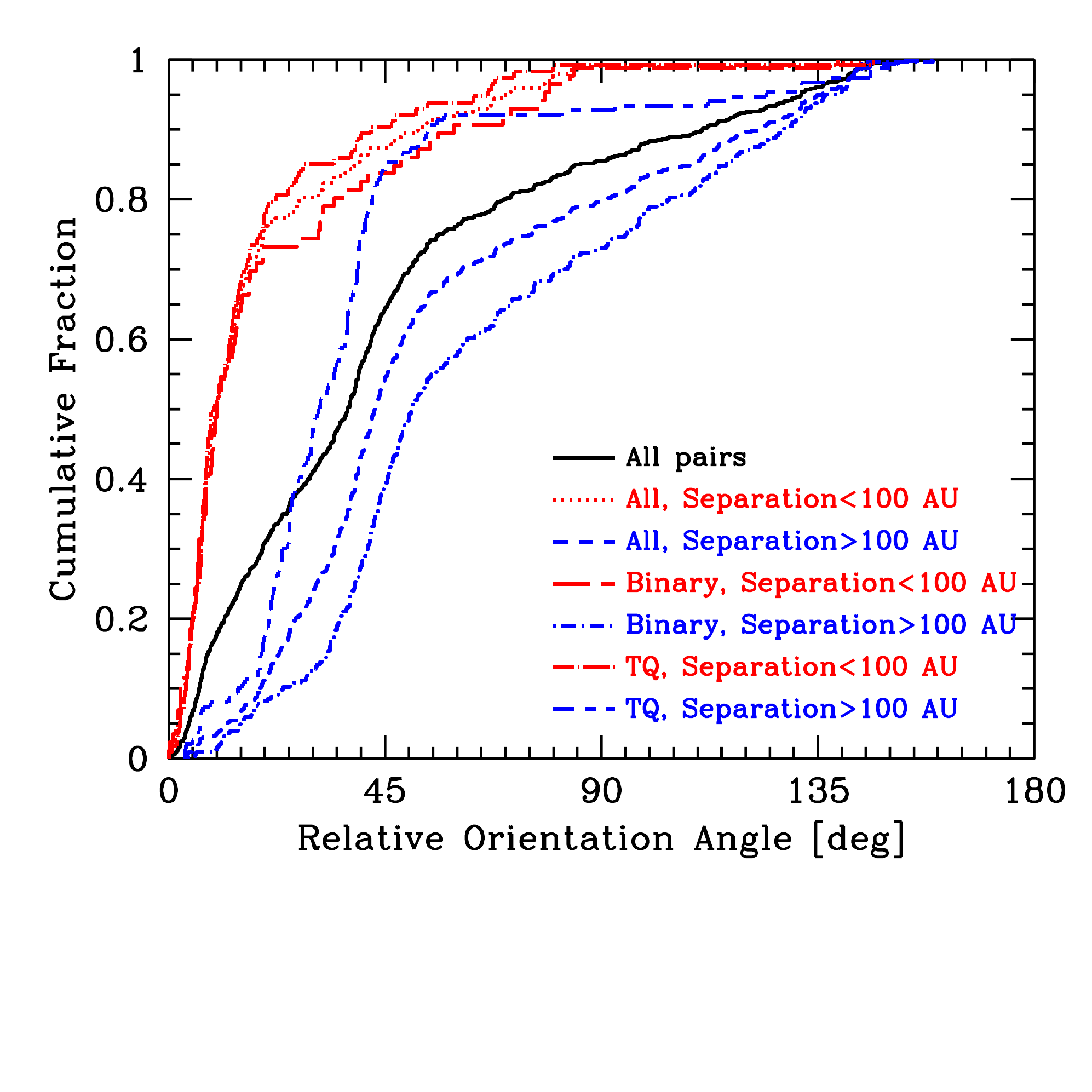} \vspace{-1.5cm}
\caption{Distributions of the relative orientation angle between the two circumstellar discs in bound protostellar pairs.  Pairs include both binaries and bound pairs in hierarchical triple or quadruple systems.  In the top left panel, we plot the relative orientation angle of each pair versus its semi-major axis, with binary systems in black, pairs in triples in red, and pairs in quadruples in blue.  In the remaining panels, we give the cumulative distributions of the orientation angles.  In the top-right panel, we give the cumulative distributions for all pairs and pairs in three different age ranges.  We also give separate distributions for binary systems, for pairs in triple or quadruple systems, and for pairs for which at least one of the components was created by disc fragmentation.  In the bottom-left panel, we plot the cumulative distributions for four ranges of semi-major axes for all pairs, and we also plot the separate distributions of binaries, and pairs that are components of triples or quadruples.  In the bottom-right panel, we plot the cumulative distributions for semi-major axes $a<100$~au and $a>100$~au for all pairs, binaries, and pairs that are components of triples or quadruples.  The circumstellar discs become more aligned with increasing age.  The circumstellar discs of pairs also tend to be more closely aligned in high-order multiple systems than in binaries.}
\label{disc_disc_cum}
\end{figure*}

\begin{figure*}
\centering \vspace{-0.5cm} \hspace{0cm}
    \includegraphics[width=8.5cm]{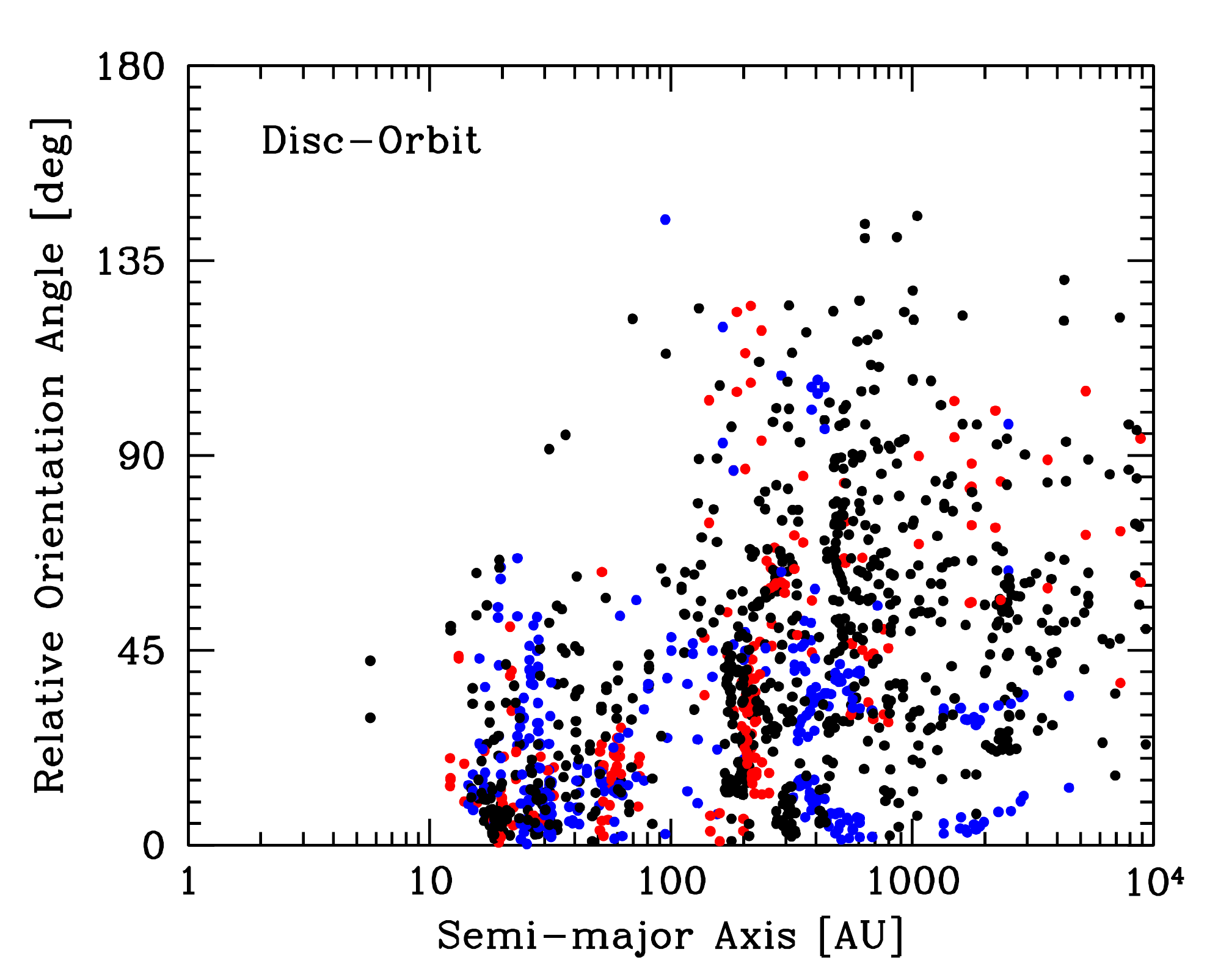} \vspace{0cm}
    \includegraphics[width=8.5cm]{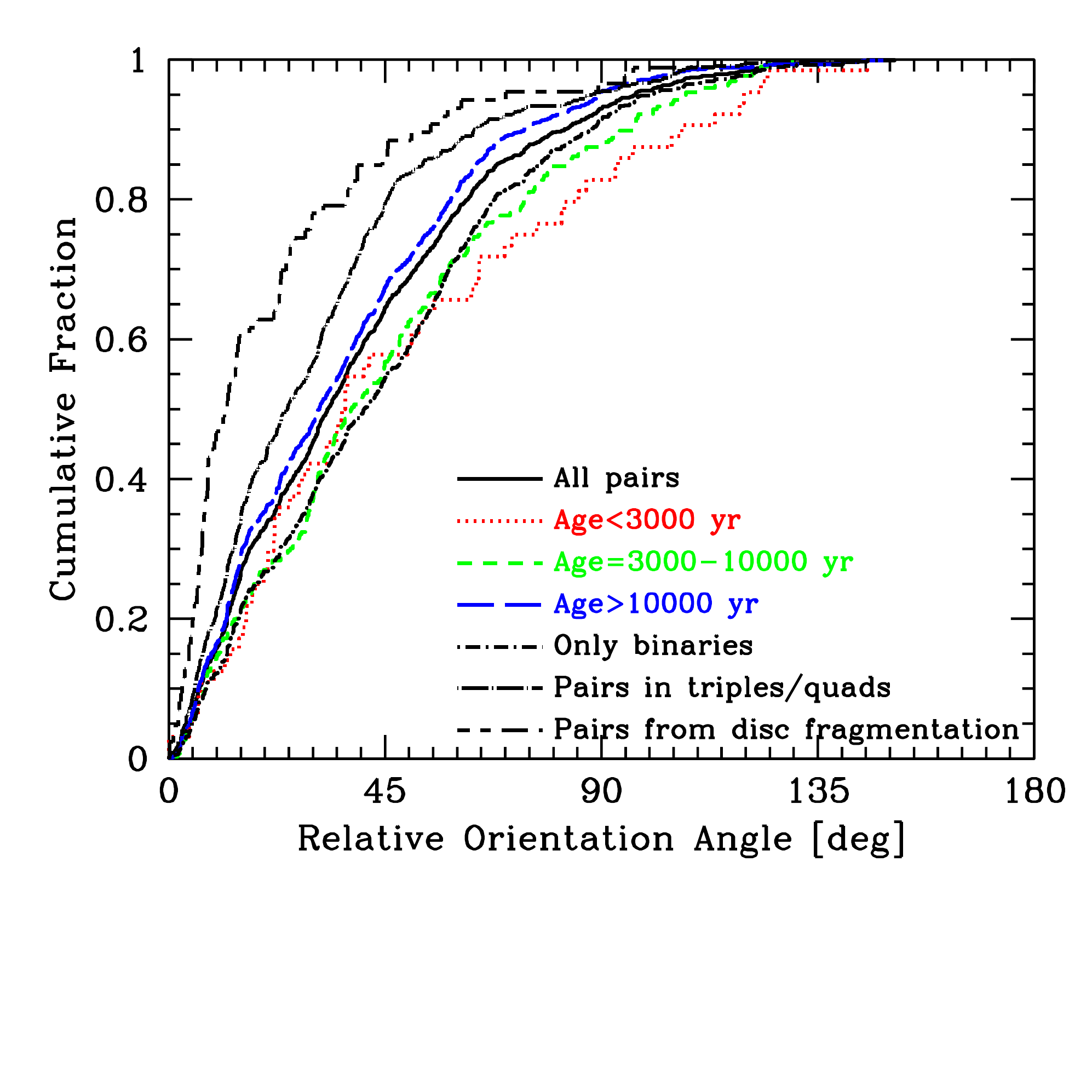} \vspace{0cm}
    \includegraphics[width=8.5cm]{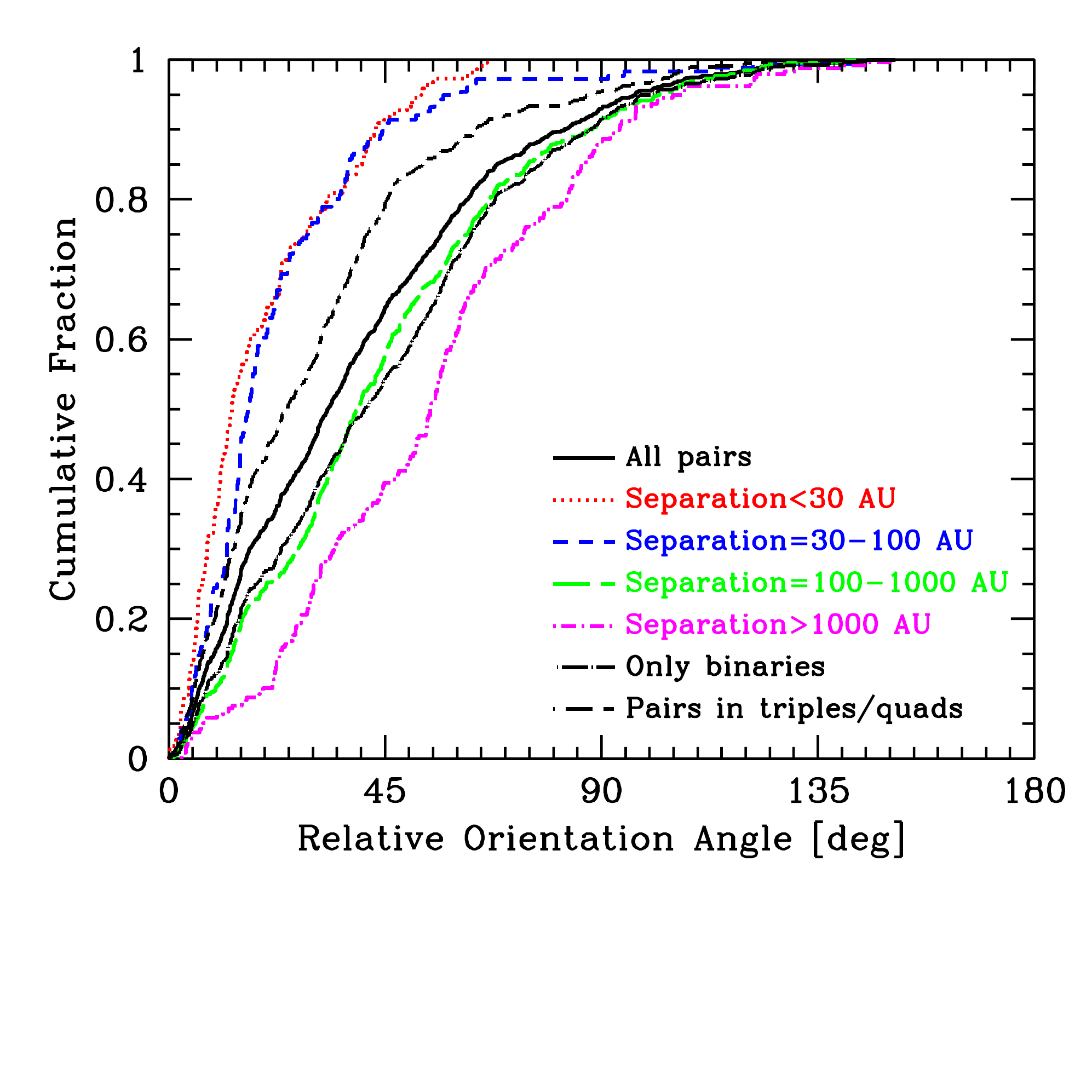} \vspace{0cm}
    \includegraphics[width=8.5cm]{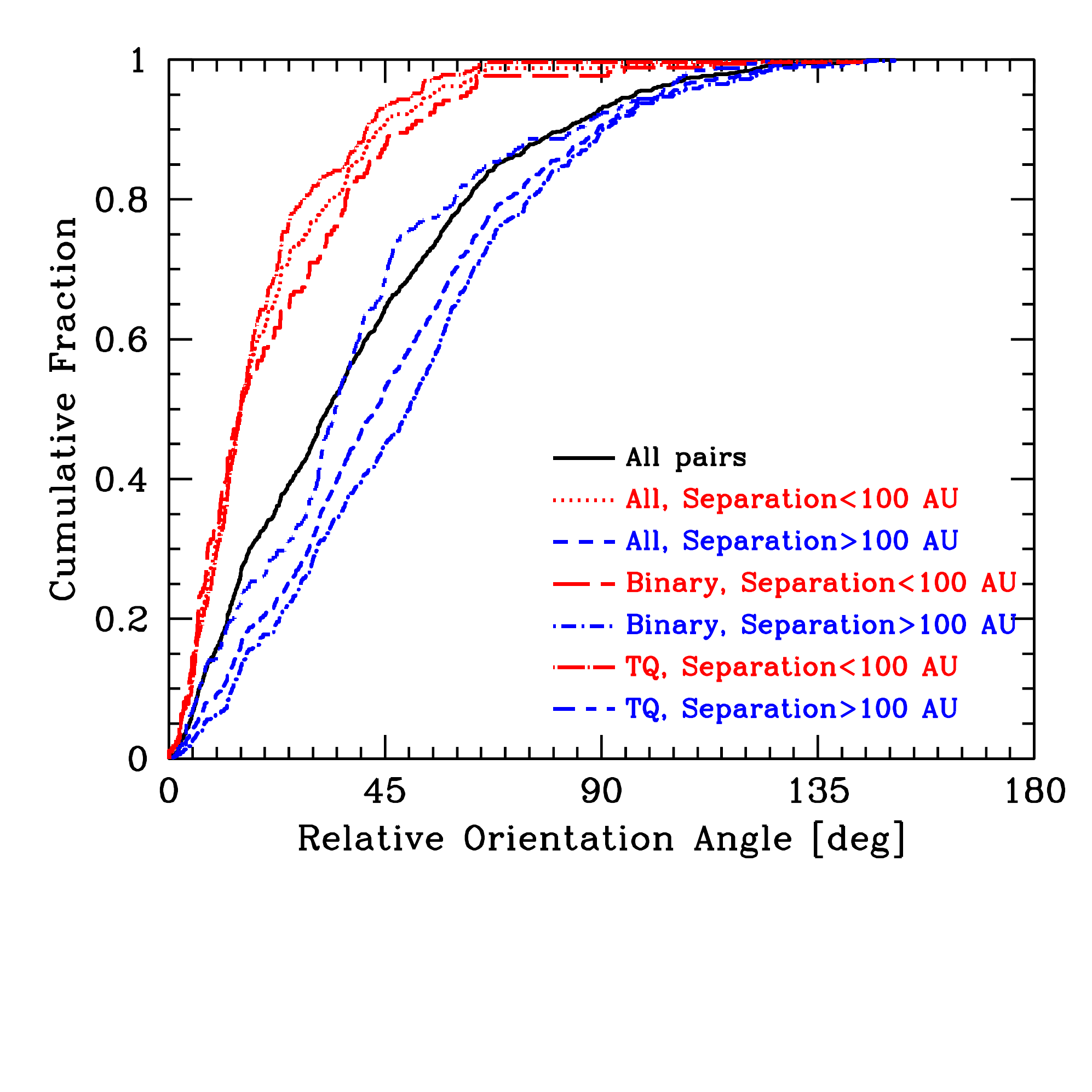} \vspace{-1.5cm}
\caption{Distributions of the relative orientation angle between each circumstellar disc and the orbital plane in bound protostellar pairs.  Pairs include both binaries and bound pairs in hierarchical triple or quadruple systems.  In the top left panel, we plot the relative orientation angle of each pair versus its semi-major axis, with binary systems in black, pairs in triples in red, and pairs in quadruples in blue.  In the remaining panels, we give the cumulative distributions of the orientation angles.  In the top-right panel, we give the cumulative distributions for all pairs and pairs in three different age ranges.  We also give separate distributions for binary systems, for pairs in triple or quadruple systems, and for pairs for which at least one of the components was created by disc fragmentation.  In the bottom-left panel, we plot the cumulative distributions for four ranges of semi-major axes for all pairs, and we also plot the separate distributions of binaries, and pairs that are components of triples or quadruples.  In the bottom-right panel, we plot the cumulative distributions for semi-major axes $a<100$~au and $a>100$~au for all pairs, binaries, and pairs that are components of triples or quadruples.  The circumstellar discs become more aligned with the orbital plane pair with increasing age.  The circumstellar discs of pairs also tend to be more closely aligned with the orbital plane in high-order multiple systems than in binaries.  However, both of these trends are weaker than when comparing the relative orientations of the two discs in a pair.}
\label{disc_orbit_cum}
\end{figure*}

\begin{figure*}
\centering \vspace{-0.5cm} \hspace{0cm}
    \includegraphics[width=8.5cm]{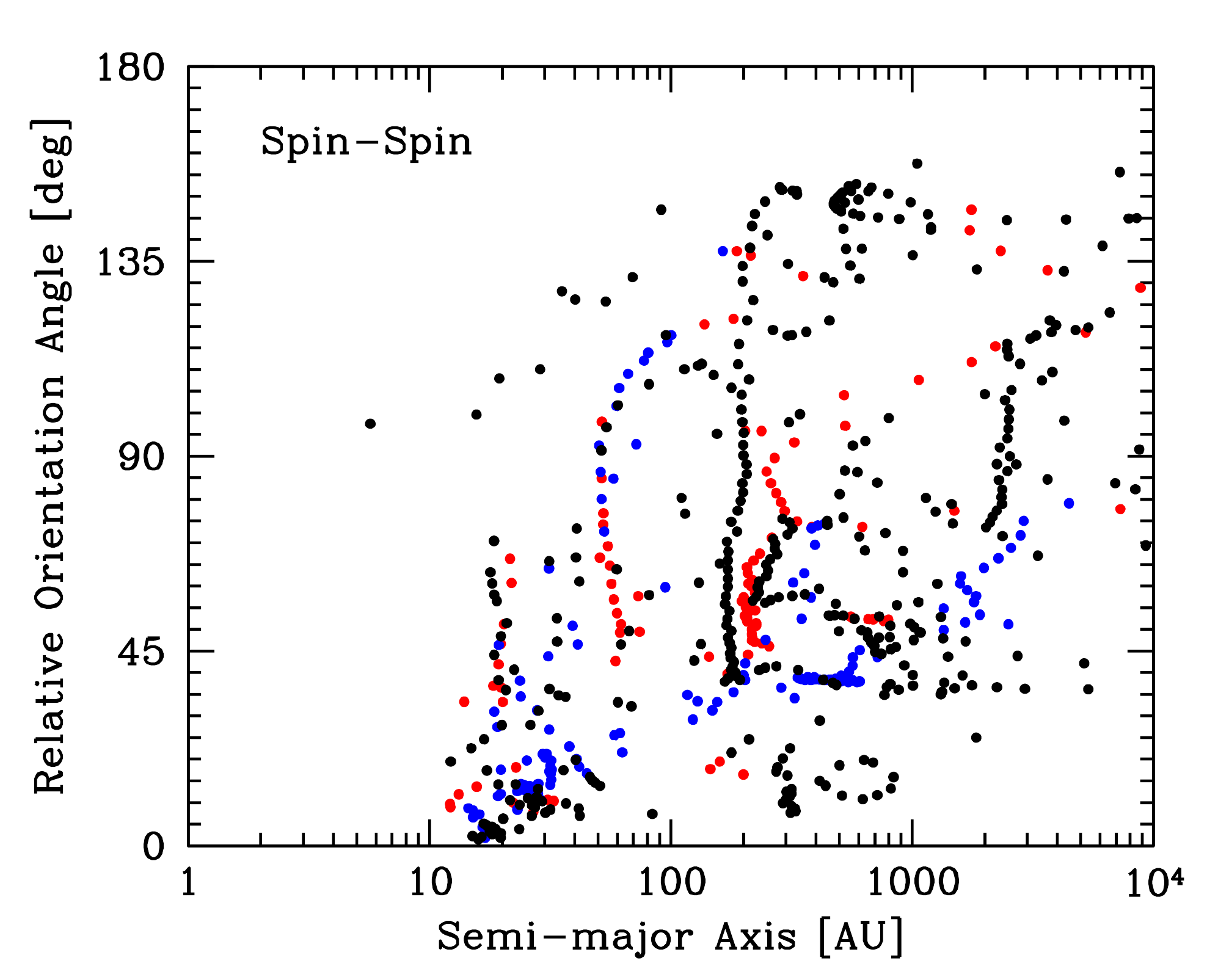} \vspace{0cm}
    \includegraphics[width=8.5cm]{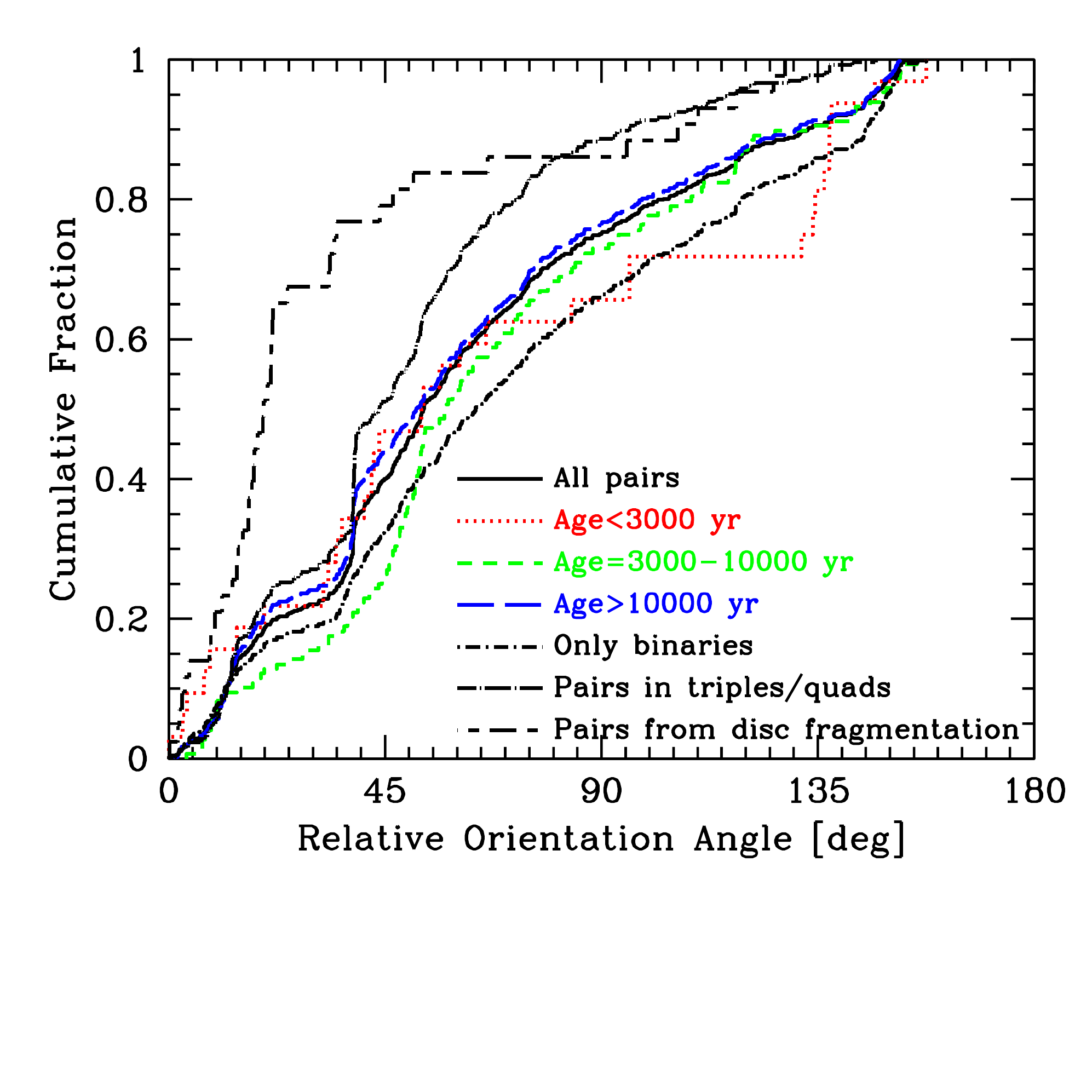} \vspace{0cm}
    \includegraphics[width=8.5cm]{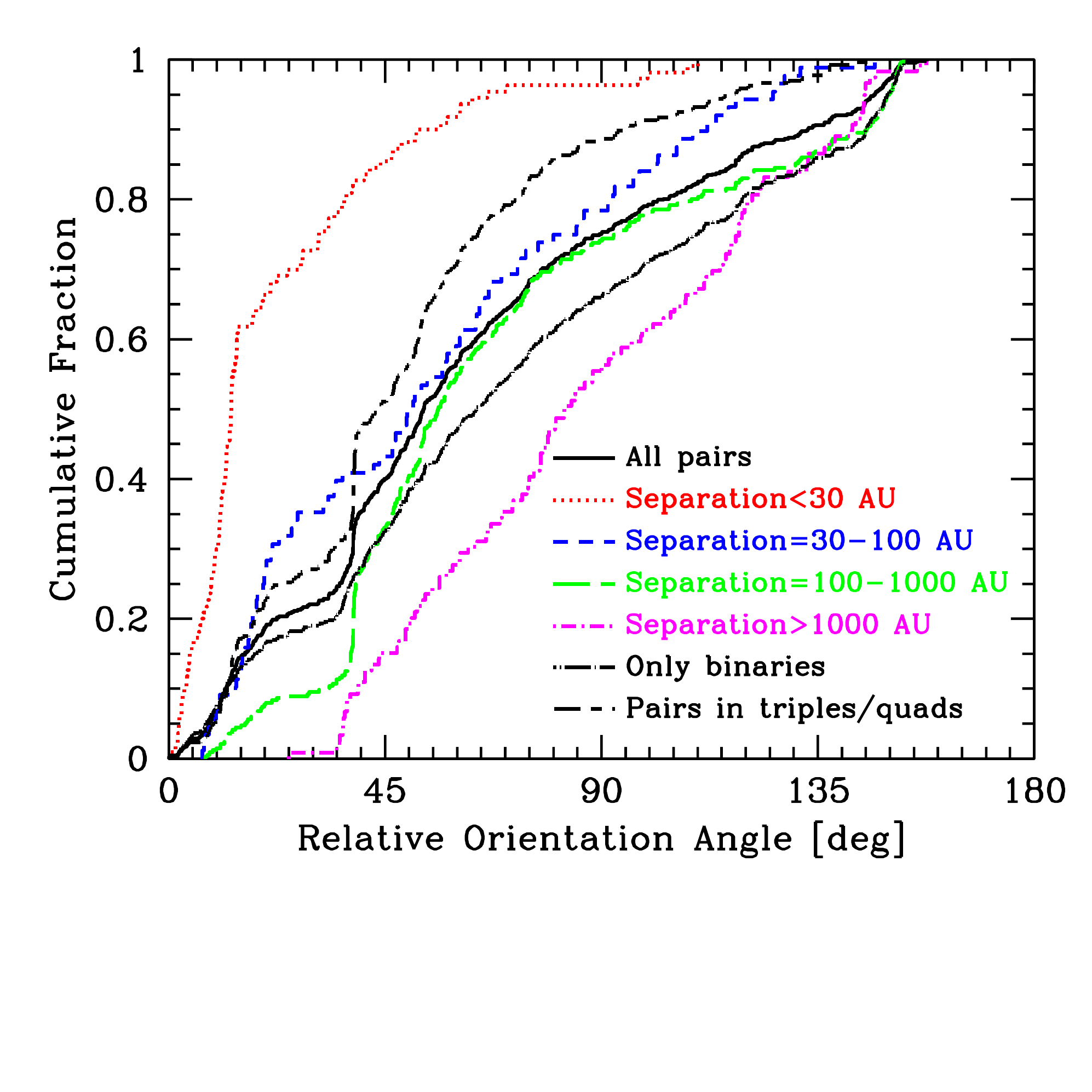} \vspace{0cm}
    \includegraphics[width=8.5cm]{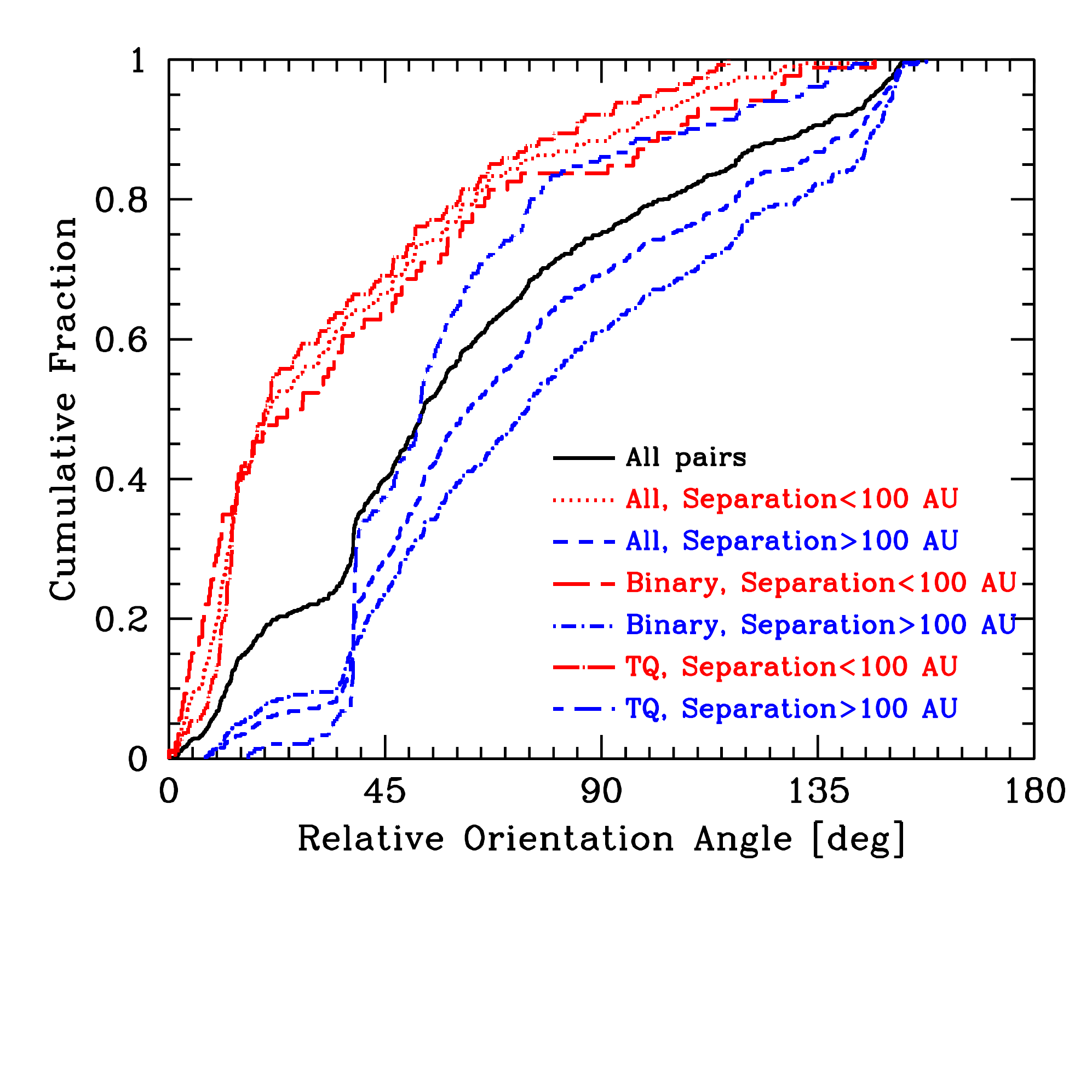} \vspace{-1.5cm}
\caption{Distributions of the relative orientation angle between the two sink particle spins in bound protostellar pairs.  Pairs include both binaries and bound pairs in hierarchical triple or quadruple systems.  In the top left panel, we plot the relative orientation angle of each pair versus its semi-major axis, with binary systems in black, pairs in triples in red, and pairs in quadruples in blue.  In the remaining panels, we give the cumulative distributions of the orientation angles.  In the top-right panel, we give the cumulative distributions for all pairs and pairs in three different age ranges.  We also give separate distributions for binary systems, for pairs in triple or quadruple systems, and for pairs for which at least one of the components was created by disc fragmentation.  In the bottom-left panel, we plot the cumulative distributions for four ranges of semi-major axes for all pairs, and we also plot the separate distributions of binaries, and pairs that are components of triples or quadruples.  In the bottom-right panel, we plot the cumulative distributions for semi-major axes $a<100$~au and $a>100$~au for all pairs, binaries, and pairs that are components of triples or quadruples.  The sink particle spins are less well aligned with each other than the circumstellar discs.  The trend of greater alignment for smaller orbital separation is stronger for sink particle spins than for the relative orientations of circumstellar discs.  However, there is less dependence on age. }
\label{spin_spin_cum}
\end{figure*}

\begin{figure}
\centering \vspace{-0.4cm} \hspace{0cm}
    \includegraphics[width=8.5cm]{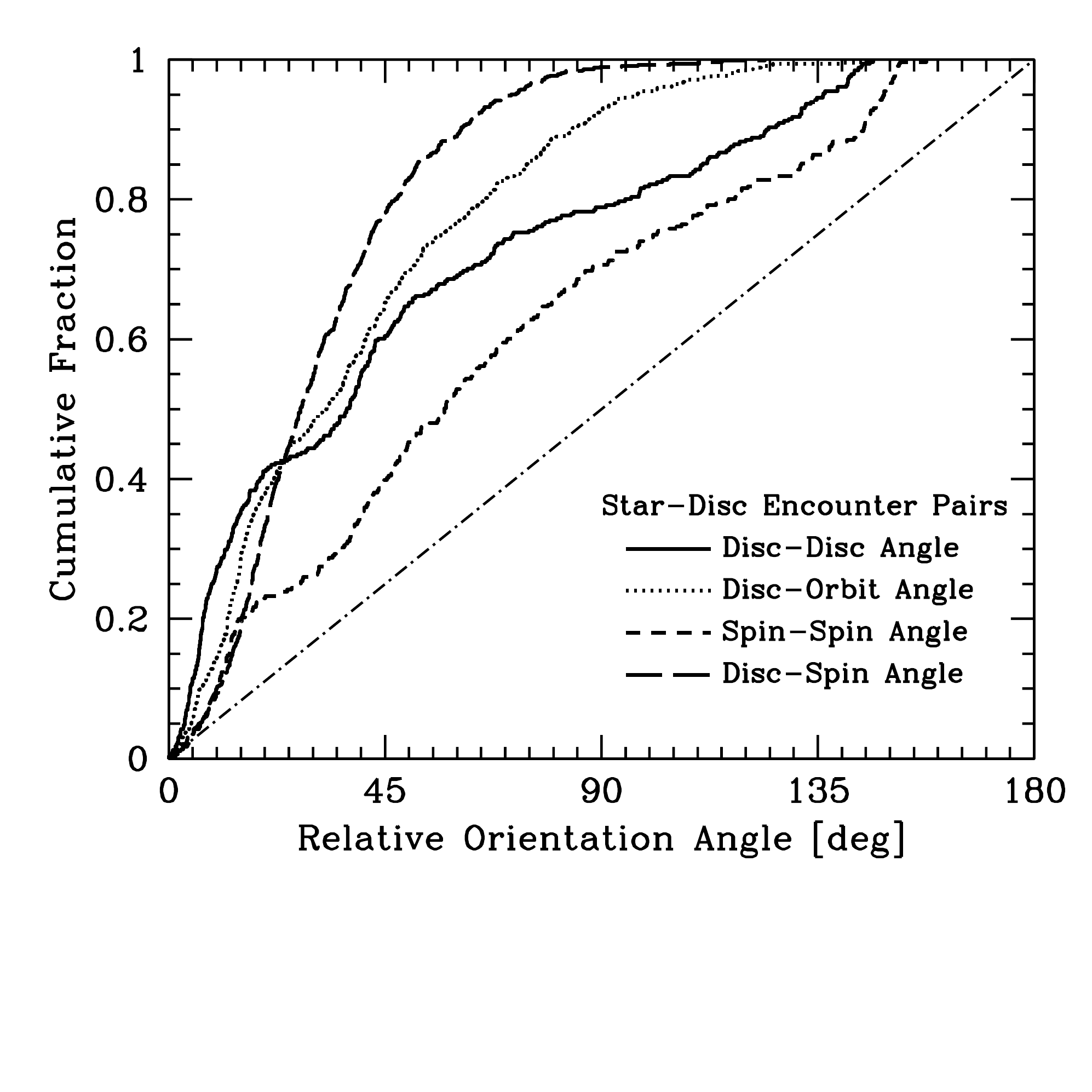} \vspace{-2cm}
\caption{For the 32 bound protostellar pairs formed by star-disc encounters, we plot the cumulative distributions of the relative orientation angle between the two circumstellar discs (disc-disc), the circumstellar discs and the orbit (disc-orbit), the two sink particle spins (spin-spin), and the circumstellar disc and spin of the associated protostar (disc-spin).  The spin-spin angles are close to randomly distributed, while other relative angles show stronger tendencies for alignment.  Circumstellar discs and protostellar spins almost always rotate in the same sense. }
\label{cum_star_disc_angles}
\end{figure}

\begin{figure*}
\centering \vspace{-0.5cm} \hspace{0cm}
    \includegraphics[width=8.5cm]{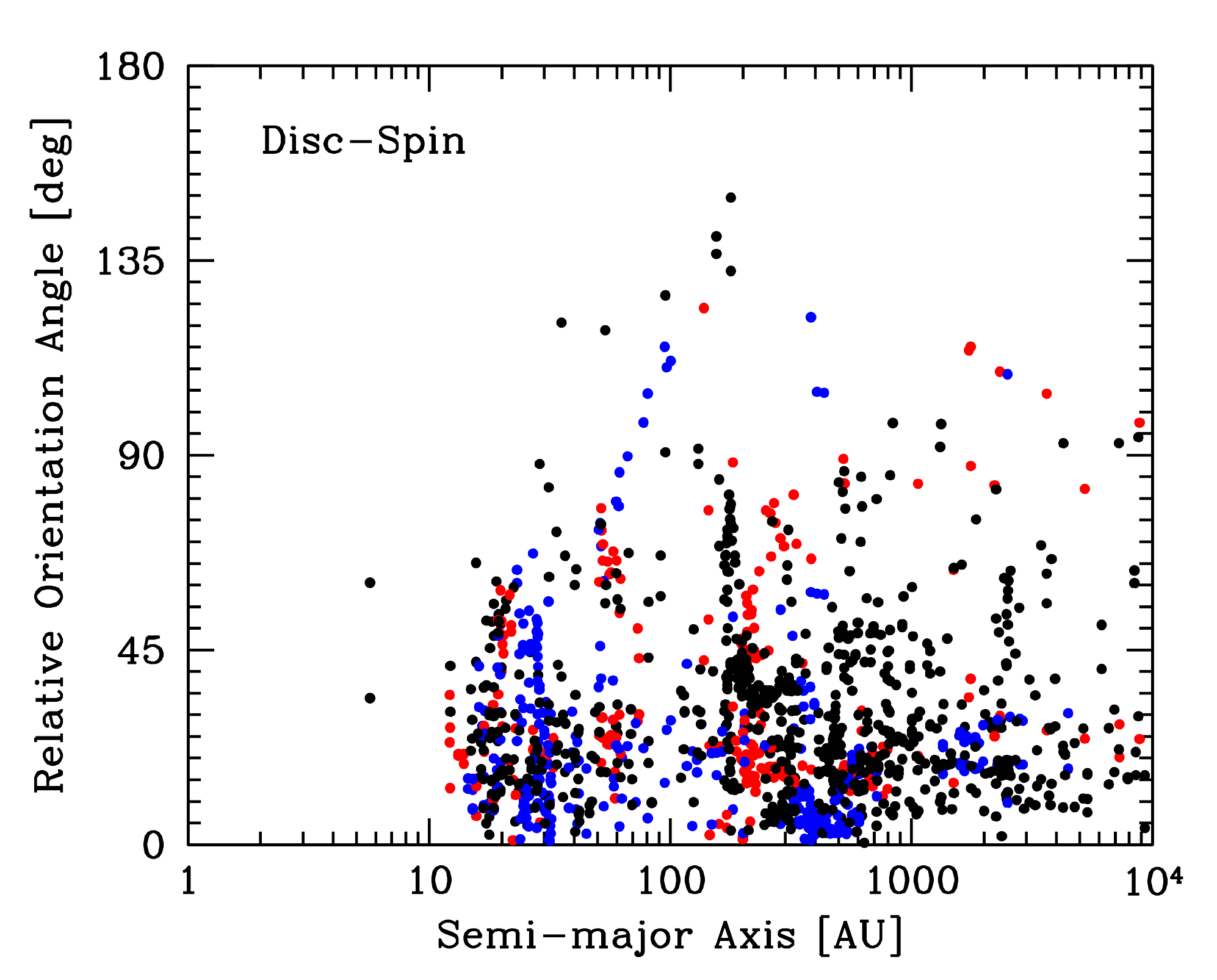} \vspace{0cm}
    \includegraphics[width=8.5cm]{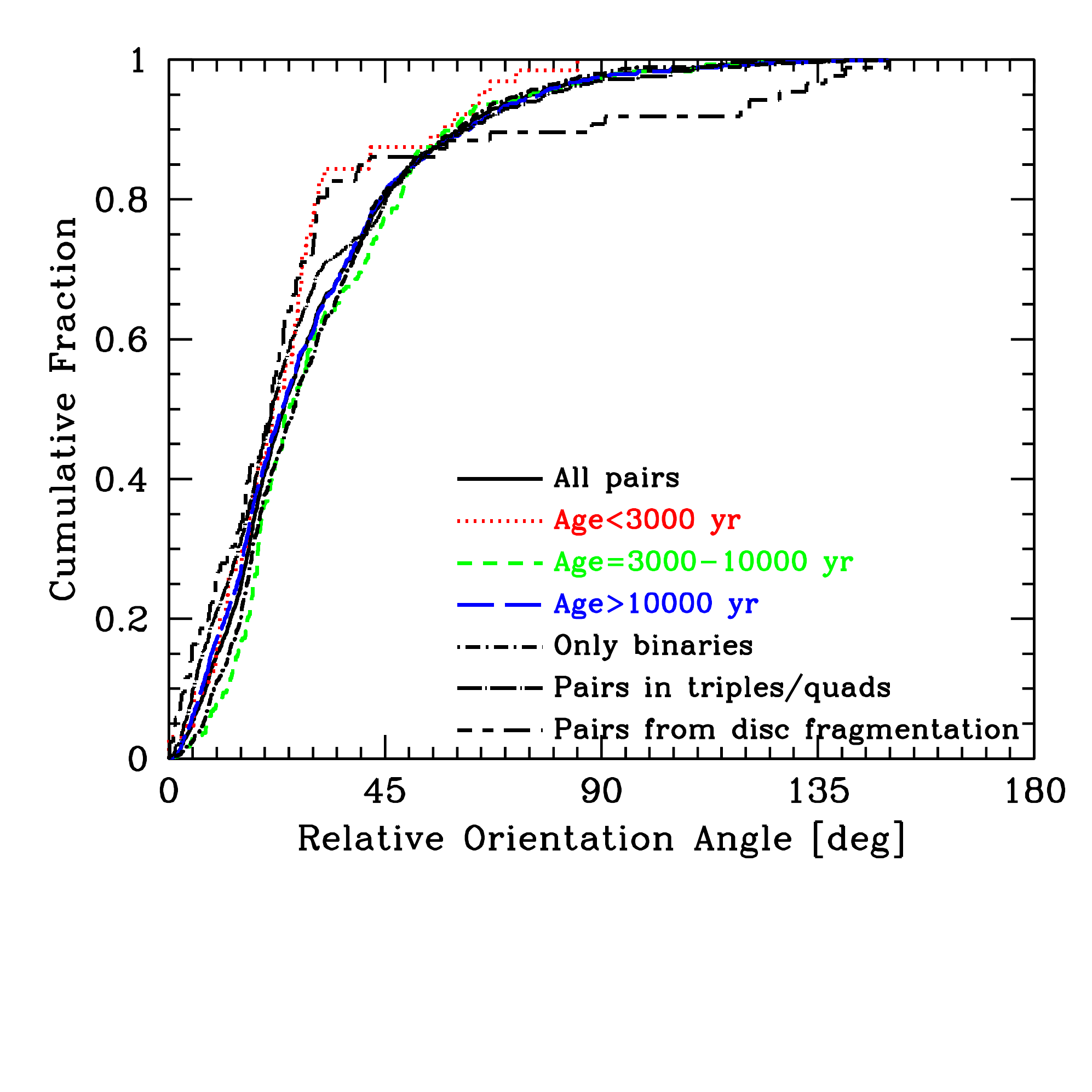} \vspace{0cm}
    \includegraphics[width=8.5cm]{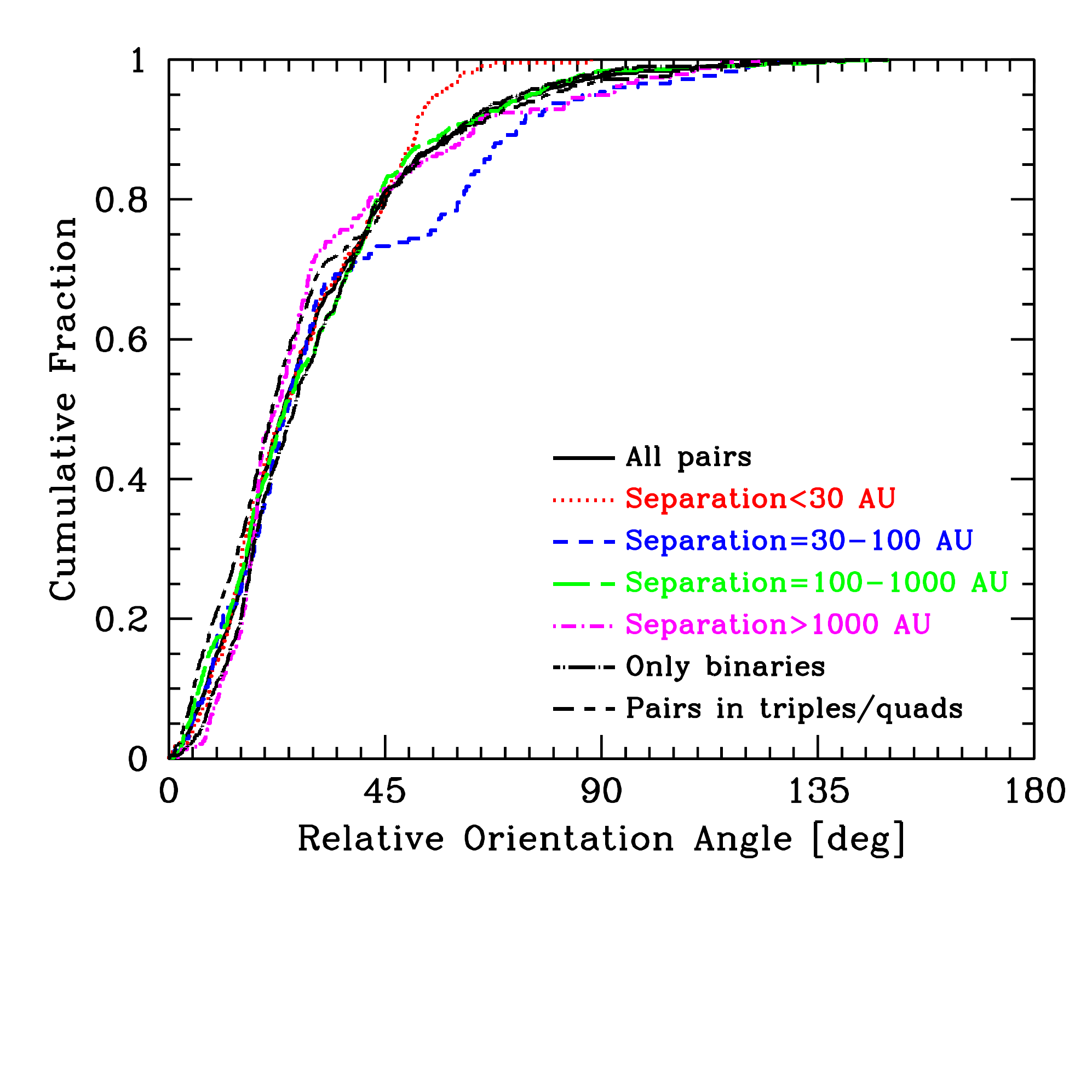} \vspace{0cm}
    \includegraphics[width=8.5cm]{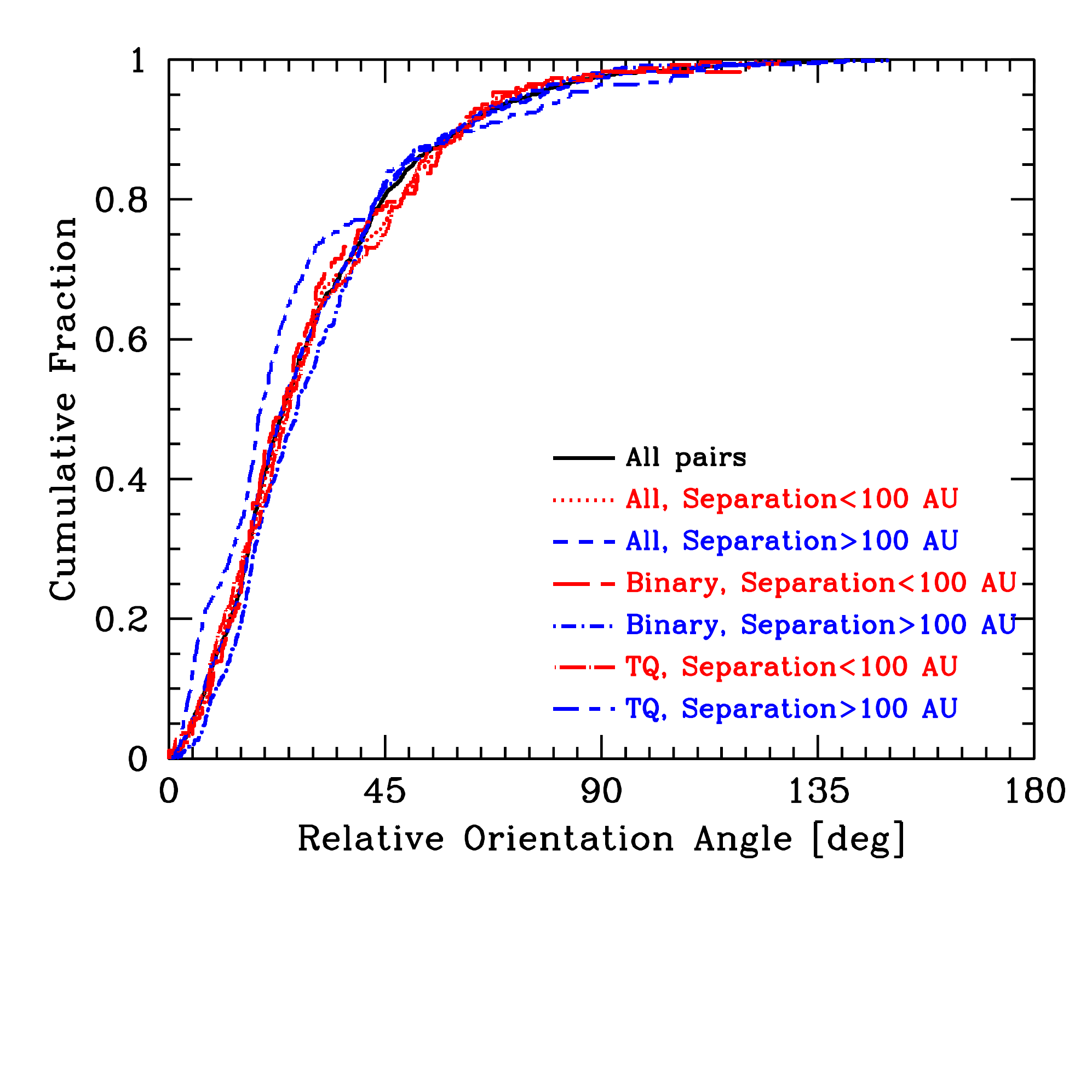} \vspace{-1.5cm}
\caption{Distributions of the relative orientation angle between the circumstellar discs and sink particle spins in bound protostellar pairs.  Pairs include both binaries and bound pairs in hierarchical triple or quadruple systems.  In the top left panel, we plot the relative orientation angle of each pair versus its semi-major axis, with binary systems in black, pairs in triples in red, and pairs in quadruples in blue.  In the remaining panels, we give the cumulative distributions of the orientation angles.  In the top-right panel, we give the cumulative distributions for all pairs and pairs in three different age ranges.  We also give separate distributions for binary systems, for pairs in triple or quadruple systems, and for pairs for which at least one of the components was created by disc fragmentation.  In the bottom-left panel, we plot the cumulative distributions for four ranges of semi-major axes for all pairs, and we also plot the separate distributions of binaries, and pairs that are components of triples or quadruples.  In the bottom-right panel, we plot the cumulative distributions for semi-major axes $a<100$~au and $a>100$~au for all pairs, binaries, and pairs that are components of triples or quadruples.  Contrary to the trends for the relative orientations circumstellar discs with discs or orbits or sink particle spins with orbits, there is very little evolution of the disc-spin cumulative distributions with age and very little dependence on the semi-major axis of the pair or whether the pair is a binary or a member of a higher-order multiple system.}
\label{disc_spin_cum}
\end{figure*}

\begin{figure*}
\centering \vspace{-0.0cm} \hspace{0cm}
    \includegraphics[width=8.5cm]{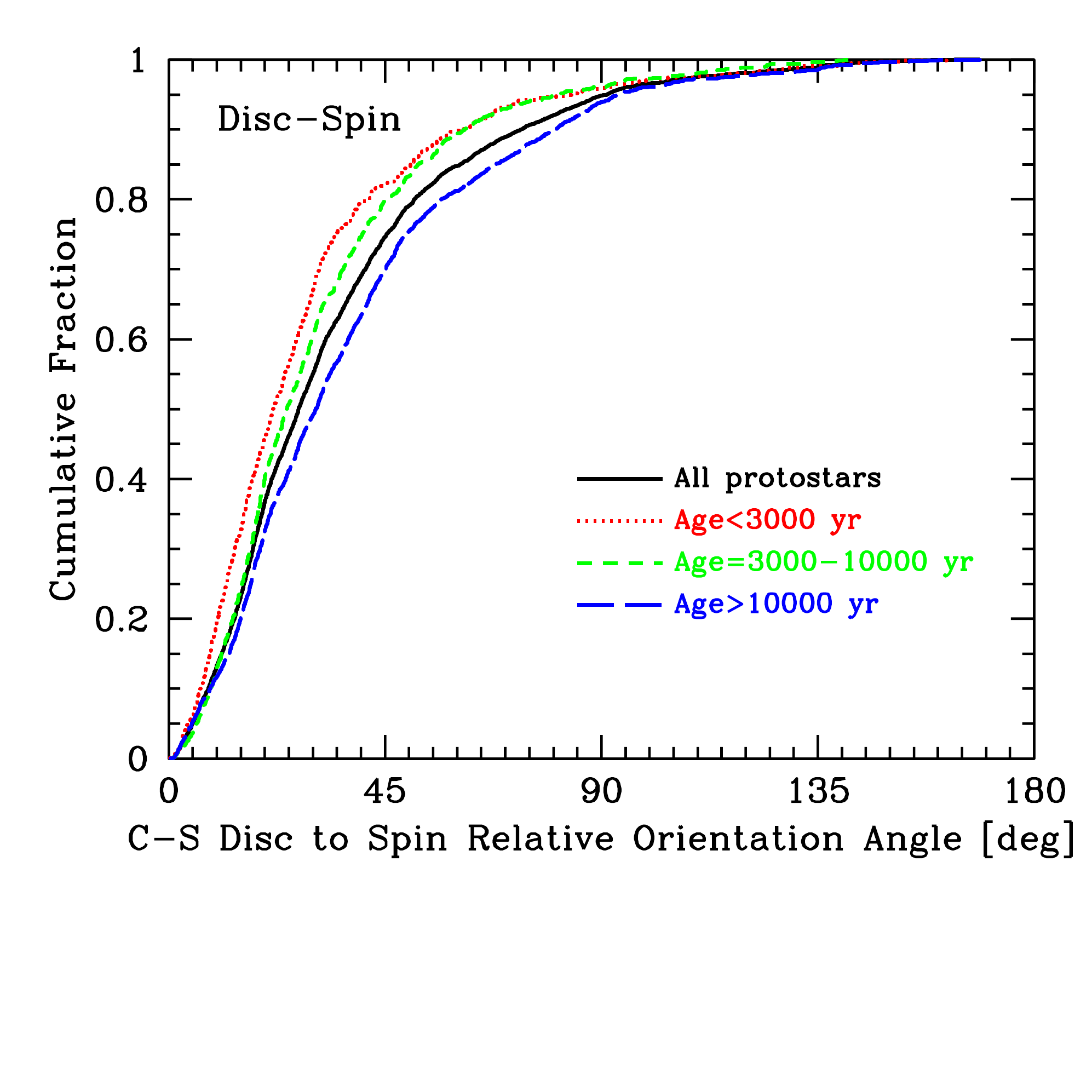} \vspace{0cm}
    \includegraphics[width=8.5cm]{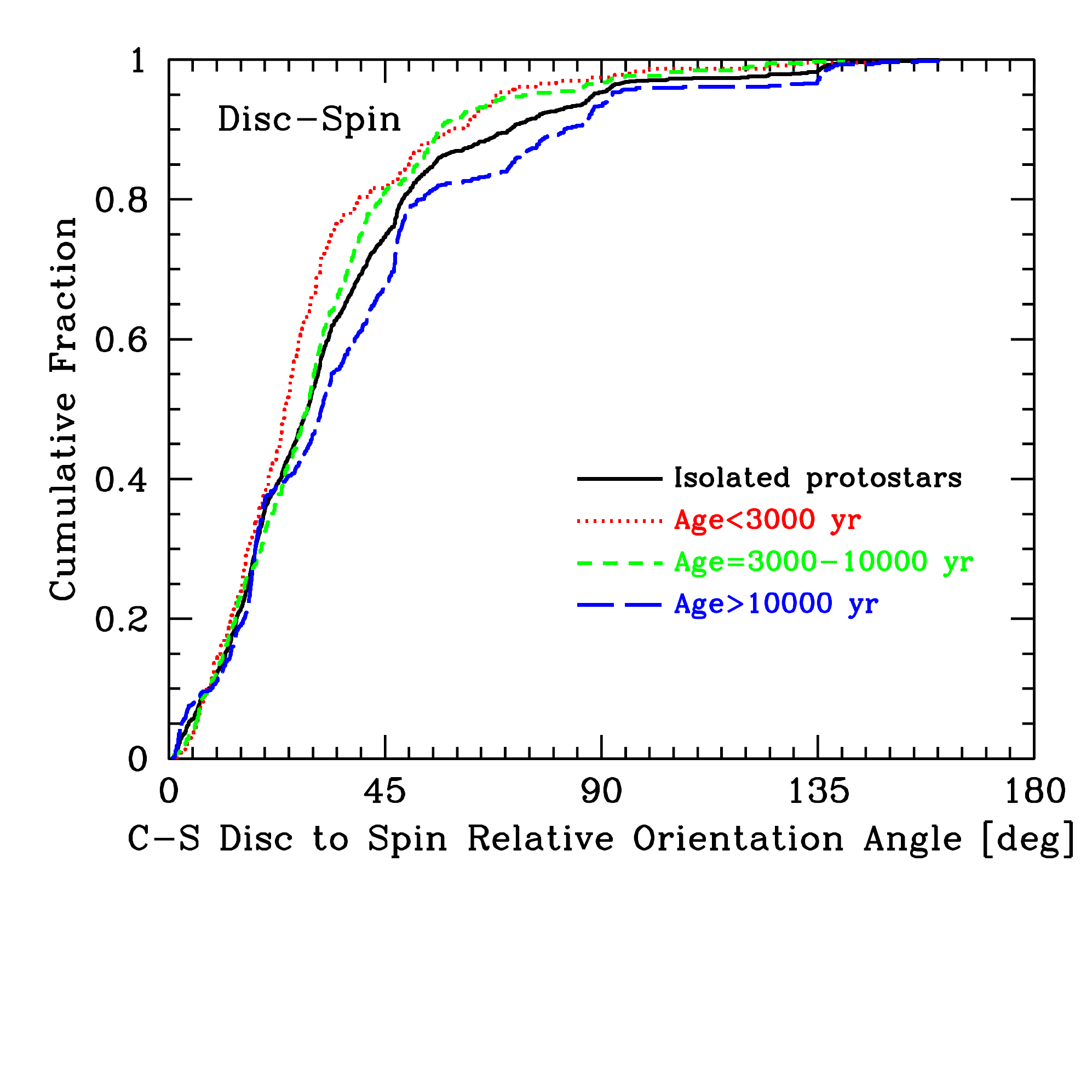} \vspace{-1.5cm}
\caption{Cumulative distributions of the relative orientation angle between the circumstellar discs and sink particle spins for protostars.    In the left panel, we give the cumulative distributions for instances for all protostars.  In the right panel, we consider only isolated protostars (those without neighbours within 2000 au). In both cases, we also consider three different age ranges.  The distributions do not depend greatly on whether all protostars, isolated protostars, or protostars in pairs are considered.  The relative orientations do depend weakly on age, with younger systems being more closely aligned than older systems.  At these young ages, even though circumstellar discs and protostellar spins have a strong tendency for alignment, 50\% are misaligned by more than $\approx 30^\circ$. }
\label{csdisc_spin}
\end{figure*}

\subsubsection{Disc orientations in binary systems}

\cite{Bate2012} examined the relative orientations of the orbits of triple systems, and also the orientations of sink particle spins relative to each other and to the orbital plane in binary systems at the end of the calculation.  Sink particle spins can be thought of as modelling the combined angular momentum of the stars themselves and the inner part of their circumstellar discs (i.e.\ radii smaller than the accretion radii $\le 0.5$~au).  \cite{Bate2012} found that the spins and orbit in binary systems tend to be aligned with each other if the semi-major axis is $a \lsim 30$~au, as is also true observationally \citep{Hale1994}.

In this section, we analyse how the relative orientations of discs, spins, and orbits depend on separation and age.  We restrict our analysis to bound pairs of protostars in which circumstellar discs have been identified around {\em both} of the protostars.  By pairs, we mean that they may be binaries, or they may be mutual closest neighbours in multiple systems (e.g. the closest pair in a triple system with a wider companion, or a pair in a quadruple system of which there may be one or two).  Each circumstellar disc must have a mass of $M_{\rm d} \geq 4.3\times 10^{-4}$~M$_\odot$ (i.e. at least 30 SPH particles).  This may seem like a low value, but all we need to determine is the angular momentum vector of the disc and 30 particles are sufficient.  The trends that we find do not change if we either decrease the limit to 10 SPH particles, or increase the limit to 100 SPH particles, but in the latter case we are left with fewer instances of discs in close pairs because the circumstellar discs in these systems are dynamically constrained to be small (and, thus, typically low mass).  In the analysis that follows, we have 653 instances of pairs with two circumstellar discs in 71 distinct systems. Of these, 390 instances are binaries in 55 different systems, and 263 are instances of 34 different pairs in higher order systems.  Note that a particular pair of protostars may be a component of a high-order system for one period of time, and a binary at a different time (e.g. the outer component of triple system may be dynamically unbound).

We begin by considering the distributions of the relative orientation angle between the two circumstellar discs. We find no significant dependence on the total mass of the protostellar system, but we do find that the relative angle depends on the separation of the pair and on the age of the system.  In Fig.~\ref{disc_disc_cum}, we plot the relative orientation angle for instances of protostellar pairs versus their separation (semi-major axis) and various cumulative distributions for all pairs and for subsets depending on their ages, separations, and separating binaries and pairs in higher-order multiple systems.  The circumstellar discs tend to be more aligned with each other in tighter pairs (left panels of Fig.~\ref{disc_disc_cum}).  A clear progression is seen in the cumulative distributions in the bottom left panel from systems with separations $a>1000$~au to $a<30$~au.  Systems with separations $a \lsim 100$~au have a strong tendency for alignment.  The dependence on separation is likely due to two main effects.  First, with typical disc sizes $a \lsim 100$~au, disc fragmentation tends to produce a larger fraction of close systems than wide systems, and it is expected that the circumstellar discs resulting from such fragmentation will be well aligned.  This is indeed the case in the simulation.  In the top right panel of Fig.~\ref{disc_disc_cum} the long-short dashed line gives the cumulative distribution of disc-disc orientation angles for pairs for which at least one of the components was created by disc fragmentation and it is clear that the vast majority of the circumstellar discs in these pairs are well aligned.  Three quarters of these pairs have separations less than 100 au, and all have separations less than 220 au. Second, the orbital timescale is much shorter for closer systems, so the gravitational torques acting on the discs that acts to align the discs with the orbit will occur on a shorter timescale. Indeed, the circumstellar discs also become more aligned with increasing age (top right panel of Fig.~\ref{disc_disc_cum}).  Not only will gravitational torques acting on the discs tend to align the discs with the orbital plane, but if a binary is formed with misaligned discs, further accretion of gas from outside the system will also tend to align the two discs.  The bottom two panels of Fig.~\ref{disc_disc_cum} show that circumstellar discs of pairs tend to be more closely aligned in high-order multiple systems than in binaries.  This is likely related to their formation.  A significant number of binary systems form via star-disc encounters (Section \ref{sec:stardisc}) in which the circumstellar discs are usually misaligned.  On the other hand, a significant number of pairs in higher-order multiple systems originate from disc fragmentation, in which it is natural for the resulting circumstellar discs to be aligned with the orbit.  The sense of this dependence of the relative orientation on whether the pair is a binary or a component of a higher-order multiple system is the same for both close systems (separations $<100$~au) or wide systems (separations $>100$~au), but it is stronger for wider systems (bottom right panel of Fig.~\ref{disc_disc_cum}).

Next we consider the distributions of the relative orientation angles of the circumstellar discs and the orbital plane of pairs.  In Fig.~\ref{disc_orbit_cum}, we plot the same quantities as we plotted for the relative orientations angles of the two discs in Fig.~\ref{disc_disc_cum}.  Note that there are two values for each pair since there are two circumstellar discs.  Compared to the disc-disc alignment, we find that the discs in close systems are slightly better aligned with each other than with the orbit, but that for wide systems there is a greater fraction of highly misaligned discs than there are discs that are highly misaligned with orbits.  For example, only $\approx 7$\% of discs are misaligned by more than $90^\circ$ relative to the orbit, while $\approx 15$\% of discs are misaligned with each other by more than $90^\circ$.  There is less dependence of the disc-orbit relative orientation angles on either age (top right panel of Fig.~\ref{disc_orbit_cum}), or separation (bottom left panel of Fig.~\ref{disc_orbit_cum}), or whether the pair is a binary or a component of a higher-order multiple system (bottom right panel of Fig.~\ref{disc_orbit_cum}) than for disc-disc alignment.  Together these relations indicate that it is probably the way the binary formed (e.g. disc fragmentation, star-disc encounter, etc) and the subsequent accretion of gas that are most important for the tendency for alignment that is seen in these young protostellar systems, rather than realignment of the discs with the orbital plane via gravitational torques.  However, realignment would be expected to have significant effects on longer timescales.

In Fig.~\ref{spin_spin_cum} we examine the relative orientation angles of the spins of the sink particles of pairs.  Recall that these can be thought of as providing the angular momenta of the protostar and the inner part of its disc.  The same dependencies on age, separation, and the multiplicity of the system are seen for spin-spin alignment as for disc-disc and disc-orbit alignment.  However, overall, the spins tend to be less well aligned with each other than the discs are with each other, or than the discs are aligned with the orbit.  For example, $\approx 25$\% of spins are misaligned by more than $90^\circ$.  The spins will tend to trace the angular momentum of the material that protostar first formed from better than either the discs or the orbital angular momentum do.  If, for example, the protostars formed in relative isolation from each other and then became bound during a star-disc encounter, the spins would generally be expected to be more misaligned with each other than the circumstellar discs because the discs would suffer gravitational torques during the encounter whereas the spins can only be affected by accretion. Thus, it is to be expected that the spins are less well aligned than the circumstellar discs.  In Fig.~\ref{cum_star_disc_angles} we check whether or not this is the case by plotting the cumulative distributions of the disc-disc, disc-orbit, spin-spin, and disc-spin orientation angles for 32 bound pairs that are formed by star-disc encounters. The relative orientations of the protostellar spins are not quite randomly distributed as there is an excess of systems with relative angles $<20^\circ$ and a deficit above $150^\circ$, but between these values the distribution is roughly uniform.  As expected the disc-disc and disc-orbit orientation angles show a greater tendency for alignment.  Finally, we note that the relative orientation angles of the circumstellar discs and the spins of the sink particles are almost all $<90^\circ$ (i.e. the discs and spins rotate in the same sense) and 80\% have relative angles $<45^\circ$.

In Fig.~\ref{disc_spin_cum} we consider the relative orientation angles of the circumstellar discs and the spins of the sink particles for all bound protostellar pairs.  It is no surprise that there is a strong preference for alignment, since the spins nominally represent the angular momentum of the protostar itself and the inner part of the disc, and the sink particles accrete from the discs.  However, in contrast to the relative orientation angles of the other components, there is essentially no dependence on age, separation, or multiplicity.  This seems to indicate that although the protostars are accreting from their discs this does not lead to appreciable realignment of the sink particle spin with the disc, at least over timescales of $\sim 10^4$~yrs.  

For comparison with Fig.~\ref{disc_spin_cum}, in Fig.~\ref{csdisc_spin} we plot the cumulative distributions of the relative orientation angles between resolved discs and sink particle spins for all circumstellar discs (4822 instances) and for circumstellar discs around isolated protostars (1226 instances).  Again we have limited the analysis to circumstellar discs that are represented by more than 30 SPH particles. The distributions are very similar to those of circumstellar discs and sink particle spins in bound pairs.  However, looking at all protostars or isolated protostars we can see some evolution with age.  In both cases, the discs and the spins are more closely aligned at {\em younger} ages and they become (slightly) less well aligned at older ages.

The results for the relative orientation angles between circumstellar discs and sink particle spins can be understood if the angular momentum of most resolved discs is incessantly being changed. The spins never `catch up' by accreting from the disc because the larger-scale disc is continually being reorientated.  In fact, statistically speaking, the discs and spins tend to be better aligned at young ages (presumably because the protostar and its young disc have originated from a relatively small, coherent volume of gas) and become less well aligned with increasing age.  Indeed, if one examines specific cases, accretion, dynamical encounters, ram-pressure stripping, etc, all act to cause the relative orientation angle to change with time.  Misalignment of the large-scale discs with respect to the inner part of the disc and the protostellar spin naturally has implications for the misalignment between planetary orbits and the rotation axes of their host stars.  \cite{BatLodPri2010} and \cite{Fielding_etal2015} have both considered the effects of accretion from turbulent clouds on the relative orientation of discs and stellar spins.  The difficulty with these studies is in trying to predict how the relative orientation angle evolves from ages of $\sim 10^5$~yrs to the ages when planets are thought to form (i.e. $\sim 10^6$~yrs).  \citealt{BatLodPri2010} highlighted the importance of reorientation of the inner disc with the star due to warp propagation and also considered the effects of dynamical encounters.  \cite{Fielding_etal2015} considered star-disc realignment due to gravitational quadrupole moments. \cite{Lai2014} has also studied the effects of magnetic star-disc interaction torques on star-disc misalignment. Neglecting these effects, \cite{Fielding_etal2015} studied 14 protostars from their hydrodynamical and magnetohydrodynamical calculations, sampled at multiple times, and found that approximately 50\% had misalignment angles in excess of $30^\circ$.  We have an almost identical result, but with an order of magnitude more protostars.

\section{Comparison with observations and further discussion}
\label{sec:discuss}

In this section, we compare the statistical properties of the discs obtained in the previous section with the statistical properties of observed discs around young stars.  In doing so, it is important to keep in mind the limitations of the calculation analysed in this paper.  We discuss two major limitations below; further limitations are discussed in Section \ref{sec:limitations}.

A major limitation is that the calculations do not follow the evolution of the discs for very long -- even the oldest discs have ages $<10^5$~yrs, and most have ages $\sim 10^4$~yrs.  Protostars with such ages are usually thought of as being Class 0 objects.  However, fundamentally, Class 0 objects are those that still have substantial envelopes \citep{AndWarBar1993} -- this will be {\em more common} for young objects than older objects, but Class 0, I, and II objects do not necessarily form a neat age sequence.  \cite{Kurosawaetal2004} found that even at an age of $\approx 10^5$~yrs, a star-forming region can have a mixture of objects ranging from Class 0 to Class III \citep[see also][]{Offneretal2012}.  Objects identified by \citeauthor{Kurosawaetal2004} as having later types had typically been involved in dynamical interactions that expelled them from dense regions of molecular gas and/or stripped their discs.   We see such effects in the calculation studied here too, with a substantial increase in the number of objects without resolved discs with increasing age.

A second limitation is that there is no accounting for the different evolution of dust and gas.  This is important because observational determinations of disc masses and radii are usually based on dust continuum emission at (sub-)mm wavelengths.  
In Class II objects where both the gas and dust are observed it is common for the radius of the gas disc to be larger than that of the (sub-)mm dust disc \citep[e.g.][]{PieDutGui2007,Pietu_etal2014,Isella_etal2007,Panic_etal2009,Andrews_etal2012,Rosenfeld_etal2013,deGregorioMonsalvo_etal2013,Walsh_etal2014,Pineda_etal2014,Cleeves_etal2016,Barenfeld_etal2017}.  For example, in the Class II object IM Lupus, the gas disc extends to $\approx 1000$~au while the mm dust disc is truncated at $\approx 300$~au \citep{Cleeves_etal2016}.  This effect is expected due to dust growth \citep{GolWar1973,Weidenschilling1980,CuzDobCha1993,DulDom2004} and inward radial migration of large grains \citep{Whipple1972,Weidenschilling1977}.  Evolutionary models of isolated dusty discs show that this has appreciable effects on the outer parts of discs on timescales of $\sim 10^{4-5}$~yrs \citep{BirAnd2014, PinLai2014,Andrews_etal2016}.

Nevertheless, the fact that all of the systems discussed in this paper are young minimises the expected differences between the distributions gas and dust.  Although the masses, radii and surface density profiles discussed in Section \ref{sec:stats} are formally those of the gas, because the systems are young (typical ages less than a few $\times 10^4$~yrs) and are often still accreting gas from the molecular cloud, the differences between the gas and dust distributions are likely much less than for a typically Class II object.  On the other hand, \cite{TsuOkuKat2017} find that even for Class 0/I objects the disc masses that are derived from dust emission (when scaled by the nominal gas-to-dust ratio of 100) may be factors of 3--5 lower than the actual gas mass.  \cite{BatLor2017} showed that if grains with sizes $>10~\mu$m are present in the pre-stellar core, the dust and gas distributions can differ even during the initial collapse before the protostar forms.  Despite these effects, deriving gas masses from dust masses currently appears to be more accurate than deriving the total gas mass from molecular emission such as CO \citep[e.g.][]{Bruderer_etal2012,Favre_etal2013, Bergin_etal2014, Kama_etal2016a,McClure_etal2016,Miotello_etal2017,Yu_etal2017}.

With these limitations in mind, in the following sections we compare the statistical properties of the discs obtained from the calculation with the statistical properties of observed discs. In Section \ref{sec:classII} we begin with observations of Class II objects because, although they will typically be much older than the discs analysed in this paper, Class II systems have been much better studied than systems with earlier types to date.  In Section \ref{sec:class0I} we examine the current status of the statistics of Class 0/I discs.  In Section \ref{sec:limitations} we discuss further limitations of the calculation analysed in this paper, and we speculate on what we may be able to learn from future studies of the statistical properties of discs.

\subsection{Comparison with Class II disc statistics}
\label{sec:classII}

Disc masses have been estimated from dust emission since the late 1980s \citep[e.g.][]{Beckwith_etal1986,Beckwith_etal1990}.  However, for a couple of decades, the only large sample of resolved circumstellar discs was that of the sillouhette discs and proplyds in the Orion Nebula Cluster \citep{OdeWenHu1993,McCODe1996,McCStaClo2000}.  The Hubble Space Telescope resolved these Class II discs down to radii of $\approx 40$.  The largest disc has a radius of $\approx 1000$~au, but the typical radius of resolved discs is $\sim 100$~au \citep{VicAlv2005}. 

Over the past decade, improvements in (sub-)millimetre resolution have allowed statistical studies of disc masses and radii to be carried out in many nearby star-forming regions, including Ophiuchus and Taurus \citep{Andrewsetal2009,Andrews_etal2010,Andrews_etal2013, Tripathi_etal2017}, Lupus \citep{Ansdell_etal2016,Tazzari_etal2017}, Upper Scopius \citep{Barenfeld_etal2016}, Chamaelon I \citep{Pascucci_etal2016}, and $\sigma$ Orionis \citep{Ansdell_etal2017}.  These studies examine Class II objects which will usually be are at more advanced evolutionary stage than the protostars we consider in this paper.  However, it is still instructive to compare these Class II populations with the sample of discs presented here.  In particular, a number of empirical trends have been found from the above studies.

\subsubsection{Disc masses}

In this section we use dust masses when referring to the masses of discs since the observations we discuss are of dust emission.  When discussing the simulated discs, we convert the gas masses into dust masses using the standard dust to gas ratio of 1:100.  

The first large sample of disc masses derived from millimetre wavelengths was carried out for the Taurus-Auriga dark clouds by \cite{Beckwith_etal1990}.  They detected dust discs around 42\% of their sample and obtained dust masses ranging from a few times $10^{-5}$~M$_\odot$ to $7 \times 10^{-3}$~M$_\odot$ (i.e. $\sim 10-2000$~M$_\oplus$) with an average mass of $\sim 10^{-4}$~M$_\odot$ (i.e. a gas mass $\sim 10^{-2}$~$M_\odot$), and disc-to-star mass ratios less than unity.  They did not find any dependence of disc mass on stellar age.

\begin{figure}
\centering \vspace{0cm} \hspace{0cm}
    \includegraphics[width=8.5cm]{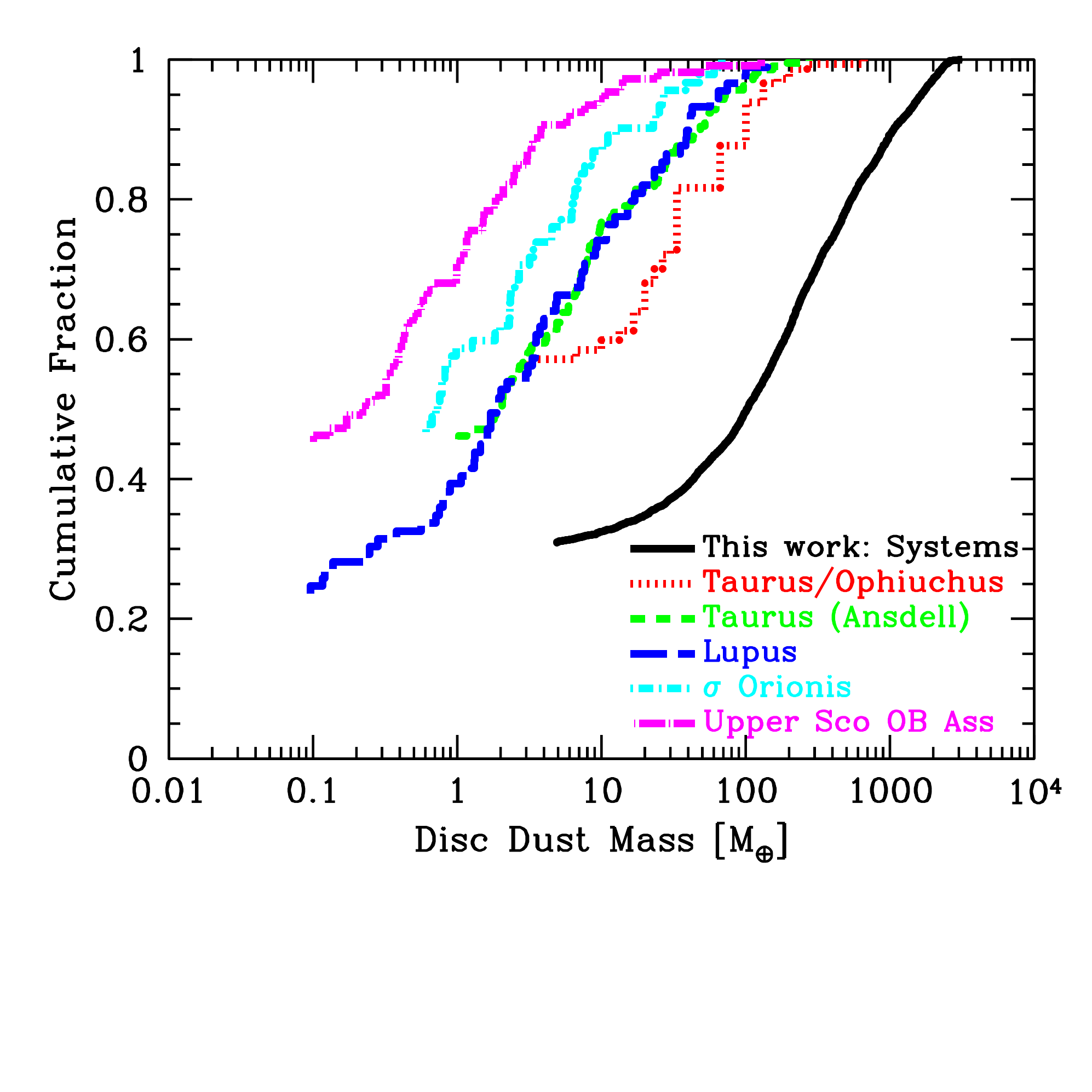} \vspace{-0.25cm}
\caption{Cumulative distributions of the disc dust mass for the discs of protostellar systems from the calculation analysed in this paper (solid line), and for discs of Class II objects observed in different star-forming regions. The observational surveys are of Taurus/Ophiuchus \citep{AndWil2007}, the reanalysis of Taurus data \citep{Andrews_etal2013} by  \citep{Ansdell_etal2016},  Lupus \citep{Ansdell_etal2016}, $\sigma$~Orionis \citep{Ansdell_etal2017}, and the Upper Scorpius OB association \citep{Barenfeld_etal2016}.  As may be expected, the young discs from the hydrodynamical simulation have higher masses than those that are typically observed in star-forming regions.  The simulated discs are approximately 1.5 orders of magnitude more massive than those in Taurus and Lupus, 2 orders of magnitude more massive than those in $\sigma$~Orionis, and 2.5 orders of magnitude more massive than those in Upper Sco. 
}
\label{obs_mass}
\end{figure}

To compare the disc masses from the hydrodynamical calculation with those of observed Class II objects, we consider the statistics from more recent surveys in different regions: the Taurus and Ophiuchus regions \citep{AndWil2007}, the Lupus star-forming region \citep{Ansdell_etal2016}, the $\sigma$ Orionis region \citep{Ansdell_etal2017}, and the Upper Scorpius OB association \citep{Barenfeld_etal2016}.  For Taurus, we also consider the reanalysis of the \cite{Andrews_etal2013} dataset by \citep{Ansdell_etal2016} because the latter used a consistent method of analysis for both the Lupus and Taurus datasets.

In Fig.~\ref{obs_mass} we plot the cumulative distributions of disc dust masses from the simulation, and from the above observational studies.  To make these cumulative distributions, we have simply taken all upper limits as being zero mass and stopped the lines at the lowest detection.  This is not the best way to treat upper limits (see any of the observational papers), but it is sufficient for our purposes and directly comparable to the way we treat discs are not resolved in the simulation.

The mean disc masses of the disc samples in Taurus, Lupus, Chamaeleon I, $\sigma$~Orionis, and Upper Scorpius are $\langle M_{\rm dust} \rangle \approx 15, 15, 13, 7, 5, $~M$_\oplus$, respectively, while the median disc masses are $\approx 3, 3, 2, 2, 0.3$~M$_\oplus$, respectively \citep{Ansdell_etal2017}.  The disc masses in Taurus, Lupus, and Chamaeleon I seem similar, with those in $\sigma$~Orionis a little lower. Upper Scorpius seems to have disc masses than are a factor of 5 lower than Taurus \citep{Barenfeld_etal2016}.  Similar results are obatained by \cite{Pascucci_etal2016}, who also find that Upper Scorpius may have a steeper $M_{\rm dust}-M_*$ relation than the other regions.  \cite{Tazzari_etal2017} reanalysed the more luminous discs in Lupus studied by \cite{Ansdell_etal2016}, excluding unresolved discs, transition discs, and known binaries.  Whereas \cite{Ansdell_etal2016} assumed a constant temperature of 20~K to derive the dust masses, \cite{Tazzari_etal2017} used a varying temperature model and obtained dust masses that were typically a factor of two higher than \cite{Ansdell_etal2016}.  This is consistent with the masses for Taurus being higher in \cite{AndWil2007} than in \cite{Ansdell_etal2016}. Similarly, using synthetic observations of protostellar disc simulations, \cite{DunVorArc2014} conclude that disc masses derived from observations at millimetre wavelengths can lead to disc mass underestimates by up to factors of two or three. On the other hand, the dust masses derived by \cite{Miotello_etal2017} tend to be $1-2$ times smaller than those of \cite{Ansdell_etal2016}.  Overall, there is currently uncertainty in dust masses derived from observations at the level of factors of a few.

From the cumulative distributions in Fig.~\ref{obs_mass}, the masses of our resolved discs are $\sim 30$ times more massive than those of the Class II discs in Taurus/Ophiuchus and Lupus.  It is not surprising that the masses are higher, since the objects from the simulation are presumably much younger than the observed discs.  In the simulation, the highest disc mass is $M_{\rm dust} \approx 3000$~M$_\oplus$, or $0.01$~M$_\odot$ (i.e. a gas mass of $\approx 1$~M$_\odot$).  Empirically, the `completeness limit' for resolved discs in the hydrodynamical calculation is $\approx 30$~M$_\oplus$ (i.e. a gas mass of $\approx 10^{-2}$~M$_\odot$, or $\approx 700$ SPH particles).  Coincidentally, these limits are similar to those in the original survey of \cite{Beckwith_etal1990}.

\subsubsection{Disc radii}
\label{sec:disc_rad}

The distributions of disc radii are more difficult to study than disc mass because high angular resolution is required.  The radii of discs of Class II objects have been studied in the Orion Nebula Cluster \citep[ONC;][]{VicAlv2005}, Ophiuchus \citep{Andrewsetal2009,Andrews_etal2010}, Lupus \citep{Tazzari_etal2017}, and the Upper Scorpius OB Association \citep{Barenfeld_etal2017}.  In addition, \cite{Tripathi_etal2017} study a collection of 50 discs that are mostly from Taurus and Ophiuchus, but with 9 that are in other regions or in isolation.  The disc radii range from $\approx 40-1000$~au, $\approx 20-200$~au, $\approx 10-400$~au, $\approx 20-200$~au, and $6-50$~au in the five samples, respectively.  All of these studies consider radii based on dust profiles, but in the ONC they are derived from optical dust absorption (discs seen in sillouhette against background nebulosity), whereas in all the other surveys they are based on millimetre dust emission.  \cite{Barenfeld_etal2017} measures both dust and gas radii for seven discs, finding that the radii of the gas discs ($30-170$~au) are larger than those measured using the dust in four of the seven cases. At face value, \citeauthor{Barenfeld_etal2017} find that the dust disc radii in Upper Scorpius are three times smaller than those found in the other regions (median radii of 21~au).

\begin{figure}
\centering \vspace{0.0cm} \hspace{0cm}
    \includegraphics[width=8.5cm]{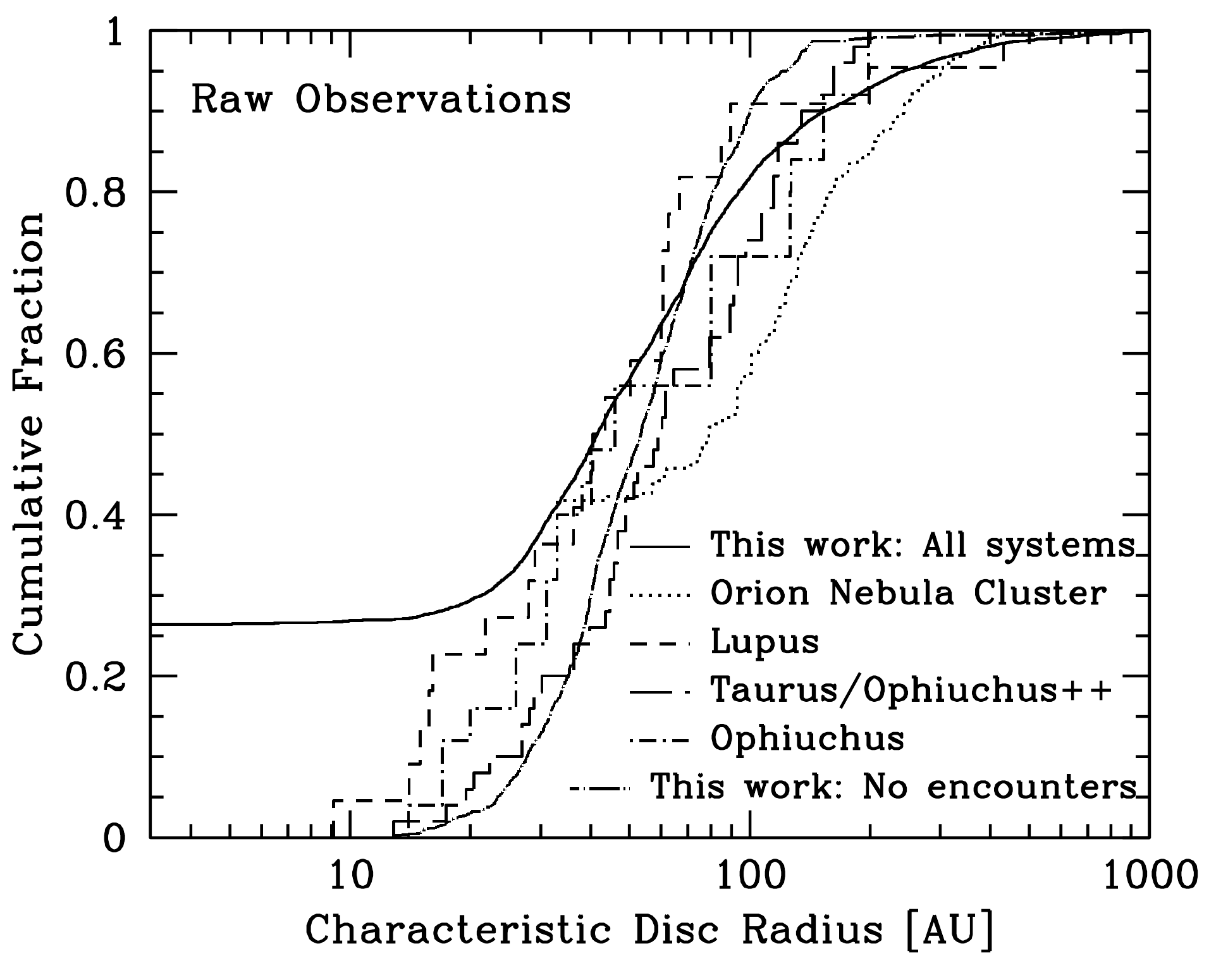} \vspace{0.0cm}
    \includegraphics[width=8.5cm]{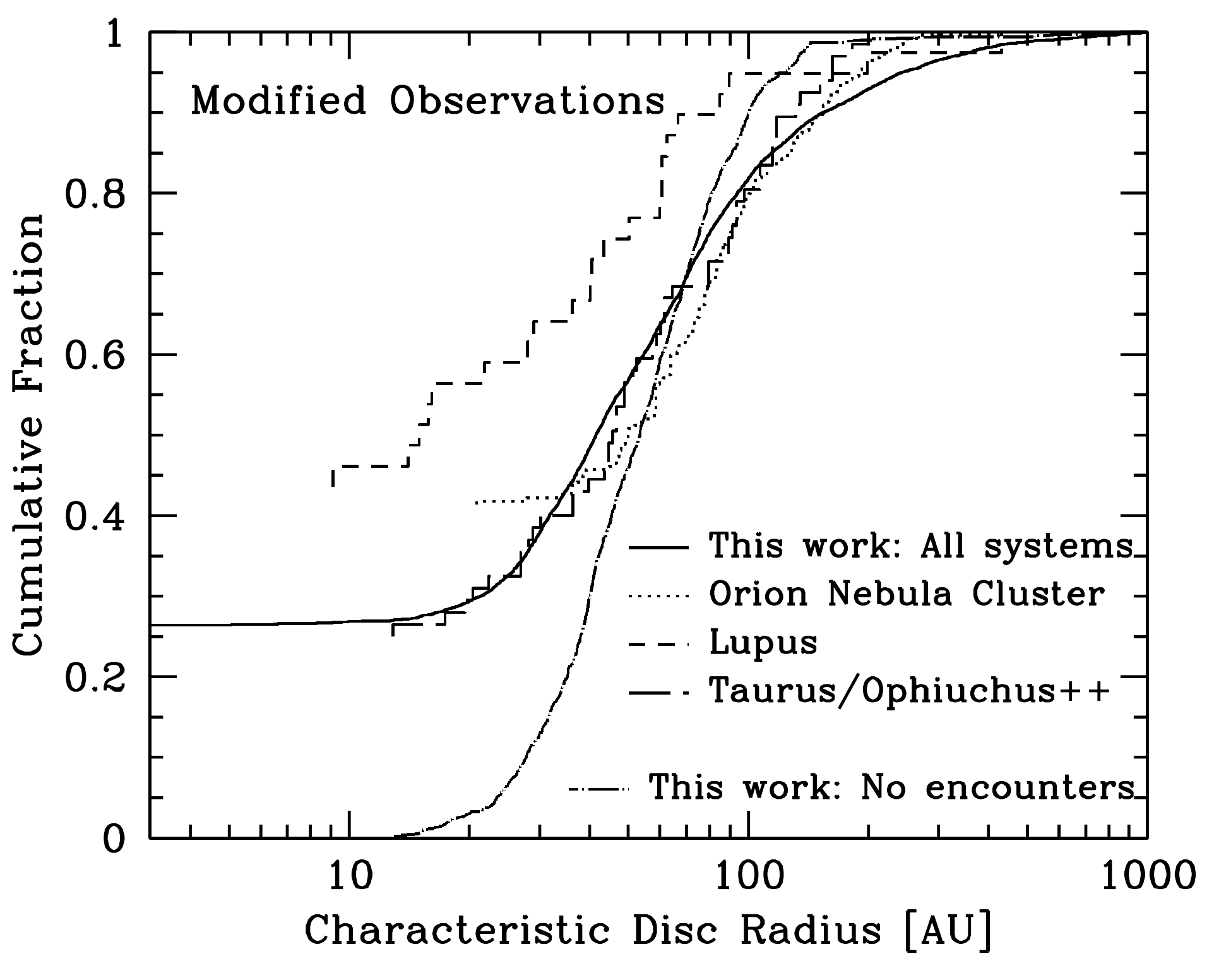} \vspace{0.0cm}
\caption{Cumulative distributions of the characteristic radii for the discs from the calculation analysed in this paper (solid line: all systems; dot-long-dashed line: protostars that have not had encounters closer than 2000 au), and for discs observed in the Orion Nebula Cluster \citep{VicAlv2005}, Lupus \citep{Tazzari_etal2017}, a sample of discs in Taurus, Ophiuchus, and other regions \citep{Tripathi_etal2017}, and Ophiuchus \citep{Andrewsetal2009,Andrews_etal2010}.  In the top panel, we give the raw observed distributions from the above papers.  In the bottom panel, we apply some corrections.  For the Orion Nebula, we scale the radii by a factor of 63.2\% to take account of the fact that the observed radii are from extinction (sihouettes) rather than from dust emission profiles.  For Lupus and Taurus, we attempt to take account of the fact that not all of the radii are able to be determined for the sample.  Our best estimate is that the observed disc have similar sizes to those produced in the numerical simulation; the discs in Orion may be up to a factor of two larger. Excluding protostars that have had encounters or have companions closer than 2000 au results in a steeper distribution.}
\label{obs_rad}
\end{figure}

In the upper panel of Fig.~\ref{obs_rad} we plot the cumulative distributions of disc sizes of the observed samples, excluding that of \cite{Barenfeld_etal2017}, and two distributions derived from the simulation. From the simulation, we plot the distribution obtained using all protostellar systems (solid line) and the distribution obtained only from protostars that have not had encounters with other protostars closer than 2000 au.  The latter is steeper as the largest discs tend to be found in multiple systems (Section \ref{sec:sys_mass_rad}), and dynamical encounters or companions are primarily responsible for producing unresolved discs (Section \ref{sec:isolated}).  At face value, the four observed distributions have median disc radii that range from 1 to 2 times the median radii of the discs of protostellar systems from the calculation (excluding unresolved discs).  The observed disc radii are also in reasonable agreement with the distribution from protostars that have not had encounters, although the latter distribution is somewhat steeper. However, the question becomes how to deal with non-detections and upper limits in the observational surveys. 

\cite{VicAlv2005} provide an estimated correction for the number of unresolved discs in the ONC which we have already used to plot the cumulative distribution in the upper panel of Fig.~\ref{obs_rad}.  But an added complication for the ONC is that the disc radii are determined from optical dust absorption which essentially give the outer radii of the discs, whereas the other surveys and the simulations measure characteristic disc radii that contain $\approx 63$\% of the disc mass.  To account for this, we can reduce the disc radii given by \cite{VicAlv2005} by a factor of 0.632.  If the disc surface density profile is $\Sigma(r) \propto r^{-1}$ (as is typical for the isolated discs in Section \ref{sec:radial}), this would give characteristic radius that contains a similar mass fraction to the other observational surveys and the simulated discs.

In Lupus, \cite{Tazzari_etal2017} give the number of systems that they are unable to determine disc radii for, but there is no indication of completeness in the studies of  \cite{Andrewsetal2009,Andrews_etal2010} and \cite{Tripathi_etal2017}.   \cite{Pietu_etal2014} performed a high angular resolution study of faint discs in the Taurus star-forming region.  They found that all of the faint discs were much smaller than the bright discs that were previously imaged.  They found that half of their discs had characteristic radii smaller than 10~au, and concluded that up to 25\% of the entire disc population of Taurus may consist of very compact dust discs.

Making these adjustments to the observational data for the ONC, Lupus, and Taurus/Ophiuchus datasets, we plot the cumulative distributions of the characteristic disc radii of modified observational data and the simulated discs in the bottom panel of Fig.~\ref{obs_rad}.  Now the characteristic radii of the discs from the simulated protostellar systems seem to be in good agreement with the disc sizes in the ONC and Taurus/Ophiuchus, but about a factor of two larger than the disc radii in the Lupus.  We note that accounting for the incompleteness of the Lupus survey, the median disc size in Lupus may be similar to that recently found by \cite{Barenfeld_etal2017} in Upper Scorpius. We also note that the results from the hydrodynamical simulation and the Lupus and Upper Scorpius results are consistent with \cite{Pietu_etal2014}'s assertion that up to 25\% of the discs in Taurus may be very compact.   The simulated distribution from protostars that have not had encounters closer than 2000 au remains too steep, implying that including multiple systems and at least some dynamical encounters is necessary to reproduce the observed disc size distribution, particularly the population of very small discs.

Given the uncertainties in the observations, particularly in terms of upper limits and sample completeness, agreement at the level of a factor of two is reasonable.  Indeed, there are several reasons why the agreement may not have been expected to be this good.  First, we know from the previous sections that the disc radii in the simulations tend to increase with age.  Second, if real discs evolve viscously, they will also grow in size.  Third, the calculations do not include magnetic fields.  Naively, magnetic fields would be expected to result in smaller discs due to magnetic braking.  We will return to this point in Section \ref{sec:limitations}.

Finally, we note from Fig.~\ref{obs_rad} that the observed distributions of disc radii for the Orion Nebula Cluster and for Taurus/Ophiuchus are very similar, despite the stellar densities being very different.  How can this be the case if dynamical interactions are important in setting disc properties?  This is possible if protostars form in small groups independent of the stellar density on larger-scales.  Then dynamical interactions between protostars will occur within the small groups as they are forming, potentially truncating discs, before the groups disperse.  Even in Taurus, many of the young stars are observed to be in groups of around a dozen protostars \citep{Gomezetal1993} which may have been more compact in the past.

\subsubsection{Disc properties versus stellar mass}

There is general agreement from studies of nearby star-forming regions that disc mass increases with stellar mass \citep[see the discussion in][]{Andrews_etal2013}, and this relation seems to extend into the sub-stellar \citep{Klein_etal2003,SchJayWoo2006,
Schaefer_etal2009,Mohanty_etal2013,Daemgen_etal2016,vanderPlas_etal2016,Testi_etal2016} and planetary-mass \citep{Bayo_etal2017} regimes.  The exact dependence, however, is model dependent, for example, whether or not disc temperature is scaled with stellar luminosity, and the assumptions made about the disc size; see, for example, \cite{Pascucci_etal2016,Hendler_etal2017}.  \cite{Andrews_etal2013} found that the millimetre flux scales as $F_{\rm mm} \propto M_*^{1.5-2.0}$ for Class II discs in the Taurus region and they argue that, accounting for dust temperature scaling, this supports a roughly linear scaling of disc mass with stellar mass (i.e. $M_{\rm d} \propto M_*$) with a dispersion of $\approx 0.7$~dex.  \cite{Ansdell_etal2016} found a slope of $M_{\rm d} \propto M_*^{1.8\pm 0.4}$ with dispersion of $0.9\pm 0.2$ for Lupus,  and $M_{\rm d} \propto M_*^{1.7\pm 0.2}$ with dispersion of $0.7\pm 0.1$ for Taurus, but a steeper slope of $M_{\rm d} \propto M_*^{2.4 \pm 0.4}$ with dispersion of $0.7\pm 0.1$ for Upper Scorpius.  \cite{Barenfeld_etal2016} obtained $M_{\rm d} \propto M_*^{1.7 \pm 0.4}$ in Upper Scorpius.  \cite{Pascucci_etal2016} derive $M_{\rm d} \propto M_*^{1.6 \pm 0.3}$ in Chamaeleon I and assert that this is similar to the relations in Taurus and Lupus, with the relation in Upper Scorpius being steeper.  

The disc masses from the hydrodynamical simulation clearly scale with stellar mass (left and right panels of Fig.~\ref{disc_systems_mass}).  The scaling appears to be roughly linear up to $M_* \approx 0.5$~M$_\odot$, with no obvious trend above this mass.  A formal fit to all systems with total protostellar masses $M_*<0.5$~M$_\odot$ and disc masses $M_{\rm d} > 0.001$~M$_\odot$ gives $M_{\rm d} \propto M_*^{0.72\pm0.03}$.  Limiting the fit to single protostars with $M_*<0.5$~M$_\odot$ gives $M_{\rm d} \propto M_*^{0.85\pm0.04}$ (i.e.\ close to a linear dependence).  For more massive systems (most of which are multiple) $M_{\rm d} \propto M_*^{-0.09\pm0.05}$ (i.e. there is no significant dependence).  We caution against over interpreting these fits because of the fact that at these young ages, both age and stellar mass matter.  There is also a very large dispersion of at least $0.6$~dex about these relations (excluding unresolved discs).  These scaling relations are broadly consistent with the observed relations for Class II objects, even though the simulated objects are much younger and the disc masses are substantially higher.  From Section \ref{sec:isolated}, we have seen that cause of much of the dispersion is due to dynamical interactions with other protostars; the dispersions of disc masses and radii of protostars that have never been within 2000~au of another protostar are significantly narrower (see Fig.~\ref{disc_cumulative}).  The implication is that the observed scaling relations of disc mass with stellar mass and their dispersion originate from the formation process, including dynamical interactions, and are not due to subsequent disc evolution.

Recent observations have found evidence that the discs of brown dwarfs may typically be smaller than the discs of more massive T Tauri stars \citep{Testi_etal2016,Hendler_etal2017}.  The first resolved observations of discs around very low mass (VLM) objects found disc sizes may range from $30-70$~au to larger than 200 au \citep{Ricci_etal2013,Ricci_etal2014}. \cite{Testi_etal2016} found evidence that two discs of VLM objects in Ophiuchus may have sharp outer disc radii of $\approx 25$~au, with three other discs having radii between 50 and 150~au, depending on model parameters. From modelling spectral energy distributions, \cite{Hendler_etal2017} find that out of 11 young stars with masses $M_*\lsim 0.2$~M$_\odot$,  7 likely have disc radii smaller than 10 au, with the remaining four objects having radii from 10--80~au.  

If it is confirmed that the discs of VLM objects are smaller than those of more massive stars, this would be consistent with the trend that we find from the hydrodynamical simulations of smaller discs around lower mass objects (e.g. the middle column of panels of Fig.~\ref{disc_systems_mass}).  We find that discs of protostellar systems with masses $M_*<0.1$~M$_\odot$ are typically half the size of systems with masses $0.1-0.3$~M$_\odot$ and 3--4 times smaller those around systems with masses $M_*>0.3$~M$_\odot$.  Taken at face value,  Fig.~\ref{disc_systems_mass} implies that roughly half of discs around systems with $M_*<0.1$~M$_\odot$ should have characteristic radii smaller than 20~au whereas for more massive systems about half should have radii smaller than 40~au.  The caveat is that, as we have seen, the disc properties of these young protostars also evolve with time and extrapolating from $\sim 10^{4-5}$~yrs to $\sim 10^6$~yrs is risky.

The observed small disc sizes would be consistent with brown dwarfs being ejected from multiple protostellar systems during their formation \citep{ReiCla2001,BatBonBro2002a,GooWhi2007,StaHubWhi2007,Bate2009a,Bate2012}.  The evidence of sharp outer radii for two objects found by \cite{Testi_etal2016} also hints at this formation mechanism.   However, it is also possible that small dust disc sizes may result from more efficient radial drift of dust in discs around low-mass objects \citep{Pinilla_etal2013}.

\begin{figure}
\centering \vspace{-0.5cm} \hspace{0cm}
    \includegraphics[width=8.5cm]{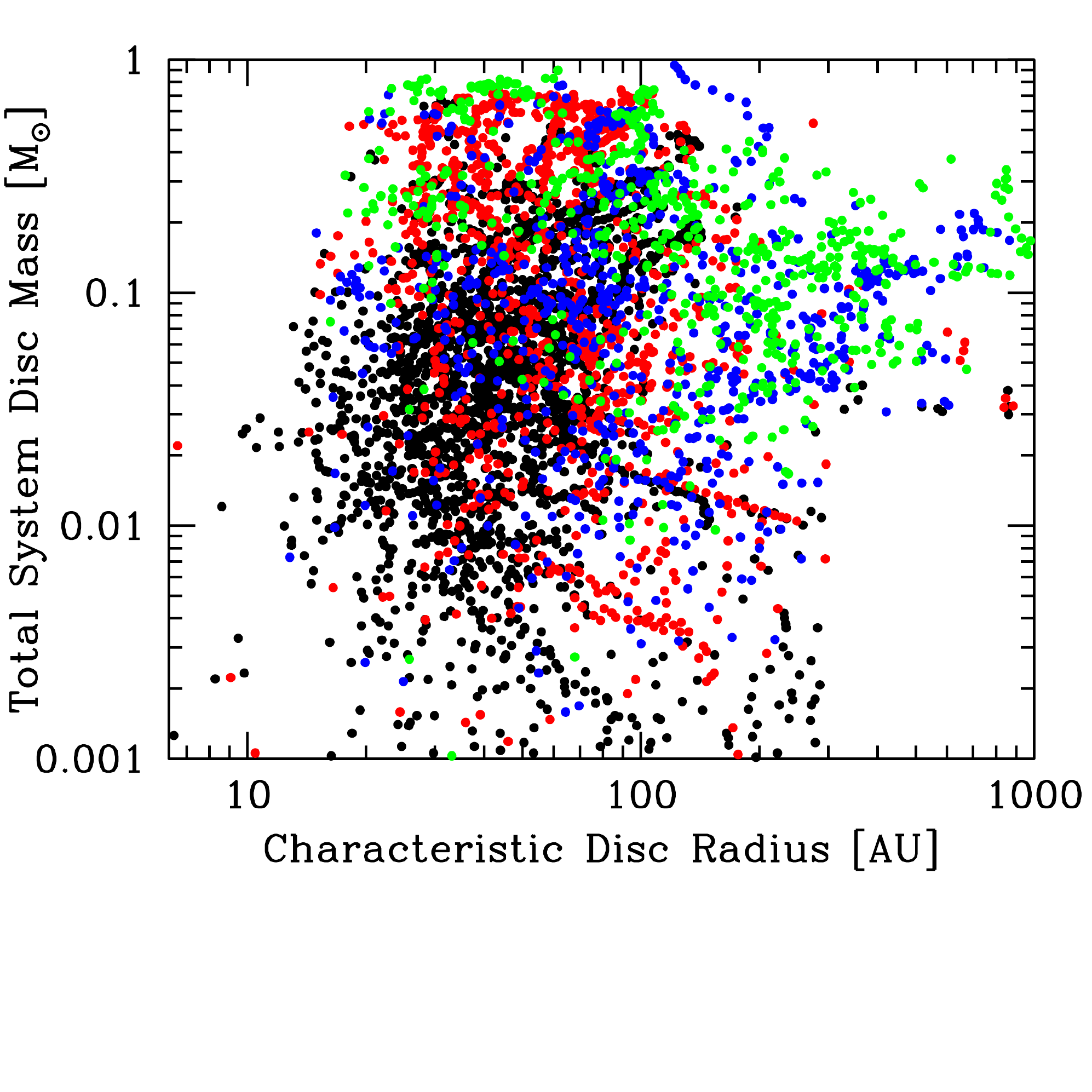}      \vspace{-2cm}
\caption{The total disc masses versus characteristic disc radii of the protostellar systems versus their total stellar mass.  Each dot represents an instance of disc(s) of a particular system, which may be a single protostar or a bound multiple protostellar system.  The colours denote the order of the system: single (black), binary (red), triple (blue), or quadruple (green).  A particular system may be represented by many instances (taken at different times). There is a weak dependence of the total mass of a disc on its characteristic radius.
}
\label{sys_mass_rad}
\end{figure}

\subsubsection{Disc radius versus disc mass}

Using 880$\mu$m observations, \cite{Andrews_etal2010} found that discs with lower luminosities are smaller, but they do not necessarily have lower surface brightnesses.  They found a relation between disc mass and characteristic radius of $M_{\rm d} \propto r_{\rm c}^{1.6 \pm 0.3}$.   \cite{Pietu_etal2014} found a similar correlation at 1.3mm, and subsequent observations in both the same and different star-forming regions have confirmed the trend \citep{Andrews2015, Tazzari_etal2017, Tripathi_etal2017}.  \cite{Tripathi_etal2017} found that the mm-luminosity scales as the square of the effective disc radius, implying that the luminosity scales linearly with the emitting area and that the average surface brightness is roughly constant for all luminosities.  \cite{Tazzari_etal2017}, who excluded binaries and transition discs, also found that the luminosity increases with effective radius, but obtained a shallower relation $L \propto R_c^{5/4}$.  \cite{Andrews_etal2010} and   \cite{Tripathi_etal2017} note that such relations are not expected from viscous evolution or photoevaporation.  Both discuss the possibility that the relation may originate from the initial angular momentum distribution in molecular clouds. \citeauthor{Tripathi_etal2017} also point out that the scaling may be due to the migration and growth of solids in discs, and/or that it may be due to unresolved optically thick (dust) emission with filling factors of a few tens of percent \cite[e.g.\ rings;][]{Pinilla_etal2012,LorBat2015a}.  

In Fig.~\ref{sys_mass_rad} we plot disc mass versus the characteristic disc radius for our protostellar systems.  Both inspection of the figure and a linear regression confirm that more massive discs do tend to be larger.  However, the relation is not as strong as that found for observed Class II objects.  If we take all protostellar systems with disc masses $M_{\rm d} >0.001$~M$_\odot$, we obtain a relation $M_{\rm d} \propto r_{\rm c}^{0.20\pm 0.03}$ with a large dispersion of 0.7 dex.  If we limit the analysis to systems with total stellar masses of $M_*<0.5$~M$_\odot$, we find the slighly stronger relation $M_{\rm d} \propto r_{\rm c}^{0.28\pm 0.04}$ and if we further consider only single protostars we obtain $M_{\rm d} \propto r_{\rm c}^{0.37\pm 0.05}$, but the dispersions increase to 0.9 and 1.1 dex, respectively.  For systems with total stellar masses $M_*>0.5$~M$_\odot$ (most of which are multiple systems), we find $M_{\rm d} \propto r_{\rm c}^{-0.12\pm0.04}$ with a dispersion of 0.6 dex (i.e. there is little dependence of disc mass on disc radius for more massive systems).  Thus, part of the observed relation may come from the initial conditions of protostellar discs, but other evolutionary effects probably dominate the observed relation (e.g. dust evolution).

\subsubsection{Young stars without discs}

By the end of the hydrodynamical calculation, there are a large number of protostars without resolved discs.  Even at ages of $\approx 10^4$~yrs, $\approx 30$\% of protostellar systems have disc masses $M_{\rm d,gas}<0.01$~M$_\odot$.  Part of this will be due to the limited numerical resolution, so the discless populations from the numerical simulations must be considered weak upper limits.  But as we have shown, most of the destruction of discs is due to dynamical interactions between protostars, with ram-pressure stripping also having a role.  Thus, even with high numerical resolution some very young protostars would be left with little or no disc material.  Is it realistic that at ages $<10^5$~yrs, a significant fraction of protostars are discless or have only very small discs?

Observationally, this is a very difficult question to answer.  The main two problems are sample selection and sensitivity.  For example, young stars are often classified as classical T Tauri stars (CTTS) and weak-line T Tauri stars (WTTS) based on H alpha emission, associated with disc accretion.  Alternatively, they may be referred to as Class II or Class III objects.  It is usually assumed that WTTS or Class III objects are `more evolved' and therefore older.  But this is not necessarily the case -- some fraction of WTTS or Class III may be young \citep{Kurosawaetal2004}.  Furthemore, some CTTS appear discless, while some WTTS have discs.  For example, in IC348 \cite{Lada_etal2006} find that $\approx 20$\% are discless while, on the other hand, 12\% of WTTS are found to have thick, primordial discs.  They also find the disc fractions peak with solar-type stars, and decline for both higher and lower masses.  It is also possible that some stars move between the CTTS and WTTS states.

For the youngest star-forming regions (a few Myr), observational determinations of disc fractions typically range from $\approx 50-85$\%  \citep{HaiLadLad2000,Lada_etal2000,HaiLadLad2001, Lada_etal2006, Balog_etal2007, Guarcello_etal2007, Hernandez_etal2007a,Hernandez_etal2007b,Hernandez_etal2008,Harvey_etal2008}.  From these studies, the highest disc fractions of $\approx 85$\% (determined using JHKL colours or {\em Spitzer} data) have been found in Orion Nebula Cluster, NGC 2024, NGC1333, and NGC2068/71. Recently, \cite{Ribas_etal2014} examined disc fractions consistently in 22 young associations using data over a wide range of wavelengths.  Their highest overall disc fractions were $60-70$\% in NGC1333 and Taurus.  Thus, there seems to be scope for a very young discless population at the level of 15-30\% and, as discussed in Section \ref{sec:disc_rad}, \cite{Pietu_etal2014} concluded that up to 25\% of the entire disc population of Taurus may consist of very compact dust discs.  But as mentioned above, we stress that because of the limited resolution in the hydrodynamical calculation, the discless population that we obtain here should be treated as a weak upper limit.

\subsection{Comparison with observed Class 0/I objects}
\label{sec:class0I}

Surveys of Class 0/I objects are not yet as extensive as for Class II objects due to the smaller numbers of objects and the difficulty of separating disc and envelope emission \citep[e.g.][]{LooMunWel2000}.  Some early studies of Class 0 and I objects inferred masses of unresolved discs ranging from 0.01 to 1.7~M$_\odot$ (dust masses 30-6000~M$_\oplus$) with typical masses of $0.05-0.2$~M$_\odot$ (dust masses 200-700~M$_\oplus$) \citep{Jorgensen_etal2009,Enoch_etal2011}.  But, only a few years ago there was much debate about whether Class 0/I objects had discs larger than $\approx 10$~au or not.  Some early observational studies found little evidence for large young discs \citep[e.g.][]{Mauryetal2010}, and this was taken as evidence that magnetic braking may inhibit large discs from forming (see Section \ref{sec:limitations}) until a later evolutionary stage. However, subsequent observations of Class 0 protostars have found a mixture of both small and large discs. B335 is estimated to have a disc radius $<5$~au \citep{Yen_etal2015}, and no Keplerian discs are detected in NGC 1333 IRAS 2A \citep{BriJorHog2009,Maret_etal2014} or L1157-mm \citep{Yen_etal2015}.  Large discs have been found in HH111 with a disc radius of $\approx 160$~au \citep{LeeHwaLi2016}, HH211 with a disc radius of $\sim 80$~au \citep{Lee_etal2009}, NGC1333 IRAS 4A2 with a disc radius of $\sim 310$~au \cite{ChoTatKan2010}, L1527 with a disc radius of $\approx 70$~au \citep{Tobinetal2012,Ohashi_etal2014,Aso_etal2017}, VLA 1623 with a disc radius of $\approx 190$~au \citep{MurLai2013,Murilloetal2013}, TMC-1A with a disc radius $\sim 100$~au \citep{Yen_etal2013,Harsono_etal2014,Aso_etal2015}, HH212 with a disc radius $\approx 60$~au \citep{Codella_etal2014, Lee_etal2017c}, and Lupus 3 MMS with a radius of $\approx 100$~au \citep{Yen_etal2017}.  Individually, all of these observed disc radii are consistent with the size distribution of protostellar discs presented in this paper.  To make further progress, we need to compare the distributions of disc properties from observation and theory.

Surveys of Class 0/I discs are now being made.  \cite{Harsono_etal2014} detected rotationally-supported discs around three out of four Class I objects.  All four objects have outer radii $\lsim 100$~au, with one object having an upper limit of 50~au.  The disc masses ranged from 0.004 to 0.033 M$_\odot$.
 \cite{Tobinetal2015} studied nine Class 0 and two Class I objects with 70-au resolution, finding flattened structures with radii $>100$~au around two sources and marginally resolved structures around three others.
 \cite{Yen_etal2017} studied three Class 0 protostars and, in addition to the 100-au resolved disc of Lupus 3 MMS mentioned above, used kinematic data to estimate disc radii of $\approx 20$ and $\approx 6$~au for the other two objects with disc masses $\sim 0.01-0.03$~M$_\odot$. First results from the VLA Nascent Disk and Multiplicity (VANDAM) survey \citep{SeguraCox_etal2016} (1601.03040) found discs radii from 10-30 au for one Class I and six Class 0 objects in Perseus.  Again these sizes are in good agreement with the typical sizes found in this paper.  It should be noted, however, that this survey was conducted at wavelengths of 8mm which may be biased towards finding small disc radii due to dust evolution (i.e. growth and radial migration).  One of their objects (Per-emb-14) that was determined to have a disc radius of $\approx 30$~au at 8mm was also resolved at a wavelength of 1.3mm by \cite{Tobinetal2015} and found to have a flattened structure with.a radius $>100$~au.  Thus care needs to be taken when interpreting the effects of wavelength and sensitivity on measurements of disc radii.  
 
\begin{figure}
\centering \vspace{-0.5cm} \hspace{0cm}
    \includegraphics[width=8.5cm]{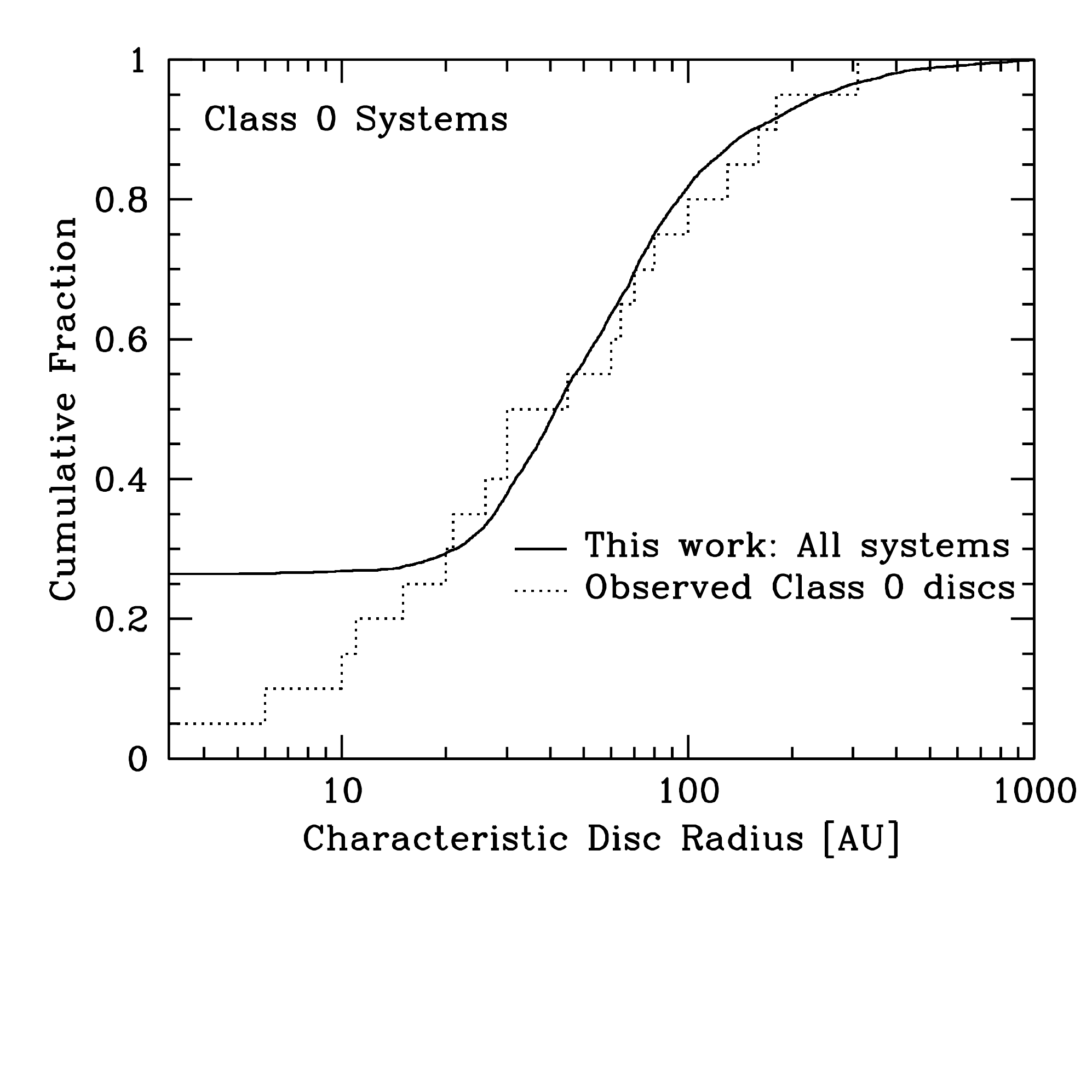} \vspace{-2cm}
\caption{Cumulative distributions of the characteristic radii for the discs from the calculation analysed in this paper (solid line), and for twenty observed discs of Class 0 objects from the literature.  The agreement is reasonable, given the uncertainties in determining Class 0 disc radii and the fact that the observations are not of a well defined sample of objects.}
\label{class0}
\end{figure}

 We note that although the values of power-law slopes of temperature, $q$, and surface density, $\lambda$, the disc radii determined by \citep{SeguraCox_etal2016} are very poorly constrained, the characteristic disc radii $r_{\rm c}$ that are determined for a particular object with different values of $q$ and $\gamma$ hardly vary at all.  This is likely because of the point we made in Section \ref{sec:characterisation}, namely that $r_{\rm c}$ simply gives the radius that contains $63.2$\% of the total disc mass (or flux) regardless of the value of $\gamma$ that is used in equation \ref{eqn:sigma} (as long as $\gamma<2$).

In Fig.~\ref{class0} we plot the cumulative distributions of characteristic disc radii from all protostellar systems in the hydrodynamical calculation, and a compilation of 20 disc radii of Class 0 objects from the papers mentioned above.  Given the small number of observed objects, the difficulties in determining the disc radii of Class 0 objects, and the fact that the sample has not been well defined, the agreement is ridiculously good.

There are not many estimates of disc masses for Class 0 protostars to date.  The unresolved observations of \cite{Jorgensen_etal2009} and \cite{Enoch_etal2011} gave estimated masses ranging from 0.01-1.7~M$_\odot$ (dust masses 30-6000~M$_\oplus$).  \cite{SeguraCox_etal2016} publish masses ranging from 0.09-0.36~M$_\odot$ (dust masses 300-1000~M$_\oplus$) for the six Class 0 objects they studied.  Tobin estimates a disc mass of 0.007~M$_\odot$  (dust mass 20~M$_\oplus$) for L1527 IRS.  These nicely span the range of disc masses that are found for the protostars in our hydrodynamical calculation (see Fig.~\ref{obs_mass}).  Thus, Class 0 disc masses are potentially $\sim 30$ times more massive than the typical Class II disc masses as suggested by Fig.~\ref{obs_mass}.

\cite{Yen_etal2017} examined the specific angular momentum profiles in 8 Class 0 objects and found signs of disc growth with disc radius increasing with protostellar mass as $R_{\rm d} \propto M_*^{0.8\pm0.14}$ or age as $R_{\rm d} \propto t^{1.09\pm0.37}$ in the Class 0 stage.  Extending the sample to include 10 Class I objects, they obtained shallower slopes of $R_{\rm d} \propto M_*^{0.24\pm0.12}$ and $R_{\rm d} \propto t^{0.18\pm0.09}$.  They speculated that this may indicate rapid growth of disc size during the Class 0 phase, and then slower growth in the Class I phase.  The characteristic disc radii from the hydrodynamical simulation clearly increase with stellar mass up to $M_* \approx 0.5$~M$_\odot$ (middle panels of Fig.~\ref{disc_systems_mass}).  A fit to all systems with total protostellar mass $M_* \approx 0.5$~M$_\odot$ gives $r_{\rm c}\propto M_*^{0.24\pm0.01}$ with a dispersion of $\approx 0.4$ dex.  Thus, the scaling is weaker than found by \citeauthor{Yen_etal2017} for Class 0 objects alone, but in good agreement with the combined sample of Class 0 and I objects. 

We caution, however, that although we have seen from the numerical calculation that the sizes of protostellar discs do typically grow with age and are larger for more massive objects (Section \ref{sec:sys_mass_rad}), it is interesting to note that the cumulative distributions of observed Class II disc radii and Class 0 disc radii in Figs. \ref{obs_rad} and \ref{class0} do not appear very different from one another (depending on how incompleteness is accounted for).  The implication is that disc radii may not differ substantially between the Class 0 and Class II phases, but their masses decrease by factors of 30 to 300.

\subsection{Limitations of the calculation and future directions}
\label{sec:limitations}

The simulation of \cite{Bate2012} from which the disc properties discussed in this paper were extracted is far from perfect.  On the positive side, it was the first hydrodynamical calculation of star cluster formation to produce more than 100 stars and brown dwarfs with a distribution of stellar masses consistent with the observed stellar initial mass function (IMF).  It also produced realistic fractions of multiple systems and the properties of those multiple systems are in reasonable agreement with those of observed multiple systems.  However, both radiative and kinetic feedback (e.g. jets and outflows) from inside sink particles were neglected.  The missing radiative feedback may have a small effect on the level of fragmentation, although because of the use of very small sink particles and the fact that only low-mass stars are produced, \citealt{Bate2012} demonstrated empirically that this effect is likely to be small.  Missing outflows are likely to result in protostellar and disc masses that are higher than they should be \citep[e.g.][]{Hansen_etal2012,KruKleMcK2012,Federrath_etal2014,Federrath2015}, but the magnitude of this effect is likely to be small ($\sim 10-20$\%) compared to the other uncertainties (i.e. disc extraction, differences in gas and dust dynamics, etc).

From the point of view of studying disc properties, apart from the obvious limitation of size of the sample, the main three limitations are: numerical resolution, the absence of differentiation between gas and dust, and the absence of magnetic fields.  The first two of these were discussed in detail at the beginning of Section \ref{sec:discuss}.  When it comes to magnetic fields, other than driving outflows, their main effects are to add additional pressure support and transport angular momentum.  Magnetic pressure support can slow down the star formation rate by factors of 2--3 compared to hydrodynamical calculations  \citep{PriBat2008,PriBat2009,Wang_etal2010,PadNor2011,FedKle2012,Myers_etal2014,Federrath2015}, but its affect on disc properties is unclear.  Magnetic angular momentum transport, however, could have a large effect.

Analytic and numerical calculations under the assumption of ideal MHD have shown that magnetic braking can stop the formation of large protostellar discs completely in simple geometries where the axis of rotation of a core is aligned with a global field that is anchored at large distances from the centre of the core \citep{AllLiShu2003,Galli_etal2006,PriBat2007,MelLi2008,HenFro2008,DufPud2009,DapBas2010,MacInuMat2011,LiKraSha2011,DapBasKun2012}.  However, various effects can reduce the effectiveness of magnetic braking.  If the magnetic field is misaligned with the rotation axis, this can reduce the braking \citep{HenCia2009,JooHenCia2012,Li_etal2014}.  In turbulent clouds, the field and rotation axis may be naturally misaligned, turbulent reconnection may reduce the field strengths, and the material at large distances is not static so the magnetic field lines can also move.  These effects all tend to reduce the effectiveness of magnetic braking \citep{Seifried_etal2012,Seifried_etal2013,Joos_etal2013,Li_etal2014,SandeGLaz2012,SandeGLaz2013}, although the discs remain smaller than those formed without magnetic fields.  Finally, the non-ideal MHD effects of ambipolar diffusion and Ohmic resistivity allow for diffusion of the magnetic field relative to the matter, and the Hall effect can cause material to either spin up or spin down depending on the relative orientation of the magnetic field and the rotation axis \citep[e.g.][]{WarNg1999,Wardle2007}.  When large discs are prevented from forming by magnetic braking in ideal MHD calculations, the effects of introducing ambipolar diffusion alone are insufficient to allow the formation of large discs \citep{DufPud2009,MelLi2009,LiKraSha2011,DapBasKun2012,TomOkuMac2015,Tsukamoto_etal2015a,WurPriBat2016}.  Similarly, introducing Ohmic diffusion only produces small discs unless an anomalously high resistivity is used \citep{Shu_etal2006,KraLiSha2010,DapBas2010,MacInuMat2011,Tomida_etal2013,WurPriBat2016}.  The Hall effect seems capable of producing large discs, but whether a large disc forms or not depends on the sign of the magnetic field \citep{BraWar2012a,BraWar2012b,KraLiSha2011,Tsukamoto_etal2015b,WurPriBat2016}.

Given the supposed importance of magnetic fields, may be surprising that the sizes of the disc produced by the hydrodynamical calculation analysed in this paper are in relatively good agreement with those that are observed (Fig.~\ref{obs_rad}).  They are certainly not too large.  What does this mean for the role of magnetic fields in disc formation?  The implication is that magnetic fields do not transport significant angular momentum during protostellar disc formation and are not important for setting disc sizes.  Although some readers may find this surprising, it may be the case, given that it has already been shown that turbulence and non-ideal MHD effects (particularly the Hall effect) can reduce the effects of magnetic braking.   Along these lines, \cite{WurPriBat2017} showed that binary star formation was primarily governed by the initial density and velocity structure of a molecular cloud core rather than magnetic effects.  The same may be true of disc formation.

Another interesting point from Figs.~\ref{obs_rad} and \ref{class0} is that there may not be much difference between the size distributions of the discs of observed Class 0 and Class II objects.  This is not expected if the typical Class 0 object is assumed to be much younger than the typical Class II object and if discs evolve in a pseudo-viscous manner; discs would be expected to get larger with increasing age.  One possibility is that the primary process(es) that drive disc evolution are not pseudo-viscous in nature.  Recent studies have suggested that accretion in protostellar discs may be driven by the loss of angular momentum in disc winds rather than by angular momentum transport within the disc itself \citep{BaiSto2013,Simon_etal2013,Suzuki_etal2016,WanGoo2017,Rafikov2017}.  This would mean that once significant envelope accretion has ceased, discs may either maintain their radius or decrease in radius with time, as opposed to increasing in radius from psuedo-viscous evolution.

Future studies will be able to explore the limitations of this calculation in more detail.  However, at present, we conclude that the disc population produced by this radiation hydrodynamical calculation of star cluster formation are in surprisingly good agreement with observed disc properties.

\section{Conclusions}
\label{conclusions}

We have presented an analysis of the protostellar discs that were produced by the radiation hydrodynamical calculation of star cluster formation first published by \cite{Bate2012}.  This can be thought of as the first attempt at protostellar disc population synthesis using a hydrodynamical calculation.  Disc evolution around 183 protostars is followed for up to 90,000 yrs, with the typical protostar being evolved for a few $\times 10^4$~yrs.  

We have shown that an enormous diversity of protostellar disc types and morphologies is to be expected around young protostars.  A particular type of system can be formed in a variety of ways.  For example, a binary system with circumstellar discs that are misaligned with the orbital plane can be produced through the  fragmentation of a laminar cylinder (i.e. filament) that is rotating about both the major and minor axes, or via fragmentation in a turbulent environment, or even from two protostars forming separately and undergoing star-disc capture (see Sections \ref{sec:misalign_csdiscs} and \ref{sec:stardisc}).  Discs with varying radial angular momenta profiles (i.e.\ warped discs) can be produced either during formation with infall whose angular momentum varies with time or by torques from companions (see Section \ref{sec:misaligned}).  Spiral density waves in discs can be self-generated via gravitational instabilities (e.g. Section \ref{sec:single}), or generated by stellar or planetary companions.  In producing the diverse range of systems, no one mechanism dominates.  Cloud/filament fragmentation, disc fragmentation, star-disc encounters, dynamical processing, accretion, and ram-pressure stripping each play significant roles.

Detailed observations of an individual system may be able to determine the processes that currently control its evolution, and the mechanisms that originally formed the system may be able to be constrained.  But in many cases their will be uncertainty over how the system originally formed (e.g. it will never be possible to distinguish whether a particular system formed from laminar or turbulent initial conditions).  In the long term, the question of how stellar systems form will only be able to be answered statistically using population synthesis and detailed observations of large samples.  This paper represents a first step in this direction, though it has severe limitations, including limited resolution, the absence of magnetic fields and protostellar feedback, and there is no accounting for the different dynamical evolution of gas and dust.  These limitations will be reduced in subsequent computations, but for the moment we have the following conclusions:
\begin{enumerate}
\item  A wide diversity of discs is already observed around young stars, but the calculation discussed here shows that the diversity is likely to be an even broader in future observations, in particular in terms of disc morphologies (e.g.\ discs in multiple systems, warped discs, eccentric discs, and other non-asymmetric disc structures).

\item We find that protostellar discs typically increase in mass with age up until $\approx 10^4$~yrs, or with protostellar mass up to $M_*\approx 0.5$~M$_\odot$.  Disc masses typically triple from ages of $\approx 10^3$~yrs to $\approx 10^4$~yrs.  Beyond this age, the typical disc mass stabilises -- while some discs continue to grow in mass, many decline due to accretion (driven by gravitational torques), fragmentation, dynamical interactions (e.g. star-disc encounters), and ram-pressure stripping.  The dependence of disc mass on protostellar mass is roughly linear ($M_{\rm d} \propto M_*^{\approx 0.85}$ for single protostars) up to $M_*\approx 0.5$~M$_\odot$, but beyond this point the disc mass has no significant dependence on protostellar mass.  There is significant dispersion in these relations, in excess of 0.6 dex.  Much of the dispersion is due to dynamical interactions between protostars (either in bound systems, or unbound encounters).  The dispersion of disc masses is significantly lower for protostars that have never had another protostar closer than 2000 au.

\item  Similarly, the characteristic radii of protostellar discs typically increase with age up until $\approx 10^4$~yrs, or with protostellar mass up to $M_*\approx 0.5$~M$_\odot$. The dependencies of disc radii on age or protostellar mass are much weaker than those for disc mass.  Disc radii typically double from ages of $\approx 10^3$~yrs to $\approx 10^4$~yrs.  Their dependence on protostellar mass scales as ($r_{\rm c} \propto M_*^{\approx 0.25}$).   The dispersion of the disc radius with protostellar mass is at the level of at least $\approx 0.4$ dex.  As with disc mass, much of this dispersion is due to interactions with other protostars; the dispersion of disc radii is smaller for protostars that have never had another protostar closer than 2000 au ($\approx 0.25$ dex).  We find that many protostellar discs are small ($r_{\rm c} \lsim 20$~au), and the fraction of small discs depends on protostellar mass.  As many as 50\% of protostars with masses $M_*<0.1$~M$_\odot$ have small discs, while for protostellar masses  $M_*>0.3$~M$_\odot$ the fraction is $\approx 10$\%.

\item  The typical disc to stellar mass ratios typically range from $M_{\rm d}/M_*\approx 0.1 - 2$ up to ages of $\lsim 10^4$~yrs, beyond which they tend to decline.  Protostars with masses $0.1 \le M_* < 0.3$~M$_\odot$ tend to have higher disc to protostar mass ratios than either lower mass or higher mass protostars.

\item  For isolated protostars the typical radial surface density profile is $\Sigma(r) \propto r^{-1}$.  This is flatter than that of the minimum solar mass nebula (MMSN) model ($\Sigma(r) \propto r^{-3/2}$), and very few of our discs have density profiles as steep as the MMSN model.

\item  We examine the relative orientations of circumstellar discs in bound protostellar pairs (both binaries and pairs in higher-order systems).  We find that the discs in closer systems tend to be preferentially more aligned, with a strong preference for alignment at orbital semi-major axes $\lsim 100$~au.  The circumstellar discs in binaries tend to be less well aligned than those of pairs in higher-order systems.  This is likely because pairs in multiple systems often originate from disc fragmentation, whereas binaries frequently originate from either disc fragmentation or star-disc encounters.  The alignment also tends to strength with increasing age.

\item  Circumstellar discs in bound protostellar pairs also have a preference for alignment with the orbit of the pair.  Compared to the disc-disc alignment, we find that the disc-orbit alignment is weaker in close systems but stronger in wide systems.  The evolution with age is not as strong as with disc-disc alignment. 

\item  In protostellar pairs, sink particle spins, which represent a combination of the angular momentum of the protostellar and inner disc (scales $\lsim 0.5$~au), show a similar tendencies for alignment with each other as the circumstellar discs.  However, the difference between binaries and pairs in higher-order multiple systems is greater.  Again this likely reflects the different dominant formation mechanisms.

\item  The relative orientations between sink particle spins and circumstellar discs in bound pairs also show a strong preference for alignment.  However, in this case, there is little variation with age, orbital separation, or the total number of protostars in the system.  Furthermore, the distribution of relative orientation angles is very similar for isolated protostars.  In all cases, around 50\% of protostellars have misalignments between their protostellar and inner disc angular momentum vectors of more than $30^\circ$.  The reason for this seems to be that the outer discs are frequently being reorientated, more quickly than the spins can `catch up' through accretion from the larger-scale disc.  This has implications for the formation of planetary systems whose orbits are misaligned with the spins of their host stars.

\item  Comparing with observations, we find that the typical disc masses at ages $\sim 10^4$~yrs are approximately $30-300$ times greater than those of observed discs of Class II protostars (depending on the star-forming region).  The range of disc masses are consistent with the few existing determinations of Class 0 objects in the literature.  The distribution of radii of the discs from the hydrodynamical simulation are also similar to those of observed Class II and Class 0 objects.  We find only a weak dependence of disc mass on disc radius.  

\end{enumerate}

Despite the absence of magnetic fields, the discs produced in the radiation hydrodynamical examined in this paper appear neither `too large' nor `too massive' when compared with the latest observations of protostellar and protoplanetary discs.  The calculation also produces a reasonable IMF and properties of multiple stellar systems.  This indicates not only that magnetic fields may have a small role to play in the formation of the IMF and multiple systems, but also that magnetic fields may have much less of an impact on the initial properties of protostellar discs that some past studies have suggested.  Although disc formation can be completely prevented by magnetic braking in idealised calculations of the collapse of isolated magnetised molecular cloud cores, over the past five years, various authors have shown that a variety of processes (i.e. misaligned magnetic fields and rotation axes, turbulence, and non-ideal MHD effects) may work together to alleviate magnetic braking.  Thus, magnetic fields may have less of an impact on the statistical properties of young protostellar discs than is often assumed.

\section*{Acknowledgements}

MRB thanks Paul Clark for encouraging him to write this paper, and Jonathan Williams, Megan Ansdell, Ilaria Pascucci, and Jim Pringle for giving comments on the original manuscript.  MRB also thanks the anonymous referee for asking a number of questions that resulted in improvements to the paper.  In particular, Appendices \ref{appendixA} and \ref{appendixB} were produced in response to the referee's questions and the discussion of the effects of the finite numerical resolution was improved.

The rendering of the calculations discussed in the Appendices were produced using SPLASH \citep{Price2007}, an SPH visualization tool publicly available at http://users.monash.edu.au/$\sim$dprice/splash.

This work was supported by the European Research Council under the European Commission's Seventh Framework Programme (FP7/2007-2013 Grant Agreement No. 339248).  The calculation discussed in this paper was performed on the University of Exeter Supercomputer, a DiRAC Facility jointly funded by STFC, the Large Facilities Capital Fund of BIS, and the University of Exeter, and on the DiRAC Complexity system, operated by the University of Leicester IT Services, which forms part of the STFC DiRAC HPC Facility (www.dirac.ac.uk). The latter equipment is funded by BIS National E-Infrastructure capital grant ST/K000373/1 and STFC DiRAC Operations grant ST/K0003259/1. DiRAC is part of the National E-Infrastructure.  The calculation was conducted as part of the award `The formation of stars and planets: Radiation hydrodynamical and magnetohydrodynamical simulations' made under the European Heads of Research Councils and European Science Foundation EURYI (European Young Investigator) Awards scheme, was supported by funds from the Participating Organisations of EURYI and the EC Sixth Framework Programme.

\section*{Supporting Information}

Additional Supporting Information may be found in the online version of this article:

{\bf Animation.} The animation shows a mosaic of 183 animations, each of which displays a region with dimensions of $400\times 400$ au centred on one of the protostars (sink particles) that is produced during the simulation.  The colour scale shows the logarithm of column density, ranging from 1 to $10^4$~g~cm$^{-2}$.  The protostars appear in the order in which the form in the radiation hydrodynamical simulation, and the animation runs from $t=0.70-1.20$~$t_{\rm ff}$, which is a period of 95,000 yrs.  The animation allows the evolution of each protostar and its disc to be followed.

{\bf Circumstellar disc data files.} We provide 183 text files, one for each sink particle, that give the time evolution of the properties of the protostar and its circumstellar disc.  The data necessary to construct Figs.~\ref{starmass} to \ref{fig:surfdens}, and Fig.~\ref{csdisc_spin} is contained in these files.  Their file names are of the format "Disc\_AAA\_UUUUUUUU.txt", where "AAA" gives the number of the sink particle in order of it formation (e.g. "001" or "183", for the first and last sink particles).  The number "UUUUUUUU" gives the unique particle identification number from the {\tt sphNG} simulation.  Each line of a file contains 26 numbers delimited by spaces that give the state of the protostar at one instance in time.  The following information is given: (1) time, (2) time of formation of the protostar, (3) mass of the protostar, (4) mass of the circumstellar disc, (5-17) 13 numbers that give the radii that contain 2, 5, 10, 20, 30, 40, 50, 63.2, 70, 80, 90, 95, and 100\% of the circumstellar disc, (18) an integer which is 1 if the protostar has no companion within 2000 au and 2 if there is at least one companion, (19-21) three numbers that give the angular momentum of the circumstellar disc, (22-24) three numbers that give the spin angular momentum of the protostar, (25) an integer whose absolute value gives the number of the nearest sink particle if (18) is equal to 2 but zero otherwise, (26) an integer whose absolute value gives the unique particle identification number of the nearest sink particle.  Integers (25) and (26) are negative if the companion is not bound to the protostar.  Time is given units of $\sqrt{(0.1~{\rm pc})^3/({\rm G~M}_\odot)} = 471300$~yrs, masses are given in M$_\odot$, radii are given in au.  Angular momentum is in units of $\sqrt{({\rm G~M}_\odot^3(0.1~{\rm pc})}$.

{\bf Data files for bound protostellar pairs.} We provide 71 text files, one for each bound pair of protostars.  The data necessary to construct Figs.~\ref{disc_disc_cum} to \ref{disc_spin_cum} is contained in these files.  Their file names are of the format "PrAg\_AAA\_BBB.txt", where "AAA" and "BBB" give the numbers of the two sink particles that form the pair.  Each line of a file contains 12 numbers delimited by spaces that give the state of the pair at one instance in time.  The following information is given: (1) the integer number of sink particles in the system containing the pair, which may be 2, 3, or 4, (2) the age of the oldest protostar in the pair, (3) time, (4) total protostellar mass of the pair, (5) the mass of the primary, (6) the semi-major axis of the pair, (7) the relative orientation angle between the two circumstellar discs, (8-9) the relative orientation angles between the primary's disc and the orbit, and the secondary's disc and the orbit, (10) the relative orientation angle between the two protostellar spins, (11-12) the relative orientation angles between the primary's disc and its spin, and the secondary's disc and its spin.  The semi-major axis is given in au, and all angles are given in degrees.

{\bf Data files for protostellar systems.} We provide 376 text files, one for each system of protostars.  The data necessary to construct Figs.~\ref{disc_systems_age}, \ref{disc_systems_mass}, and \ref{sys_mass_rad} is contained in these files.  Their file names are of the format "SysDMR\_N(\_AAA).txt", where "N" gives the number of protostars in the system and there is one occurrence of "\_AAA" for each protostar to give the numbers of the sink particles (e.g. SysDMR\_1\_001.txt or SysDMR\_4\_117\_114\_145\_137.txt).  Each line of a file contains 8 numbers delimited by spaces that give the state of the system at one instance in time.  The following information is given: (1) an integer giving the number of protostars in the system which may be 1, 2, 3, or 4, (2) time, (3) the age of the oldest protostar in the system, (4) total protostellar mass of the system, (5) the mass of the primary, (6) the total mass in all of the system's discs, (7) the characteristic disc radius that contains 50\% of the total disc mass, (8) the characteristic disc radius that contains 63.2\% of the total disc mass.  The units are the same as those used in the other data files.

{\bf SPH output files.} Finally, the data set consisting of the output from the calculation of \cite{Bate2012} that is analysed in this paper is available from the University of Exeter's Open Research Exeter (ORE) repository and can be accessed via the handle: http://hdl.handle.net/10871/14881.

\bibliography{mbate}

\appendix

\section{}
\label{appendixA}

\begin{figure*}
\centering \vspace{-0.5cm} \hspace{0.0cm}
    \includegraphics[width=17.5cm]{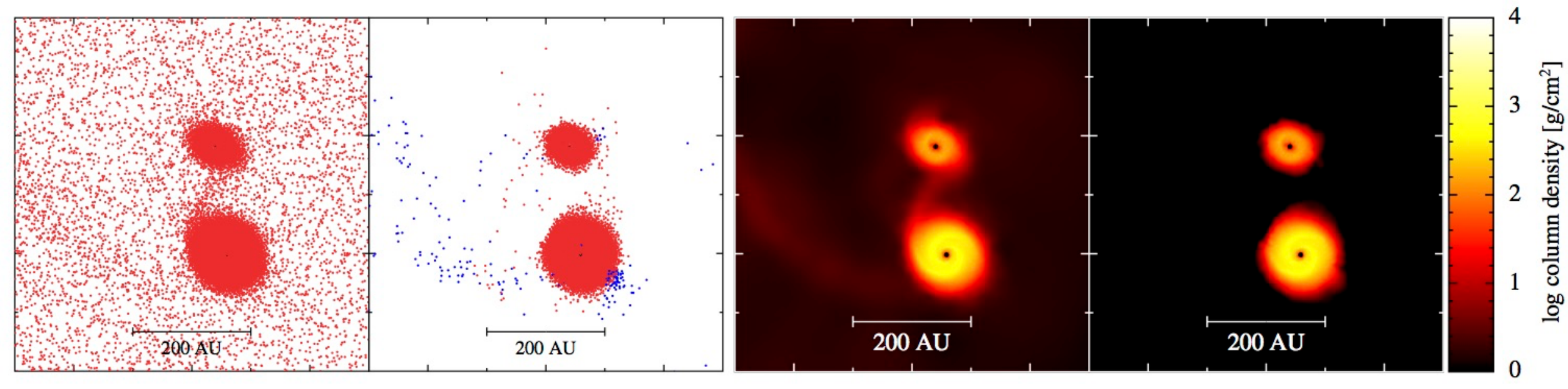} \vspace{-0.25cm}
\caption{An example of how the disc extraction algorithm extracts the discs of binary system (21,2) which is also shown in the middle row of Fig.~\ref{csdiscs}.  The panels measure 600 AU across.  The left-most panel shows all gas SPH particles (red) in a projection through the cloud.  The second panel shows the particles that have been determined to be in circumstellar discs (red) or circum-multiple discs (blue, almost none).  The right panels give the associated column-density plots, using all particles (third panel) and using only disc particles (right panel).  The sink particles are shown as black dots.  Most of the cloud material and the `bridge' of material between the two protostars has clearly been removed, leaving only the two circumstellar discs.  }
\label{A1}
\end{figure*}

\begin{figure*}
\centering \vspace{-0.5cm} \hspace{0.0cm}
    \includegraphics[width=17.5cm]{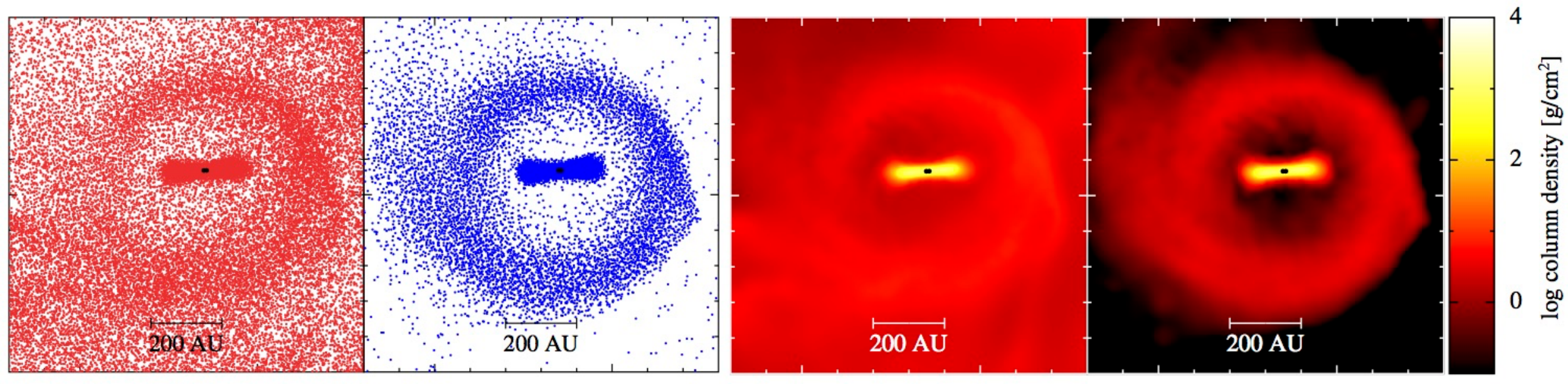} \vspace{-0.25cm}
\caption{An example of how the disc extraction algorithm extracts the discs of close binary system (6,13) which has two distinct circumbinary discs where the inner and outer discs are misaligned.  This system is also shown in Fig.~\ref{misaligned}.  The panels measure 1000 AU across.  The left-most panel shows all gas SPH particles (red) in a projection through the cloud.  The second panel shows the particles that have been determined to be in circumbinary discs (blue).  The right panels give the associated column-density plots, using all particles (third panel) and using only disc particles (right panel).  The sink particles are shown as black dots.  }
\label{A2}
\end{figure*}

\begin{figure*}
\centering \vspace{-0.5cm} \hspace{0.0cm}
    \includegraphics[width=17.5cm]{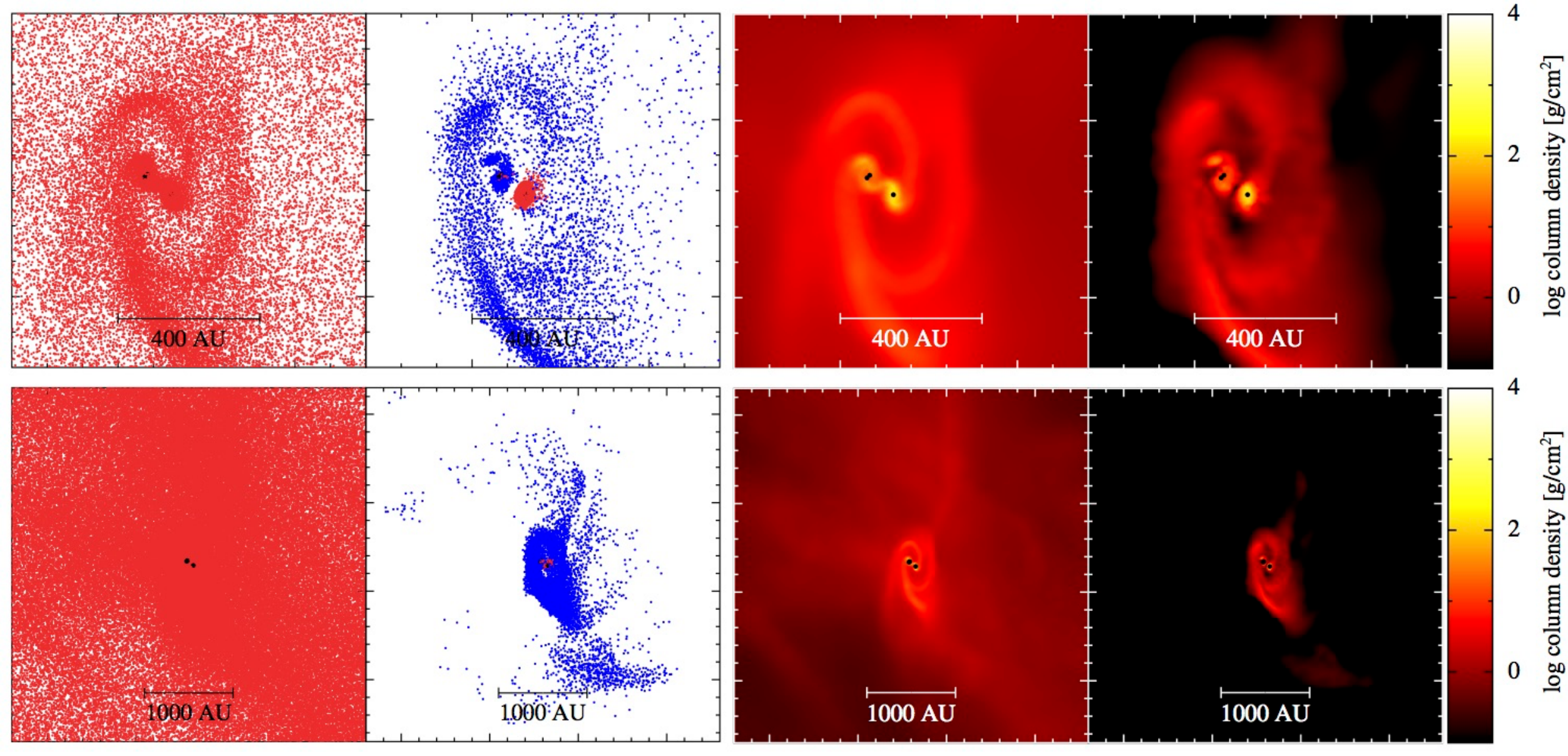} \vspace{-0.25cm}
\caption{An example of how the disc extraction algorithm extracts the discs of triple system ((27,19),22), which is also shown in the top-left panel of Fig.~\ref{multdiscs}.  The panels in the upper row measure 1000 AU across, while the panels in the lower row measure 4000 AU across because particles up to 2000 AU from a protostellar system are tested to determine whether they are part of a disc.  The left-most panels show all gas particles (red), while the second row of panels shows the particles that have been determined to be in circumstellar discs (red, mostly surrounding protostar number 22) or circum-multiple discs (blue).  The sink particles are shown as black dots.  The right panels give the associated column-density plots, using all particles (third column) and using only disc particles (right panels).  The cloud material has clearly been removed, leaving only the discs.  }
\label{A3}
\end{figure*}

In this Appendix we provide some examples of how the disc extraction algorithm that is described in Section \ref{sec:characterisation} operates.  The algorithm essentially identifies SPH particles that have would have ballistic orbits with eccentricities less than 0.3 as belonging to a disc.  Because many of the discs are self-gravitating, the computation of the ballistic orbit takes into account the mass of previously identified disc particles that are closer to the protostar than the particle being considered.  Circum-multiple discs are identified for systems consisting of up to four protostars.

In Fig.~\ref{A1} we show the discs that are extracted for the comparatively simple binary system (21,2) which is also shown in Fig.~\ref{csdiscs}.  The algorithm eliminates most of the surrounding cloud material and the faint `bridge' of material that lies between the two discs.  The discs have masses of 0.05~M$_\odot$ (3404 SPH particles) and 0.33~M$_\odot$ (23212 SPH particles), top to bottom.  Almost no circumbinary disc material is identified in this case (0.006~M$_\odot$, or 426 particles), and most of this gas is associated with the outer parts of the two circumstellar discs.

In Fig.~\ref{A2} we show the discs that are extracted for binary system (6,13) which has a misaligned circumbinary disc (also shown in Fig.~\ref{misaligned}).  The algorithm has no problem extracting the disc, even though the outer part of the disc has a different orientation from the inner part of the disc.  The binary is so close in this case that there are no circumstellar disc particles.  The circumbinary disc mass is 0.49~M$_\odot$ (34247 SPH particles).

Finally, in Fig.~\ref{A3} we show the discs that are extracted for the embedded triple system ((27,19),22) which is also shown in the top-right panel of Fig.~\ref{multdiscs}.  In this case, the circumstellar disc surrounding protostar number 22 is clearly identified and has a mass of 0.021~M$_\odot$ (1459 red particles, in the top panel of the second column).  Neither of protostars in the tight pair (27,19) has much of a circumstellar disc (fewer than 20 particles).  Instead, the gas close to the pair is identified as being circumbinary (0.006~M$_\odot$, 427 blue particles).  The majority of the disc mass of the triple system is contained in a large circumtriple disc which has strong spiral arms (0.10~M$_\odot$, 7167 blue particles).  The extraction algorithm does a good job of differentiating between the circumtriple disc and the surrounding cloud material.

\section{}
\label{appendixB}

To investigate the effects of the limited numerical resolution on the evolution of the discs, we performed two much smaller star formation calculations at varying resolutions.  The first case was the collapse of a rotating molecular cloud core with an initial Bonnor-Ebert density profile to form a single protostar with a disc.  The second case was the collapse of a rotating molecular cloud core with an $m=2$ density perturbation to form a binary system.  The methods used to perform these calculations and characterise the disc properties were identical to the methods employed in the main part of in this paper, including setting the size of the sink particle accretion radii to 0.5~AU.

\begin{figure}
\centering \vspace{-0.3cm} \hspace{0cm}
    \includegraphics[width=15cm]{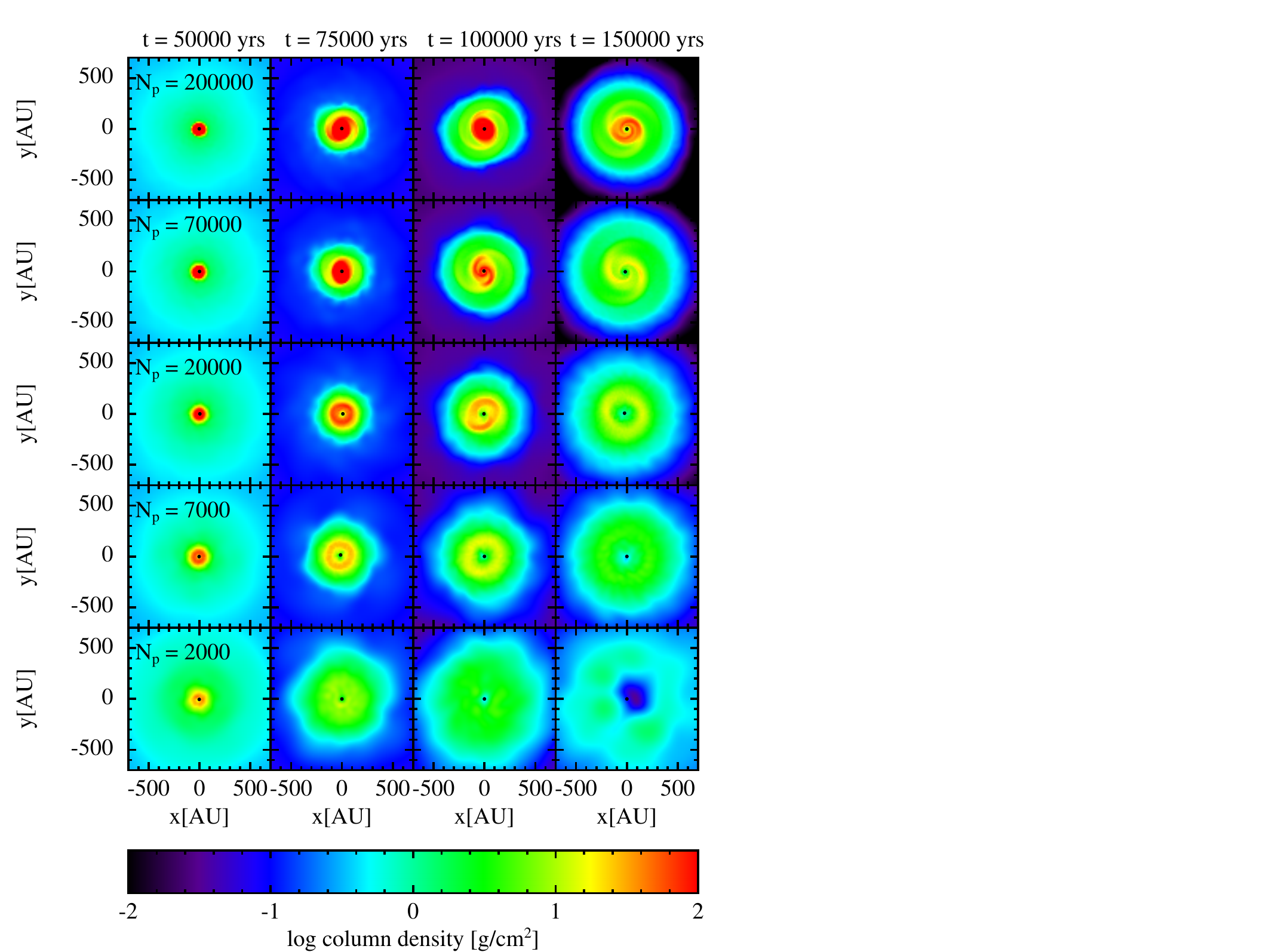}      \vspace{-0.5cm}
\caption{The evolution of the disc with time in calculations of the collapse of a 1-M$_\odot$ rotating Bonnor-Ebert sphere to form a single protostar with a disc. The calculations are performed using five different numerical resolutions: $N_{\rm p}=2000$, 7000, $2\times 10^4$, $7\times 10^4$,  $2\times 10^5$ SPH particles.  With low resolution ($N_{\rm p}< 20000$) the spiral arms do not develop and the disc accretes and spreads much more rapidly.  A reasonable level of convergence is obtained using $\gsim 2\times 10^4$ particles per solar mass.  The main calculation in this paper uses $7\times 10^4$ particles per solar mass.
}
\label{appB1}
\end{figure}

\begin{figure}
\centering \vspace{0.0cm} \hspace{0cm}
    \includegraphics[width=8.5cm]{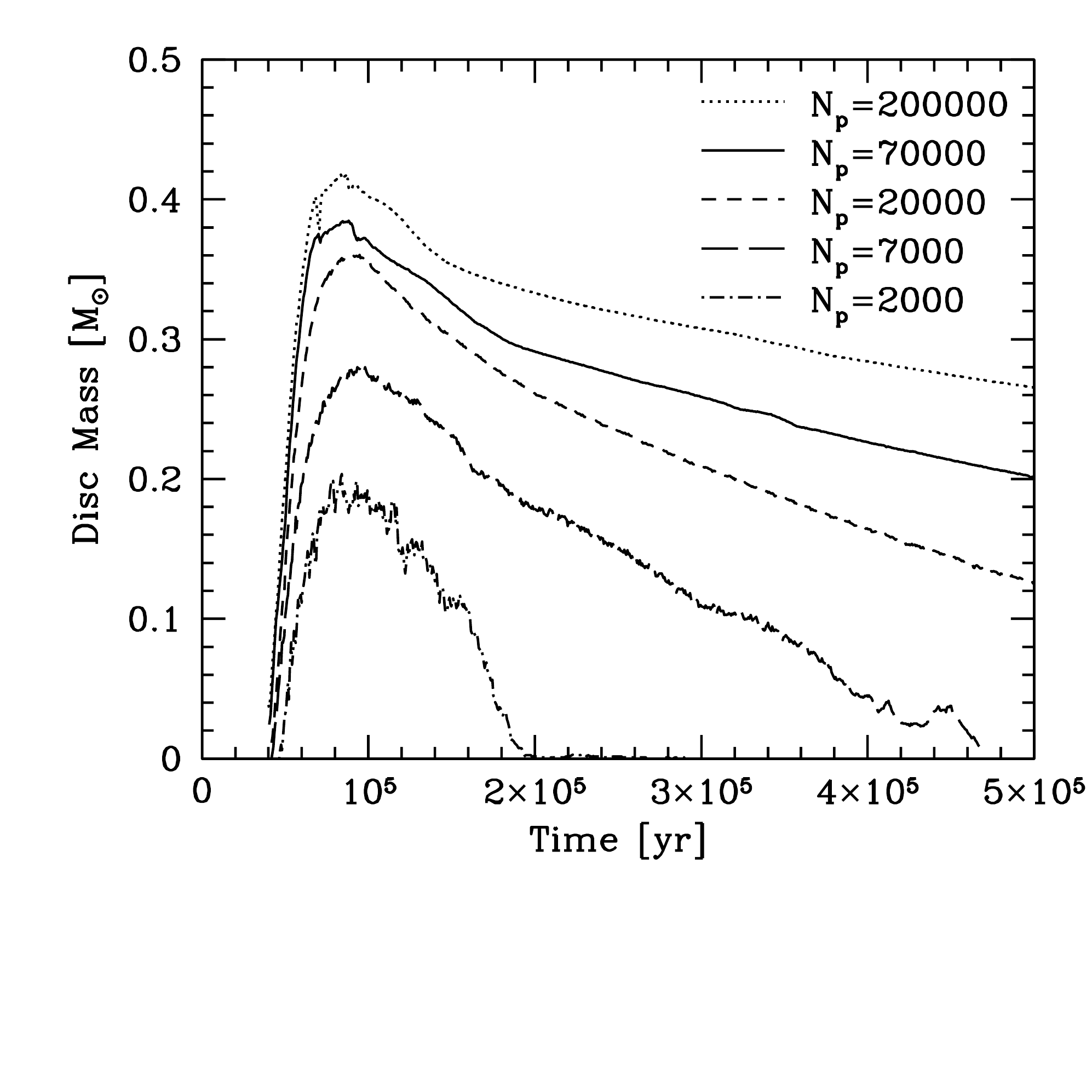}      \vspace{-0.2cm}
\caption{The evolution of the disc mass with time in calculations of the collapse of a 1-M$_\odot$ rotating Bonnor-Ebert sphere to form a single protostar with a disc. The calculations are performed using five different numerical resolutions: 2000, 7000, $2\times 10^4$, $7\times 10^4$,  $2\times 10^5$ SPH particles.  Convergence of the peak disc mass to the level of $\approx 20$\% is obtained using $\gsim 2\times 10^4$ particles per solar mass.  The main calculation in this paper uses $7\times 10^4$ particles per solar mass.  With low resolution the disc mass is underestimated, and when the number of particles modelling the disc drops below $\sim 400$ the disc quickly drains away.  
}
\label{appB2}
\end{figure}

\begin{figure*}
\centering \vspace{0cm} \hspace{0cm}
    \includegraphics[width=17.5cm]{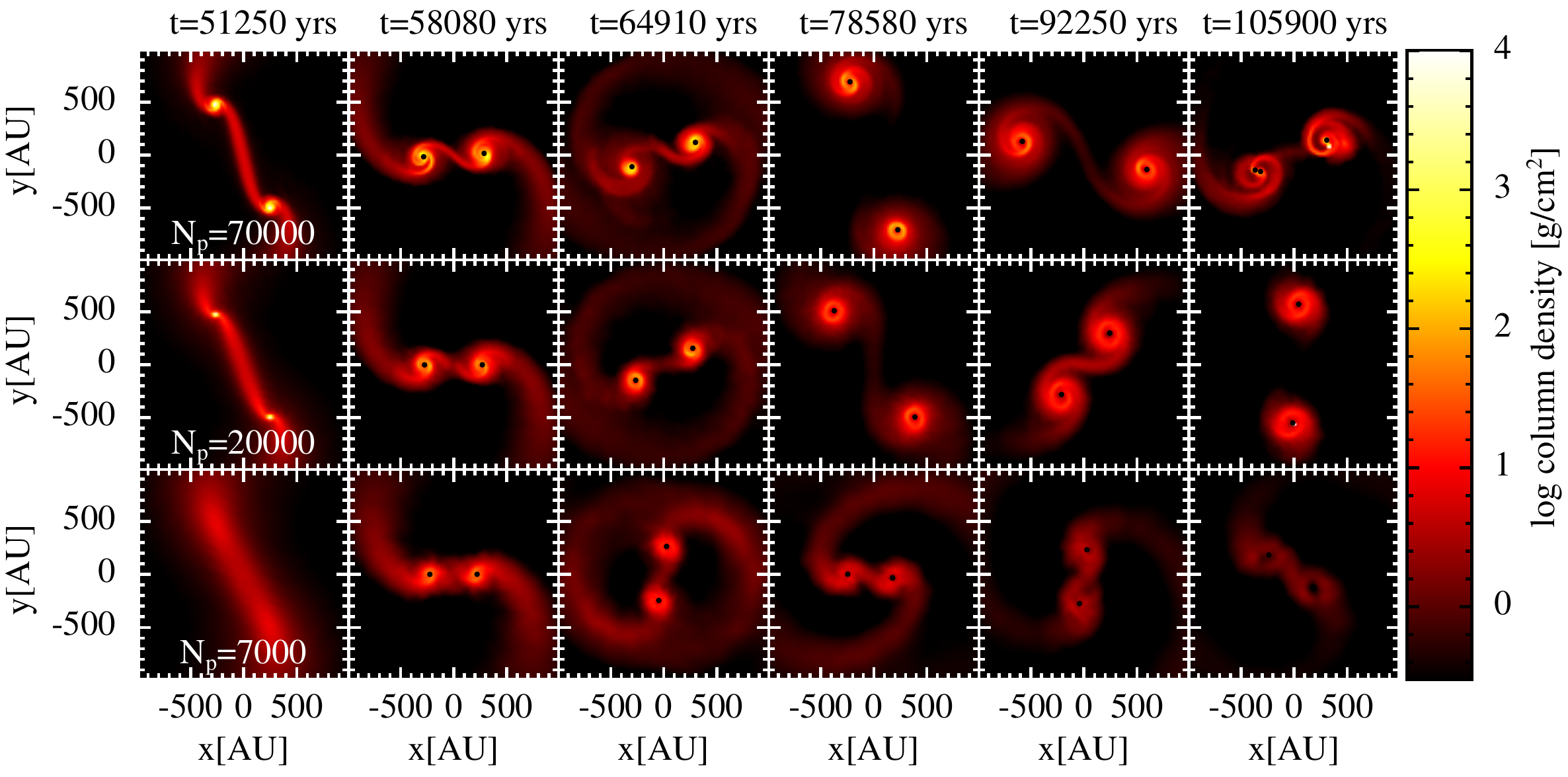}      \vspace{0cm}
\caption{The evolution of a binary system formed from the collapse of a 1-M$_\odot$ rotating molecular cloud core. The calculations are performed using three different numerical resolutions: 7000, $2\times 10^4$, and $7\times 10^4$ SPH particles.  The two highest resolution calculations clearly resolve the circumstellar discs, but with the lowest resolution the discs are quickly accreted onto the sink particles of the binary.
}
\label{appB3}
\end{figure*}

\subsection{A single protostellar disc}

The initial conditions were a 1-M$_\odot$ spherical molecular cloud core with a Bonnor-Ebert density profile for which the ratio of density between the centre and the outer edge of the cloud was 20:1.  
The initial radius of the cloud was 4800 AU, the initial temperature was 10~K, and the cloud was placed in solid-body rotation with an angular velocity of $\Omega = 1.38 \times 10^{-13}$~rad~s$^{-1}$.  This gives ratios of the thermal and rotational energies to the magnitude of the gravitational potential energy of $\alpha = 0.39$ and $\beta = 0.010$, respectively.  Calculations were performed using 2000, 7000, $2\times 10^4$, $7\times 10^4$ (the resolution of the main calculation), and $2\times 10^5$ SPH particles.

The cloud collapses to form a single protostar with a circumstellar disc that initially has an outer radius of $\approx 100$~AU, but grows as gas with greater specific angular momentum falls in.  Fig.~\ref{appB1} shows some snapshots of the evolution.  The disc becomes quite massive and develops spiral arms but does not fragment.  Thus, in addition to the infall of gas with more specific angular momentum, the disc grows in size due to the action of gravitational torques from spiral arms, and the action of numerical viscosity.  Fig.~\ref{appB2} shows the evolution of the disc mass versus time for each of the calculations.  The disc grows in mass from $t=40,000$ to $\approx 85000$~yrs.  Its disc mass then declines as it accretes onto the central protostar.  Lower numerical resolution gives lower masses throughout the evolution.  The discs grow less rapidly, and their peak masses are lower.  As expected, the discs modelled with lower numerical resolution also accrete more quickly due to the increased numerical viscosity.  The disc mass evolution shows signs of convergence when using $>2\times 10^4$ particles.  When using $7\times 10^4$ particles (the resolution used for the main calculation discussed in this paper), the mass differs from that obtained using $2\times 10^5$ particles by less than 10\% to $t\approx 200000$~yrs.  For the calculations using $2\times 10^4$ particles or more, a change in the slope of the decay can be seen at $t\approx 160000-180000$~yrs.  Before this time, the accretion is primarily driven by gravitational torques.  After this time, the accretion is primarily viscous and, thus, depends on the resolution (the slope is shallower with higher resolution).  Also note that the discs rapidly disappear when the resolution drops to below $\approx 400$ particles ($0.2$~M$_\odot$ for $N_{\rm p}=2000$ particles, or $0.06$~M$_\odot$ for $N_{\rm p}=7000$ particles).

\begin{figure}
\centering \vspace{-0.5cm} \hspace{0cm}
    \includegraphics[width=8.5cm]{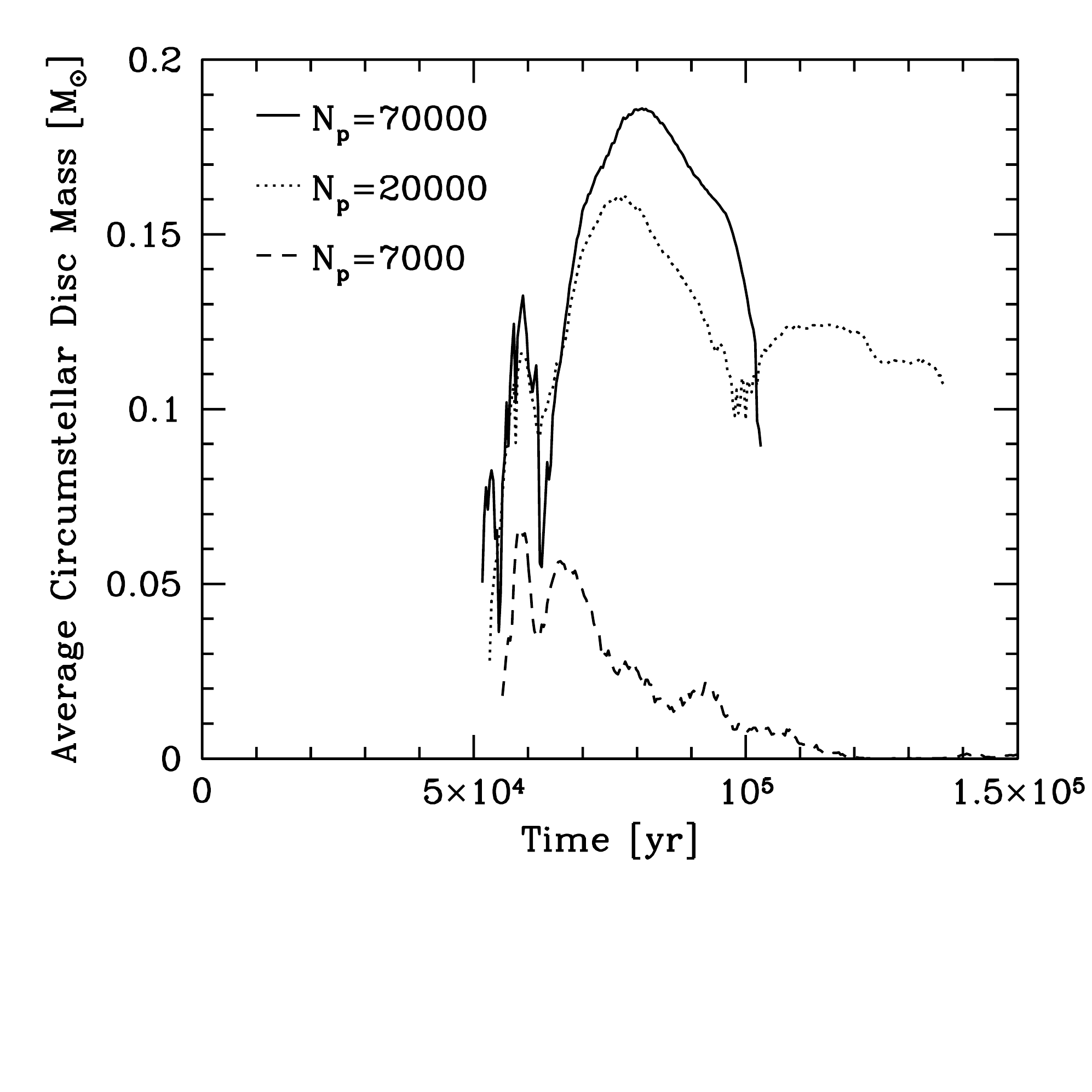}      \vspace{-2cm}
\caption{The time evolution of the average circumstellar disc mass for the two protostars formed in the binary system depicted in Fig.~\ref{appB3}.  The calculations are performed using three different numerical resolutions: 7000, $2\times 10^4$, and $7\times 10^4$ SPH particles.  The disc masses have minimums at the periastron passages of the binary (approximately $t=$62000 and $t=$100000~yrs).  The circumstellar disc masses are well resolved for 20000 SPH particles per solar mass and higher, but with only 7000 SPH particles the discs quickly accrete onto the sink particles.  The difference in the disc mass at ages $t=75000-100000$~yrs for the two highest resolution calculations is primarily due to the separation of the binary being slightly smaller when the calculation is performed using fewer particles.
}
\label{appB4}
\end{figure}

\subsection{A binary protostellar system}

The initial conditions for the calculation of binary formation were based on those of \cite{BosBod1979} and \cite{BatBonPri1995}, but with slightly different values.  A 1-M$_\odot$ spherical molecular cloud core with a nominal uniform density has an $m=2$ density perturbation applied such that $\rho = \rho_0[1 + 0.5 \cos(2\phi)]$, where $\rho_0$ is a constant and $\phi$ is the azimuthal angle. The initial radius of the cloud was $5\times 10^{16}$~cm, the initial temperature was 12~K, and the cloud was placed in solid-body rotation with an angular velocity of $\Omega = 8.0 \times 10^{-13}$~rad~s$^{-1}$.  This gives ratios of the thermal and rotational energies to the magnitude of the gravitational potential energy of $\alpha = 0.39$ and $\beta = 0.20$, respectively.  Calculations were performed using $N_{\rm p}=7000$, $2\times 10^4$, and $7\times 10^4$ SPH particles, with the latter being the resolution of the main calculation.

The cloud collapses to form an equal-mass binary protostellar system that has a mildly eccentric orbit.  Each protostar has a circumstellar disc, and there is also a weak circumbinary disc later in the calculation. Fig.~\ref{appB3} shows some snapshots of the evolution for each of the three numerical resolutions.  In Fig.~\ref{appB4} we plot the average masses of the two circumstellar discs (extracted using the same disc extraction method as that used in the rest of this paper) as a function of time for all three calculations.  With fewer SPH particles, the collapse takes slightly longer and the resulting binary is slightly tighter.  The circumstellar discs are resolved in all calculations initially, but with the lowest resolution the initial disc masses ($\approx 0.07$~M$_\odot$) are approximately half those that are obtained in the other two calculations ($\approx 0.13$~M$_\odot$).  Moreover, in the lowest resolution calculation each circumstellar disc initially contains only $\approx 500$ SPH particles.  These poorly resolved discs quickly evolve viscously and after $\approx 50,000$~yrs of evolution (i.e. at $t=100,000$~yrs) the circumstellar discs have almost disappeared.  By contrast, using either $N_{\rm p}=20000$ or $N_{\rm p}=70000$, the disc masses are well resolved well beyond $t=100,000$~yrs.  Note that the difference in the average circumstellar disc mass during the period $t=75000-100000$~yrs between the two highest resolution calculations is primarily due to the different binary separation (the orbit of the binary is tighter in the $N_{\rm p}=20000$ calculation than in the $N_{\rm p}=70000$ calculation, so the circumstellar discs are slightly smaller and less massive after the first periastron passage).   The comparison after $t=103,000$~yrs is complicated by the fact that one of the circumstellar discs fragments when using $N_{\rm p}=70000$ (hence the solid line in Fig.~\ref{appB4} is not plotted beyond this point).  

\subsection{Summary of the resolution tests}

In both of the resolution tests, it is found that discs modelled by less than $\approx 500$ SPH particles tend to suffer rapid viscous evolution.  Their masses should, therefore, be treated as lower limits.  It is important to note, however, that none of the protostars studied in the main calculation of this paper have ages $>9\times 10^4$~yrs, and most have substantially younger ages.  In the resolution tests, an isolated disc modelled by 400 SPH particles (i.e. the $N_{\rm p}=2000$ calculation) still has half its peak mass at an age of $\approx 10^5$~yrs, while in the binary test the circumstellar discs whose peak mass is modelled by $\approx 400$ SPH particles retain half their peak mass to ages of $\approx 2 \times 10^4$~yrs (in the $N_{\rm p}=7000$ calculation).

In both of the test calculations, using $N_{\rm p}=7 \times 10^4$ SPH particles per solar mass means that the circumstellar discs last well in excess of $10^5$~yrs.  Thus, if similar systems form within the main calculation, their disc properties should be well characterised.  However, the number of SPH particles that make up a disc depends on the mass of the disc.  The initial masses in the two test calculations range from 0.4 to 0.13~M$_\odot$.  Only about 10\% of the instances of protostars studied in the main calculation have such high masses, or about 20\% of isolated protostars.  Discs with lower masses will evolve more rapidly.

The timescale over which a disc accretes also depends on the size of the disc, which also evolves with time.  For an isolated disc, viscous evolution and/or gravitational torques result in spreading of the disc as it accretes.  The characteristic viscous timescale for a disc $\tau \sim r^2/\nu$ where $\nu$ is the kinematic viscosity.  The numerical viscosity scales proportional to the SPH smoothing length, $h$ \citep*{Monaghan1985,Pongracic1988,MegWicBic1993,MerBat2012}.  For a well resolved disc (in which the vertical scale height is resolved)  $h \propto r N_{\rm p}^{-1/3}$, but for a poorly resolved disc the scaling will be more like $h \propto r N_{\rm p}^{-1/2}$.  Thus, one expects that the viscous timescale will scale as $\tau \propto r N_{\rm p}^{1/2}$ for poorly resolved discs.  Thus, for the same number of SPH particles, smaller discs will evolve more rapidly than larger discs.  

The evolution of the disc also depends on its circumstances.  An isolated disc gets larger as it evolves and, therefore, although the number of particles decreases, the dependence on radius of the viscous timescale lengthens its lifetime.  This effect can be seen in the mass evolution of the $N_{\rm p}=20000$ calculation in Fig.~\ref{appB2} with the slightly concave shape of the mass versus time curve between $t=200000$ and $500000$~yrs.  But for a circumstellar disc in a binary system, its outer radius is constrained by the gravitational torques from the companion (i.e. it is truncated).  Thus, for the same initial state (mass, radius, number of SPH particles), an isolated disc will last longer than one in a binary \cite[the same way that the viscous disc of the primary in an unequal-mass binary will last longer than that of the secondary;][]{ArmClaTou1999}.

Overall, we expect that typical instances of discs in the main calculation that have masses $\gsim 0.03$~M$_\odot$ (2000 SPH particles) will be well modelled in terms of both their mass and radius for the ages of the protostars from the main calculation analysed in this paper.  This is based on the fact that discs in these test calculations that have this number of particles last well in excess of $10^5$~yrs (e.g. the $N_{\rm p}=7000$ case in Fig.~\ref{appB2} and the $N_{\rm p}=20000$ case in Figs.~\ref{appB4}).  Very small discs in close multiple systems (radii $\lsim 20$~AU) may not survive for long even if they are initially resolved by more than 2000 SPH particles (because the viscous timescale scales as $\tau \propto r N_{\rm p}^{1/2}$).  But such small, comparatively massive, circumstellar discs are apparently rare (c.f.\ Fig.~\ref{sys_mass_rad}, which plots disc mass versus radius for protostellar systems of all ages).  Below $\approx 0.03$~M$_\odot$, some discs will have significantly underestimated masses because they will evolve viscously on timescales comparable to the ages of their protostar.  This is consistent with the forms of the cumulative disc mass distributions found in Section \ref{sec:stats} of this paper (e.g. Figs.~\ref{eccentricitytest} and \ref{disc_cumulative}), which are found to flatten below this disc mass.

\end{document}